\documentclass[12pt]{article}
\usepackage{epsfig}
\usepackage{a4}
\usepackage{latexsym}
\usepackage{cite}

\usepackage{pslatex}
\usepackage[latin1]{inputenc}
\usepackage[T1]{fontenc}

\usepackage{color,colordvi}
\def\colour4colour#1{\Blue{#1}}

\renewcommand{\theequation}{\thesection.\arabic{equation}}

\newcommand{\lsim}{\raisebox{-0.07cm}{$\:\:\stackrel{<}{{\scriptstyle
 \sim}}\:\: $} }
\newcommand{\hspn}{{\hspace{-4mm}}}
\newcommand{\hspp}{{\hspace{5mm}}}
\newcommand{\beq}{\begin{equation}}
\newcommand{\eeq}{\end{equation}}
\newcommand{\bea}{\begin{eqnarray}}
\newcommand{\eea}{\end{eqnarray}}
\newcommand{\nn}{\nonumber}
\newcommand{\MSb}{$\overline{\mbox{MS}}$}
\newcommand{\as}{\alpha_{\rm s}}
\newcommand{\ar}{a_{\rm s}}
\newcommand{\ra}{\rightarrow}
\newcommand{\DD}{{\cal D}}
\newcommand{\ep}{\epsilon}

\newcommand{\qV}{\mbox{\boldmath $q$}}
\newcommand{\FV}{\mbox{\boldmath $F$}}
\newcommand{\CV}{\mbox{\boldmath $C$}}
\newcommand{\PV}{\mbox{\boldmath $P$}}
\newcommand{\ZV}{\mbox{\boldmath $Z$}}
\newcommand{\KV}{\mbox{\boldmath $K$}}

\begin{document}
\setlength{\parskip}{0.3cm}
\setlength{\baselineskip}{0.55cm}

\def\Ftwo{{F_{\:\! 2}}}
\def\FL{{F_{\:\! L}}}
\def\F3{{F_{\:\! 3}}}
\def\Qs{{Q^{\, 2}}}
\def\GeV2{{\mbox{GeV}^{\:\!2}}}
\def\x1{{(1 \! - \! x)}}
\def\DDk{{{\cal D}_{\:\! k}^{}}}
\def\DD#1{{{\cal D}_{\:\! #1}^{}}}
\def\z#1{{\zeta_{\:\! #1}}}
\def\zss{{\zeta_{2}^{\,2}}}
\def\zst{{\zeta_{3}^{\,2}}}
\def\zts{{\zeta_{2}^{\,3}}}
\def\ca{{C_A}}
\def\cas{{C^{\: 2}_A}}
\def\cat{{C^{\: 3}_A}}
\def\cf{{C_F}}
\def\cfs{{C^{\: 2}_F}}
\def\cft{{C^{\: 3}_F}}
\def\cff{{C^{\: 4}_F}}
\def\nf{{n^{}_{\! f}}}
\def\nfs{{n^{\,2}_{\! f}}}
\def\nft{{n^{\,3}_{\! f}}}
\def\caf{{C_{AF}}}
\def\cafs{{C_{AF}^{\: 2}}}
\def\caft{{C_{AF}^{\: 3}}}
\def\dabc2{{d^{\:\!abc}d_{abc}}}
\def\dabcnc{{{d^{\:\!abc}d_{abc}}\over{n_c}}}
\def\fl11{fl_{11}}
\def\fl02{fl_{02}}
\def\b#1{{{\beta}_{#1}}}
\def\bb#1#2{{{\beta}_{#1}^{\,#2}}}

\def\pqq(#1){p_{\rm qq}(#1)}
\def\pqg(#1){p_{\rm qg}(#1)}
\def\pgq(#1){p_{\rm gq}(#1)}
\def\pgg(#1){p_{\rm gg}(#1)}

\def\H(#1){{\rm{H}}_{#1}}
\def\Hh(#1,#2){{\rm{H}}_{#1,#2}}
\def\THh(#1,#2){{\widetilde{\rm{H}}}_{#1,#2}}
\def\Hhh(#1,#2,#3){{\rm{H}}_{#1,#2,#3}}
\def\THhh(#1,#2,#3){{\widetilde{\rm{H}}}_{#1,#2,#3}}
\def\Hhhh(#1,#2,#3,#4){{\rm{H}}_{#1,#2,#3,#4}}
\def\Hhhhh(#1,#2,#3,#4,#5){{\rm{H}}_{#1,#2,#3,#4,#5}}

\begin{titlepage}
\noindent
LTH 857 \hfill {\tt arXiv:0912.0369 [hep-ph]}\\
DESY 09-211 \\
SFB/CPP-09-119 \\
NIKHEF 09-031 
\vspace{1.1cm}
\begin{center}
\Large
{\bf On Higgs-exchange DIS, physical evolution kernels} \\
\vspace{0.15cm}
{\bf and fourth-order splitting functions at large \boldmath $x$} \\
\vspace{1.6cm}
\large
G. Soar$^{\, a}$, S. Moch$^{\, b}$, J.A.M. Vermaseren$^{\, c}$ 
and A. Vogt$^{\, a}$\\
\vspace{1cm}
\normalsize
{\it $^a$Department of Mathematical Sciences, University of Liverpool \\
\vspace{0.1cm}
Liverpool L69 3BX, United Kingdom}\\
\vspace{0.5cm}
{\it $^b$Deutsches Elektronensynchrotron DESY \\
\vspace{0.1cm}
Platanenallee 6, D--15735 Zeuthen, Germany}\\
\vspace{0.5cm}
{\it $^c$NIKHEF Theory Group \\
\vspace{0.1cm}
Science Park 105, 1098 XG Amsterdam, The Netherlands} \\
\vfill
\large
{\bf Abstract}
\vspace{-0.2cm}
\end{center}
We present the coefficient functions for deep-inelastic scattering (DIS) via 
the exchange of a scalar $\phi$ directly coupling only to gluons, such as the 
Higgs boson in the limit of a very heavy top quark and $\nf$ effectively 
massless light flavours, to the third order in perturbative QCD.
The two-loop results are employed to construct the next-to-next-to-leading 
order physical evolution kernels for the system $(\,\Ftwo, F_{\phi\,}$) of 
flavour-singlet structure functions. The practical relevance of these kernels 
as an alternative to \MSb\ factorization is bedevilled by artificial double 
logarithms at small values of the scaling variable $x$, where the large 
top-mass limit ceases to be appropriate.
However, they show an only single-logarithmic enhancement at large $x$.
Conjecturing that this feature persists to the next order also in the present 
singlet case, the three-loop coefficient functions facilitate exact predictions
(backed up by their particular colour structure) of the double-logarithmic 
contributions to the fourth-order singlet splitting functions, i.e., of the 
terms $\x1^a \ln^{\,k} \! \x1$ with $k = 4,\,5,\,6$ and $k = 3,\,4,\,5$, 
respectively, for the off-diagonal and diagonal quantities to all powers $a$ in
$\x1$. 
\vspace*{0.2cm}
\end{titlepage}
%
%
\section{Introduction}
\label{sec:intro}
%
%
Structure functions in lepton-nucleon deep-inelastic scattering are 
classic observables probing Quantum Chromodynamics (QCD), the theory of the
strong interaction, and in particular the structure of the nucleon. Indeed,
structure function measurements provide the backbone of our knowledge of the
quark and gluon longitudinal momentum distributions in the proton to this day,
40 years after the pioneering measurements of DIS at SLAC \cite{SLAC-DIS}, see 
also Refs.~\cite{Nobel90}. 
In turn these distributions, in short referred to as the parton distributions 
or parton densities of the nucleon, are indispensable for the description and 
analysis of hard scattering processes at the proton colliders forming the 
high-energy frontier of particle physics, i.e., the {\sc Tevatron} and the LHC.
See, e.g., Ref.~\cite{PDG08} for introductory overviews and the present status 
of structure functions and QCD.

The asymptotic freedom of QCD \cite{AsymptF} facilitates a partly perturbative
description of hard hadron processes. At all practically relevant scales, 
however, the strong coupling constant $\as$ is much larger than its 
electroweak counterparts. Hence calculations to a higher order are required in
perturbative QCD in order in arrive at precise predictions. Also in this 
respect structure functions form a special tool, since they can be expressed 
in terms of the operator-product expansion \cite{OPEstr} which provides a 
framework that renders higher-order calculations considerably more accessible.

In a series of previous articles, three of us have exploited this fact to 
derive the exact third-order QCD corrections to the splitting functions 
governing the scale dependence of the parton distributions \cite{MVV3,MVV4} 
-- and thus the transfer of information from DIS at moderate scales to the 
high scales involved in Higgs, gauge boson, top quark and new physics 
processes at {\sc Tevatron} and the LHC -- 
and to the partonic cross sections (coefficient functions) for the most 
important structure functions \cite{MVV5,MVV6,MVV10}.
As all modern higher-order calculations in perturbative QCD, those 
calculations have been performed in dimensional regularization \cite{DimReg}, 
and the splitting functions and coefficient functions have been determined in 
the modified~\cite{BBDM78} minimal subtraction \cite{MS} scheme, \MSb.

Consequently the higher-order parton densities are not defined in terms of any 
physical process, but by a mathematical description employed for isolating the
mass singularities which lead to their scale dependence. The \MSb\ scheme has 
definite advantages, e.g., it leads to a stable (order-independent) functional
form of the dominant diagonal (quark-quark and gluon-gluon) splitting functions
in the limit of large momentum fractions $x$ \cite{Korchemsky:1989si}, see also
Ref.~\cite{DMS05}, a feature that assists a stable evolution of the parton 
densities over the wide range of scales mentioned above. 

An alternative approach has been suggested already long ago in 
Ref.~\cite{FP82}, see also Refs.~\cite{physanom}:
As discussed below, the parton densities can be eliminated from the description
of the structure functions (and other hard processes). The dependence of the 
observables on the physical hard scale (in DIS $\Qs = -q^2$, where $q$ is the 
four-momentum of the exchanged electroweak gauge boson) is then given in terms 
of so-called physical evolution kernels or physical anomalous dimensions.
While the direct relation between different observables via the universal 
parton densities is obscured in this approach, it can be practically useful 
for analyses of selected observables such as the determination of $\as$ 
from the scaling violations of structure functions \cite{Vogt:1999ik}.

The construction of physical evolution kernels is particularly simple (and 
unique) for flavour non-singlet, i.e., gluon insensitive observables. It turns 
out that these kernels show an only single-logarithmic enhancement at large 
values of the scaling variable $x$ for a wide range of non-singlet quantities 
in DIS, semi-inclusive $e^+e^-$ annihilation (SIA) and the Drell-Yan process 
\cite{MV3,MV5}.
This general behaviour is in contrast to that of all corresponding \MSb\ 
coefficient functions which receive double-logarithmic contributions. As a
result the physical kernels provide an -- at present not yet formally proven
-- all-order exponentiation of the highest $\ln \x1$ terms beyond the leading 
$\x1^{-1}$ contributions covered by the standard threshold resummation
\cite{SoftGlue} to all powers in $\x1$. 

In the present article we address the flavour singlet case which requires two
observables and a $2\!\times\! 2$ matrix evolution kernel. Unlike Refs.~\cite
{physanom,Vogt:1999ik} we study the ideal -- for theoretical purposes -- 
gluonic complement to the most important gauge-boson exchange structure 
function $\Ftwo$, i.e., DIS by the exchange of a scalar $\phi$ coupling 
directly only to gluons via an additional term $\phi\,G^{\,\mu\nu\!}G_{\mu\nu}$
in the Lagrangian, where $G^{\,\mu\nu}$ represents the gluon field strength 
tensor. Such a term, already suggested as a trick in Ref.~\cite{FP82}, is of 
course not present in the fundamental Standard Model Lagrangian. However, it 
does occur effectively for the Higgs boson in the limit of a very heavy top 
quark \cite{HGGeff}. 
 
This interaction was included, in order to directly access also the quark-gluon
and gluon-gluon splitting functions, in the calculations of Ref.~\cite{MVV4} 
and, in fact, its precursors \cite{Mom3loop,Moch:1999eb}.
Hence the one- and two-loop coefficients functions for the resulting structure 
function $F_\phi$ were calculated, but not published, quite a while ago. 
The more recent, but also unpublished three-loop results have already been 
employed to determine the third-order gluon jet function in threshold 
resummation \cite{MVV7} and the pole terms of the three-loop gluon form factor
\cite{MVV9} -- and from the latter the next-to-next-to-next-to-leading 
logarithmic (N$^3$LL) threshold resummation for Higgs production in the 
heavy-top limit \cite{MV1}.
These coefficient functions also provide, via an analytic continuation to 
timelike kinematics, a strong check of the next-to-next-to-next-to-leading 
order (N$^3$LO) computation in Ref.~\cite{Baikov:2006ch} of the top-induced
Higgs decay into hadrons \cite{MV2}. The first- and second-order coefficient 
functions for $F_\phi$ have now been derived independently in a completely 
different manner~\cite{DGGLcphi}.

The outline of this article is as follows. In Section~2 we introduce the 
formalism used for calculating the coefficient functions for $F_\phi$ to order 
$\as^{\,3}$, briefly discuss the diagram calculation and its checks, and 
provide the transformation to the physical kernels to the fourth order. 
In Section~3 we then present the exact expressions for the coefficient 
functions and address their size and end-point behaviour. The one- and two-loop
results are then combined in Section~4 with the known second- and third-order
splitting functions to derive the next-to-leading order (NLO) and next-to-%
next-to-leading order (NNLO) physical kernels for the system $(\Ftwo,F_\phi)$.
Also these kernels show an only single-logarithmic large-$x$ enhancement to all
powers in $\x1$, a feature that we conjecture to hold also at the next order.
In Section~5 this conjecture and the three-loop coefficient functions of 
Section 3 are then employed to predict the coefficients of the highest three 
powers of $\ln \x1$ in all four singlet splitting functions at order 
$\as^{\,4}$ (N$^3$LO).
We summarize our results and give a brief outlook in Section~6. 
Appendices A and B contain the partly very long expressions, respectively, 
for the new third-order coefficient functions and the elements of the NNLO 
physical kernel matrix. In Appendix C we finally provide the large-$x$ limit of
the gluon coefficient functions for $\Ftwo$. 
%
%
\section{Formalism and calculation} 
\label{sec:formalism}
%
%
We consider inclusive deep-inelastic lepton-nucleon scattering through the 
exchange of a  scalar (Higgs) boson $\phi$, which proceeds through the reaction
\beq
\label{eq:dis}
  \phi(q) \:+\: {\rm nucl}\:\! (p) \:\:\ra\:\: X\; .
\eeq
Here $X$ denotes all hadronic final states allowed by quantum number 
conservation. The boson $\phi$ transfers a space-like momentum $q$ 
(i.e., $\Qs \equiv -q^2 > 0$), while the nucleon carries momentum~$p$. 
The scaling variable of the reaction is defined as usual as 
$x=\Qs / (2p\cdot q)$ with $0 < x \leq 1$.

In complete analogy to ordinary (gauge-boson exchange) DIS, the cross section 
for this process can be parametrized in terms of a structure function $F_\phi$.
Through the optical theorem the total cross section (hence the structure
function $F_\phi$) is related to the imaginary part of the forward amplitude 
${\cal T}_\phi$ for the scattering process in Eq.~(\ref{eq:dis}) of a virtual 
Higgs boson off the nucleon,
\beq
\label{eq:T}
  {\cal T}_\phi(p,q) \; = \; i \! \int \! d^{\, 4}z \: e^{\,iqz} \,
  \langle {\rm nucl,}\, p \vert \, T \big(J(z) \, J(0)\big) \, 
  \vert {\rm nucl,}\, p \rangle \; .
\eeq
This quantity contains the time-ordered product of two scalar currents (see 
below) to which the standard operator-product expansion (OPE) can be applied 
in the Bjorken limit (large $\Qs$ for fixed~$x$).
The relevant steps have been already discussed in previous lower-order 
and fixed Mellin-$N$ third-order calculations, see Refs.~\cite
{Mom3loop,Moch:1999eb}, hence we can be brief in recalling some key issues.

The OPE decomposes the current product in Eq.~(\ref{eq:T}) in terms of the 
standard set of spin-averaged matrix elements of the (renormalized) spin-$N$ 
twist-two irreducible flavour-singlet quark and gluon operators
\bea
\label{eq:OME}
  \langle {\rm nucl,}\, p \vert \, O_{i}^{\{\mu^{\,}_1,...,\,\mu^{\,}_N\}}
  \,\vert {\rm nucl,}\, p \rangle 
  \; = \; p^{\{\mu^{\,}_1}...\,p^{\,\mu^{\,}_N\}}\, A_{i,\rm nucl}(N,\mu^{\,2})
\eea
and the respective hard scattering parton coefficient functions $C_{\phi,i}$ 
with $i= \mbox{q, g}$, where $\mu$ stands for the renormalization scale.
The operators $O_{\rm q}$ and $O_{\rm g}$ in Eq.~(\ref{eq:OME}) arise from 
the symmetric and traceless part of $N$ covariant derivatives $D^{\,\mu}$,
respectively acting on the quark ($\psi$) and gluon fields ($G^{\,\mu\nu\,}$).
Neglecting $1/\Qs$ power corrections, the OPE applied to Eq.~(\ref{eq:T}) 
allows us to express the Mellin-$N$ moments of the structure function $F_\phi$ 
in terms the matrix elements $A_{i,\rm nucl}$ and coefficient functions 
$C_{\phi,i}$ as
\bea
\label{eq:Fphimellin}
  F_\phi(N,\Qs) & = & \int_0^1 \! dx\: x^{\,N-1} \, F_\phi(x,\Qs) \, ,
\\
&=&
\label{eq:OPE-C*A}
\sum_{i = {\rm q, g}}
  C_{\phi,i} \left(N, \frac{\Qs}{\mu^{\,2}}, \as \right) 
  A_{i,\rm nucl}(N,\mu^{\,2}) \; .
\eea
Here the first line~(\ref{eq:Fphimellin}) fixes our conventions for Mellin 
moments $N$ and the second relation~(\ref{eq:OPE-C*A}) holds (due to symmetry 
properties for $x \to -x$) for even-integer values of $N$.
All (complex) moments, and the complete $x$-dependence, are uniquely fixed 
by analytic continuation though.

The renormalization of the singlet operators $O_{\rm q}$ and $O_{\rm g}$ in 
Eq.~(\ref{eq:OME}) in terms of their bare counterparts proceeds as
\beq
\label{eq:Oren}
  O_i \; = \; Z_{\,ik}\,O_k^{\,\rm bare} \; ,
\eeq
where the renormalization factors $Z_{ik}$ are matrix-valued (i.e., summation 
of $k$ is understood) and the dependence on $N$ has been suppressed here (and 
in the following) for brevity. 
The well-known anomalous dimensions $\gamma_{\,ik}$ arise from the 
renormalization factors $Z_{\,ik}$ as 
\beq
\label{eq:gamZ}
 \gamma_{\,ik} \: = \: 
 -\,\left( \frac{d }{d\ln\mu^{\,2} }\: Z_{\,ij} \right) (Z^{-1})_{\!jk} \; ,
\eeq
and the scale dependence of the operators $O_{i}$ is given by
\beq
\label{eq:gamma}
  \frac{d}{d \ln \mu^{\,2} }\: O_i 
  \; = \; - \,\gamma_{\,ik}\, O_k
  \; \equiv \; P_{\,ik}\, O_k  
  \; .
\eeq
Eq.~(\ref{eq:gamma}) also recalls the conventional relation between the 
anomalous dimensions and the moments of the splitting functions $P_{\,ik}(x)$. 

Using dimensional regularization~\cite{DimReg} in $ D = 4 - 2\ep $ dimensions 
and the modified minimal subtraction scheme \MSb\ \cite{BBDM78,MS}, the 
renormalization factors $Z_{\,ik}$ in Eq.~(\ref{eq:Oren}) are a given by a 
series of poles in $1/\ep$ which can be expressed in terms of the perturbative 
expansion coefficients $\gamma^{\,(l)}$ of the anomalous dimensions and the 
coefficients $\beta_{\rm n}$ governing the running coupling.
In a power expansion in $\as$ the former can be written as
\beq
\label{eq:gam-exp}
  \gamma_{\,ik}^{} \; = \; 
  \sum_{l=0}^{\infty}\, a_{\rm s}^{\,l+1}\, \gamma_{\,ik}^{\,(l)} \; ,
\eeq
while the latter are given by 
\beq
\label{eq:arun}
  \frac{d}{d \ln \mu^{\,2}}\: \frac{\as}{4\pi} \:\: \equiv \:\: 
  \frac{d\,\ar}{d \ln \mu^{\,2}} \:\: = \:\: - \ep\, \ar 
  - \beta_0\, \ar^{\,2} - \beta_1\, \ar^{\,3} 
  - \beta_2\, \ar^{\,4} - \ldots \:\: .
\eeq
Here $\beta_{1, \ldots, 4 }$ denote the known four-dimensional expansion 
coefficients of the beta function of QCD \cite{AsymptF,beta1,beta2,beta3}, 
$\,\beta_0 = 11/3\:\ca - 2/3\:\nf\,$ etc, with $\nf$ representing the number of 
active quark flavours and $\ca=3$ (and $\cf=4/3$) denoting the usual 
SU($n_c\!=\!3$) colour factors.

The N$^{\,l}$LO anomalous dimensions $\gamma^{\,(l)}$ are related to the 
$1/\ep$ single poles of $Z_{\,ik}$ in Eq.~(\ref{eq:Oren}) at order 
$a_{\rm s}^{\,l+1}$. The $D$-dimensional coefficient functions 
$\widetilde{C}_{\phi,i}$, on the other hand, have an expansion in non-negative 
powers of $\ep$, viz
\beq
\label{eq:cf-exp}
  \widetilde{C}_{\phi,i} \; = \; \delta_{\,i \rm g} \,
    + \,\sum_{l=1}^{\infty} \, \ar^{\, l} \left( c_{\phi,i}^{\,(l)} 
    + \ep\:\! a_{\phi,i}^{\,(l)} + \ep^2 b_{\phi,i}^{\,(l)} + \ldots \right)
\eeq
where $i$ = q, g as before, and we have again suppressed the dependence on $N$ 
(and $\Qs/\mu^{\,2}$). Recall that contributions of order $\ep^{\,k>0}$ in 
Eq.~(\ref{eq:cf-exp}) enter the extraction of the anomalous dimensions and 
coefficient functions at higher orders in $\as$. For example, as shown in 
Eq.~(\ref{eq:Tphi3}) below, the determination of the third-order coefficient 
functions $c_{\phi,i}^{\,(3)}$ requires the two-loop $\ep^1$ quantities 
$a_{\phi,i}^{\,(2)}$.

For the perturbative determination of splitting functions and coefficient 
functions the nucleons in Eqs.~(\ref{eq:T}) and (\ref{eq:OME}) can be replaced 
by partons. Thus we need to compute all Feynman diagrams for the forward 
scattering amplitudes which contribute at leading-twist accuracy to the singlet
structure functions $F_\phi$ in partonic DIS, 
\beq
\label{eq:Tpart}
  T_{\phi, k}(p,q) \; = \; i \! \int \! d^{\, 4}z \: e^{\,iqz} \, 
  \langle \, k, p \vert \, T \big(J(z) \, J(0) \big) \, \vert k, p \rangle 
  \, , \quad k = {\rm q,\,g} 
  \; , 
\eeq
i.e., we are considering the reactions,
\beq
\label{eq:qaqa}
 \mbox{parton}\,(p)+\mbox{scalar}\,(q) \:\:\rightarrow\:\:
 \mbox{parton}\,(p)+\mbox{scalar}\,(q) 
\end{equation}
where the scalar couples directly only to gluons via a $\phi\,G^{\,\mu\nu\!}
G_{\mu\nu}$ contribution to the Lagrangian. Such an interaction is effectively 
included in the Standard Model for the Higgs boson, in the limit of a heavy top
quark and negligible Yukawa couplings to all other quark flavours \cite{HGGeff}.
Actually, the resulting coefficient functions for the Higgs boson differ from 
those of a scalar with a generic $\phi\,G^{\,\mu\nu\!}G_{\mu\nu}$ coupling by a
perturbative prefactor, known to N$^3$LO \cite{Chetyrkin:1997un}, which is 
however irrelevant for our considerations. 
We will return to the limitations of the heavy-top limit in Sections 3 and 4.

The above scalar current in Eq.~(\ref{eq:Tpart}) requires an additional, 
well-known renormalization~\cite{GGrenorm},
\bea
\label{eq:scalarRen}
\left( G^{\,\mu\nu} G_{\mu\nu} \right) \;=\; 
 Z_{G^{\,2}} \left( G^{\,\mu\nu} G_{\mu\nu} \right)^{\rm{bare}} 
 \, +\; \dots \;\; , \qquad
 Z_{G^{\,2}} \;=\; \frac{1}{1-\beta(\ar)/(\ep \ar)} \;\; ,
\eea
which is calculable in terms of the coefficients $\beta_{\rm n}$ of the QCD 
beta function~(\ref{eq:arun}). Moreover the field strength in 
Eq.~(\ref{eq:scalarRen}) is subject to operator mixings which, however, give 
either vanishing contributions to the on-shell matrix elements considered in 
Eq.~(\ref{eq:qaqa}) or, as in case of a quark mass term 
$m_{\rm{q}}\bar{\psi} \psi$, vanish in the present limit of massless quarks.

In order to determine the coefficient functions $C_{\phi,i}$ to order $\as^3$,
we thus `only' need to evaluate the forward scattering amplitudes
(\ref{eq:qaqa}) to three loops. The corresponding Feynman diagrams have been 
generated automatically with {\sc Qgraf} \cite{Nogueira:1991ex}. The resulting
number of diagrams is shown in Table~\ref{tab:diagrams}. Note that, as already
in Refs.~\cite{Mom3loop}, the generation and counting of the diagrams are 
non-standard, as some tricks (e.g., using symmetries valid only at the relevant
even values of $N$) have been employed to reduce the number of diagrams. 
\begin{table}[htp]
\vspace{-3mm}
\begin{center}
\begin{tabular}{c r r r r}\\
\hline & & & & \\[-3mm]
{process} &{tree} &1-loop &2-loop &3-loop \\[1mm]
\hline & & & & \\[-3mm]
$q\,\phi\:\ra\: q\,\phi$ &   &  1 &  23 &   696 \\
$g\,\phi\:\ra\: g\,\phi$ & 1 &  8 & 218 &  6378 \\
$h\,\phi\:\ra\: h\,\phi$ &   &  1 &  33 &  1184 \\[1mm]
\hline & & & & \\[-2mm]
 sum & 1 & 10 & 274 & 8258 \\[1mm] \hline
\end{tabular}
\end{center}
\vspace{-3mm}
\caption{ \label{tab:diagrams}
 The number of diagrams for the forward amplitudes employed in the calculation 
 of the structure function $F_\phi$ to three loops. $h$ stands for the external
 ghost discussed in the text.}
\vspace{-3mm}
\end{table}
 
For all external partons states in Eq.~(\ref{eq:qaqa}) we need to project on 
their physical polarizations. To that end it has proven efficient to contract 
the external gluon lines only with the metric tensor $-g^{\alpha\beta}$ 
instead of a physical projector
\begin{eqnarray}
  \label{eq:gluon-pol-proj}
  -g^{\alpha\beta} 
  \, + \, \frac{p^\alpha q^\beta + p^\beta q^\alpha}{p \cdot q} 
  \, - \, \frac{p^\alpha p^\beta q^2}{(p \cdot q)^2}
  \; ,
\end{eqnarray}
matching the kinematics of Eq.~(\ref{eq:qaqa}) with $p^2 = 0$, $q^2= - \Qs < 0$.
To compensate the additional unphysical degrees of freedom we have to consider 
an extra class of diagrams with external ghosts instead of external gluons, 
i.e., the process $h\,\phi \ra h\,\phi$ listed in the third line of
Table~\ref{tab:diagrams}. In particular the resulting absence of higher tensor 
integrals from the numerator of Eq.~(\ref{eq:gluon-pol-proj}) leads to a 
vital simplification of the calculations. For the same reason our all-$N$
computations have been performed in the Feynman gauge. As before 
the gauge independence has been verified for a few low values of $N$ along the
lines of Refs.~\cite{Mom3loop} where also the projector 
(\ref{eq:gluon-pol-proj}) has been employed for checks.

The calculation of the Mellin-$N$ projection of the forward-scattering 
integrals for all values of $N$ has been discussed extensively, if not 
exhaustively, in previous publications~\cite{Moch:1999eb,MVV2,MVV3,MVV4,MVV6}.
Also the present calculation has been carried out using version 3 of the
symbolic manipulation program {\sc Form} \cite{Vermaseren:2000nd} to which new 
features, such as an efficient database facility for huge numbers of 
intermediate integrals, were added in order to make such computations feasible
\cite{Vermaseren:2002rp}. We would like to stress again that the calculation
has been set up in such a manner that checks of fixed low moments could be 
performed at all intermediate stages using the program {\sc Mincer} 
\cite{Gorishnii:1989gt,Larin:1991fz} for massless three-loop self-energy 
integrals. Note also that the recent computations of fixed moments of 
heavy-quark structure functions \cite{Bierenbaum:2009mv} and of the three-loop 
quark and gluon form factors \cite{FF3loop} provide strong independent checks 
of different aspects of our calculations. 

Having calculated the perturbative corrections to the individual parton 
contributions to $T_{\phi, k}$ in Eq.~(\ref{eq:Tpart}) it remains (first) to 
perform the renormalization of $\alpha_s$ and the scalar current in 
Eq.~(\ref{eq:scalarRen}), and (second) to disentangle the coefficients of the 
Laurent expansion in terms of anomalous dimensions and coefficient functions.
The second task also known as mass factorization is non-trivial, requires 
an explicit representation for the matrix $Z_{ik}$ in Eq.~(\ref{eq:Oren}) 
and introduces the dependence on a specific scheme, i.e., \MSb\ in our case.
To that end, we set the factor $\exp ( \ep \{\ln(4\pi)-\gamma_{\rm e}\} ) = 1$ 
where $\gamma_{\rm e}$ is the Euler-Mascheroni constant and choose, without
loss of generality, the renormalization and mass factorization scales as 
$\mu^2 = \Qs$.

A suitable normalization, already used in Eq.~(\ref{eq:cf-exp}) above, at order
$\as^0$ implies  
\beq
\label{eq:Tphi0}
  T_{\phi,\rm q}^{(0)} \; = \; c_{\phi,\rm q}^{(0)} \; = \; 0 
  \, , \qquad\qquad
  T_{\phi,\rm g}^{(0)} \; = \; c_{\phi,\rm g}^{(0)} \; = \; 1 
  \; .
\eeq
At the first order in $\as$ the respective forward amplitudes need to be 
calculated up to order $\ep^2$ for our purposes, yielding
\beq
\label{eq:Tphi1}
  T_{\phi, \rm p}^{(1)} \; = \; 
  \frac{1}{\ep}\, \gamma_{\,\rm gp}^{\,(0)} \: + \: c_{\phi,\rm p}^{(1)} 
  \: + \: \ep\, a_{\phi,\rm p}^{(1)} \: + \: \ep^2 b_{\phi,\rm p}^{(1)} 
  \; ,
\eeq 
with $\rm p = q,\, g$. 
Correspondingly the two-loop contributions are required up to order $\ep$. 
These quantities are given by
\bea
\label{eq:Tphi2}
  T_{\phi,\rm p}^{(2)} 
  &\!=\!& 
  \frac{1}{2 \ep^2}\, \bigg\{
  \left( \gamma^{\,(0)}_{\,\rm{gi}} - \beta_0 \delta_{\,\rm gi} \right) 
    \gamma^{\,(0)}_{\,\rm{ip}} \bigg\}
  \, + \, \frac{1}{2 \ep}\, \bigg\{ \gamma^{\,(1)}_{\,\rm{gp}} 
     + 2 c^{(1)}_{\phi,{\rm{i}}} \gamma^{\,(0)}_{\,\rm{ip}} 
    \bigg\}
  \\[1mm]
  & &
  \, +\,  c^{(2)}_{\phi,{\rm{p}}} 
  \, +\,  a^{(1)}_{\phi,{\rm{i}}} \gamma^{\,(0)}_{\,\rm{ip}} 
  \, +\,  \ep\, \bigg\{
  a^{(2)}_{\phi,{\rm{p}}} 
  + b^{(1)}_{\phi,{\rm{i}}} \gamma^{\,(0)}_{\,\rm{ip}} 
  \bigg\}
  \; ,
  \nn
\eea
where $\delta_{\,\rm ip}$ is the Kronecker symbol and summation over repeated
indices is understood. Finally we are ready to write down the third-order 
coefficients $T_{\phi,k}^{(3)}$ in Eq.~(\ref{eq:Tpart}), 
\bea
\label{eq:Tphi3}
  T_{\phi,\rm p}^{(3)}   &\!=\!& 
  \frac{1}{6 \ep^3} \,
  \bigg\{ \gamma^{\,(0)}_{\,\rm{gi}} \gamma^{\,(0)}_{\,\rm{ik}} 
       \gamma^{\,(0)}_{\,\rm{kp}} 
     - 3 \beta_0 \gamma^{\,(0)}_{\,\rm{gi}} \gamma^{\,(0)}_{\,\rm{ip}} 
     + 2 \beta_0^{\,2}\, \gamma^{\,(0)}_{\,\rm{gp}} 
  \bigg\} 
\nn \\[1mm] & & \mbox{}
  + \frac{1}{6 \ep^2}\, 
  \bigg\{ \gamma^{\,(0)}_{\,\rm{gi}} \gamma^{\,(1)}_{\,\rm{ip}} 
     + 2\, \gamma^{\,(1)}_{\rm{gi}} \gamma^{\,(0)}_{\,\rm{ip}}
     - 2 \beta_0 \gamma^{\,(1)}_{\,\rm{gp}} 
     - 2 \beta_1 \gamma^{\,(0)}_{\,\rm{gp}}
     + 3 c^{(1)}_{\phi,{\rm{i}}} \left( \gamma^{\,(0)}_{\,\rm{ik}} 
       - \beta_0 \delta_{\,\rm ik} \right) \gamma^{\,(0)}_{\,\rm{kp}} 
  \bigg\}
\nn \\[1mm] & & \mbox{}
  + \frac{1}{6 \ep}\, 
  \bigg\{2\, \gamma^{\,(2)}_{\,\rm{gp}} 
    + 3 c^{(1)}_{\phi,{\rm{i}}} \gamma^{\,(1)}_{\,\rm{ip}} 
    + 6 c^{(2)}_{\phi,{\rm{i}}} \gamma^{\,(0)}_{\,\rm{ip}} 
    + 3 a^{(1)}_{\phi,{\rm{i}}} \left( \gamma^{\,(0)}_{\,\rm{ik}} 
      - \beta_0 \delta_{\,\rm ik} \right) \gamma^{\,(0)}_{\,\rm{kp}} 
  \bigg\}
\nn \\[1mm] & & \mbox{}
  + c^{(3)}_{\phi,{\rm{p}}} 
  + {1 \over 2} a^{(1)}_{\phi,{\rm{i}}} \gamma^{\,(1)}_{\,\rm{ip}} 
  + a^{(2)}_{\phi,{\rm{i}}} \gamma^{\,(0)}_{\,\rm{ip}} 
  + {1 \over 2} b^{(1)}_{\phi,{\rm{i}}} \left( \gamma^{\,(0)}_{\,\rm{ik}} 
    - \beta_0 \delta_{\,\rm ik} \right) \gamma^{\,(0)}_{\,\rm{kp}} 
  \; ,
\eea
from which the N$^3$LO coefficient functions $c^{(3)}_{\phi,{\rm{p}}}$ are
extracted.

As mentioned above, the mass factorization in Eqs.~(\ref{eq:Tphi1}) -- 
(\ref{eq:Tphi3}) provides the splitting functions and coefficient functions
in the \MSb\ scheme at the scale $\mu^2 = \Qs$. The full scale dependence,
including the case of unequal renormalization and factorization scales, can be
reconstructed from these results using, e.g., Eqs.~(2.16) -- (2.18) of Ref.\
\cite{NV2}.

For transformations to other schemes it is convenient to combine the present
`gluon' structure function $F_\phi$, recall Eq.~(\ref{eq:Tphi0}), with a 
corresponding `quark' observable such as (the flavour-singlet part of$\,$) 
$\Ftwo$ in photon-exchange DIS \cite{MVV6}, with $c_{2,\rm q}^{(0)} = 1$ and 
$c_{2,\rm g}^{(0)} = 0$. Hence we define the two-dimensional vector of the
moments of the singlet quark and gluon distributions,
\beq
\label{eq:QSvec}
  \qV \;=\; \left( \begin{array}{c} \!q_{\rm s}^{}\! \\ g  \end{array} \right)
  \quad \mbox{ with } \quad
  q_{\rm s}^{} \; = \; \sum_{i=1}^{\nf} \left( q_i^{} + \bar{q}_i^{} \right) 
\eeq
and the corresponding $N$-dependent quantities (recall Eq.~(\ref{eq:gamma})) 
\beq
\label{eq:FPCmat}
  \FV \;=\; \left( \begin{array}{c} \!\Ftwo\! \\ \!F_\phi\! \end{array} \right) 
  \; , \quad
  \PV \;=\; \left( \begin{array}{cc} 
            \!  P_{\rm qq} & P_{\rm qg} \! \\
            \!  P_{\rm gq} & P_{\rm gg} \! \end{array} \right)
  \; , \quad
  \CV \;=\; \left( \begin{array}{cc}
            \!  C_{\,\rm 2,q} \! & C_{\,\rm 2,g} \! \\
            \!  C_{\,\rm \phi,q} \! & C_{\,\rm \phi,g} \! \end{array} \right)
\eeq
at $\mu^2 = \Qs$.
Any scheme transformation can now be performed by inserting a suitable (finite,
no relation to Eq.~(\ref{eq:Oren}) above) transformation matrix $\ZV$ with 
\beq
\label{eq:Zmat}
  Z_{\,\rm ik} \;=\; \delta_{\,\rm ik} \, + \,
  \sum_{l=1}^{\infty} \, \ar^{\, l} \, Z_{\,\rm ik}^{\,(l)}
\eeq 
which results in transformed coefficient functions $\CV^{\,\prime}$ and parton
distributions $\qV^{\,\prime}$ via
\beq
\label{eq:Cqnew}
  \FV \:\:=\:\: \CV \cdot \qV 
      \:\:=\:\: \CV\, \ZV^{\,-1} \cdot \ZV\, \qV
      \:\:=\:\: \CV^{\,\prime} \cdot \qV^{\,\prime}
  \; .
\eeq
 
The corresponding transformation of the splitting functions reads
\beq
\label{eq:Pnew}
 \PV^{\,\prime} \; = \; \PV \:+\: \Big( \beta\, \frac{d\:\! \ZV}{d \ar} 
 \:+\: [\ZV,\PV\,] \Big) \,\ZV^{\,-1}
\eeq
where $\beta$ represents the right-hand-side of Eq.~(\ref{eq:arun}), and
$[\ZV,\PV\,]$ the standard commutator of the matrices $\ZV$ and $\PV$.
To order $\as^{\,4}$ (N$^3$LO) the insertion of the perturbative expansions 
thus leads to
\bea
\label{eq:Pexp}
  \PV^{\,\prime} &\!\! =\!\! & \; \ar\: \PV_0
  \nn \\[1mm] & & \mbox{\hspn}
  + \,\ar^{\,2} \Big( \PV^{\,(1)} + [\,\ZV^{(1)\!},\PV^{(0)\,}] 
    - \beta_0\ZV^{(1)} \Big)
 \nn  \\[1mm] & & \mbox{\hspn}
  + \,\ar^{\,3} \Big( \PV^{\,(2)} + [\,\ZV^{(2)\!},\PV^{(0)\,}]
    + [\,\ZV^{(1)\!},\PV^{(1)\,}] - [\,\ZV^{(1)\!},\PV^{(0)\,}] \,\ZV^{(1)}
    + \beta_0 \Big( \{\ZV^{(1)}\}^2 - 2 \ZV^{(2)} \Big) 
    - \beta_1 \ZV^{(1)} \Big)
 \nn \\[1mm] & & \mbox{\hspn}
  + \,\ar^{\,4} \Big( \PV^{\,(3)} + [\,\ZV^{(3)\!},\PV^{(0)\,}]
    + [\,\ZV^{(2)\!},\PV^{(1)\,}] + [\,\ZV^{(1)\!},\PV^{(2)\,}]
    - \Big( [\,\ZV^{(2)\!},\PV^{(0)\,}] + [\,\ZV^{(1)\!},\PV^{(1)\,}] \Big)
      \,\ZV^{(1)} 
 \nn \\ & & \mbox{\hspp}
    + [\,\ZV^{(1)\!},\PV^{(0)\,}] \Big( \{\ZV^{(1)}\}^2 - 2 \ZV^{(2)} \Big)
    - \beta_0 \Big( \{\ZV^{(1)}\}^3 - \ZV^{(1)} \ZV^{(2)} 
      - 2 \ZV^{(2)} \ZV^{(1)} +3 \ZV^{(3)}  \Big)
 \nn \\ & & \mbox{\hspp}
    + \beta_1 \Big( \{\ZV^{(1)}\}^2 - 2 \ZV^{(2)} \Big)
    - \beta_2 \ZV^{(1)} \Big) 
    \;\; + \;\; {\cal O} (\ar^{\,5\,})
 \; .
\eea
In order to transfer the momentum sum rule, $q_{\rm s}^{} + g \,=\, 1$ at 
$N=2$ for all scales, from \MSb\ to the transformed parton densities 
$q^{\,\prime}$ in Eq.~(\ref{eq:Cqnew}), the coefficients in Eq.~(\ref{eq:Zmat})
need to satisfy the relations
\beq
\label{eq:Zmomsum}
  Z_{\,\rm qq}^{\,(l)}(N\!=\!2) \,+\, Z_{\,\rm gq}^{\,(l)}(N\!=\!2) \; = \; 0 
  \;\; , \quad
  Z_{\,\rm qg}^{\,(l)}(N\!=\!2) \,+\, Z_{\,\rm gg}^{\,(l)}(N\!=\!2) \; = \; 0
  \;\; . 
\eeq

Eqs.~(\ref{eq:Zmat}) -- (\ref{eq:Pexp}) can be employed to transform to the 
physical evolution equations for the system $F = (\Ftwo, F_\phi)$ by choosing
\beq
\label{eq:F2Htrf}
  \left( \begin{array}{cc}
       \!  Z_{\,\rm qq}^{\,(l)} & Z_{\,\rm qg}^{\,(l)} \! \\[2mm]
       \!  Z_{\,\rm gq}^{\,(l)} & Z_{\,\rm gg}^{\,(l)} \! 
         \end{array} \right)
  \; = \;
  \left( \begin{array}{cc}
       \!  c_{\,\rm 2,q}^{\,(l)} \! & c_{\,\rm 2,g}^{\,(l)} \! \\[2mm]
       \!  c_{\,\rm \phi,q}^{\,(l)} \! & c_{\,\rm \phi,g}^{\,(l)} \!
         \end{array} \right) 
\eeq
in Eq.~(\ref{eq:Zmat}). This leads to
\beq
\label{eq:Kab}
  \frac{d}{d \ln \Qs }\: F \; = \; \KV\, F
  \; \equiv \; \sum_{l=0}^{\infty} \: \ar^{\:l+1\!} 
    \left( \begin{array}{cc}
       \!  K_{\, 22}^{\,(l)} \! & K_{\, 2\phi}^{\,(l)} \! \\[2mm]
       \!  K_{\, \phi 2}^{\,(l)} \! & K_{\,\phi\phi}^{\,(l)} \!
         \end{array} \right)
  \cdot
  \left( \begin{array}{c} \!\Ftwo\! \\ \!F_\phi\! \end{array} \right)
  \; ,
\eeq
where the matrix elements of the physical evolution kernel $\KV$ are given by 
Eq.~(\ref{eq:Pexp}) after inserting Eq.~(\ref{eq:F2Htrf}). As far as we know
this transformation has first been suggested (at NLO) in Ref.~\cite{FP82}.
The above relations refer to the choice $\mu^2 = \Qs$ of the renormalization
scale -- the mass factorization scheme and scale have now been eliminated from
the problem. From these results the physical kernel at $\mu^2 \neq \Qs$ (or in 
other renormalization schemes) can be reconstructed in the usual way.
As we shall see below, Eq.~(\ref{eq:F2Htrf})  does not represent a `normal' 
scheme transformation of the quark and gluon distributions, since the \MSb\ 
coefficient functions for $\Ftwo$ and $F_\phi$ on the right-hand-side do not 
fulfill the momentum sum rule constraints (\ref{eq:Zmomsum}).
%
%
\setcounter{equation}{0}
\section{Coefficient functions for Higgs-exchange DIS}
\label{sec:coeffs}
%
%
In this section we present and discuss the $x$-space results, obtained from our
above $N$-space calculations by a by now standard inverse Mellin transformation 
\cite{Moch:1999eb,Remiddi:1999ew}, for the previously unpublished \MSb-scheme 
coefficient functions $c_{\,\rm \phi,q}^{\,(l)}$ and 
$c_{\,\rm \phi,g\,}^{\,(l)}$, $l = 1,\,2,\,3$, at the standard choice 
$\mu^2 = \Qs$ of the renormalization and factorization scale.

We express our results in terms of the harmonic polylogarithms 
$H_{m_1,...,m_w}(x)$ with $m_j = 0,\pm 1$. Our notation for these functions 
follows Ref.~\cite{Remiddi:1999ew} to which the reader is referred for a 
detailed discussion. 
For completeness we recall the basic definitions: The lowest-weight ($w = 1$) 
functions $H_m(x)$ are given by
\beq
\label{eq:hpol1}
  H_0(x)       \: = \: \ln x \:\: , \quad\quad
  H_{\pm 1}(x) \: = \: \mp \, \ln (1 \mp x) \:\: .
\eeq
The higher-weight ($w \geq 2$) functions are recursively defined as
\beq
\label{eq:hpol2}
  H_{m_1,...,m_w}(x) \: = \:
    \left\{ \begin{array}{cl}
    \displaystyle{ \frac{1}{w!}\,\ln^w x \:\: ,}
       & \quad {\rm if} \:\:\: m^{}_1,...,m^{}_w = 0,\ldots ,0 \\[2ex]
    \displaystyle{ \int_0^x \! dz\: f_{m_1}(z) \, H_{m_2,...,m_w}(z)
       \:\: , } & \quad {\rm else}
    \end{array} \right.
\eeq
with
\beq
\label{eq:hpolf}
  f_0(x)       \: = \: \frac{1}{x} \:\: , \quad\quad
  f_{\pm 1}(x) \: = \: \frac{1}{1 \mp x} \:\: .
\eeq
For chains of indices zero we employ the abbreviated notation
\beq
\label{eq:habbr}
  H_{{\footnotesize \underbrace{0,\ldots ,0}_{\scriptstyle m} },\,
  \pm 1,\, {\footnotesize \underbrace{0,\ldots ,0}_{\scriptstyle n} },
  \, \pm 1,\, \ldots}(x) \: = \: H_{\pm (m+1),\,\pm (n+1),\, \ldots}(x)
\eeq
and usually suppress the argument $x$. 

The $l$-th order coefficient functions involve harmonic polylogarithms up to 
weight $2l\!-\!1$. Hence only the results up to two loops, with $w \leq 3$,
can be expressed in terms of standard polylogarithms. A complete list can be 
found in appendix A of Ref.~\cite{Moch:1999eb}. 
A {\sc Fortran} program for the harmonic polylogarithms including weight $w=4$ 
has been published in Ref.~\cite{Gehrmann:2001pz}. Its extension to $w=5$,
required for the third-order coefficient functions, is also available~\cite
{Gehrmann:pc}.

In this notation the first-order coefficient functions are given by
\bea
\label{eq:cphiq1}
 c^{\,(1)}_{\rm \phi\, , q}(x) &\! =\! &
        \colour4colour{ \cf\, } 
          \* \Big( 2\, \* \pgq(x)\, \* ( - \H(0) - \H(1) )\,
        - 3\, \* x^{\,-1}\, + 2\, \* x\, \Big)
\:\: ,
\\[2mm]
\label{eq:cphig1}
 c^{\,(1)}_{\rm \phi\, , g}(x) &\! =\! &
        \colour4colour{ \ca } \, \* \Big( 4\, \* \pgg(x) \* (
          - \H(0)
          - \H(1) )\,
          - 11/3\: \* ( \x1^{-1} + x^{-1}\, )\,
          + ( 67/9 - 4\, \* \z2\, )\: \* \delta \x1 \Big)\,
\nn \\[0.5mm] & & \mbox{\hspn}
       + \colour4colour{ \nf } \, \* \Big(2/3\: \* ( \x1^{-1} + x^{\,-1}\ )\,
       - 10/9\: \* \delta \x1 \Big)
\:\: .
\eea
Here and below $\zeta_{\:\!n}$ represents the Riemann zeta-function, and as 
above $\nf$ denotes the number of effectively massless flavours.  $\ca$ and 
$\cf$ are the usual QCD colour factors specified below Eq.~(\ref{eq:arun}).
Finally we have employed the abbreviations 
\pagebreak

\vspace*{-1.5cm}
\bea
\label{eq:p0gqgg}
  p_{\rm{gq}}(x) &\! =\! & 2\:\! x^{\,-1} -2 + x  \nn \\[0.5mm]
  p_{\rm{gg}}(x) &\! =\! & (1-x)^{-1} + x^{\,-1} - 2 + x - x^{\,2}
  \:\: .
\eea

\noindent
The corresponding two-loop (NLO) coefficient functions read
\bea
\label{eq:cphiq2}
 \lefteqn{c^{\,(2)}_{\rm \phi\, , q}(x) \; = \;
%
%
       \colour4colour{ \cfs }\, \* \Big(
        4\, \* \pgq(x)\,  \*  (
          - 6\, \* \Hh(1,1)\,
          - 2\, \* \Hhh(1,1,0)\,
          - \Hhh(1,1,1)\,
          - 2\, \* \Hh(1,2)\,
          - 2\, \* \Hh(2,0)\,
          - 2\, \* \Hh(2,1)\,
          - 2\, \* \H(3)\,
          )\,
 }
   \nn \\[-0.5mm] & & \mbox{}
       + 8\, \* \pgq( - x)\,  \*  (
          - \H(-1)\, \* \z2\,
          - 2\, \* \Hhh(-1,-1,0)\,
          + \Hhh(-1,0,0)\,
          - \H(0)\, \* \z2\,
          )\,
       - 16\, \* ( 1\, + x\, )\,  \*  \Hh(-1,0)\,
   \nn \\[0.5mm] & & \mbox{}
       - 4\, \* \z3\, \* (2\, + 8\, \* x^{-1}\, + 3\, \* x\, )\,
       + 2\, \* (2 - x)  \*  (
            5\, \* \Hhh(0,0,0)\,
          + 2\, \* \Hh(2,0)\,
          + 2\, \* \Hh(2,1)\,
          + 4\, \* \H(3)\,
          )\,
   \nn \\[0.5mm] & & \mbox{}
       + 4\, \* (5\, - 9\, \* x^{-1}\,
       - 6\, \* x\,)\,  \*  \H(2)\,
       - 4\, \* \z2\, \* (5\, - 9\, \* x^{-1}\, - 2\, \* x\,)\,
       - 8\, \* \z2\, \* (6\, - x\,)\,  \*  \H(0)\,
       - (12\, - x\,)\,  \*   \Hh(0,0)\,
   \nn \\[0.5mm] & & \mbox{}
       + (15\, - 10\, \* x^{-1}\,
       + 18\, \* x\,)\,  \*   \H(0)\,
       + 2\, \* (16\, - 18\, \* x^{-1}\, - 7\, \* x\,)\,  \*  \Hh(1,0)\,
       + (56\, - 64\, \* x^{-1}\, + 3\, \* x\,)\,  \*   \H(1)\,
   \nn \\ & & \mbox{}
       + 1/2\: \* (208\, - 59\, \* x^{-1}\,
       - 30\, \* x\,)\,
       - 32\, \* \Hh(-2,0)\,
       - 4\, \* \Hh(1,1)\,
                        \Big)\,
   \nn \\[-0.5mm] & & \mbox{\hspn}
   +   \colour4colour{ \ca\, \* \cf }\, \* \Big(
         20\, \* \pgq(x)\,  \*  (
            \z2\, \* \H(1)\,
          - \Hhh(1,0,0)\,
          - \Hhh(1,1,0)\,
          - \Hhh(1,1,1)\,
          - \Hh(1,2)\,
           )\,
   \nn \\[-0.5mm] & & \mbox{}
       + 4\, \* \pgq( - x)  \*  (
          - 2\, \* \H(-1)\, \* \z2\,
          + 3\, \* \Hhh(-1,0,0)\,
          + 2\, \* \Hh(-1,2)\,
          )\,
       + 16\, \* (1\, + 2\, \* x^{-1}\, + 2\, \* x\,)\,  \*  (
           \z2\, \* \H(0)\,
          - \H(3)\,
          )\,
   \nn \\[0.5mm] & & \mbox{}
       + 4\, \* (2\, - 10\, \* x^{-1}\, - 7\, \* x\,)\, \* \Hh(2,1)\,
       - 4\, \* (10\, + 4\, \* x^{-1}\, + 7\, \* x\,)\,  \*  \Hhh(0,0,0)\,
       - 4\, \* \z3\, \* (16\, - 18\, \* x^{-1}\, - 11\, \* x\,)\,
   \nn \\[0.5mm] & & \mbox{}
       + 4/3\: \* (24\, + 22\, \* x^{-1}\,
       + 9\, \* x\, + 4\, \* x^{2}\,)\,  \*  \Hh(-1,0)\,
       + 4/3\: \* (46\, - 53\, \* x^{-1}\, 
       + x\, + 4\, \* x^{2}\,)\, \* \Hh(1,0)\,
   \nn \\[0.5mm] & & \mbox{}
       - 1/9\: \* (60\, + 1362\, \* x^{-1}\, + 513\, \* x\, 
       + 352\, \* x^{2}\,)\,  \*  \H(0)\,
       + 2/3\: \* (70\, - 84\, \* x^{-1}\, + 13\, \* x\, 
       + 8\, \* x^{2}\,)\,  \*  \Hh(1,1)\,
   \nn \\[0.5mm] & & \mbox{}
       - 4/3\: \* \z2\, \* (88\, - 53\, \* x^{-1}\, + 7\, \* x\, 
       + 12\, \* x^{2}\,)\,
       + 4/3\: \* (88\, - 31\, \* x^{-1}\, + 16\, \* x\, 
       + 12\, \* x^{2}\,)\,  \*  \H(2)\,
   \nn \\[0.5mm] & & \mbox{}
       + 1/9\: \* (242\, - 288\, \* x^{-1}\, - 127\, \* x\, 
       - 176\, \* x^{2}\,)\,  \*  \H(1)\,
       + 2/3\: \* (268\, - 106\, \* x^{-1}\, + 13\, \* x\, 
       + 32\, \* x^{2}\,)\,  \*  \Hh(0,0)\,
   \nn \\ & & \mbox{}
       + 1/54\: \* (6172\, - 13457\, \* x^{-1}\, 
           + 76\, \* x\,+ 1216\, \* x^{2}\,)\,
       - 8\, \* (4\, \* x^{-1}\, + 3\, \* x\,)\,  \*  \Hh(2,0)\,
       - 16\, \* \Hh(-2,0)\,
                              \Big)\,
   \nn \\[-0.5mm] & & \mbox{\hspn}
   +    \colour4colour{ \cf\, \* \nf }\, \* \Big(
           2/9\: \* \pgq(x)\,  \*  (
           24\, \* \Hh(0,0)
          + 29\, \* \H(1)
          + 12\, \* \Hh(1,0)
          + 6\, \* \Hh(1,1)
          + 12\, \* \H(2)
          )\,
       + 8/3\: \* \z2\, \* (2\, - 2\, \* x^{-1} - x\,)
   \nn \\[-1.5mm] & & \mbox{}
       - 2/3\: \* (26 - 32\, \* x^{-1} - 11\, \* x\,)\, \*  \H(0)\,
       - 1/27\: \*  (332\, - 737\, \* x^{-1}\, - 28\, \* x\,)\,
                \Big)\,
\eea
and
\bea
\label{eq:cphig2}
 \lefteqn{c^{\,(2)}_{\rm \phi\, , g}(x) \; = \;
%
%
     \colour4colour{ \cas } \, \*  \Big(
         \pgg(x)\,  \*  (
          - 2570/27\:
          + 68\, \* \z3\,
          + 176/3\: \* \z2\,
          + 24\, \* \Hh(-2,0)\,
          - 778/9\: \* \H(0)\,
          + 56\, \* \H(0)\, \* \z2\,
 }
   \nn \\[-0.5mm] & & \mbox{}
          - 44\, \* \Hh(0,0)\,
          - 28\, \* \Hhh(0,0,0)\,
          - 778/9\: \* \H(1)\,
          + 40\, \* \H(1)\, \* \z2\,
          - 176/3\: \* \Hh(1,0)\,
          - 40\, \* \Hhh(1,0,0)\,
          - 176/3\: \* \Hh(1,1)\,
   \nn \\[0.5mm] & & \mbox{}
          - 56\, \* \Hhh(1,1,0)\,
          - 48\, \* \Hhh(1,1,1)\,
          - 56\, \* \Hh(1,2)\,
          - 176/3\: \* \H(2)\,
          - 48\, \* \Hh(2,0)\,
          - 56\, \* \Hh(2,1)\,
          - 56\, \* \H(3)\,
          )
   \nn \\[0.5mm] & & \mbox{}
       + 4\, \* \pgg(-x)\,  \*  (
            7\, \* \z3\,
          + 6\, \* \Hh(-2,0)\,
          - 8\, \* \H(-1)\, \* \z2\,
          - 8\, \* \Hhh(-1,-1,0)\,
          + 10\, \* \Hhh(-1,0,0)\,
          + 4\, \* \Hh(-1,2)\,
          + 2\, \* \H(0)\, \* \z2\,
   \nn \\[0.5mm] & & \mbox{}
          - 3\, \* \Hhh(0,0,0)\,
          - 2\, \* \H(3)\,
          )
       + 16\, \* (1\, + x\, )\,  \*  (
            2\, \* \z3\,
          + 4\, \* \H(0)\, \* \z2\,
          - 5\, \* \Hhh(0,0,0)\,
          - 2\, \* \Hh(2,0)\,
          - 2\, \* \Hh(2,1)\,
          - 4\, \* \H(3)\,
          )
   \nn \\[0.5mm] & & \mbox{}
       + 8/3\: \* (6\, + 11\, \* x^{-1}
       + 6\, \* x + 11\, \* x^{2})  \*  \Hh(-1,0)\,
       + 1/27\: \* (10\, - 5659\, \* x^{-1} - 1916\, \* x + 3177\, \* x^{2})
   \nn \\[0.5mm] & & \mbox{}
       + 4/3\: \* (14\, - 44\, \* x^{-1}
            - 25\, \* x + 33\, \* x^{2}) \, \*  (
            \Hh(1,0)\,
          + \Hh(1,1)\,
          )
       - 4/3\: \* \z2\, \* (45\, - 44\, \* x^{-1} - 24\, \* x + 77\, \* x^{2})
   \nn \\[0.5mm] & & \mbox{}
       + 4/3\: \* (45\, - 22\, \* x^{-1} - 12\, \* x 
         + 77\, \* x^{2}) \, \*  \H(2)\,
       + 2/3\: \* (171\, - 44\, \* x^{-1} - 33\, \* x 
         + 220\, \* x^{2}) \, \*  \Hh(0,0)\,
   \nn \\[0.5mm] & & \mbox{}
       - 2/9\: \* (182\, + 121\, \* x^{-1} - 58\, \* x) \, \*  \H(1)\,
       - 1/9\: \* (1107\, + 778\, \* x^{-1} + 699\, \* x 
       + 536\, \* x^{2}) \, \*  \H(0)\,
   \nn \\ & & \mbox{}
       +  (
            30425/162\:
          - 242/3\: \* \z3\,
          - 778/9\: \* \z2\,
          + 101/5\: \* \zss\,
          )\, \*  \delta \x1
               \Big)\,
   \nn \\[-0.5mm] & & \mbox{\hspn}
      + \colour4colour{ \ca\, \*  \nf } \, \*  \Big(
         \pgg(x)\,  \*  (
            224/9\:
          - 32/3\: \* \z2\,
          + 56/3\: \* \H(0)\,
          + 8\, \* \Hh(0,0)\,
          + 56/3\: \* \H(1)\,
          + 32/3\: \* 
            ( \Hh(1,0)\,
            + \Hh(1,1) )
   \nn \\[-0.5mm] & & \mbox{}
          + 32/3\: \* \H(2)\,
          )
       + 4\, \* (1\, + x\, )\,  \*  \Hh(0,0)\,
       + 8/3\: \* (2\, - x + x^{2})  \*  (
            \Hh(1,0)\,
          + \Hh(1,1)\,
          )
       + 8/3\: \* (3\, + x^{2})  \*  (
            \H(2)\,
          - \z2
          )
   \nn \\[0.5mm] & & \mbox{}
       + 2/9\: \* (40\, + 48\, \* x^{-1} + 7\, \* x - 4\, \* x^{2}) \,
          \*  \H(1)\,
       + 2/9\: \* (47\, + 48\, \* x^{-1} + 14\, \* x - 4\, \* x^{2}) \,
         \*  \H(0)\,
   \nn \\[0.5mm] & & \mbox{}
       + 2/27\: \* (389\, + 320\, \* x^{-1} - 46\, \* x + 59\, \* x^{2})
       -   (
            4112/81\:
          + 28/3\: \* \z3\,
          - 56/3\: \* \z2\,
          ) \, \*  \delta \x1
               \Big)\,
   \nn \\[-0.5mm] & & \mbox{\hspn}
      +  \colour4colour{ \cf\, \*  \nf } \, \*  \Big(
         2\, \* \pgg(x)\,
       + 4\, \* (1\, + x\, )\, \*  (
          - 2\, \* \z3\,
          - 4\, \* \H(0)\, \* \z2\,
          + 5\, \* \Hhh(0,0,0)\,
          + 2\, \* \Hh(2,0)\,
          + 2\, \* \Hh(2,1)\,
          + 4\, \* \H(3)\,
          )
   \nn \\[-0.5mm] & & \mbox{}
       + 4/3\: \* (3\, + 4\, \* x^{-1} - 3\, \* x - 4\, \* x^{2}) \*  (
            \Hh(1,0)\,
          + \Hh(1,1)\,
          )
       + 4/3\: \* (9\, + 4\, \* x^{-1} + 6\, \* x - 4\, \* x^{2}) \*  (
            \H(2)\,
          - \z2
          )
   \nn \\[0.5mm] & & \mbox{}
       + 2/3\: \* (21\, + 8\, \* x^{-1} + 15\, \* x 
         - 8\, \* x^{2}) \, \*  \Hh(0,0)\,
       + 2/9\: \* (93\, + 32\, \* x^{-1} - 57\, \* x 
         - 68\, \* x^{2}) \, \*  \H(1)\,
   \nn \\[0.5mm] & & \mbox{}
       + 2/9\: \* (129\, + 32\, \* x^{-1} - 3\, \* x 
         - 68\, \* x^{2}) \, \*  \H(0)\,
       + 1/27\: \* (1671\, + 307\, \* x^{-1} - 1212\, \* x - 442\, \* x^{2})
   \nn \\ & & \mbox{}
       -  (
            63/2\:
          - 24\, \* \z3\,
          ) \, \*  \delta \x1
               \Big)\,
   \nn \\[-0.5mm] & & \mbox{\hspn}
      +  \colour4colour{ \nfs } \, \ \*  \Big(
         8/27\: \* \pgg(x)\,  \*  (
          - 5\,
          - 3\, \* \H(0)\,
          - 3\, \* \H(1)\,
          )
       + 8/27\: \* (2\, - x + x^{2})  \*  (
          - 5\,
          - 3\, \* \H(0)\,
          - 3\, \* \H(1)\,
          )
   \nn \\[-1.5mm] & & \mbox{}
       + (
            100/81\:
          - 8/9\, \* \z2\,
          ) \, \*  \delta \x1
               \Big)
 \:\: .
\eea
As mentioned in the introduction, the above results have also been derived, in 
a quite different and completely independent manner, in Ref.~\cite{DGGLcphi}.
As their counterparts for photon- and $W\!$-$\:\!$exchange DIS presented in 
Refs.~\cite{MVV6,MVV10}, the full third-order expressions, not obtained by any 
other group so far, for the present Higgs-exchange coefficient functions are 
exceedingly long. Therefore these expressions are deferred to Appendix A.

Before we illustrate the numerical size and perturbative stability of these
coefficient functions, it is instructive to discuss their behaviour close to 
the endpoints $x=1$ and $x=0$. For this we will use to abbreviations
\beq
\label{abbrev}
  x_1^{} \; = \; 1-x            \:\: ,\quad
  L_0    \; = \; \ln\, x        \:\: ,\quad
  L_1    \; = \; \ln\, x_1^{}   \:\: ,\quad
  \DDk   \; = \; [\, x_1^{-1} L_1^{\,k\,}]_+ \:\: .
\eeq
As usual, the +-distributions are defined by
\beq
\label{plus}
  \int_0^1 \! dx \: a(x)_+ f(x) \; = \; \int_0^1 \! dx \: a(x)
  \,\{ f(x) - f(1) \}
\eeq
for regular functions $f(x)$. It is understood that all $1/\x1$ poles in Eqs.\
(\ref{eq:cphig1}), (\ref{eq:cphig2}) and (\ref{eq:cphig3}) have to be read as
+-distributions.

The leading large-$x$ contributions to the gluon coefficient functions 
$\,c_{\rm \phi\, , g}^{\,(n)}(x)\,$ are given by soft-gluon emission 
contributions $\DDk$ with $k \,=\, 0,\: \ldots, \: 2n\!-\!1$. At the first 
order these and all other endpoint contributions can be read off directly 
from Eqs.~(\ref{eq:cphiq1}) and (\ref{eq:cphig1}) with the help of 
Eq.~(\ref{eq:hpol1}).
The +-distribution coefficients of the two-loop gluon coefficient function 
read
\bea
\label{eq:cphig2D3}
 c_{\rm \phi\, , g}^{\,(2)} \Big|_{\,\DD3} &\! = \! &
            8\, \* \cas
 \; ,
 \\[1mm]
 \label{eq:cphig2D2}
 c_{\rm \phi\, , g}^{\,(2)} \Big|_{\,\DD2} &\! = \! &
         - {88 \over 3}\: \* \cas
         + {16 \over 3}\: \* \ca\, \* \nf
 \; ,
 \\[1mm]
 \label{eq:cphig2D1}
 c_{\rm \phi\, , g}^{\,(2)} \Big|_{\,\DD1} &\! = \! &
        \cas \* \Big( \,
           {778 \over 9}\,
         - 40\, \* \z2 \Big)\,
         - {56 \over 3}\: \* \ca\, \* \nf\,
         + {8 \over 9}\: \* \nfs
 \; ,
 \\[1mm]
 \label{eq:cphig2D0}
 c_{\rm \phi\, , g}^{\,(2)} \Big|_{\,\DD0} &\! = \! &
   - \cas \* \Big( \,
           {2570 \over 27}\,
         - 32\, \* \z3\,
         - {176 \over 3}\: \* \z2 \Big)\,
   + \ca\, \* \nf\, \* \Big( \,
           {224 \over 9}\,
         - {32 \over 3}\: \* \z2 \Big)\,
         + 2\, \* \cf\, \* \nf
         - {40 \over 27}\: \* \nfs
\; , \;\qquad
\eea
and the corresponding contributions at the third-order are given by
\bea
\label{eq:cphig3D5}
 c_{\rm \phi\, , g}^{\,(3)} \Big|_{\,\DD5} &\! = \! &
          8\, \* \cat
 \; ,
 \\[1mm]
 \label{eq:cphig3D4}
 c_{\rm \phi\, , g}^{\,(3)} \Big|_{\,\DD4} &\! = \! & \mbox{}
         - {550 \over 9}\: \* \cat\,
         + {100 \over 9}\: \* \cas\, \* \nf
 \; ,
 \\[1mm]
 \label{eq:cphig3D3}
 c_{\rm \phi\, , g}^{\,(3)} \Big|_{\,\DD3} &\! = \! &
    \cat\, \* ( 340\,
         - 128\, \* \z2 ) \,
         - {256 \over 3}\: \* \cas\, \* \nf\,
         + {16 \over 3}\: \* \ca \* \nfs
 \; ,
 \\[1mm]
 \label{eq:cphig3D2}
 c_{\rm \phi\, , g}^{\,(3)} \Big|_{\,\DD2} &\! = \! & \mbox{}
    - \cat\, \* \Big( \,
           {9623 \over 9}\,
         - {1892 \over 3}\: \* \z2 \,
         - 256\, \* \z3 \Big)\,
    + \cas\, \* \nf\, \* \Big( \,
           {3106 \over 9}\:
         - {344 \over 3}\: \* \z2 \Big)\,
   \nn \\ & & \mbox{}
         + 16\, \* \cf\, \* \ca\, \* \nf\,
         - {292 \over 9}\: \* \ca\, \* \nfs\,
         + {8 \over 9}\: \* \nft
 \; ,
 \\[2mm]
 \label{eq:cphig3D1}
 c_{\rm \phi\, , g}^{\,(3)} \Big|_{\,\DD1} &\! = \! & \mbox{}
   \cat\, \* \Big( \,
           {192268 \over 81}\,
         - {16268 \over 9}\: \* \z2\,
         - {2816 \over 3}\: \* \z3\,
         + {1316 \over 5}\: \* \zss \Big)\,
   \nn \\ & & \mbox{}
 - \cas\, \* \nf\, \* \Big( \,
           {67730 \over 81}\,
         - {4192 \over 9}\: \* \z2\, 
         - {128 \over 3}\: \* \z3 \Big)\
 - \cf\, \* \ca\, \* \nf\, \* \Big( \,
           {598 \over 3}\:
         - 128\, \* \z3  \Big)\,
   \nn \\[1mm] & & \mbox{}
 + \ca\, \* \nfs\, \* \Big( \,
           {6652 \over 81}\:
         - {272 \over 9}\: \* \z2 \Big)\,
       + {20 \over 3}\: \* \cf\, \* \nfs\,
       - {80 \over 27}\: \* \nft
 \; ,
 \\[2mm]
 \label{eq:cphig3D0}
 c_{\rm \phi\, , g}^{\,(3)} \Big|_{\,\DD0} &\! = \! & \mbox{}
   - \cat\, \* \Big( \,
           {1616486 \over 729}\,
         - {169910 \over 81}\: \* \z2\,
         - {40454 \over 27}\: \* \z3\,
         + {7931 \over 15}\: \* \zss\,  
         + {1280 \over 3}\: \* \z2\, \* \z3\,
         - 80\, \* \z5
     \Big)\,
   \nn \\ & & \mbox{}
     + \cas\, \* \nf\, \* \Big( \,
           {1234307 \over 1458}\,
         - {55388 \over 81}\: \* \z2\,
         - {1816 \over 9}\: \* \z3\,
         + {1538 \over 15}\: \* \zss
     \Big)\,
     - \cfs\, \* \nf\,
   \nn \\[1mm] & & \mbox{}
     + \cf\, \* \ca\, \* \nf\, \* \Big( \,
           {7810 \over 27}\,
         - 32\, \* \z2\,
         - {1624 \over 9}\: \* \z3\,
         - {32 \over 5}\: \* \zss
     \Big)\,  
     - \cf\, \* \nfs\, \* \Big( \,
           {350 \over 9}\,
         - {80 \over 3}\: \* \z3 \Big)\,
   \nn \\[1mm] & & \mbox{}
     - \ca\, \* \nfs\, \* \Big( \,
           {138493 \over 1458}\,
         - {584 \over 9}\: \* \z2\,  
         + {152 \over 27}\: \* \z3
    \Big)\,
     + \nft\, \* \Big( \,
           {200 \over 81}\,
         - {16 \over 9}\: \* \z2 \Big)\,
\:\: .
\eea

Together with the $\delta \x1$ contributions arising from soft-gluon emissions
and virtual corrections -- which can be read off directly from 
Eqs.~(\ref{eq:cphig1}), (\ref{eq:cphig2}) and (\ref{eq:cphig3}) -- 
these results have be employed in Ref.~\cite{MVV7} to obtain the soft-gluon
exponentiation of $C_{\rm \phi\, , g}$ to the next-to-next-to-next-to-leading
logarithmic (N$^{\,3\:\!}$LL) accuracy. Consequently the highest seven 
+-distributions are known at the fourth and all higher orders in $\as$,
which the exception of the (almost certainly numerically irrelevant) 
contribution of the four-loop cusp anomalous dimension to the coefficient of 
$\as^{\,n} {\cal D}_{\,2n-7\,}$.

As discussed in Refs.~\cite{MVV8,MVV9}, the computation of the forward 
amplitudes $T_{2,\rm q}^{(l)}$ and $T_{\phi,\rm g}^{(l)}$ in photon- and 
Higgs-exchange DIS to order $\as^{\,l} \,\ep^{\,3-l}$ with $l\leq 3$, in the
above soft$\,$+$\,$virtual limit, facilitates the determination of all $1/\ep$ 
pole terms of the three-loop $\gamma^{\,\ast\!} qq$ and $Hgg$ form factors 
(the latter, of course, in the heavy-top limit). 
The corresponding results, including the additional $\ep^{\,0\,}\nf$ 
contributions in the $\gamma^{\,\ast\!}qq$ case, have been verified recently in
direct calculations of these form factors \cite{FF3loop}, thus providing a 
check of Eqs.~(\ref{eq:cphig3D5}) -- (\ref{eq:cphig3D0}) and the $\delta \x1$ 
in Eq.~(\ref{eq:cphig3}) and the corresponding results in 
Refs.~\cite{MVV3,MVV4,MVV6} including, in particular, the full results for the 
three-loop quark and gluon cusp anomalous dimensions.
Those results, in turn, involve a considerable part of the three-loop
forward-scattering diagrams and integrals entering the complete (all-$x$) 
calculations of the third-order splitting functions and coefficient functions.

Returning to the large-$x$ behaviour of the gluon coefficient functions
$c^{\,(n)}_{\rm \phi\, , g}(x)$ we note that, as Eqs.~(\ref{eq:cphig2D3}) --
(\ref{eq:cphig2D1}) and (\ref{eq:cphig3D5}) -- (\ref{eq:cphig3D3}) above, 
the coefficients of the highest three subleading $\ln^{\,k} \x1$ terms,
$k = 3,\,2,\,1$ for $c^{\,(2)}_{\rm \phi\, , g}(x)$ and 
$k = 5,\,4,\,3$ for $c^{\,(2)}_{\rm \phi\, , g}(x)$, do not include 
contributions with the colour factor $C_F$, i.e., gluon emission from quarks. 
Hence they are guaranteed to originate from `non-singlet like' diagrams with a 
(modulo self-energy insertions) unbroken gluon line connecting the incoming 
gluon to the scalar $\phi$. Consequently non-singlet considerations and 
structures hold for these contributions, cf.~Refs.~\cite{DMS05,MV2,MV3}, and 
these subleading logarithms can be predicted to all orders $n$ along the lines 
of Ref.~\cite{MV5}. 

We now turn to the quark coefficient functions $c^{\,(n)}_{\rm \phi\, , q}(x)$ 
where the integrable logarithms $L_1^{\,k} \equiv \ln^{\,k}\! \x1$ with 
$k\,=\, 1,\:\ldots,\: 2n\!-\!1$ form the leading large-$x$ terms.
Again the first-order coefficient is obvious from Eq.~(\ref{eq:cphiq1}), and
the second-order contributions are given by
\bea
\label{eq:cphiq2L3}
 c_{\rm \phi\, , q}^{\,(2)} \Big|_{\,L_1^3} &\! = \! &
           {10 \over 3}\: \* \cf\, \* \ca\,
         + {2 \over 3}\: \* \cfs\,
 \; ,
 \\[1mm]
\label{eq:cphiq2L2}
 c_{\rm \phi\, , q}^{\,(2)} \Big|_{\,L_1^2} &\! = \! &
           {7 \over 3}\: \* \cf\, \* \ca\,
         - 14\, \* \cfs\,
         + {2 \over 3}\: \* \cf\, \* \nf
 \; ,
 \\[1mm]
\label{eq:cphiq2L1}
 c_{\rm \phi\, , q}^{\,(2)} \Big|_{\,L_1} &\! = \! &
         \cf\, \* \ca\, \* \Big( \,
           {349 \over 9}\:
         - 20\, \* \z2 
         \Big)\,
         + 5\, \* \cfs\,
         - {58 \over 9}\: \* \cf\, \* \nf
 \; .
\eea
The corresponding three-loop results read
\bea
\label{eq:cphiq3L5}
 c_{\rm \phi\, , q}^{\,(3)} \Big|_{\,L_1^5} &\! = \! &
           {10 \over 3}\, \* \cf\, \* \cas\,
         + {2 \over 3}\, \* \cft
 \; ,
 \\[1mm]
 \label{eq:cphiq3L4}
 c_{\rm \phi\, , q}^{\,(3)} \Big|_{\,L_1^4} &\! = \! & \mbox{}
         - {61 \over 54}\, \* \cf\, \* \cas\,
         - {442 \over 27}\, \* \cfs\, \* \ca\,
         - {157 \over 18}\, \* \cft\,
         +  {47 \over 27}\, \* \cf\, \* \ca\, \* \nf\,
         + {13 \over 27}\, \* \cfs\, \* \nf
 \; ,
 \\[1mm]
 \label{eq:cphiq3L3}
 c_{\rm \phi\, , q}^{\,(3)} \Big|_{\,L_1^3} &\! = \! &
      \cf\, \* \cas\, \* \Big( \,
           {5894 \over 81}\,
         - {368 \over 9}\, \* \z2 \Big)\,
       + \cfs\, \* \ca\, \* \Big( \,
           {5161 \over 81}\,
         - {296 \over 9}\, \* \z2 \Big)\,
       + \cft\, \* \Big( \,
           {74 \over 3}\,
         + {88 \over 9}\, \* \z2 \Big)\,
   \nn \\ & & \mbox{}
         - {1148 \over 81}\, \* \cf\, \* \ca\, \* \nf\,
         - {1018 \over 81}\, \* \cfs\, \* \nf\,
         + {16 \over 27}\, \* \cf\, \* \nfs
 \; ,
 \\[2mm]
 \label{eq:cphiq3L2}
 c_{\rm \phi\, , q}^{\,(3)} \Big|_{\,L_1^2} &\! = \! & \mbox{}
       - \cf\, \* \cas\, \* \Big( \,
           {8213 \over 81}\,
         - {296 \over 9}\, \* \z2\, 
         - 84\, \* \z3\, \Big)\,
       - \cfs\, \* \ca\, \* \Big( \,
           {40486 \over 81}\,
         - {608 \over 3}\, \* \z2\,
         - {316 \over 3}\, \* \z3\, \Big)
   \nn \\ & & \mbox{}
      + \cft\, \* \Big( \,
           {295 \over 3}\,
         + {104 \over 3}\, \* \z2\,   
         - {184 \over 3}\, \* \z3 \Big)\,
      + \cf\, \* \ca\, \* \nf\, \* \Big( \,
           {3236 \over 81}\,
         - {248 \over 9}\, \* \z2 \Big)\,
   \nn \\[1mm] & & \mbox{}
  +    \cfs\, \* \nf\, \* \Big( \,
           {6616 \over 81}\,
         + {4 \over 3}\, \* \z2 \Big)\,
      - {104 \over 27}\, \* \nfs\, \* \cf\,
 \; ,
 \\[2mm]
 \label{eq:cphiq3L1}
 c_{\rm \phi\, , q}^{\,(3)} \Big|_{\,L_1} &\! = \! & \mbox{}
         \cf\, \* \cas\, \* \Big( \,
           {227065 \over 243}\,
         - {17384 \over 27}\, \* \z2\,
         - {590 \over 3}\, \* \z3\,
         + {368 \over 5}\, \* \zss
   \Big)\,
       + \cfs\, \* \ca\, \* \Big( \,
           {524873 \over 972}\,
   \nn \\ & & \mbox{}
         - {638 \over 27}\, \* \z2\,
         - {2954 \over 9}\, \* \z3\,
         + {2674 \over 15}\, \* \zss
    \Big)\,
       - \cft\, \* \Big( \,
           {2927 \over 12}\,
         + 170\, \* \z2\,
         - {328 \over 3}\, \* \z3\,
         + {1804 \over 15}\, \* \zss
    \Big)\,
   \nn \\[1mm] & & \mbox{}
       - \cf\, \* \ca\, \* \nf\, \* \Big( \,
           {75052 \over 243}\,
         - {3260 \over 27}\, \* \z2 \,
         - {52 \over 3}\, \* \z3 \Big)\,
       - \cfs\, \* \nf\, \* \Big( \,
           {70747 \over 486}\,
         - {812 \over 27}\, \* \z2\, 
         - {272 \over 9}\, \* \z3 
           \Big)\,
   \nn \\[1mm] & & \mbox{}
       + \cf \, \* \nfs\, \* \Big( \,
           {544 \over 27}\,
         - {40 \over 9}\, \* \z2 
           \Big)\,
 \; .
\eea
The two-loop coefficients (\ref{eq:cphiq2L3}) -- (\ref{eq:cphiq2L1}) contribute 
to the large-$x$ behaviour of the NNLO physical kernel $K_{\,\rm \phi 2}^
{\,(2)}(x)$, recall Eqs.~(\ref{eq:F2Htrf}) and (\ref{eq:Kab}) above, which will
be discussed in the next section. The third-order results (\ref{eq:cphiq3L5}) 
-- (\ref{eq:cphiq3L3}) will enter our large-$x$ predictions for the four-loop 
splitting function $P_{\rm gq}^{\,(3)}(x)$ in Section 5.

At small $x$ both the coefficient functions $c^{\,(n)}_{\rm \phi\, , p}(x)$,
$ \rm p = q,\,g$, show a double-logarithmic enhancement, i.e., 
terms $x^{\,-1} \ln^{\,k} x$ contribute with 
$k \,=\, 0,\: \ldots,\: 2n\!-\!1$. The second-order coefficients~are
\bea
\label{eq:cphiq2L03}
 xc_{\rm \phi\, , q}^{\,(2)} \Big|_{\,L_0^3} &\! = \! & \mbox{}
          - {8 \over 3}\: \* \cf\, \* \ca\,
 \; ,
 \\[1mm]
 \label{eq:cphiq2L02}
 xc_{\rm \phi\, , q}^{\,(2)} \Big|_{\,L_0^2} &\! = \! & \mbox{}
          - {106 \over 3}\: \* \cf\ \* \ca\,
          + {16 \over 3}\: \* \cf\, \* \nf
 \; ,
 \\[1mm]
 \label{eq:cphiq2L01}
 xc_{\rm \phi\, , q}^{\,(2)} \Big|_{\,L_0} &\! = \! &
  - \cf\, \* \ca\, \* \Big( \,
            {454 \over 3}\,
          - 32\, \* \z2 \Big)\,
   - \cfs\, \* ( 
            10\,
          - 16\, \* \z2 )\,
   + {64 \over 3}\: \* \cf\, \* \nf
 \; ,
 \\[1mm]
 \label{eq:cphiq2L00}
 xc_{\rm \phi\, , q}^{\,(2)} \Big|_{\,L_0^0} &\! = \! & \mbox{}
     - \cf\, \* \ca\, \* \Big( \,
            {13457 \over 54}\,
          - {212 \over 3}\: \* \z2\,
          - 72\, \* \z3 \Big)\, 
     - \cfs\, \* \Big( \,
            {59 \over 2}\,
          - 36\, \* \z2\,
          + 32\, \* \z3 \Big)\,
   \nn \\[1mm] & & \mbox{}
     + \cf\, \* \nf\, \* \Big( \,
            {737 \over 27}\,
          - {16 \over 3}\: \* \z2 \Big)\,
\eea
and
\bea
 \label{eq:cphig2L03}
 xc_{\rm \phi\, , g}^{\,(2)} \Big|_{\,L_0^3} &\! = \! & \mbox{}
          - {8 \over 3}\: \* \cas\,
 \; ,
 \\[1mm]
 \label{eq:cphig2L02}
 xc_{\rm \phi\, , g}^{\,(2)} \Big|_{\,L_0^2} &\! = \! & \mbox{}
          - {110 \over 3}\: \* \cas\,
          + 4\, \* \ca\, \* \nf\,
          + {8 \over 3}\: \* \cf\, \* \nf
 \; ,
 \\[1mm]
 \label{eq:cphig2L01}
 xc_{\rm \phi\, , g}^{\,(2)} \Big|_{\,L_0} &\! = \! & \mbox{}
    - \cas\, \* \Big( \,
            {1556 \over 9}\,
          - 48\, \* \z2 \Big)\,
    + {88 \over 3}\: \* \ca\, \* \nf\,
    + {64 \over 9}\: \* \cf\, \* \nf\,
    - {8 \over 9}\: \* \nfs
 \; ,
 \\[2mm]
 \label{eq:cphig2L00}
 xc_{\rm \phi\, , g}^{\,(2)} \Big|_{\,L_0^0} &\! = \! & \mbox{}
   - \cas\, \* \Big( \,
            {2743 \over 9}\,
          - {352 \over 3}\: \* \z2\,
          - 40\, \* \z3 \Big)\,
   + \ca\, \* \nf\, \* \Big( \,
            {1312 \over 27}\,
          - {32 \over 3}\: \* \z2 \Big)\,
   \nn \\ & & \mbox{}
   + \cf\, \* \nf\, \* \Big( \,
            {361 \over 27}\,
          - {16 \over 3}\: \* \z2 \Big)\,
   - {40 \over 27}\: \* \nfs
 \; .
\eea
Their three-loop counterpart are found to be 
\bea
\label{eq:cphiq3L05}
 xc_{\rm \phi\, , q}^{\,(3)} \Big|_{\,L_0^5} &\! = \! &
          - {8 \over 15}\: \* \cf\, \* \cas\,
 \; ,
 \\[1mm]
 \label{eq:cphiq3L04}
 xc_{\rm \phi\, , q}^{\,(3)} \Big|_{\,L_0^4} &\! = \! & \mbox{}
          - 24\, \* \cf\, \* \cas\,
          + {28 \over 9}\: \* \cf\, \* \ca\, \* \nf\,
          + {8 \over 9}\: \* \cfs\, \* \nf
 \; ,
 \\[1mm]
\label{eq:cphiq3L03}
 xc_{\rm \phi\, , q}^{\,(3)} \Big|_{\,L_0^3} &\! = \! &
      - \cf\, \* \cas\, \* \Big( \,
            {1010 \over 3}\,
          - 32\, \* \z2 \Big)\,
     - \cfs\, \* \ca\, \* \Big( \,
            {20 \over 3}\,
          - {32 \over 3}\: \* \z2  \Big)\,
          + {2348 \over 27}\: \* \nf\, \* \cf\, \* \ca\,
   \nn \\ & & \mbox{}
          + {40 \over 27}\: \* \cfs\, \* \nf\,
          - {16 \over 3}\: \* \cf\, \* \nfs
 \; ,
 \\[2mm]
 \label{eq:cphiq3L02}
 xc_{\rm \phi\, , q}^{\,(3)} \Big|_{\,L_0^2} &\! = \! & \mbox{}
       - \cf\, \* \cas\, \* \Big( \,
            {16793 \over 9}\,
          - {1288 \over 3}\: \* \z2\,  
          - {208 \over 3}\: \* \z3
          \Big)\,
     - \cfs\, \* \ca\, \* \Big( \,
            {833 \over 3}\,
          - {992 \over 3}\: \* \z2\, 
          + 16\, \* \z3
          \Big)\,
   \nn \\ & & \mbox{}
     + \cft\, \* (
            52\,
          - 48\, \* \z2 \,
          + 32\, \* \z3 )\,
     + \cf\, \* \ca\, \* \nf\, \* (
            546\,
          - 48\, \* \z2 )\,
   \nn \\[1mm] & & \mbox{}
     + \cfs\, \* \nf\, \* \Big( \,
            {1654 \over 27}\,
          - 64\, \* \z2  \Big)\,
     - {964 \over 27}\: \* \nfs\, \* \cf\,
 \; ,
 \\[3mm]
 \label{eq:cphiq3L01}
 xc_{\rm \phi\, , q}^{\,(3)} \Big|_{\,L_0} &\! = \! &
    - \cf\, \* \cas\, \* \Big( \,
            {353878 \over 81}\,
          - {49580 \over 27}\: \* \z2\,
          - {15104 \over 9}\: \* \z3\,
          + {5584 \over 15}\: \* \zss  \Big)\,
   \nn \\[1mm] & & \mbox{}
     - \cfs\, \* \ca\, \* \Big( \,
            {14701 \over 9}\,
          - {15052 \over 9}\: \* \z2\,
          + {1448 \over 3}\: \* \z3\,
          + {208 \over 5}\: \* \zss  \Big)\,
     + \cft\, \* ( \,
            450\,
          - 224\, \* \z2\,
   \nn \\[1mm] & & \mbox{}
          + 144\, \* \z3\,
          - 128\, \* \zss  )\,
     + \: \cf\, \* \ca\, \* \nf\, \* \Big( \,
            {122170 \over 81}\,
          - {10600 \over 27}\: \* \z2\,
          - {400 \over 9}\: \* \z3
          \Big)\,
   \nn \\[1mm] & & \mbox{}
     + \cfs\, \* \nf\, \* \Big( \,
            {102137 \over 243}\,
          - {7024 \over 27}\: \* \z2\,  
          - {1376 \over 9}\: \* \z3
          \Big)\,
     - \cf\, \* \nfs\, \* \Big( \,
            {7672 \over 81}\,
          - {128 \over 9}\: \* \z2  \Big)\,
 \; ,
 \\[3mm]
 \label{eq:cphiq3L00}
 xc_{\rm \phi\, , q}^{\,(3)} \Big|_{\,L_0^0} &\! = \! & \mbox{}
    \cf \* \cas\, \* \Big( \,
            {19967 \over 36}\,
          + {149944 \over 81}\: \* \z2
          + {111202 \over 27}\: \* \z3
          - {55748 \over 45}\: \* \zss  
          - {608 \over 3}\: \* \z2 \* \z3
          - {4184 \over 3}\: \* \z5\,
          \Big)\,
%
   \qquad
%
   \nn \\ & & \mbox{}
     - \cfs\, \* \ca\, \* \Big( \,
            {327439 \over 108}\,
          - {82576 \over 27}\: \* \z2
          + {3088 \over 3}\: \* \z3
          + {1528 \over 5}\: \* \zss  
          + {592 \over 3}\: \* \z2 \* \z3
          + {448 \over 3}\: \* \z5
          \Big)\,
   \nn \\[1mm] & & \mbox{}
     + \cft\, \* \Big( \,
            {6601 \over 6}\,
          - 544\, \* \z2\,
          - 76\, \* \z3\,
          - {1384 \over 5}\: \* \zss 
          + 64\, \* \z2 \* \z3\,
          + 560\, \* \z5
          \Big)\,
   \nn \\[1mm] & & \mbox{}
     + \cf\, \* \ca\, \* \nf\, \* \Big( \,
            {398771 \over 243}\,
          - {17104 \over 27}\: \* \z2\,
          - {13204 \over 27}\: \* \z3\,
          + {896 \over 15}\: \* \zss  \Big)\,
     + \cfs\, \* \nf\, \* \Big( \,
            {421208 \over 729}\,
   \nn \\[1mm] & & \mbox{}
          - {33896 \over 81}\, \* \z2\,
          - {3824 \over 27}\: \* \z3\,
          + {656 \over 15}\: \* \zss  \Big)\,
%
%
     - \cf\, \* \nfs\, \* \Big( \,
            {25291 \over 243}\,
          - {848 \over 27}\: \* \z2 
          - {16 \over 3}\: \* \z3\,
          \Big)
\eea
and
\bea
 \label{eq:cphig3L05}
 xc_{\rm \phi\, , g}^{\,(3)} \Big|_{\,L_0^5} &\! = \! & \mbox{}
          - {8 \over 15}\: \* \cat\,
 \; ,
 \\[1mm]
 \label{eq:cphig3L04}
 xc_{\rm \phi\, , g}^{\,(3)} \Big|_{\,L_0^4} &\! = \! & \mbox{}
          - {220 \over 9}\: \* \cat\,
          + {8 \over 3}\: \* \cas\, \* \nf\,
          + {16 \over 9}\: \* \cf\, \* \ca\, \* \nf
 \; ,
 \\[2mm]
 \label{eq:cphig3L03}
 xc_{\rm \phi\, , g}^{\,(3)} \Big|_{\,L_0^3} &\! = \! & \mbox{}
      - \cat\, \* \Big( \,
            {9698 \over 27}\,
          - {128 \over 3}\: \* \z2 \Big)\,
          + {2032 \over 27}\: \* \cas\, \* \nf\,
          + {296 \over 9}\: \* \cf\, \* \ca\, \* \nf\,
          - {88 \over 27}\: \* \ca\, \* \nfs
    \nn \\ & & \mbox{}
          - {112 \over 27}\: \* \nfs\, \* \cf\,
 \; ,
 \\[3mm]
 \label{eq:cphig3L02}
 xc_{\rm \phi\, , g}^{\,(3)} \Big|_{\,L_0^2} &\! = \! & \mbox{}
     - \cat\, \* \Big( \,
            {61454 \over 27}\,
          - {2200 \over 3}\: \* \z2\, 
          - {256 \over 3}\: \* \z3 \Big)\,
     + \cas\, \* \nf\, \* \Big( \,
            {16882 \over 27}\,
          - 80\, \* \z2 \Big)
   \nn \\ & & \mbox{}
     + \cf\, \* \ca\, \* \nf\, \* \Big( \,
            {5872 \over 27}\,
          - {128 \over 3}\: \* \z2 \Big)\,
     + \cfs\, \* \nf\, \* \Big( \,
            {44 \over 3}\,
          - {32 \over 3}\: \* \z2 \Big)\,
     - {1184 \over 27}\: \* \ca\, \* \nfs\,
   \nn \\[1mm] & & \mbox{}
     - {760 \over 27}\, \* \cf\, \* \nfs
     + {8 \over 9}\, \* \nft\,
 \; ,
 \\[3mm]
 \label{eq:cphig3L01}
 xc_{\rm \phi\, , g}^{\,(3)} \Big|_{\,L_0} &\! = \! & \mbox{}
      - \cat\, \* \Big(
            {507719 \over 81}\,
          - {95912 \over 27}\: \* \z2\,
          - {4136 \over 3}\: \* \z3\,
          + {8128 \over 15}\: \* \zss \Big)\,
   \nn \\[1mm] & & \mbox{}
     + \cas\, \* \nf\, \* \Big( \,
            {59272 \over 27}\,
          - {19648 \over 27}\: \* \z2\, 
          + {16 \over 3}\: \* \z3
          \Big)\,
     + \cfs\, \* \nf\, \* \Big( \,
            {568 \over 9}\,
          - {544 \over 9}\: \* \z2\, 
          + {128 \over 3}\: \* \z3
          \Big)\,
   \nn \\[1mm] & & \mbox{}
     + \cf \* \ca\, \* \nf\, \* \Big( \,
            {181109 \over 243}\,
          - {5296 \over 27}\: \* \z2\, 
          - {2560 \over 9}\: \* \z3
          \Big)\,
     - \ca\, \* \nfs\, \* \Big( \,
            {13988 \over 81}\,
          - {256 \over 9}\: \* \z2 \Big)\,
   \nn \\[1mm] & & \mbox{}
     - \cf\, \* \nfs\, \* \Big( \,
            {7370 \over 81}\,
          - {160 \over 9}\: \* \z2 \Big)\,
          + {80 \over 27}\: \* \nft\,
 \; ,
 \\[3mm]
 \label{eq:cphig3L00}
 xc_{\rm \phi\, , g}^{\,(3)} \Big|_{\,L_0^0} &\! = \! & \mbox{}
    -  \cat\, \* \Big( \,
            {57560 \over 27}\,
          - {407248 \over 81}\: \* \z2\,
          - {87922 \over 27}\: \* \z3\,
          + {9724 \over 5}\: \* \zss\,
          + 336\, \* \z2 \* \z3\,
          + 984\, \* \z5
          \Big)\,
   \nn \\ & & \mbox{}
     + \cas\, \* \nf\, \* \Big( \,
            {4398355 \over 1458}\,
          - {123284 \over 81}\: \* \z2\,
          - {7568 \over 27}\: \* \z3\,
          + {5864 \over 45}\: \* \zss \Big)\,
   \nn \\[1mm] & & \mbox{}
     + \cf\, \* \ca\, \* \nf\, \* \Big( \,
            {881945 \over 1458}\,
          - {8228 \over 27}\: \* \z2\,
          - {6752 \over 9}\: \* \z3\,
          + {704 \over 9}\: \* \zss \Big)\,
   \nn \\[1mm] & & \mbox{}
     + \cfs\, \* \nf\, \* \Big( \,
            {3053 \over 54}\,
          - {3112 \over 27}\: \* \z2\,
          + {880 \over 9}\: \* \z3\,
          + {832 \over 45}\: \* \zss \Big)\,
     - \ca\, \* \nfs\, \* \Big( \,
            {344365 \over 1458}\,
          - {2504 \over 27}\: \* \z2\, 
   \nn \\[1mm] & & \mbox{}
          + {296 \over 27}\: \* \z3
          \Big)\,
     - \cf\, \* \nfs\, \* \Big( \,
            {105398 \over 729}\,
          - {1264 \over 27}\: \* \z2\, 
          - {1552 \over 27}\: \* \z3
          \Big)\,
     + \nft\, \* \Big( \,
            {200 \over 81}\,
          - {16 \over 9}\: \* \z2 \Big)\,
\; . \;\qquad
\eea
 
This behaviour is different from that of the (spacelike) splitting functions 
and gauge-boson exchange coefficient functions which receive an only 
single-logarithmic (`BFKL') enhancement of the leading $1/x$ contributions to 
all orders in $\as$ \cite{BFKL,Catani:1990eg,Catani:1994sq}. In fact, this
seemingly surprising feature of the $\phi\, G^{\,\mu\nu} G_{\mu\nu}$ probe of
the hadronic system was already briefly stated in Ref.~\cite{Catani:1990eg},
and the leading double-logarithmic terms were explicitly calculated in 
Ref.~\cite{Hautmann:2002tu} for the related case of Higgs production in 
proton-proton collisions in the heavy-top limit.
 
However, the functions $c^{\,(n)}_{\rm \phi\, , p}(x)$ are probed at very 
small values of $x$ only at very high centre-of-mass (CM) energies $\sqrt{s}$.
Given the finite physical top mass, the large top-mass approximations for Higgs
exchange$/$production breaks down in this region. Indeed, as explicitly shown 
recently for Higgs production in Ref.~\cite{Marzani:2008az}, the full 
finite-$m_{\,\rm top}$ coefficient functions receive only single logarithmic 
contributions at small $x$. Consequently the effective-theory coefficient 
functions $c^{\,(n)}_{\rm \phi\, , p}(x)$ cease to provide useful 
approximations for Higgs-exchange DIS in this limit. Where exactly in $x$ this 
occurs cannot be established without the very non-trivial calculation of at 
least all NNLO $1/x$ terms in the full theory. This point has not been reached 
yet even in the phenomenologically important case of Higgs production in 
proton$\:\!$--$\:\!$(anti-)$\:\!$proton collisions, for the present status 
see~Refs.~\cite{Harlander:2009mq}.

We are now ready to discuss the numerical size of the coefficient functions
(\ref{eq:cphiq1}) -- (\ref{eq:cphig2}), (\ref{eq:cphiq3}) and (\ref{eq:cphig3}).
In Figs.~1 and 2 their cumulative effect is shown for $\nf = 4$ quark flavours
and an order-independent value $\as (\mu_{\,\rm r}^{\,2}\!=\! 
\mu_{\,\rm f}^{\,2}\!=\! Q_0^{\,2})\:=\: 0.2$ for the strong coupling constant 
corresponding to a scale of $\,Q_0^{\,2} \approx 20 \ldots 50\:\GeV2$. The same
reference point was used in Refs.~\cite{MVV5,MVV6}, hence the present results
can be compared directly to the photon-exchange case discussed in those 
articles. Note that Fig.~2 does not provide the full information about the
effect of the gluon coefficient function $C_{\rm \phi, g}(x)$ due to presence
of +-distributions, recall Eq.~(\ref{plus}), and $\delta \x1$ contributions.

As shown in the right parts of both figures, the perturbative expansions of
$C_{\rm \phi, p}(x)$ are rather stable in the mid-$x$ region, but,
unsurprisingly, large beyond-NLO corrections are found in the soft-gluon 
large-$x$ and the high-energy small-$x$ regions. While in the former the NNLO 
and N$^3$LO are similar, the perturbative expansion appears to break down below
$x \approx 10^{\,-3}$ and $x \approx 10^{\,-4}$ for $C_{\rm \phi, q}(x)$ and 
$C_{\rm \phi, g}(x)$, respectively. 
This breakdown thus occurs in a region where this coefficient functions are 
definitely not expected to represent Higgs exchange anymore, as a CM energy of 
about $m_{\,\rm top}$ is required to access $x = 10^{\,-3}$ at the chosen scale
$\Qs\approx 30$ GeV$^2$.

Obviously the (perturbative) coefficient functions enter physical quantities 
such as the structure function $F_\phi$ only through convolutions such as
\beq
\label{eq:Mconv}
  [\, C_{\rm \phi, q} \otimes q \,](x) \; = \; \int_x^1 \! \frac{dy}{y} 
  \: C_{\rm \phi, q}(x) \: q \!\left( \frac{x}{y} \right)
\eeq
with the (partly non-perturbative) quark and gluon distributions which, due
to the general shape of $q_{\rm s}^{}(x)$ and $g(x)$, shift the onset of the 
small-$x$ instability to considerably lower values of $x$ than in Figs.1 and 2.
The resulting contributions of Eqs.~(\ref{eq:cphiq1}) -- (\ref{eq:cphig2}), 
(\ref{eq:cphiq3}) and (\ref{eq:cphig3}) to $F_\phi$ are illustrated in Fig.~3 
for the schematic, but sufficiently realistic order-independent input 
\bea
\label{eq:p-sg}
  xq_{\rm s}(x,Q_0^{\,2}) &\! = \! &
  0.6\: x^{\, -0.3} (1-x)^{3.5}\, (1 + 5.0\: x^{\, 0.8\,}) \:\: ,
\nn \\[0.5mm]
  xg (x,Q_0^{\,2})\:\: &\! = \! &
  1.6\: x^{\, -0.3} (1-x)^{4.5}\, (1 - 0.6\: x^{\, 0.3\,})
\eea
already employed in Refs.~\cite{MVV4,MVV5,MVV6}. In reality both $\as$ and
the parton distributions do depend on the perturbative order in a manner that
reduces the relative higher-order corrections in particular at large $x$.
Keeping this in mind, the medium-$\Qs$ perturbative stability in Fig.~3 is 
satisfactory over a wide range in $x$ despite, as expected for a 
gluon-dominated quantity, the presence of considerably larger $\as$ corrections
than found for the standard-DIS structure function $\Ftwo\,$.

\begin{figure}[p]
\vspace*{-1mm}
\centerline{\epsfig{file=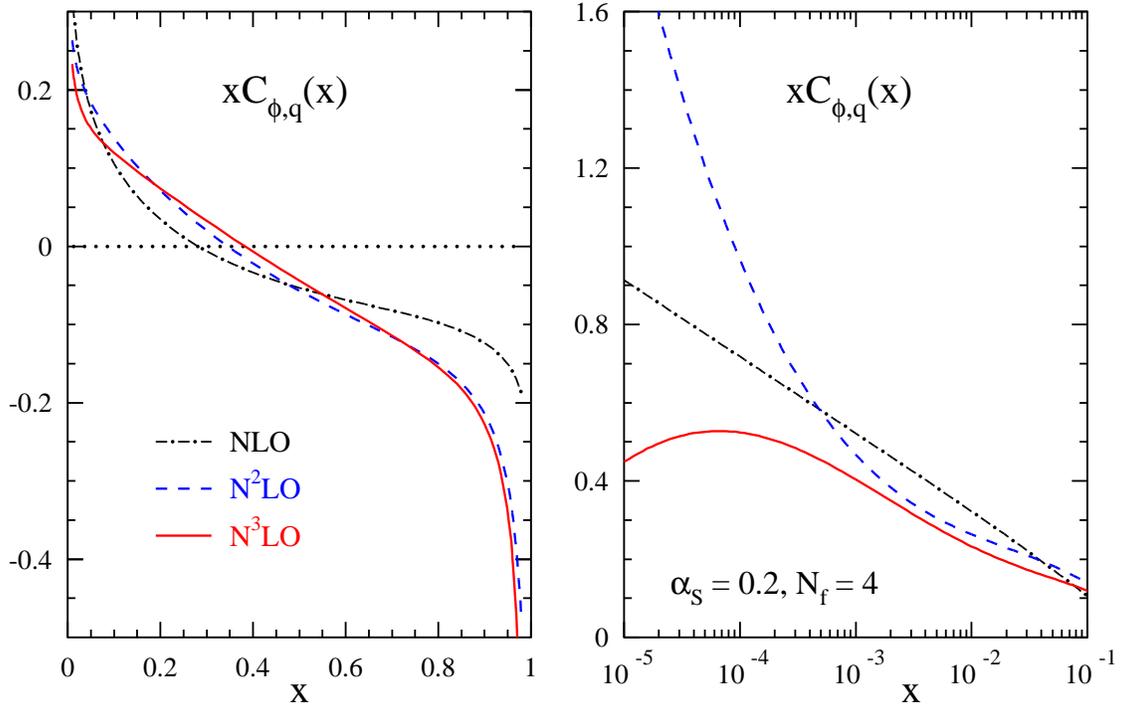,width=15.0cm,angle=0}}
\vspace{-2mm}
\caption{\label{pic:fig1}
 The perturbative expansion to order $\as^{\,3}$ (N$^3$LO) of the quark 
 coefficient function $C_{\rm \phi, q}(x)$ for DIS via the exchange of a scalar
 $\phi$ with a $\phi\, G^{\,\mu\nu} G_{\mu\nu}$ coupling to gluons. The results
 are shown for a standard medium-scale reference point and have been multiplied
 by $x$ for display purposes.
 }
\end{figure}
\begin{figure}[p]
\centerline{\epsfig{file=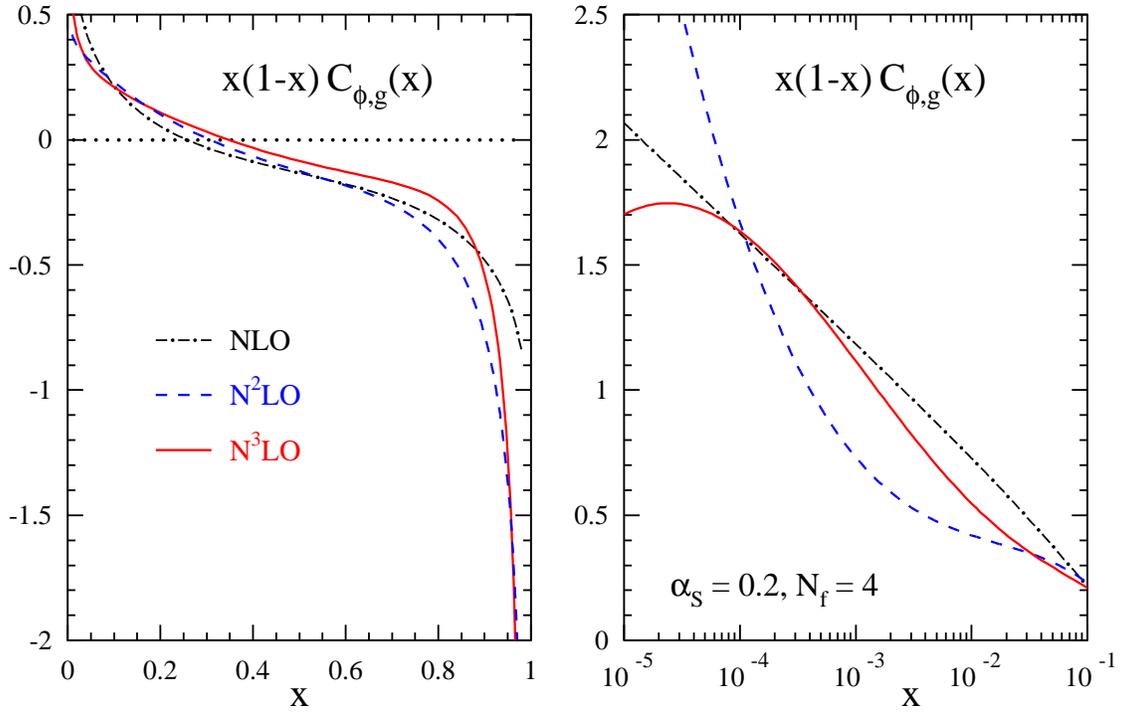,width=15.0cm,angle=0}}
\vspace{-2mm}
\caption{\label{pic:fig2}
 As Fig.~\ref{pic:fig1}, but for the gluon coefficient function 
 $C_{\rm \phi, g}(x)$ at $x < \!1$, multiplied by an additional factor $\x1$
 compensating the large $x$-dependence due to the +-distribution contributions. 
 }
\vspace{-1mm}
\end{figure}

\begin{figure}[thb]
\vspace*{-2mm}
\centerline{\epsfig{file=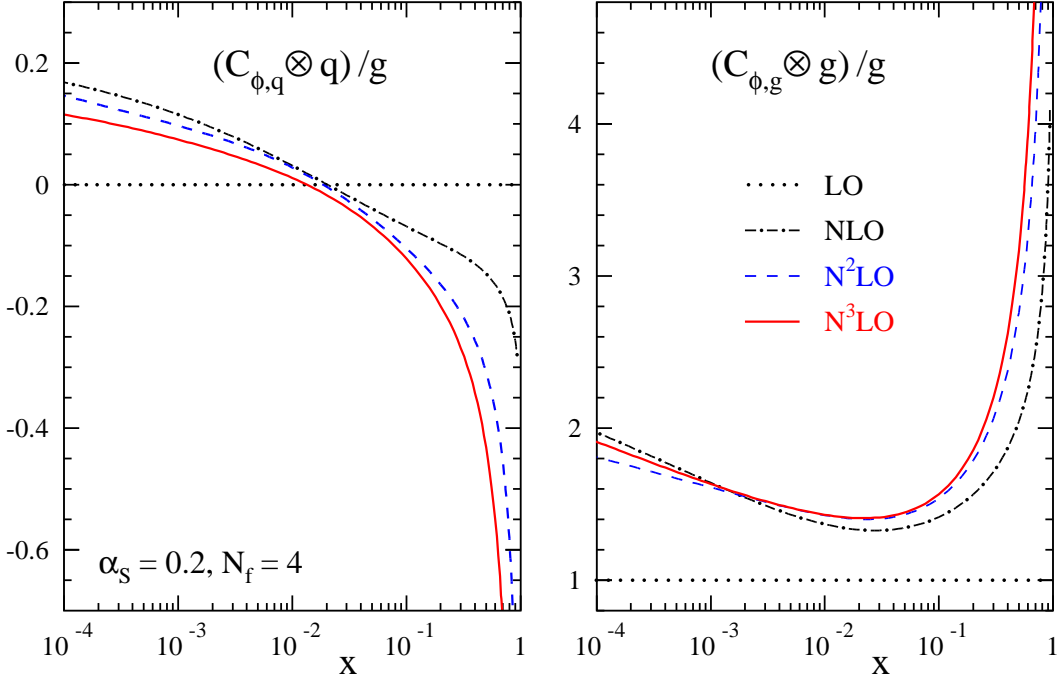,width=14.5cm,angle=0}}
\vspace{-3mm}
\caption{\label{pic:fig3}
 The perturbative expansion of the quark (left) and gluon (right) contributions
 to the structure function $F_\phi$, normalized to the lowest-order result
 $F_\phi(x,\Qs) = x g(x,\Qs)$, for the order-independent input (\ref{eq:p-sg}) 
 at our reference point corresponding to a scale $\,\Qs \approx 20 \ldots 50\: 
 \GeV2$.
\vspace{1mm}
 }
\end{figure}

Finally we need to address the Mellin moments of the new coefficient functions
$c^{\,(n)}_{\rm \phi\, , p}(x)$ for $n\,=\, 1,\:2,\:3$ ($\rm p = q,\,g$).  
The second moments for the quark case are given by
\bea
\label{eq:cq1N2}
 c_{\rm \phi\, , q}^{\,(1)} (N\!=\!2) &\! = \! &
        - {4 \over 3}\: \* \cf
 \:\;\cong\:\;  - 1.77778
 \:\: , \\[2mm] 
\label{eq:cq2N2}
 c_{\rm \phi\, , q}^{\,(2)} (N\!=\!2) &\! = \! & \mbox{}
       -  \cf\,  \*  \ca\,   \*  \Big( \,
            {6224 \over 81}
          - 32\, \*  \z3
          \Big)
       +  \cfs\,  \*  \Big( \,
            {1237 \over 81}
          - 32\, \*  \z3
          \Big)
       + {44 \over 3}\:  \* \cf\,  \*  \nf\,
 \nn \\[2mm]
   &\!\cong\! & - 194.729 + 19.5556\: \nf
 \:\: , \\[4mm]
\label{eq:cq3N2}
 c_{\rm \phi\, , q}^{\,(3)} (N\!=\!2) &\! = \! & \mbox{}
       -  \cf\,  \*  \cas\,  \*  \Big( \,
            {10744957 \over 4374}\,
          - {121232 \over 81}\: \*  \z3
          + {64 \over 15}\: \*  \zss
          + {640 \over 3}\: \*  \z5
          \Big)
   \nn \\ & & \mbox{}
       +  \cfs\, \*  \ca\,   \*  \Big( \,
            {4294603 \over 4374}\,
          - {6488 \over 9}\: \*  \z3
          + {64 \over 5}\: \*  \zss
          - {1280 \over 3}\: \*  \z5
          \Big)
   \nn \\ & & \mbox{}
       -  \cft\,  \*  \Big( \,
            {137462 \over 729}\,
          + {41792 \over 81}\: \*  \z3
          + {128 \over 15}\: \*  \zss
          - {2560 \over 3}\: \*  \z5
          \Big)
   \nn \\ & & \mbox{}
       -  \cfs\, \*  \nf\,   \*  \Big( \,
            {62201 \over 2187}\,
          - {11264 \over 81}\: \*  \z3
          + {128 \over 15}\: \*  \zss
          \Big)
       -  \cf\,  \*  \nfs\,  \*  \Big( \,
            {33854 \over 729}\,
          - {64 \over 9}\: \*  \z3
          \Big)
   \nn \\ & & \mbox{}
       +  \cf\,  \*  \ca\,  \*  \nf\,   \*  \Big( \,
            {1616063 \over 2187}\,
          - {22768 \over 81}\, \*  \z3
          + {128 \over 15}\: \*  \zss
          \Big)
 \nn \\[2mm]
 &\!\cong\! & - 12116.73 + 1902.16\: \nf - 50.5213\: \nfs
\eea
where the second (approximate) equalities are obtained by inserting $C_A=3$, 
$C_F=4/3$ and the numerical values of the $\zeta$-function.
The results for the gluon coefficient function read
\bea
\label{eq:cg1N2}
 c_{\rm \phi\, , g}^{\,(1)} (N\!=\!2) &\! = \! &
          {203 \over 18}\: \*  \ca
        - {10 \over 9}\: \* \nf
 \:\;\cong\:\; 33.8333 - 1.11111\: \nf
 \:\: , \\[2mm]
\label{eq:cg2N2}
 c_{\rm \phi\, , g}^{\,(2)} (N\!=\!2) &\! = \! &
         \cas\, \*  \Big( \,
            {23473 \over 108}
          - 66\, \*  \z3
          \Big)
       -  \ca  \*  \nf   \*  \Big( \,
            {8215 \over 162}
          + 12\, \*  \z3
          \Big)
    \nn \\[1mm] & & \mbox{}
       -  \cf  \*  \nf   \*  \Big( \,
            {1726 \over 81}
          - 24\, \*  \z3
          \Big)
       + {100 \over 81}\: \*  \nfs
 \nn \\[2mm]
 &\!\cong\! &   1242.06 - 185.349\: \nf  + 1.23457\: \nfs
\:\: , \\[4mm] 
\label{eq:cg3N2}
 c_{\rm \phi\, , g}^{\,(3)} (N\!=\!2)  &\! = \! &
          \cat\,  \*  \Big( \,
            {27878711 \over 5832}\,
          - {66970 \over 27}\: \*  \z3
          + {880 \over 3}\: \*  \z5
          \Big)
   \nn \\ & & \mbox{}
       -  \cf\,  \*  \ca\,  \*  \nf\,   \*  \Big( \,
            {2824243 \over 2916}
          - 652\, \*  \z3
          - {104 \over 15}\: \*  \zss
          - {760 \over 3}\: \*  \z5 
          \Big)
   \nn \\ & & \mbox{}
       -  \cas\, \*  \nf\,   \*  \Big( \,
            {3723227 \over 2187}\,
          - {4162 \over 81}\: \*  \z3
          + {24 \over 5}\: \*  \zss
          - {280 \over 3}\: \*  \z5 
          \Big)
   \nn \\ & & \mbox{}
       +  \cfs\, \*  \nf\,   \*  \Big( \,
            {40081 \over 486}\,
          + {24668 \over 81}\: \*  \z3 
          - {32 \over 15}\: \*  \zss
          - {1360 \over 3}\: \*  \z5 
          \Big)
       - {1000 \over 729}\: \* \nft
   \nn \\ & & \mbox{}
       +  \cf\,  \*  \nfs\,  \*  \Big( \,
            {53945 \over 486}\,
          - {8360 \over 81}\: \*  \z3
          \Big)
       +  \ca\,  \*  \nfs\,  \*  \Big( \,
            {1155203 \over 8748}\,
          + {3652 \over 81}\: \*  \z3
          \Big)
 \qquad \nn \\[2mm]
 &\!\cong\! &   56778.8 \: - 13673.8\:\nf + 541.328\: \nfs - 1.37174\: \nft
 \:\: .
\eea
Note the conspicuous absence of coefficient linear in $\z2$ from the above 
second- and third-order results, a feature that has been observed and discussed
before for the corresponding moments of the gauge-boson exchange structure 
functions in Refs.~\cite{Mom3loop}.

For $\nf=4$ flavours Eqs.~(\ref{eq:cq1N2}) -- (\ref{eq:cg3N2}) yield the rather
benign expansions
\bea
\label{CphiqN2}
 C_{\rm \phi,\, q}(N\!=\!2) &\!\cong\!& 
  \!\quad \:-\: 0.1415\,\as \:-\: 0.7378\,\as^{\,2}
          \:-\: 2.6791\,\as^{\,3} \:+\; \ldots
\:\: , \\[1mm]
\label{CphigN2}
 C_{\rm \phi,\, g}(N\!=\!2) &\!\cong\!&  
      1   \:+\: 2.3387\,\as \:+\: 3.2956\,\as^{\,2}
          \:+\: 5.3704\,\as^{\,3} \:+\; \ldots
\eea
which nevertheless, as expected, are rather different from their counterparts
for the photon-exchange structure function $\Ftwo$ given by \cite{Mom3loop}
\bea
\label{C2qN2}
 C_{\,\rm 2,\, q}(N\!=\!2) &\!\cong\!&
      1   \:+\: 0.0354\,\as \:-\: 0.0785\,\as^{\,2}
          \:-\: 0.1986\,\as^{\,3} \:+\; \ldots
\:\: , \\[1mm]
\label{C2gN2}
 C_{\,\rm 2,\, g}(N\!=\!2) &\!\cong\!&
 \!\quad \:-\: 0.1592\,\as \:-\: 0.2259\,\as^{\,2}
         \:-\: 0.0274\,\as^{\,3} \:+\; \ldots
\:\: .
\eea
Consequently, as already indicated at the end of Section 2, the 
scheme-transformation relation (\ref{eq:Zmomsum}) is not fulfilled with the 
choice (\ref{eq:F2Htrf}) involving Eqs.~(\ref{CphiqN2}) -- (\ref{C2gN2}).

The $N$-dependences of $C_{\rm \phi,\, q}$ and $C_{\rm \phi,\, g}$ are shown 
graphically, again for $\as = 0.2$, in Fig.~4. The corresponding results for 
$\Ftwo$ and $\FL$ in standard DIS have been presented in Figs.~11 and 12 of 
Ref.~\cite{MVV6}. The present pattern is rather similar to that for $\Ftwo$
(with the quark and gluon coefficient functions interchanged), albeit with 
considerably larger $\as$ corrections. In fact, the $N$-dependent values
of the strong coupling,
\beq
\label{eq:ashat}
  \widehat{\alpha}_{\:\!\phi,i}^{\,(n)}(N) \:\: = \:\: 4\:\!\pi\;
  \frac{c_{\phi,i}^{(n-1)}(N)}{2\,c_{\phi,i}^{(n)}(N)} \:\: ,
\eeq
for which the $n$-th order corrections are half as large as those of the 
previous order 
-- recall that the coefficient functions are expressed in terms of the small 
expansion parameter $\,\ar\equiv\as/(4\pi)\,$ of Eq.~(\ref{eq:arun}) --
shown in Fig.~5 are rather more similar to those for $F_L$ than for $F_2$ which 
can be found, respectively, in Figs.~13 and 14 of Ref.~\cite{MVV6}. Note,
however, that also in Fig.~5 the values for $\widehat{\alpha}^{\,(n)}(N)$ 
increase, rather than decrease, with the perturbative order $n$. This behaviour
indicates that also here we have not yet reached the expected asymptotic regime 
of the expansion in powers of $\as$ with a factorial-type growth 
$c^{\,(n)} \sim n!$ of the expansion coefficients.

\begin{figure}[p]
\vspace*{-1mm}
\centerline{\epsfig{file=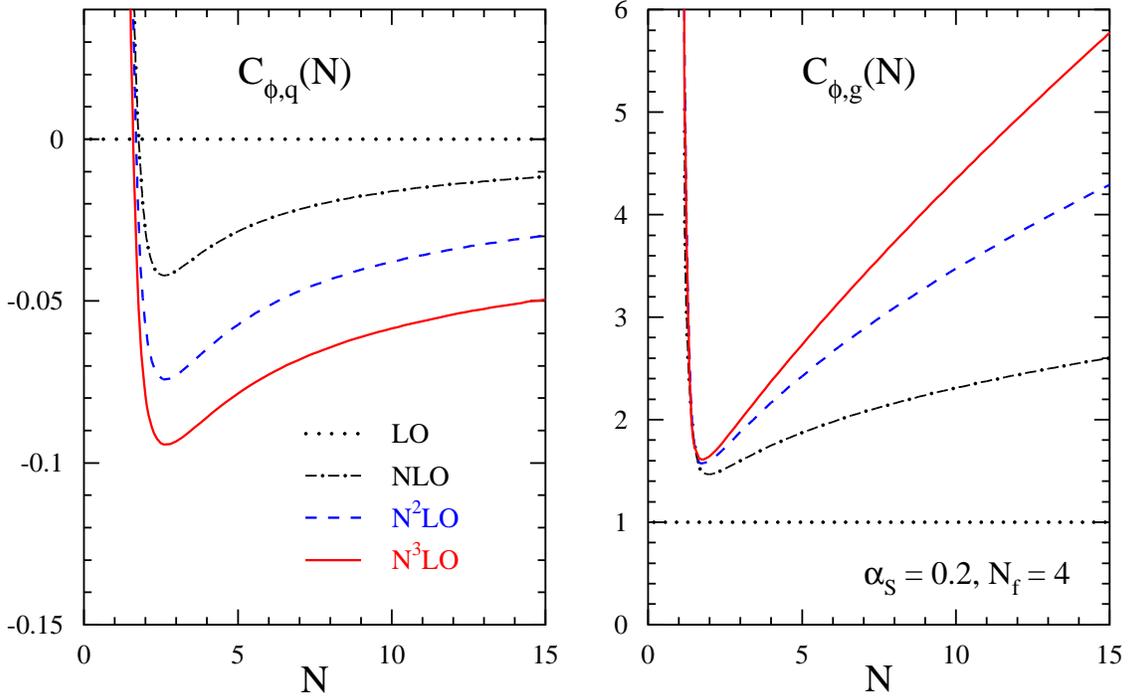,width=15.0cm,angle=0}}
\vspace{-2mm}
\caption{\label{pic:fig4}
 The perturbative expansion to order $\as^{\,3}$ (N$^3$LO) of the quark and 
 gluon coefficient functions for the structure function $F_\phi$ in Mellin-$N$
 space, shown for four effectively massless flavours and the order-independent 
 value of $\as$ specified in the right panel.
 }
\end{figure}
\begin{figure}[p]
\centerline{\epsfig{file=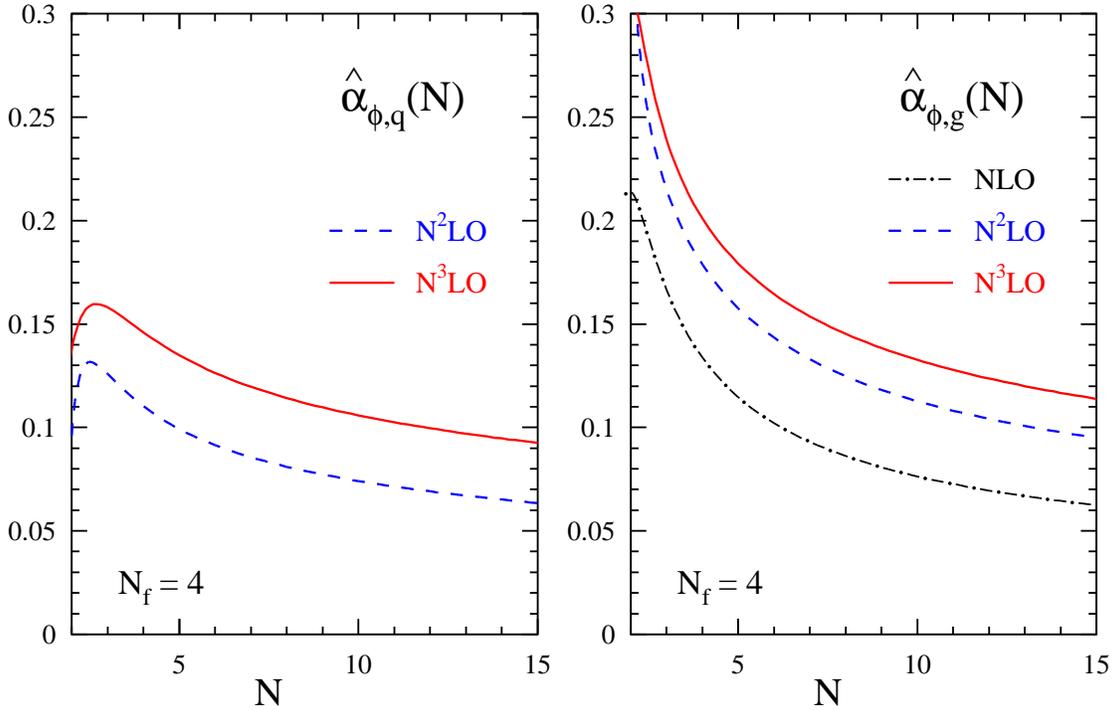,width=15.0cm,angle=0}}
\vspace{-2mm}
\caption{\label{pic:fig5}
 The $N$-dependent values (\ref{eq:ashat}) of $\as$ at which the effect of the 
 $n$-th order (N$^n$LO) quark and gluon coefficient functions for $F_\phi$ is 
 half as large as that of the previous order. A NLO curve can only be shown for
 the gluon case (right) since only here the LO contribution does not vanish.
 }
\vspace{-1mm}
\end{figure}
%
%
\setcounter{equation}{0}
\section{Physical evolution kernel for $\Ftwo$ and $F_{\phi}$}
\label{sec:kernel}
%
%
We are now in a position to write down and discuss the physical kernels for the
coupled flavour-singlet evolution of the structure functions $\Ftwo$ and 
$F_{\phi}$ as defined at the end of Section 2. To NNLO ($\,l = 2$) these 
kernels,
\beq 
\label{eq:Kabexp} 
  K_{\,ab} \:\: = \:\: \sum_{l=0} \ar^{\,l+1}\, K_{\,ab}^{\,(l)} 
  \quad \mbox{ with} \quad a,\, b \; =\; 2,\, \phi
  \:\: ,
\eeq
are specified in terms of the three-loop splitting functions \cite{MVV3,MVV4}
together with the second-order coefficient functions for $\Ftwo$, first 
calculated in Refs.~\cite{ZvN-F2}, and $F_\phi$ provided in the previous 
section. Also these results will by given in terms of harmonic polylogarithms 
\cite{Remiddi:1999ew} as summarized in Eqs.\ (\ref{eq:hpol1}) -- (\ref{eq:habbr})
above. From $\,l = 1$ we will present $K_{\,22}$ in the form
\beq
\label{eq:K22nsps}
   K_{\,22}^{\,(l)} \:\: = \:\: K_{\,2,\rm ns }^{\,(l)} \,+\,
   K_{\,22, \rm ps}^{\,(l)} \; ,
\eeq
i.e., we separately provide the scalar evolution kernel for the non-singlet 
structure function $F_{\,2,\rm ns}$ (addressed, but not written down in this 
form, before in Refs.~\cite{NV3,MV5}), which is of direct phenomenological 
interest for, e.g., the determination of $\as$ from scaling violations. As
above, our results are given in the \MSb\ renormalization scheme for the 
scale $\,\mu = Q$.

At leading order (LO) the elements of the physical kernel are identical to
the splitting functions, recall the first line of Eq.~(\ref{eq:Pnew}), i.e.,
\bea
 \label{eq:K220}
 K_{\,22}^{\,(0)} &\! = \!&
  \colour4colour{\cf} \* \big( \,
          2 \* \pqq(x)
          + 3\* \delta(1 - x)
          \,\big)
%
 \; ,\\[1mm]
 \label{eq:K2p0}
 K_{\,2\phi}^{\,(0)} &\! = \!&
          2\, \* \colour4colour{\nf} \, \* \pqg(x)
 \; , \\[1mm]
 \label{eq:Kp20}
 K_{\,\phi 2}^{\,(0)} &\! = \!&
          2\, \* \colour4colour{\cf} \, \* \pgq(x)
 \; ,\\[1mm]
 \label{eq:Kpp0}
 K_{\,\phi \phi}^{\,(0)} &\! = \!&
         \colour4colour{\ca}  \*  \bigg(
            4 \* \pgg(x)
          + {11 \over 3} \* \delta(1 - x)
          \bigg)
          - {2 \over 3} \, \* \colour4colour{\nf} \, \* \delta(1 - x)
 \; ,
\eea
where we have used Eq.~(\ref{eq:p0gqgg}) and the additional abbreviations
\bea
\label{eq:p0qqqg}
  p_{\rm{qq}}(x) &\!=\!& 2\, (1 - x)^{-1} - 1 - x
  \nn \\
  p_{\rm{qg}}(x) &\! =\! & 1\, -\, 2x\, +\, 2x^{\,2} 
  \:\: .
\eea

\noindent
The next-to-leading order (NLO) contributions to these kernels are given by
\bea
 \label{eq:Kns1}
 \lefteqn{ K^{\,(1)}_{\,2,\rm ns}(x)  \; = \; 
%
       \colour4colour{\cfs}\, \* \Big( \pqq(x)\, \* [
          - 6\, \* \H(0)\,
          + 8\, \* \Hh(1,0)\,
          + 8\, \* \H(2)\,
          ]\,
       + 8\, \* \pqq(-x)\, \* [
          - \z2\,
          - 2\, \* \Hh(-1,0)\,
          + \Hh(0,0)\,
          ]\,
%
 }
%
   \nn \\[-0.5mm] & & \mbox{}
       +  2\, \* (
          - 2\,
          + 2\, \* x\,
          + \H(0)\,
          - 3\, \* x\,\* \H(0)\,
          - 2\, \* \Hh(0,0)\,
          - 2\, \* x\, \* \Hh(0,0)\,
             )
          + [ \, 3/2 - 12\, \* \z2 + 24\, \* \z3 ] \: \* \delta \x1
          \Big)
   \nn \\[-0.5mm] & & \mbox{\hspn}
       + \colour4colour{\ca\, \* \cf}\, \* \Big( \pqq(x)\, \* [ \,
            367/18\:
          + 2/3\: \* (
          - 6\, \* \z2\,
          + 22\: \* \H(0)\,
          + 6\, \* \Hh(0,0)\,
          + 11\: \* \H(1)\,
            )
          ]\,
   \nn \\[-0.5mm] & & \mbox{}
       + 4\, \* \pqq(-x)\, \* [ \,
            \z2\,
          + 2\, \* \Hh(-1,0)\,
          - \Hh(0,0)\,
                 ]\,
          + 1/6\: \* ( 13\,
          - 167\,\* x\,
          + [ \, 215 + 176\, \* \z2 - 72\, \* \z3 ] \, \* \delta \x1
                  )
          \Big)
    \nn \\[-0.5mm] & & \mbox{\hspn}
         + \colour4colour{ \cf\, \* \nf} \, \* \Big( \pqq(x)\, \* [
          - 29/9
          - 8/3\, \* \H(0)\,
          - 4/3\, \* \H(1)\,
          ]\,
          + 1/3\: \* (1\,
          + 13\, \* x\,
          - [ 19 + 16\, \* \z2 ]\, \* \delta \x1
            )
          \Big)
 \:\: ,
 \\[2mm]
 \label{eq:Kps1}
 \lefteqn{ K^{\,(1)\,}_{\,22,\,\rm ps}(x) \; = \;  
         \colour4colour{ \cf\, \* \nf} \, \* \Big(
            2/3\: \* ( 32\, \* x^{-1}
          - 69\,
          + 45\, \* x\,
          - 8\, \* x^2\, )\,
          + 4/3\: \* (4\, \* x^{-1}\,
          - 9\, \* x\,
          + 8\, \* x^2\, )\, \* \H(0)\,
%
  }
%
   \nn \\[-0.5mm] & & \mbox{}
          - 8\, \* ( 1\, + x )\, \* \Hh(0,0)\,
          \Big)
 \:\: ,
 \\[2mm]
 \label{eq:Kqg1}
 \lefteqn{ K^{\,(1)\,}_{\, 2\phi}(x) \; = \; 
        \colour4colour{ \ca\, \* \nf} \, \*  \Big(
         8\, \* \pqg(x)\, \* \H(2)\,
       - 8\, \* \pqg(-x)\, \* \Hh(-1,0)\,
       + 2/3\: \* (1\, - 50\, \* x + 50\, \* x^2)\, \* \H(1)\,
%
  }
%
   \nn \\ & & \mbox{}
       - 8\, \* (1\, + 2\, \* x)\, \* \Hh(0,0)\,
       - 8\, \* \z2\, \* (1\, + 2\, \* x^2 \,)
       + 2/3\: \* ( 8\, \* x^{-1} - 11\, - 98\, \* x + 66\, \* x^2)\, 
         \* \H(0)\,
   \nn \\[0.5mm] & & \mbox{}
       + 2/9\: \* ( 100\, \* x^{-1}\, - 322\, + 155\, \* x + 18\, \* x^2)
         \Big)\,
       \:\: + \:\: \colour4colour{ \cf\, \* \nf} \, \*  \Big(
        4\, \* \pqg(x)\, \*  [ \,
            2\, \* \Hh(0,0)\,
          - 3\, \* \H(1)\,
          + 2\, \* \Hh(1,0)\,
          ]\,
   \nn \\[-0.5mm] & & \mbox{}
       - 4\, \* (1\, - 2\, \* x)\, \* \Hh(0,0)\,
       + 2\, \* (1\, + 4\, \* x - 12\, \* x^2)\, \* \H(0)\,
       + 2\, \* (2\, - 7\, \* x + 4\, \* x^2 \,)
       + 16\, \* \H(1)\,          
              \Big)
   \nn \\[-0.5mm] & & \mbox{\hspn}
        + \colour4colour{ \nfs} \, \*  \Big(
        4/3\: \* \pqg(x)\,  \*  [ \,
            \H(0)\,
          + \H(1)\,
          ]\,
       + 4/9\: \* (- 2\, \* x^{-1} + 11\, - 25\, \* x + 21\, \* x^2 \,)
                \Big)
\:\: ,
 \\[3mm]
 \label{eq:Kgq1}
 \lefteqn{ K^{\,(1)\,}_{\,\phi 2}(x) \; = \;
    \colour4colour{ \cfs} \, \* \Big(
        8\, \* \pgq(x)\,  \*  \H(2)\,
       + 8\, \* \z2\, \* (- 2\, \* x^{-1} + 2\, - x)
       - 4\, \* (2\, - x)\,  \*  \Hh(0,0)\,
       + 2\, \* (2\, + 3\, \* x)\,  \*  \H(0)\,
%
  }
%
   \nn \\[-0.5mm] & & \mbox{}
       + ( 10\, \* x^{-1} - 18\, - x)
       + 12\, \* x^{-1}  \*  \H(1)\,
             \Big)
   \:\: + \:\: \colour4colour{\ca\, \* \cf} \, \* \Big(
        \pgq(x)\, \*  [ \,
            4/3\: \* \H(1)\,
          + 8\, \* \Hh(1,0)\,
          ]\,
   \nn \\ & & \mbox{}
       - 8\, \* \pgq(-x)\, \*  \Hh(-1,0)\,
       - 6\, \* (2\, - x)\,\*  \H(1)\,
       + 8\, \* (2\, + x)\,\*  \Hh(0,0)\,
       + 1/3\: \* (  87\, \* x^{-1} - 22\, + 83\, \* x + 64\, \* x^2 \,)
   \nn \\[0.5mm] & & \mbox{}
       + 2/3\: \* ( 22\, \* x^{-1} - 138\, + 15\, \* x - 16\, \* x^2 \,) \,\* 
         \H(0)\,
       + 8\, \* \z2\, \* (2\, \* x^{-1} + x)
              \Big)
   \nn \\[-0.5mm] & & \mbox{\hspn}
   + \colour4colour{ \cf\, \* \nf} \, \* \Big(
        \pgq(x)\,  \*  [
          - 4\, \* \H(0)\,
          - 4/3\: \* \H(1)\,
          ]\,\,
       - 2/3\: \* ( 29\, \* x^{-1} - 26\, + 13\, \* x)
             \Big )
\:\: ,
 \\[3mm]
 \label{eq:Kgg1}
 \lefteqn{ K^{\,(1)\,}_{\,\phi\phi}(x) \; = \; 
%
   \colour4colour{ \cas} \, \* \Big(
        \pgg(x)\,  \*  [ \,
           389/9\:
          - 8\, \* \z2\,
          + 44/3\: \* \H(0)\,
          + 8\, \* \Hh(0,0)\,
          + 44/3\: \* \H(1)\,
          + 16\, \* \Hh(1,0)\,
          + 16\, \* \H(2)\,
          ]\,
%
  }
%
   \nn \\ & & \mbox{}
       - 8\, \* \pgg(-x)\,  \*  [ \,
            \z2\,
          + 2\, \* \Hh(-1,0)\,
          - \Hh(0,0)\,
          ]\,
       + 32\, \* (1+x)\, \*  \Hh(0,0)\,
       - 4/3\: \* (25\, - 11\, \* x + 44\, \* x^2)\, \*  \H(0)\,
   \nn \\[0.5mm] & & \mbox{}
       - 1/9\: \* ( 268\, \* x^{-1} - 728\, + 607\, \* x - 389\, \* x^2 \,)
       -  (
            449/27\:
          - 44/3\: \* \z2\,
          - 12\, \* \z3\,
          )\: \* \delta \x1
                \Big)
   \nn \\[-0.5mm] & & \mbox{\hspn}
   + \colour4colour{ \ca\, \* \nf} \, \* \Big(
          4/3\: \* \pgg(x)\,  \*  [ 
          - 7
          - 2\, \* \H(0)\,
          - 2\, \* \H(1)\,
          ]\,
       - 8/3\: \* (1+x)\, \*  \H(0)\,
       - 4/9\: \* (13\, \* x^{-1} + 13\, - 2\, \* x - 2\, \* x^2 \,)
   \nn \\[-0.5mm] & & \mbox{}
       + 4/27\: \* (
           43
          - 18\, \* \z2\,
          )\: \* \delta \x1
             \Big)
   \:\: + \:\: \colour4colour{ \cf\, \* \nf} \, \* \Big(
       - 8\, \* (1+x)\, \*  \Hh(0,0)\,
       - 4/3\: \* (4\, \* x^{-1} + 6\, - 9\, \* x)\, \*  \H(0)\,
   \nn \\ & & \mbox{}
       - 2/9\: \* ( 44\, \* x^{-1} - 27\, - 45\, \* x + 28\, \* x^2)
       - 2\, \* \delta \x1
             \Big)
   \nn \\[-0.5mm] & & \mbox{\hspn}
    + \colour4colour{ \nfs} \, \* \Big(
       4/9\:  \* \pgg(x)\,
       + 4/9\: \* (2\, - x + x^2)
       - 20/27\: \* \delta \x1
             \Big)
 \:\: .
\eea
The five corresponding third-order (NNLO) corrections fill about 10 pages (in 
total) and therefore are deferred to Appendix B.

The leading large-$x$ contributions to the evolution kernels (\ref{eq:Kns1})
-- (\ref{eq:Kgg1}) and (\ref{eq:Kns2}) -- (\ref{eq:Kgg2}) are given by terms
of the form $\DDk$ (defined in Eq.~(\ref{abbrev})) for the diagonal 
entries $K^{\,(n)}_{\,22}$ and $K^{\,(n)}_{\,\phi\phi}$, and $L_1^{\,k} \equiv 
\ln^{\,k} \! \x1$ for their off-diagonal counterparts $K^{\,(n)}_{\,2\phi}$
and $K^{\,(n)}_{\,\phi 2 }$. It turns out that in all cases the higher-order
enhancement (as far as known, obviously) is only single-logarithmic, $k \,=\,
0,\: \ldots,\: n$. This is in contrast to the behaviour of the off-diagonal 
\MSb\ splitting functions, cf.~Eqs.~(4.17) -- (4.19) of Ref.~\cite{MVV4}, and
of all coefficient functions for $\Ftwo$ and $F_\phi$ as discussed in 
Ref.~\cite{MVV6} and Section~3 and Appendix C of the present article. Hence the
key non-singlet observation of Refs.\mbox{\cite{MV3,MV5}} generalizes to the 
present flavour-singlet case. We will utilize this fact in the next section.

The NLO and NNLO large-$x$ coefficients for $\,K_{\,22}(x)\,$ are given by
\bea
\label{eq:K221D1}
 K_{\,22}^{\,(1)} \Big|_{\,\DD1} &\! = \! & \mbox{}
%
         - {44 \over 3}\: \* \cf \* \ca\,
         + {8 \over 3}\: \* \cf\, \* \nf
 \:\: ,
 \\[1mm]
\label{eq:K221D0}
 K_{\,22}^{\,(1)} \Big|_{\,\DD0} &\! = \! & \mbox{}
%
      \Big( \, {367 \over 9} - 8\, \* \z2 \Big)\, \* \cf \* \ca\,
      - {58 \over 9}\: \* \cf\, \* \nf
\eea
and
\bea
\label{eq:K222D2}
 K_{\,22}^{\,(2)} \Big|_{\,\DD2} &\! = \! & \mbox{}
%
           {484 \over 9}\: \*  \cf \*  \cas\,
         - {176 \over 9}\: \*  \cf \*  \ca\, \*  \nf\,
         + {16 \over 9}\: \, \*  \cf\, \*  \nfs\,
  \:\: ,
 \\[2mm]
\label{eq:K222D1}
 K_{\,22}^{\,(2)} \Big|_{\,\DD1} &\! = \! & \mbox{}
%
         - \Big( \, 
             {9298 \over 27} - {176 \over 3}\: \* \z2
           \Big)\, \*  \cf \*  \cas\,
         + \Big( \, 
             {3104 \over 27} - {32 \over 3}\: \*  \z2
           \Big)\, \*  \cf \*  \ca\, \*  \nf\,
   \nn \\ & & \mbox{}
         + 8\, \*  \cfs\, \*  \nf\,
         - {232 \over 27}\: \*  \cf\, \*  \nfs
  \:\: ,
 \\[3mm]
\label{eq:K222D0}
 K_{\,22}^{\,(2)} \Big|_{\,\DD0} &\! = \! & \mbox{}
%
         \Big( \,
           {50689 \over 81}
         - {680 \over 3}\: \*  \z2
         - 264\, \* \z3\,
         + {176 \over 5}\: \*  \zss\,
         \Big)\, \*  \cf \*  \cas\,
   \nn \\ & & \mbox{\hspn}
       + ( \,
           11
         - 88\, \*  \z2
         + 176\, \*  \z3
         )\, \*  \cfs\, \*  \ca\,
       - \Big( \,
           {15062 \over 81}
         - 16\, \* \z3\,
         - {512 \over 9}\: \* \z2
         \Big)\, \* \cf \*  \ca\, \*  \nf\,
   \nn \\[1mm] & & \mbox{\hspn}
       - \Big( \,
           {134 \over 3}
         - 16\, \* \z2
         \Big)\, \* \cfs\, \*  \nf\,
      +  \Big( \,
           {940\over 81} 
         - {288 \over 81}\: \* \z2
         \Big)\, \*  \cf\, \*  \nfs
\:\: .
\eea
The corresponding terms for $\,K^{\,(1)}_{\,\phi\phi}(x)\,$ and 
$\,K^{\,(2)}_{\,\phi\phi}(x)\,$ read
\bea
\label{eq:Kpp1D1}
 K_{\,\phi\phi}^{\,(1)} \Big|_{\,\DD1} &\! = \! & \mbox{}
%
         - {44 \over 3}\: \* \cas\,
         + {8 \over 3}\: \* \ca\, \* \nf
 \:\: ,
 \\[1mm]
\label{eq:Kpp1D0}
 K_{\,\phi\phi}^{\,(1)} \Big|_{\,\DD0} &\! = \! & \mbox{}
%
           \Big( \, {389 \over 9} - 8\, \* \z2 \Big)\, \* \cas\,
         - {28 \over 3}\: \* \ca\, \* \nf\,
         + {4 \over 9}\: \* \nfs
\eea
and
\bea
\label{eq:Kpp2D2}
 K_{\,\phi\phi}^{\,(2)} \Big|_{\,\DD2} &\! = \! & \mbox{}
%
           {484 \over 9}\: \*  \cat\,
         - {176 \over 9}\: \*  \cas\, \*  \nf\,
         + {16 \over 9}\: \, \* \ca\, \*  \nfs\,
  \:\: ,
 \\[2mm]
\label{eq:Kpp2D1}
 K_{\,\phi\phi}^{\,(2)} \Big|_{\,\DD1} &\! = \! & \mbox{}
%
       - \Big( \, 
           {9782 \over 27}
         - {176 \over 3}\: \* \z2 
           \Big)\, \* \cat\,
       + \Big( \,
           {3764 \over 27}
         - {32 \over 3}\: \*  \z2 
           \Big)\, \* \cas\, \* \nf\,
   \nn \\ & & \mbox{}
         + 8\, \*  \cf\, \*  \ca\, \* \nf\,
         - {424 \over 27}\: \*  \ca\, \*  \nfs\,
         + {16 \over 27}\: \* \nft
  \:\: ,
 \\[3mm]
\label{eq:Kpp2D0}
 K_{\,\phi\phi}^{\,(2)} \Big|_{\,\DD0} &\! = \! & \mbox{}
%
         \Big( \,
           18974/27
         - {3008 \over 9}\: \* \z2\,
         - 88\, \*  \z3\,
         + {176 \over 5}\: \*  \zss 
           \Big) \, \* \cat\,
   \nn \\ & & \mbox{}
       -  \Big( \,
           20858/81\:
         - 96\, \*  \z2\,
         + 16\, \*  \z3 
           \Big) \, \*  \cas\, \* \nf\,
       -  \Big( \,
           {176 \over 3}
         - 32\, \* \z3
           \Big) \, \*  \cf \*  \ca\, \* \nf\,
   \nn \\[1mm] & & \mbox{}
       + \Big( \,
           {2284 \over 81}
         - {64 \over 9}\: \*  \z2
           \Big) \, \*  \ca\, \*  \nfs\,
         + 4\, \* \cf\, \*  \nfs\,
         - {80 \over 81}\: \*  \nft\,
\:\: .
\eea
Note that the leading large-$x$ coefficients at all three orders are simply
given by $(-\beta_0)^n \, A_{1,\,\rm p}$, where $A_{1,\,\rm p}$ denotes the
appropriate one-loop quark or gluon cusp anomalous dimension, i.e., the 
coefficient of $\x1^{-1}$ in Eq.~(\ref{eq:K220}) or (\ref{eq:Kpp0}). 

The log-enhanced large-$x$ contributions to $\,K^{\,(1)}_{\,2\phi}(x)\,$ and 
$\,K^{\,(2)}_{\,2\phi}(x)\,$ are
\bea
\label{eq:K2p1L1}
 K_{\,2\phi}^{\,(1)} \Big|_{\,L_1} &\! = \! & \mbox{}
%
         -{2 \over 3}\:\* \ca\, \* \nf\,
         -4\,\* \cf\, \* \nf\,
         -{4 \over 3}\: \* \nfs\,
\eea
and
\bea
\label{eq:K2p2L2}
 K_{\,2\phi}^{\,(2)} \Big|_{\,L_1^{\,2}} &\! = \! & \mbox{}
%
       - \Big( \,
           {118 \over 3}\:
         - 16\, \*  \z2\
         \Big)\, \*  \cf \* \ca\, \*  \nf\,
       + \Big( \, {100 \over 3} - 16\, \*  \z2
         \Big)\, \*  \cas\, \*  \nf\,
   \nn \\ & & \mbox{}
         + \, 20\, \*  \cfs\, \*  \nf\,
         + {8 \over 3}\: \*  \ca\, \*  \nfs\,
         + {4 \over 3}\: \*  \cf\, \*  \nfs
  \:\: ,
 \\[2mm]
\label{eq:K2p2L1}
 K_{\,2\phi}^{\,(2)} \Big|_{\,L_1} &\! = \! & \mbox{}
       - ( 
           2\,
         - 64\, \* \z2
         + 96\, \* \z3\,
         )\, \* \cfs\, \* \nf\,
       - \Big(
           {110 \over 3}
         + {64 \over 3}\: \* \z2\,
         - 96\, \* \z3
         \Big)\, \* \cf \* \ca\, \* \nf\,
   \nn \\ & & \mbox{}
         - 24\, \* \cas\, \* \nf\,
       + \Big(\,
           {8 \over 3}
         + {16 \over 3}\: \* \z2
         \Big)\, \* \cf\, \* \nfs\,
 \:\: .
\eea
The corresponding terms for $\,K_{\,\phi 2}(x)\,$ are given by
\bea
\label{eq:Kp21L1}
 K_{\,\phi 2}^{\,(1)} \Big|_{\,L_1} &\! = \! & \mbox{}
%
           {14 \over 3}\: \* \cf\, \* \ca\,
         - 12\, \* \cfs\,
         + {4 \over 3}\: \* \cf\, \* \nf
\eea
and
\bea
\label{eq:Kp22L2}
 K_{\,\phi 2}^{\,(2)} \Big|_{\,L_1^{\,2}} &\! = \! & \mbox{}
%
         ( 16\, + 32\, \* \z2 )\, \*  \cft\,
       + ( 34\, - 64\, \* \z2 )\, \*  \cfs\, \*  \ca\,
       - \Big( \,
           {208 \over 9}
         - 32\, \* \z2
         \Big)\, \*  \cf\, \*  \cas\,
   \nn \\ & & \mbox{}
         - 12\, \*  \cfs\, \*  \nf\,
         + {20 \over 9}\: \*  \cf\, \*  \ca\, \*  \nf\,
         + {8 \over 9}\: \*  \cf\, \*  \nfs\,
  \:\: ,
 \\[3mm]
\label{eq:Kp22L1}
 K_{\,\phi 2}^{\,(2)} \Big|_{\,L_1} &\! = \! & \mbox{}
%
         ( \, 
           106\,
         + 24\, \* \z2 
         - 48\, \* \z3\,
         )\, \* \cft\,
       - \Big( \,
           {2024 \over 3}\:
         + {212 \over 3}\: \* \z2
         - 120\, \* \z3\,
           \Big)\, \* \cfs\, \* \ca\,
   \nn \\ & & \mbox{}
       + \Big( \, 
           {6308 \over 27}
         + {236 \over 3}\: \* \z2
         - 72\, \* \z3\,
         \Big)\, \* \cf \* \cas\,
       + \Big( \, 
           {272 \over 3}
         + {32 \over 3}\: \* \z2
         \Big)\, \* \cfs\, \* \nf\,
   \nn \\[1mm] & & \mbox{}
       + \Big( \,
           {620 \over 27}
         - {56 \over 3}\: \* \z2
         \Big)\, \* \cf \* \ca\, \* \nf\,
         - {232 \over 27}\: \* \cf\, \* \nfs
 \:\: .
\eea
Thus already the leading large-$x$ terms are of a more complicated form 
in the off-diagonal cases.

At small-$x$ these physical evolution kernels, unlike the \MSb\ splitting 
functions \cite{BFKL}, include double-logarithmic $1/x$-contributions 
originating, via Eqs.~(\ref{eq:Pexp}) and (\ref{eq:F2Htrf}), in the 
corresponding behaviour for the coefficient functions 
$c_{\rm \phi\,,p}^{\,(n)}$ discussed in the previous section.  
Specifically, the quantities $K_{\,\phi 2}^{\,(n)}(x)$ and 
$K_{\,\phi\phi}^{\,(n)}(x)$ are found to include terms up to $x^{\,-1} 
\ln^{\,2n-1}\! x$ for $n = 1,\: 2,\: 3$. The corresponding leading small-$x$
contributions to the N$^2$LO and N$^3$LO kernels $K_{\,22}^{\,(n)}(x)$ and 
$K_{\,2\phi}^{\,(n)}(x)$ are of the form $x^{-1}\ln^{\,2n-2} \! x$. Since also
these terms are artifacts of the large-$m_{\,\rm top}$ approximation discussed 
above, we will not write them down here.

Instead we directly proceed to a brief discussion of the numerical implications
of Eqs.\ (\ref{eq:Kns1})~-- (\ref{eq:Kgg1}) and (\ref{eq:Kns2}) -- 
(\ref{eq:Kgg2}). The resulting perturbative expansions (\ref{eq:Kabexp}) of the 
elements $K_{\,ab}$ of this evolution kernel are illustrated in Figs.~6 -- 9.
Here the initial conditions $\Ftwo(x,Q_0^{\,2})$ and $F_\phi(x,Q_0^{\,2})$ are 
physical quantities and consequently, unlike the \MSb\ parton densities in 
Section 3, in principle independent of the perturbative order. 
Hence it is easier here to provide a realistic (and not only schematic, as in 
the previous section) account of the perturbative (in-)$\:\!$stability of the 
quantities under consideration.

The only remaining order-dependent quantity is the \MSb\ coupling at our chosen reference scale. Here we employ the NLO and NNLO values obtained from a
schematic analysis of non-singlet scaling violation in Refs.~\cite{NV3,MVV6},
supplemented by a corresponding LO result,  
\bea
\label{eq:als-sec4}
  \as (Q_0^{\,2})_{\,\rm LO}^{}  \;\;\;\;    &\! =\! & 0.255 \nn \\
  \as (Q_0^{\,2})_{\,\rm NLO}^{}  \;\;       &\! =\! & 0.208 \nn \\
  \as (Q_0^{\,2})_{\,\rm NNLO}^{}            &\! =\! & 0.201
\:\: .
\eea
These values are consistent with a NLO$\,-\,$NNLO difference of about 0.002
to 0.003 at the mass of the $Z$-boson as found in recent fits of structure 
functions and related data \cite{asfits}. The actual scale $Q_0^{\,2}$ in 
Eq.~(\ref{eq:als-sec4}) depends, of course, on the value of $\as(M_Z^{\,2})$
with, at present, $20\:\GeV2\! \lsim Q_0^{\,2} \lsim  50\:\GeV2\!$.

For the illustrations of the resulting scaling violations in Figs.~10 and 11
we use the sufficiently characteristic form (\ref{eq:p-sg}) for the structure 
functions at the reference scale $Q_0^{\,2}$, i.e., we employ
\bea
\label{eq:F2Hinp}
  \Ftwo(x,Q_0^{\,2}) &\! = \! &
  0.6\: x^{\, -0.3} (1-x)^{3.5}\, (1 + 5.0\: x^{\, 0.8\,}) \:\: ,
\nn \\[0.5mm]
  F_\phi (x,Q_0^{\,2}) &\! = \! &
  1.6\: x^{\, -0.3} (1-x)^{4.5}\, (1 - 0.6\: x^{\, 0.3\,}) \:\: .
\eea
Note that, as implied by the identity of Eqs.~(\ref{eq:p-sg}) and 
(\ref{eq:F2Hinp}), we have absorbed the average $\langle e^2 \rangle$ of the
quark charges into the definition of the singlet structure function $\Ftwo$, 
which is thus given at leading order by $\Ftwo \,=\, xq_{\rm s}$ instead of 
the standard normalization $\Ftwo \,=\, \langle e^2 \rangle\, xq_{\rm s}$.

At medium to large $x$ especially the diagonal entries $K_{\,22}$ and 
$K_{\,\phi\phi}$, respectively shown in Figs.~6 and 7, are rather stable, with
the relative NNLO corrections not exceeding 10\% in the range $0.2 < x < 0.9$ 
in both cases. Larger higher-order effects are found in this region for the 
(absolutely smaller) off-diagonal kernels displayed in Figs.~8 and 9.
In all cases the perturbative expansion appears ill-behaved at small values of
$x$, starting from about $x \simeq 10^{\,-1}$ for $K_{\,\phi\phi}$ to 
$x \simeq 10^{\,-3}$ for $K_{\,\phi 2}$. We expect that a (physically realistic)
inclusion of the finite-$m_{\,\rm top}$ effects as discussed below Eq.~(\ref
{eq:cphig3L00}) would fundamentally change these small-$x$ results.

Due to the ubiquitous Mellin convolutions (\ref{eq:Mconv}) -- and the diagonal
distribution effects not visible in Figs.~6 and 7 -- even the present
evolution of the structure functions $\Ftwo$ and $F_\phi$, illustrated in 
Figs.~10 and 11, is perfectly stable down to about $x = 10^{\,-2}$ and rather
acceptable at $x$-values as low as $x \simeq 10^{\,-3}$. At large $x$ these 
physical scaling violations are, unsurprisingly, much stronger (by a factor
between 2 and 3) for the gluonic observable $F_\phi$ than for $\Ftwo$. Here
the relative NNLO corrections exceed 5\% only at $x > 0.7$ for the scale
derivative of $\Ftwo$, and at $x \geq 0.6$ for that of $F_\phi$.

\begin{figure}[p]
\vspace*{-1mm}
\centerline{\epsfig{file=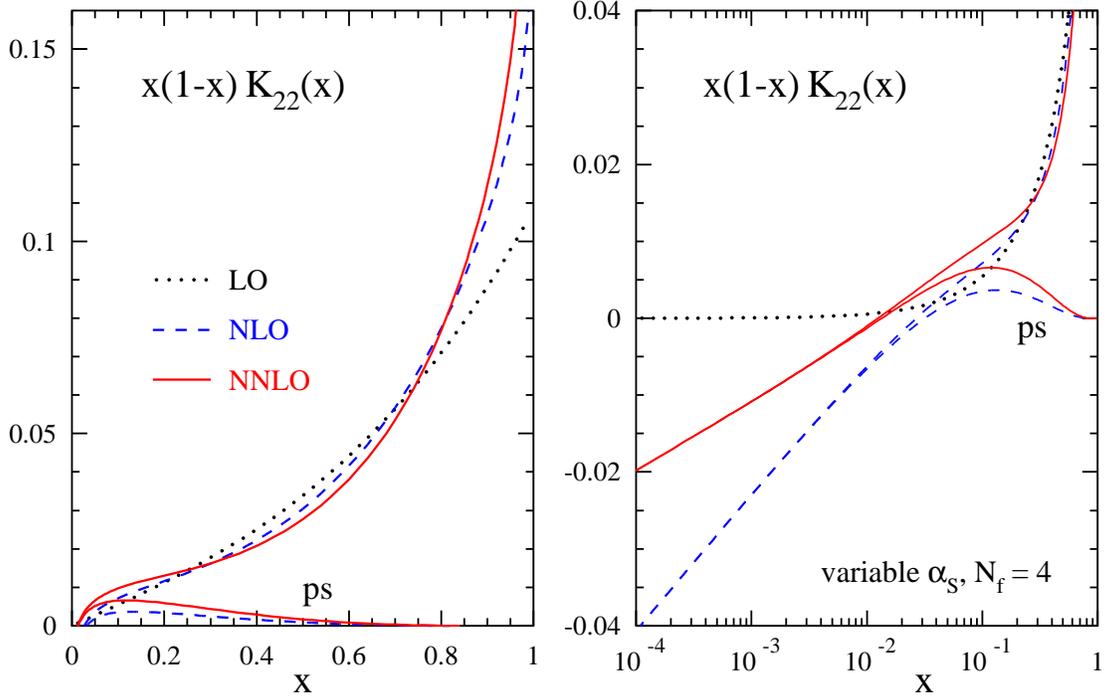,width=15.0cm,angle=0}}
\vspace{-2mm}
\caption{\label{pic:fig6}
 The perturbative expansion to order $\as^{\,3}$ (NNLO) of the diagonal element
 $K_{\,22}(x)$ of the physical evolution-kernel matrix for the system 
 ($\,\Ftwo$, $F_\phi\,$) of flavour-singlet DIS structure functions at $x < 1$.
 The poles at $x=0$ and $x=1$ have been suppressed by multiplication with 
 $x \x1$. Also shown is the pure-singlet contribution $K_{\,22,\,\rm ps}$ 
 defined in Eq.~(\ref{eq:K22nsps}).
 }
\end{figure}
\begin{figure}[p]
\centerline{\epsfig{file=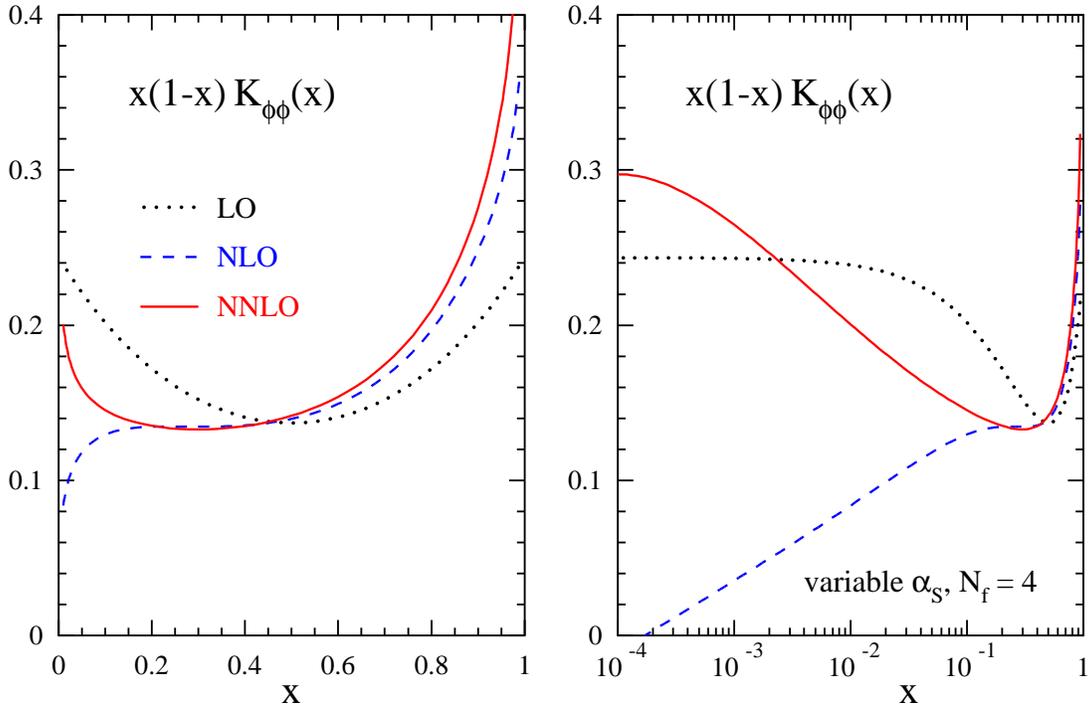,width=15.0cm,angle=0}}
\vspace{-2mm}
\caption{\label{pic:fig7}
 As Fig.~\ref{pic:fig6}, but for the second diagonal element 
 $K_{\,\phi\phi}(x)$. As in all figures in this section, the order-dependent
 values (\ref{eq:als-sec4}) are employed for the strong coupling $\as$.
 }
\vspace{-1mm}
\end{figure}

\begin{figure}[p]
\vspace*{-1mm}
\centerline{\epsfig{file=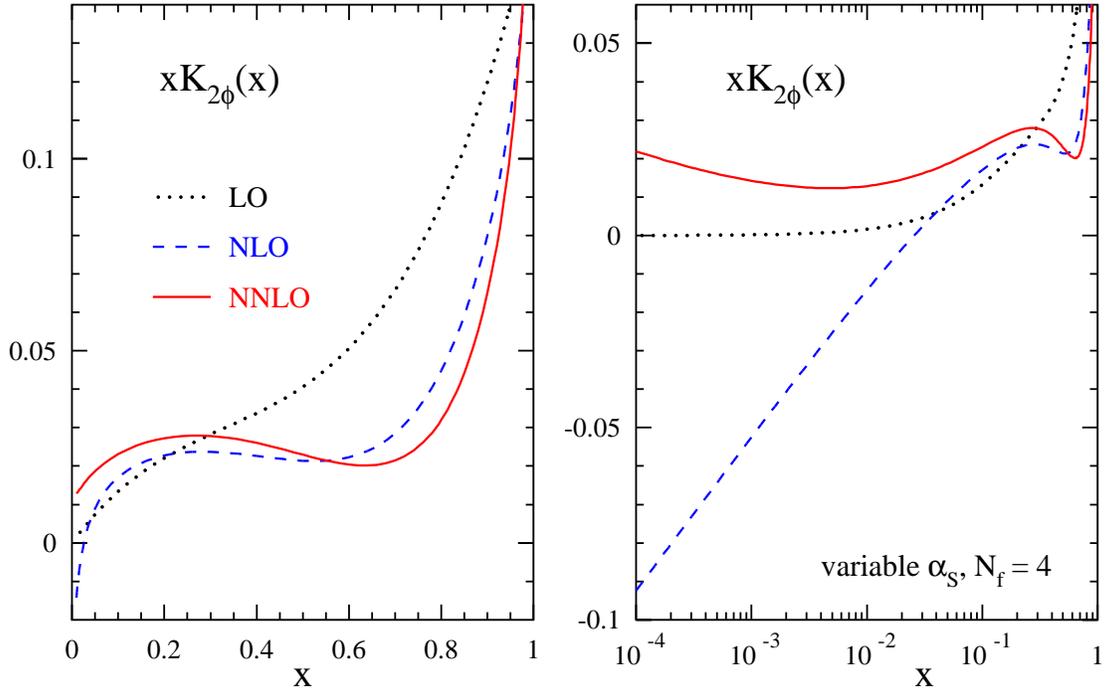,width=15.0cm,angle=0}}
\vspace{-2mm}
\caption{\label{pic:fig8}
 The perturbative expansion to order $\as^{\,3}$ (NNLO) of the off-diagonal
 element $K_{\,2\phi}(x)$ of the physical evolution-kernel matrix for the 
 system ($\Ftwo$, $F_\phi$) of singlet structure functions. 
 }
\end{figure}
\begin{figure}[p]
\centerline{\epsfig{file=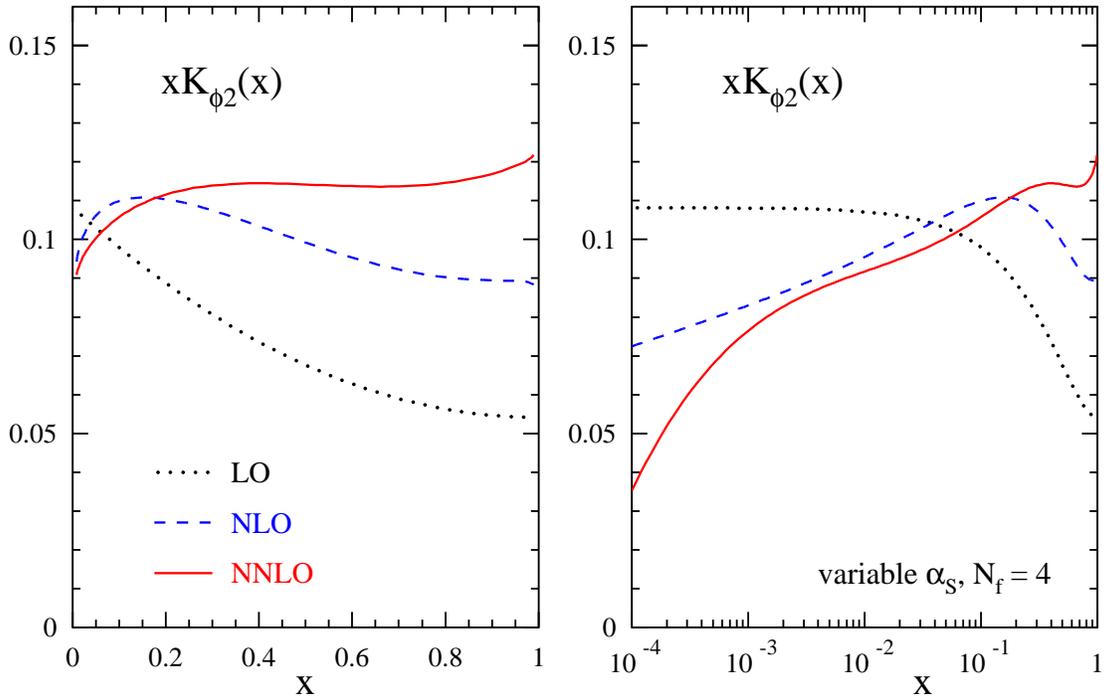,width=15.0cm,angle=0}}
\vspace{-2mm}
\caption{\label{pic:fig9}
 As Fig.~\ref{pic:fig8}, but for the second off-diagonal element
 $K_{\,\phi 2}(x)$. In both cases the curves have multiplied by a factor of 
 $x$ compensating the main $x$-dependence.
 }
\vspace{-1mm}
\end{figure}

\begin{figure}[p]
\vspace*{-1mm}
\centerline{\epsfig{file=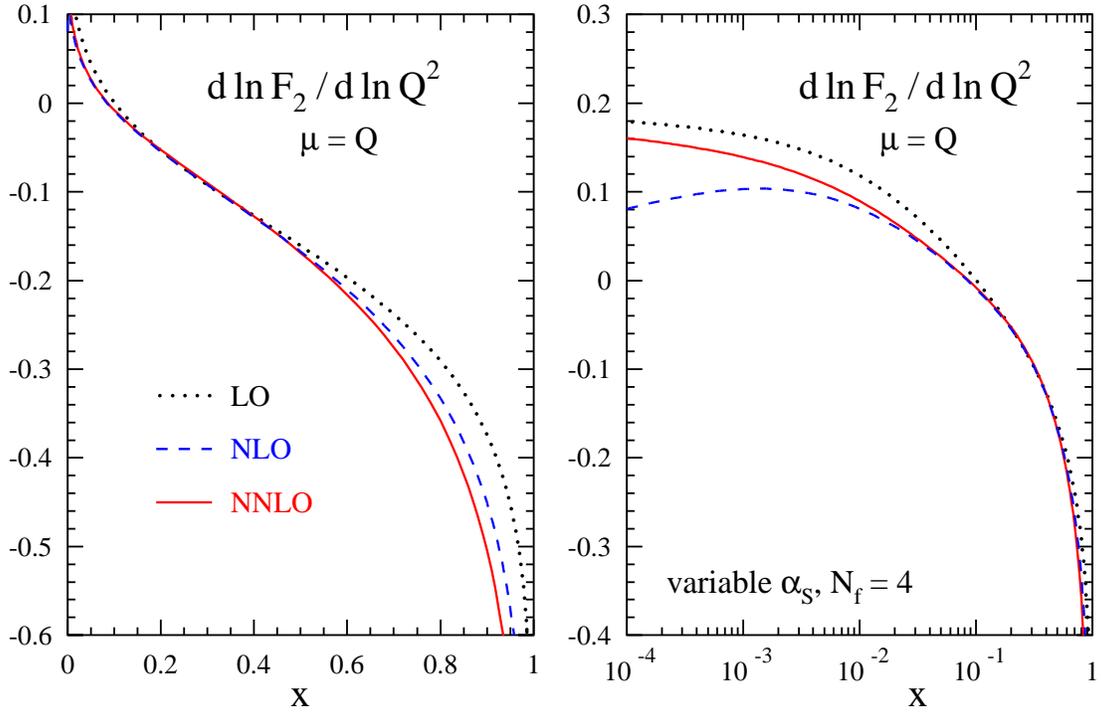,width=15.0cm,angle=0}}
\vspace{-2mm}
\caption{\label{pic:fig10}
 The perturbative expansion (\ref{eq:Kab}) to $l=2$ (NNLO) of the normalized
 scale derivative $d \ln \Ftwo / d \ln \Qs$ of the photon-exchange singlet
 structure function $\Ftwo$ for the initial conditions (\ref{eq:F2Hinp}) at a 
 mid-scale reference point with the coupling constants (\ref{eq:als-sec4}).
 }
\end{figure}
\begin{figure}[p]
\centerline{\epsfig{file=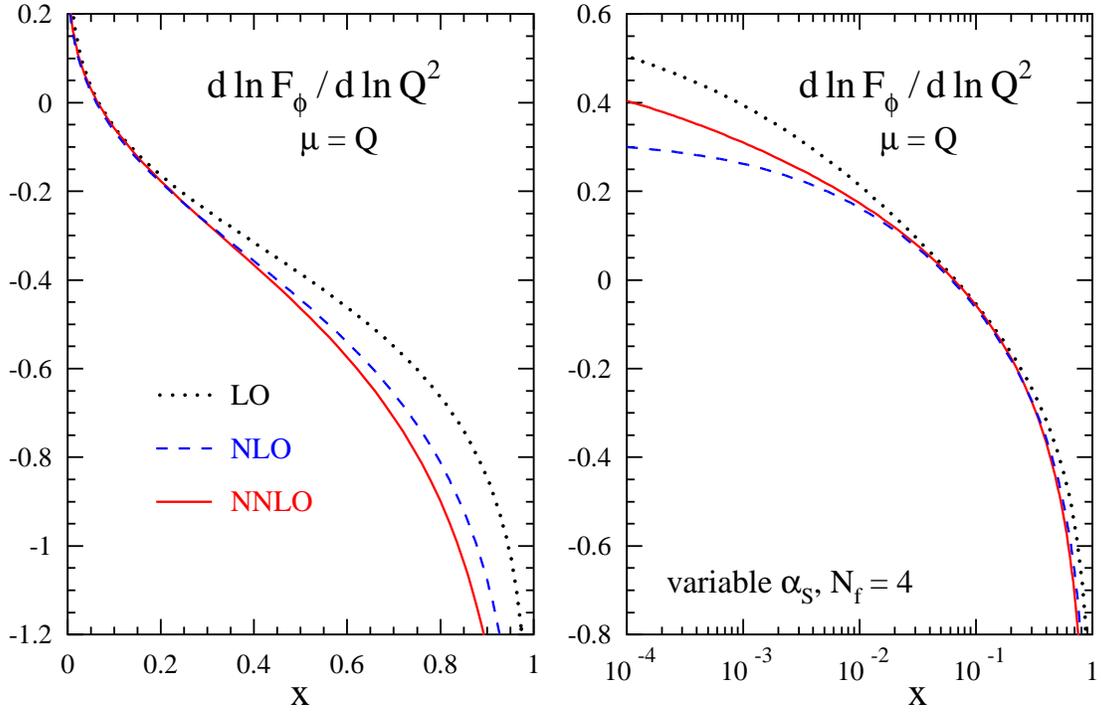,width=15.0cm,angle=0}}
\vspace{-2mm}
\caption{\label{pic:fig11}
 As Fig.~\ref{pic:fig10}, but for the scaling violations of the
 scalar-exchange structure function $F_\phi\,$.
 }
\vspace{-1mm}
\end{figure}
%
%
\setcounter{equation}{0}
\section{Large-$x$ behaviour and four-loop predictions}
\label{sec:fourloop}
%
%
We now return to the expansion of the physical kernels and the \MSb\ splitting
functions in powers of $L_1 \,\equiv\, \ln \x1$. The leading large-$x$ terms,
$\x1^{-1} L_1^{\,k}$ for $K_{\,22}$ and $K_{\,\phi\phi}$ and $\x1^{0}L_1^{\,k}$
for $K_{\,2\phi}$ and $K_{\,\phi 2}$, have been given to order $\as^{\,3}$ 
(NNLO) in Eqs.\ (\ref{eq:K221D1}) -- (\ref{eq:Kp22L1}). 
In fact, the only single-logarithmic enhancement -- $k\leq n$ at N$^{\:\! n}$LO
-- does not only hold for these leading contributions, but to all orders in 
$\x1$ for all matrix entries. Thus the physical evolution kernels fulfill
\beq
\label{KoffdL}
  K_{\, ab}^{\,(n)} \Big|_{L_1^{\,n+k} f(x)} \; = \; 0 
  \quad \mbox{ for } \quad k \,\geq\: 1 \:\: ,
\eeq
to (at least) $n=2$, where $f(x)$ denotes functions with a non-logarithmic
expansion at $x=1$.
 
Already the leading $L_1^{\,n} f(x)$ terms are not particularly simple for the 
off-diagonal evolution kernels, hence we refrain from writing them down here
for brevity. The observation made below Eq.~(\ref{eq:Kpp2D0}) for the diagonal 
entries, on the other hand, also generalizes to all orders in $\x1$, i.e., 
the leading terms are those of the non-singlet kernels, and therefore given by 
\cite{MV5}
\bea
\label{K22L1}
  K_{\, 22}^{\,(1)}(x) &\!\! = \!&
  \ln \x1 \: \* \pqq(x) \* \left[
      -2\, \* \cf \* \beta_0 \:-\: 8\, \* \cfs\, \* \H(0) \right]
%
      \;+\; {\cal O} \! \left( \:\!\ln^{\,0} \!\x1 \:\!\right) \:\: ,
 \\[2mm]
\label{K22L2}
 K_{\, 22}^{\,(2)}(x) &\!\! = \!&
      \ln^{\,2} \x1 \, \* \pqq(x) \* \left[
      2\, \* \cf \* \beta_0^{\,2} \:+\: 12\, \* \cfs\, \* \beta_0\, \* \H(0)
      + 32\, \* \cft\: \* \Hh(0,0) \right]
%
     \;+\; {\cal O} \! \left( \:\!\ln \x1 \:\!\right) 
 \;\;
\eea
and
\bea
\label{KppL1}
  K_{\, \phi\phi}^{\,(1)}(x) &\!\! = \!&
  \ln \x1 \: \* 2 \* \pgg(x) \* \left[
      -2\, \* \ca \* \beta_0 \:-\: 8\, \* \cas\, \* \H(0) \right]
%
      \;+\; {\cal O} \! \left( \:\!\ln^{\,0} \!\x1 \:\!\right) \:\: ,
 \\[2mm]
\label{KppL2}
 K_{\, \phi\phi}^{\,(2)}(x) &\!\! = \!&
      \ln^{\,2} \x1 \, \* 2 \* \pgg(x) \* \left[
      2\, \* \ca \* \beta_0^{\,2} \:+\: 12\, \* \cas\, \* \beta_0\, \* \H(0)
      + 32\, \* \cat\: \* \Hh(0,0) \right]
%
     \;+\; {\cal O} \! \left( \:\!\ln \x1 \:\!\right)
 \:\: .
\eea
Note that the additional factor of 2 in the latter two equations arises from 
a mismatch, in terms of the normalization of the leading $\x1^{-1}$ terms, of
the definitions of $\pqq(x)$ and $\pgg(x)$ in Eqs.~(\ref{eq:p0gqgg}) and 
(\ref{eq:p0qqqg}). These results imply that even the $L_1^{\,n} f(x)$ 
contributions to the pure-singlet quark and gluon kernels (recall the 
discussion in the paragraph above Eq.~(\ref{eq:cphiq2L3})) vanish to at least
$n=2$,
\beq
\label{KdiagL}
 K_{\,\rm 22, ps }^{\,(n)} \Big|_{L_1^{\,n} f(x)}    \; = \;
 K_{\,\phi\phi }^{\,(n)} \Big|_{\cf L_1^{\,n} f(x) } \; = \; 0
\:\: .
\eeq

The above behaviour of the physical evolution kernels is in striking contrast 
to that of the \MSb-scheme splitting functions, which do include 
double-logarithmic contributions, $\ln^{\,k}\! \x1$ with $k > n \geq 1$ at 
N$^{\:\!n}$LO, at all orders in the expansion at $x=1$.
At NLO these contributions, and the single-logarithmic terms of the
diagonal splitting functions, are given by
\bea
\label{Pqg1DL}
  P_{\rm qg}^{\,(1)}(x) &\!\! = \!&
%
    \ln^{\,2}\! \x1 \: \* \caf \* \nf \: \* \{\, -\,4\,\* \pqg(x) \} 
    \; + \; {\cal O}(\ln \x1)
  \:\: ,
\\[1mm]
\label{Pgq1DL}
 P_{\rm gq}^{\,(1)}(x) &\!\! = \!& 
%
    \ln^{\,2}\! \x1 \: \* \caf \* \cf \: \* \{\, +\,4\,\* \pgq(x) \} 
    \; + \; {\cal O}(\ln \x1)
\eea
with $\caf \,\equiv\, \ca - \cf$, and
\bea
\label{Pqq1SL}
P_{\rm qq}^{\,(1)}(x) &\!\! = \!&
%
  \ln \x1 \: \* \cfs \; \* \{\; -\,8\,\* \pqq(x) \}\; \* \H(0)
    \; + \; {\cal O}(\ln \x1^{0})
 \:\: ,
\\[1mm]
\label{Pgg1SL}
 P_{\rm gg}^{\,(1)}(x) &\!\! = \!&
%
  \ln \x1 \: \* \cas \: \* \{\, -\,16\,\* \pqq(x) \}\, \* \H(0)
    \; + \; {\cal O}(\ln \x1^{0}) 
 \:\: .
\eea

\noindent
The corresponding NNLO results read 
\bea
\label{Pqg2DL}
  P_{\rm qg}^{\,(2)}(x) &\!\! = \!&
%
   \ln^{\,4}\! \x1 \: \* \cafs\, \* \nf\: \* 4/3\: \* \pqg(x) 
   \nn \\ & & \mbox{\hspn}
 + \ln^{\,3}\! \x1 \* \Big[ \, \cafs\, \* \nf \:\* 
   \{ 
        - (34
         +16/x
         -92\, \* x
         +64\, \* x^2\,) / 9
         - 8/3\: \* ( 1 + 4\, \* x )\, \* \H(0)
   \} 
   \nn \\ & & \mbox{}
   + \, \caf \* \cf\, \* \nf \:\* 
   \{
         -40/9
         -16/(9\* x)
         -28/9\: \* x
         +80/9\: \* x^2
         + (4 - 24\, \* x + 32/3\: \* x^2\, )\, \* \H(0)
   \}
   \nn \\ & & \mbox{}
   + \, \caf\* \nfs\: \* 4/9\: \* \pqg(x) 
   \Big]
   \; + \; {\cal O}(\ln^{\,2}\! \x1)
  \:\: ,
\\[1mm]
\label{Pgq2DL}
 P_{\rm gq}^{\,(2)}(x) &\!\! = \!&
%
      \ln^{\,4}\! \x1 \: \* \cafs \* \cf\: \* 4/3\: \* \pgq(x) 
   \nn \\[0.5mm] & & \mbox{\hspn}
      + \ln^{\,3}\! \x1 \: \* \Big[ \, \cafs \* \cf\, \: \* 
   \{ \,
         ( 344/x
         - 316
         + 170\, \* x
         - 16\, \* x^2\, ) / 9
         - 8/3\: \* (4/x - 8 + x)\: \* \H(0)
   \} 
   \nn \\ & & \mbox{}
      + \caf \* \cfs\, \: \*
   \{ 
           128/(9 \* x)
         - 76/9
         - 16/9\: \* (x + x^2\,)
         - ( 32/(3 \* x) - 24 + 4\, \* x)\, \* \H(0)
   \} 
   \nn \\[0.5mm] & & \mbox{}
      + \caf \* \cf\, \* \nf\, \* (-20/9\: \* \pgq(x))
   \Big]
   \; + \; {\cal O}(\ln^{\,2}\! \x1)
\eea
and
\bea
\label{Pqq2DSL}
  P_{\rm qq}^{\,(2)}(x) &\!\! = \!&
%
  \ln^{\,3}\! \x1 \* \Big[ \, \caf \* \cf \* \nf \: \*
   \{
         - 32/(9\, \* x\,)
         - 8/3
         + 8/3\: \*  x\,
         + 32/9\: \*  x^2
         - 16/3\: \*  (1\,
         +  x\,)\, \*  \H(0)
   \}
        \Big]
   \nn \\ & & \mbox{\hspn}
+  \ln^{\,2}\! \x1 \* \Big[ \, \cfs\, \* \nf \: \*
   \{
           2/3\: \*  (16/x
         + 7
         - 7\, \*  x\,
         - 16\, \*  x^2\,)
         + 8\, \*  (3
         + 4\, \*  x
         + 4/3\: \*  x^2\,)\, \*  \H(0)
   \}
   \nn \\ & & \mbox{}
+  \cf \* \ca \* \nf \: \*
   \{
         - 8/(9 \* x)
         - 6
         + 6\, \*  x\,
         + 8/9\: \*  x^2
         + 4/3\: \*  (4/x
         - 11
         - 17\, \*  x
         - 4\, \*  x^2\,)\, \*  \H(0)
    \}
   \nn \\[1.0mm] & & \mbox{}
+  \cf\, \* \nfs \: \*
   \{
           16/(9 \* x)
         + 4/3
         - 4/3\: \*  x
         -16/9\: \*  x^2
         + 8/3\: \*  (1 + x)\, \*  \H(0)
   \}
   \nn \\ & & \mbox{}
+  \caf\, \* \cf\, \* \nf \: \*
   \{ 
      8\, \* (1 + x) \, \* ( \Hh(0,0) - 2\,\* \THh(1,0) )
   \}
       \Big]
  \: + \: \,\cft\; \* 32\, \* \pqq(x)\, \* \H(0,0)
   \nn \\ & & \mbox{\hspn} 
  \; + \; {\cal O}(\ln \x1)
  \:\: ,
 \\[3mm]
\label{Pgg2DSL}
  P_{\rm gg}^{\,(2)}(x) &\!\! = \!&
%
   \ln^{\,3}\! \x1 \* \Big[ \, \caf \* \cf\, \* \nf \: \*
   \{
           32/(9\, \* x\,)
         + 8/3
         - 8/3\: \*  x
         - 32/9\: \*  x^2
         + 16/3\: \*  (1 +  x)\, \*  \H(0)
   \}
        \Big]
   \nn \\ & & \mbox{\hspn}
 +  \ln^{\,2}\! \x1 \* \Big[ \, \cfs\, \* \nf \: \*
   \{
         - 2/3\: \*  (16/x
         + 7
         - 7\, \*  x
         - 16\, \*  x^2\,)
         - 8\, \*  (3
         + 4\, \*  x
         + 4/3\: \*  x^2\,)\, \*  \H(0)
   \}
   \nn \\ & & \mbox{}
+  \cf \* \ca\, \* \nf \: \*
   \{
           8/(9\, \* x\,)
         + 6
         - 6\, \*  x
         - 8/9\: \*  x^2
         - 4/3\: \*  (4/x
         - 11
         - 17\, \*  x
         - 4\, \*  x^2\,)\, \*  \H(0)
   \}
   \nn \\[1.0mm] & & \mbox{}
+  \cf\, \* \nfs \: \*
   \{
         - 16/(9\, \* x)
         - 4/3
         + 4/3\: \*  x
         +16/9\: \*  x^2
         - 8/3\: \*  (1 + x) \*  \H(0)
   \}
   \nn \\ & & \mbox{}
+  \caf\, \* \cf\, \* \nf \: \*
   \{
      - 8\, \* (1 + x) \, \* ( \Hh(0,0) - 2\,\* \THh(1,0) )
   \}
       \Big]
  \: + \: \,\cat\; \* 64\, \* \pgg(x)\, \* \H(0,0)
   \nn \\ & & \mbox{\hspn}
  \; + \; {\cal O}(\ln \x1)
  \:\: .
\eea
The respective final terms of the last two equations represent the non-singlet
contributions directly corresponding to the final terms in Eqs.~(\ref{K22L2})
and (\ref{KppL2}).

Eqs.~(\ref{Pqg1DL}) -- (\ref{Pgg2DSL}) show a couple of interesting features:
The double-logarithmic contributions, unlike the single-logarithmic terms, all
vanish for the case $\cf = \ca$ which forms a subset of the colour factor 
conditions leading to an ${\cal N} \!=\! 1$ supersymmetric theory. In fact, the
coefficients of the leading, next-to-leading etc double logarithms successively
include terms of the form \mbox{$(\ca - \cf)^{\,n-1}$}, $(\ca - \cf)^{\,n-2}$ 
etc. Moreover, to the accuracy of Eqs.~(\ref{Pqq2DSL}) and (\ref{Pgg2DSL}),
the pure-singlet parts of $P_{\,\rm qq}^{\,(2)}$ and $P_{\,\rm gg}^{\,(2)}$ 
turn out to be completely identical up to an overall sign.

Both the single logarithmic enhancement of the physical kernels also in the
present flavour-singlet case, and the above features are non-trivial (and not
obviously related) properties. In the non-singlet case the former property is
known to hold also at N$^3$LO \cite{MV5}. Recall also that it is implied 
\cite{NV3} for the leading $\x1^{-1}$ by the soft-gluon exponentiation 
\cite{SoftGlue}, and established at all powers of $\x1$ in the large-$\nf$ 
limit by Refs.~\cite{Gracey:1994nn,Mankiewicz:1997gz}. 

Hence it appears very natural to conjecture that also Eqs.~(\ref{KoffdL}) and
Eqs.~(\ref{KdiagL}) are valid at N$^3$LO as well. As the three-loop coefficient
functions are known from Refs.~\cite{MVV6} and Section~3, this implies that the
known double logarithms originating from the matrix (\ref{eq:F2Htrf}) in 
Eq.~(\ref{eq:Pexp}) at order $\as^{\,4}$ have to be compensated by the 
(hitherto unknown) corresponding contributions to the four-loop splitting 
functions $P^{\,(3)}$. These terms can therefore be predicted from the above
conjecture, and checked against the colour-factor pattern expected from 
Eqs.~(\ref{Pqg1DL}) -- (\ref{Pgg2DSL}).

For the off-diagonal splitting functions these predictions are found to be
\bea
\label{Pqg3DL}
  P_{\rm qg}^{\,(3)}(x) &\!\! = \!&
%
  \ln^{\,5}\! \x1 \* \Big[ \, \caft\, \* \nf \:\*
   \{
          32/(27\* x)
         + 46/27
         + 52/27\: \* x
         - 4\, \* x^2
         + 16/9\: \* ( 1 + 4\, \* x)\, \* \H(0)
   \}
   \nn \\ & & \mbox{}
      + \cafs \* \cf\, \* \nf \: \*
   \{
          32/( 27 \* x)
         + 58/27
         + 100/27\: \* x
         - 68/9\: \* x^2
         + 8/9\: \* (1 + 10\, \* x)\, \* \H(0)
   \}
   \nn \\ & & \mbox{}
      - \cafs\, \* \nfs\: \* \{\, 4/27\: \* \pqg(x) \}
      \Big]
   \nn \\ & & \mbox{\hspn}
      + \ln^{\,4}\! \x1 \: \* \Big[ \, \caft\, \* \nf\, \: \*
    \{
          152/(27\* x)
         - 106/9
         + 1178/27\: \* x
         - 719/27\: \* x^2
   \nn \\ & & \mbox{\hspp}
         - 2/9\: \* ( 8/(3\* x) - 1 + 16\, \* x - 1114/3\: \* x^2)\: \* \H(0)
         - 80/9\: \* ( 1 + 2\* x + 2\* x^2\,)\: \* \THh(-1,0)
   \nn \\[1mm] & & \mbox{\hspp}
         - (104/9 + 160/9\: \* x ) \: \* \H(0,0)
         + 8/9\: \* (11 + 38\, \* x + 2 \* x^2\, )\, \* \THh(1,0)
         - 80/9\: \* \z2\, \* \pqg(x)
    \} 
   \nn \\[1.5mm] & & \mbox{}
+ \, \cafs \* \cf\, \* \nf\, \: \* 
    \{
         - 176/(81\,\*x)
         + 685/162
         - 1426/81\: \* x
         + 1166/27\: \* x^2
   \nn \\[0.5mm] & & \mbox{\hspp}
         - 1/27\: \* (32/x
         - 71
         + 742\, \* x
         - 2156\, \* x^2\, )\,  \* \H(0)
         - 80/9\: \* (1
         + 2\, \* x
         + 2\, \* x^2\, )\,  \* \THh(-1,0)
   \nn \\[0.5mm] & & \mbox{\hspp}
         + 8/9\: \* (31
         - 2\, \* x
         + 52\, \* x^2\, )\,  \* \THh(1,0)
         + 16/9\: \* (7\, \* x
         + 10\, \* x^2\, )\,  \* \Hh(0,0)
         - 8\, \* \z2\, \* \pqg(x)
    \}
   \nn \\[1.5mm] & & \mbox{}
+ \, \cafs \* \nfs\, \: \* 
    \{
           8/(27\,\*x)
         - 98/81
         + 184/81\: \* x
         - 226/81\: \* x^2
         + 4/9\: \* (1 + 4\, \* x )\,  \* \H(0)
    \}
   \nn \\[1.5mm] & & \mbox{}
+ \, \caf \* \cfs\, \* \nf\, \: \* 
    \{
         - 632/(81\,\*x)
         - 877/81
         + 1349/81\: \* x
         + 49/27\: \* x^2
   \nn \\[0.5mm] & & \mbox{\hspp}
         - 1/27\: \* (16/x
         + 211
         + 28\, \* x
         + 552\, \* x^2\, )\,  \* \H(0)
         - (10\,
         - 220/3\: \* x
         + 80/3\: \* x^2\, )\,  \* \Hh(0,0)
    \}
   \nn \\[1.5mm] & & \mbox{}
+ \, \caf \* \cf\, \* \nfs\, \: \*
    \{
         - 40/(81\,\*x)
         + 719/81
         - 1060/81\: \* x
         + 398/81\: \* x^2
   \nn \\[0.5mm] & & \mbox{\hspp}
         + 2/27\: \* (29
         + 50\, \* x
         - 52\, \* x^2\, )\,  \* \H(0)
         + 8/9\: \* (2 - 29\, \* x )\,  \* \Hh(0,0)
    \}
   \nn \\ & & \mbox{}
- \, \caf \* \nft\, \: \*
    \{
         4/81\: \* \pqg(x)
    \}
      \Big]
  \; + \; {\cal O}(\ln^{\,3} \! \x1)
\eea
and
\bea
\label{Pgq3DL}
  P_{\rm gq}^{\,(3)}(x) &\!\! = \!&
%
  \ln^{\,5}\! \x1 \* \Big[ \, \caft\, \* \cf \: \*
      \{
          (292/x
         - 236
         + 46\, \*  x
         - 32\, \*  x^2\,)\, / 27
         + 16/9\: \*(4 +  x )\, \*  \H(0)
      \}
   \nn \\ & & \mbox{}
+ \, \cafs\, \* \cfs\, \: \* 
      \{
           (148/x
         - 44
         - 86\, \*  x
         - 32\, \*  x^2\,)\, / 27
         + 8/9\: \* (10 + x )\, \*  \H(0)
     \}
   \nn \\ & & \mbox{}
- \, \cafs\, \* \cf\, \* \nf\, \: \* 
     \{
           4/27\: \*  \pgq(x)
     \}
         \Big]
   \nn \\ & & \mbox{}
 + \ln^{\,4}\! \x1 \* \Big[ \, \caft\, \* \cf \: \*
      \{
           3029/(81\, \*  x)
         - 428/81
         + 295/81\: \*  x
         + 128/27\: \*  x^2
   \nn \\ & & \mbox{\hspp}
         - (1168/x
         - 2300
         - 68\, \*  x
         + 176\, \*  x^2\,) /27\: \*  \H(0)
         + (80/(3\, \*  x) - 400/9 + 16\, \*  x)\, \*  \Hh(0,0)
   \nn \\[1mm] & & \mbox{\hspp}
         + 80/9\: \*  (2/x + 2 + x)\, \*  \THh(-1,0)
         + 8/9\: \*  (26/x + 14 + 23 \*  x)\, \*  \THh(1,0)
         + 16/9\: \*  \z2\, \*  \pgq(x)
      \}
   \nn \\[1.5mm] & & \mbox{}
+ \, \cafs\, \* \cfs\, \: \* 
      \{
           7142/(81\, \*  x)
         - 1418/27
         - 505/54\: \*  x
         + 736/81\: \*  x^2
         - (1472/(27\, \*  x)
   \nn \\[0.5mm] & & \mbox{\hspp}
         - 1022/9
         - 89/9\: \*  x
         + 64/9\: \*  x^2\,)\, \*  \H(0)
         + (160/(3 \*  x) - 128 + 280/9\: \*  x)\, \*  \Hh(0,0)
   \nn \\[0.5mm] & & \mbox{\hspp}
         + 80/9\: \*  (2/x + 2 + x)\, \*  \THh(-1,0)
         + 8/9\: \*  (52/x - 2 + 31\, \*  x)\, \*  \THh(1,0)
         - 8\, \*  \z2\, \*  \pgq(x)
      \}
   \nn \\[1.5mm] & & \mbox{}
+ \, \cafs\, \* \cf\, \* \nf \: \* 
      \{
         - 2290/(81\, \*  x)
         + 656/27
         - 70/9\: \*  x
         + 184/81\: \*  x^2
   \nn \\[0.5mm] & & \mbox{\hspp}
         + (8/(3 \*  x) - 440/27 - 56/27\: \*  x)\, \*  \H(0)
      \}
   \nn \\[1.5mm] & & \mbox{}
+ \, \caf\, \* \cft\, \: \* 
      \{
           47/(3\, \*  x)
         - 1855/81
         + 185/81\: \*  x
         + 352/81\: \*  x^2
   \nn \\[0.5mm] & & \mbox{\hspp}
         - (304/x
         - 424
         + 17\, \*  x
         + 16\, \*  x^2\,) /27\: \*  \H(0)
         + (80/(3\, \*  x) - 220/3 + 10\, \*  x)\, \*  \Hh(0,0)
      \}
   \nn \\[1.5mm] & & \mbox{}
+ \, \caf \* \cfs\, \* \nf \: \* 
      \{
         - (1474/x
         - 1460
         + 217\, \*  x
         - 248\, \*  x^2\,) / 81 
   \nn \\[1.5mm] & & \mbox{\hspp}
         + (8/x
         - 532
         + 62\, \*  x\,) / 27\:  \*  \H(0)
         + 16/9\: \*(2 -  x)\, \*  \Hh(0,0)
      \}
   \nn \\[1.5mm] & & \mbox{}
+ \, \caf\, \* \cf\, \* \nfs\, \: \* 
      \{
           32/81\: \*  \pgq(x)
      \}
          \Big]
  \; + \; {\cal O}(\ln^{\,3} \! \x1)
  \:\: .
\eea
Here, and in Eqs.~(\ref{Pqq3DSL}) and (\ref{Pgg3DSL}) below, we have employed
the modified basis of Ref.~\cite{MV5} for the harmonic polylogarithms, in which 
all logarithms and $\zeta$-functions are made explicit in the expansion at 
$x=1$ to all orders. The corresponding functions entering the above expression 
are
\beq
\label{HPLmod1}
  \THh(-1,0)(x) \; = \; \Hh(-1,0)(x) \:+\: \z2/2 \;\; , \quad
  \THh(1,0)(x)  \; = \; \Hh(1,0)(x) \:+\: \z2
\eeq
and, e.g., $\H(0,1)(x)$ has been expressed by
\beq
\label{HPLmod2}
  \H(0,1)(x) \; = \; - \,\THh(1,0)(x) \:-\: \ln \x1 \: \H(0)(x) \:-\: \z2
\:\: .
\eeq

The corresponding predictions for the four-loop pure-singlet splitting 
functions (which, as expected, are suppressed by another power of $\x1$ for
$x \ra 1$) read
\bea
\label{Pqq3DSL}
  P_{\rm qq,ps}^{\,(3)}(x) &\!\! = \!&
%
  \ln^{\,4}\! \x1\, \* \Big[\, \cafs \* \cf\, \*  \nf\, \*
      \{
         - 728/(81\, \*  x)
         - 602/27
         + 122/27\: \*  x
         + 2168/81\: \*  x^2
   \nn \\ & & \mbox{\hspp}
         + 4/27\: \*  (40/x
         - 121
         - 241\, \*  x
         - 40\, \*  x^2\, )\, \*  \H(0)
         + 80/9\: \*  (1 + x )\, \*  \Hh(0,0)
      \}
   \nn \\[1.5mm] & & \mbox{}
+ \,  \caf\,  \* \cfs\,  \* \nf\,  \: \* 
      \{
         - 248/(81\, \*  x)
         - 62/27
         + 62/27\: \*  x
         + 248/81\: \*  x^2
   \nn \\[0.5mm] & & \mbox{\hspp}
         + 4/27\: \*  (40/x
         - 1
         - 61\, \*  x
         - 40\, \*  x^2\, )\, \*  \H(0)
         + 160/9\: \*  (1 + x )\, \*  \Hh(0,0)
      \}
   \nn \\ & & \mbox{}
+ \,  \caf  \* \cf\,  \* \nfs\,  \: \* 
      \{
           176/(81\, \*  x)
         + 44/27\: \* ( 1 - x)
         - 176/81\: \*  x^2 + 88/27\: \*  (1 + x )\, \*  \H(0)
      \}
               \Big]
   \nn \\ & & \mbox{\hspn}
+  \ln^{\,3}\! \x1\, \* \Big[ \,  \cft\,  \* \nf \: \*
      \{
           1072/(27\, \*  x)
         + 610/9
         - 2182/27\: \*  x
         - 80/3\: \*  x^2
   \nn \\ & & \mbox{\hspp}
         - (320/(27\,\*  x)\:
         + 80/9\: \* ( 1 - x)
         - 320/27\: \*  x^2)\, \* \z2
         + (2252/27
         - 884/27\: \*  x
   \nn \\[0.5mm] & & \mbox{\hspp}
         - 2560/27\: \*  x^2
         - 160/9\: \* \z2\, \* (1 + x) )\, \* \H(0)
         - 8/3\: \* (5
         + 25/3\: \*  x
         + 152/9\: \*  x^2\, )\, \*  \Hh(0,0)
   \nn \\[0.5mm] & & \mbox{\hspp}
         + (1216/(27\, \*  x)
         - 112/3
         - 1264/9\: \*  x
         - 1216/27\: \*  x^2\, )\, \*  \THh(1,0)
   \nn \\[0.5mm] & & \mbox{\hspp}
         - 320/9\: \*  (1 + x )\, \*  \THhh(1,0,0)
         + 32\, \*  (1 + x )\, \*  \THhh(0,1,0)
         + 560/9\: \*  (1 + x )\, \*  \Hhh(0,0,0)
      \}
   \nn \\[1.5mm] & & \mbox{}
+ \,  \cfs\,  \* \ca\,  \* \nf\,  \: \* 
      \{
         - 24232/(243\, \*  x)
         + 8954/81
         - 16730/81\: \*  x
         + 47560/243\: \*  x^2
   \nn \\[0.5mm] & & \mbox{\hspp}
         + (64/(27\, \*  x)\:
         + 16/9\: \* ( 1  - x)
         - 64/27\: \*  x^2)\, \* \z2
         - (4864/(81\, \*  x)
         + 2584/27
   \nn \\[0.5mm] & & \mbox{\hspp}
         - 4952/27\: \*  x
         - 6944/81\: \*  x^2
         - 32/9\: \* (1 + x)\, \* \z2 )\, \* \H(0)
         - 16/27\: \*  (142 
   \nn \\[0.5mm] & & \mbox{\hspp}
           + 301\, \*  x )\, \*  \Hh(0,0)
         + (640/(27\, \*  x)
         - 160/9\: \* ( 1 + x )
         + 640/27\: \*  x^2\, )\, \*  \THh(-1,0)
   \nn \\[0.5mm] & & \mbox{\hspp}
         - 16/27\: \*  (88/x
         - 205
         - 457\, \*  x
         - 88\, \*  x^2\, )\, \*  \THh(1,0)
         - 320/9\: \*  (1 - x )\, \*  \THhh(0,-1,0)
   \nn \\[0.5mm] & & \mbox{\hspp}
         - 128/3\: \*  (1 + x )\, \*  \THhh(0,1,0)
         - (176/3\: - 208/9\: \*  x )\, \*  \Hh(0,0,0)
      \}
   \nn \\[1.5mm] & & \mbox{}
+ \,  \cf\,  \* \cas\,  \* \nf\,  \: \* 
      \{
           1580/(27\, \*  x)
         - 13028/81
         + 23804/81\: \*  x
         - 1724/9\: \*  x^2
   \nn \\[0.5mm] & & \mbox{\hspp}
         + (256/(27\, \*  x)
         + 64/9\: \* ( 1 - x ) 
         - 256/27\: \*  x^2)\, \* \z2
         + (2720/(81\, \*  x)
         + 632/27
   \nn \\[0.5mm] & & \mbox{\hspp}
         - 3656/27\: \*  x
         + 3872/81\: \*  x^2
         + 128/9\: \* (1 + x)\, \* \z2 ) \, \*  \H(0)
         - (832/(27\, \*  x)
         - 1376/27
   \nn \\[0.5mm] & & \mbox{\hspp}
         - 4064/27\: \*  x
         - 128/9\: \*  x^2\, )\, \*  \Hh(0,0)
         - 160/9\: \* (4/(3 \*  x)\:
         - 1 - x 
         + 4/3\: \*  x^2\, )\, \*  \THh(-1,0)
   \nn \\[0.5mm] & & \mbox{\hspp}
         + (64/(9\, \*  x)\:
         - 2272/27\:
         - 3520/27\: \*  x
         - 64/9\, \*  x^2\, )\, \*  \THh(1,0)
         + 32/3\: \*  (1 + x )\, \*  \THhh(0,1,0)
   \nn \\[0.5mm] & & \mbox{\hspp}
         + 320/9\: \*  (1 - x )\, \*  \THhh(0,-1,0)
         + 320/9\: \*  (1 + x )\, \*  \THhh(1,0,0)
         - (32/9 + 256/3\: \*  x )\, \*  \Hh(0,0,0)
      \}
   \nn \\[1.5mm] & & \mbox{}
+ \,  \cfs\,  \* \nfs\,  \: \* 
      \{
         - (1504/x
         + 1820
         - 860\, \*  x
         - 2464\, \*  x^2) / 81
         - (368/9\:
         + 160/3\: \*  x
   \nn \\[0.5mm] & & \mbox{\hspp}
         + 704/81\: \*  x^2\, )\, \*  \H(0)
         + 16/27\: \*  (1 + x )\, \*  \Hh(0,0)
         - 352/27\: \*  (1 + x )\, \*  \THh(1,0)
      \}
   \nn \\[1.5mm] & & \mbox{}
+ \,  \cf\,  \* \ca\,  \* \nfs  \: \* 
      \{
           3280/(243\, \*  x)
         + 1472/81
         - 512/81\: \*  x
         - 6160/243\: \*  x^2
   \nn \\[0.5mm] & & \mbox{\hspp}
         - (704/(81\, \*  x)
         - 832/27
         - 448/9\: \*  x
         - 704/81\: \*  x^2\, )\, \*  \H(0)
   \nn \\[0.5mm] & & \mbox{\hspp}
         - 32/27\: \*  (13 + 10\, \*  x )\, \*  \Hh(0,0)
         + 352/27\: \*  (1 + x )\, \*  \THh(1,0)
      \}
   \nn \\[1.5mm] & & \mbox{}
+ \,  \cf\,  \* \nft\,  \: \* 
      \{
         - 128/(81\, \*  x)\:
         - 32/27\: \* (1 - x)
         + 128/81\: \*  x^2
         - 64/27\: \*  (1 + x )\, \*  \H(0)
      \}
        \Big]
   \nn \\[1mm] & & \mbox{\hspn}
  \; + \; {\cal O}(\ln^{\,2} \! \x1)
\eea
and, in perfect agreement with the NNLO observation below Eq.~(\ref{Pgg2DSL}),
\beq
\label{Pgg3DSL}
  \left. P_{\rm gg}^{\,(3)}(x) \right|_{\cf} \;\; = \;\;
    - \, P_{\rm qq, ps}^{\,(3)}(x) 
  \; + \; {\cal O} \!\left(\ln^{\,2} \! \x1 \, f(x) \right) 
\;\; . \qquad\qquad
\eeq
Eq.~(\ref{Pqq3DSL}) includes three more modified basis functions,
\bea
  \THhh(0,-1,0)(x) & = & \Hhh(0,-1,0)(x) \:+\: \H(0)(x)\: \z2/2 \:+\: 3\,\z3/2
\:\: , \nn \\[1mm]
  \THhh(0,1,0)(x)\; & = & \Hhh(0,1,0)(x) \:+\: \H(0)(x)\: \z2 \:+\: 2\,\z3
\:\: , \nn \\[1mm]
  \THhh(1,0,0)(x)\; & = & \Hhh(1,0,0)(x) \:-\: \z3
\:\: .
\eea
 
Both the full quark-quark and gluon-gluon splitting functions differ from the 
above results by a term with one unknown coefficient, respectively given by 
 $ \,\xi_{P^{(3)}} \cff \ln^{\,3} \! \x1 \,    \pqq(x)\, \H(0,0,0)(x)$ and 
 $ \,\xi_{P^{(3)}} C_A^{\,4} \ln^{\,3} \! \x1 \, 2\,\pgg(x)\, \H(0,0,0)(x)$,
which directly enters the large-$x$ behaviour of the diagonal physical kernels,
given by \cite{MV5}
\beq
 K_{\,22}^{\,(3)} \Big|_{L_1^3 \, f(x)} \; = \; 
      \; \pqq(x) \* \left[
      - 2\,\cf \* \beta_0^{\,3} \:-\: 44/3\: \cfs\,\beta_0^{\,2}\: \H(0)
      - 64\,\cft\,\beta_0\: \Hh(0,0) + \xi_{P_3}^{}\, \cff\, \Hhh(0,0,0)
      \right]
\; , \;
\eeq
and correspondingly
\beq
  K_{\,\phi\phi}^{\,(3)} \Big|_{L_1^3 \, f(x)} \; = \;
      2 \pgg(x) \* \left[
      - 2\,\ca \* \beta_0^{\,3} \:-\: 44/3\: \cas\,\beta_0^{\,2}\: \H(0)
      - 64\,\cat\,\beta_0\: \Hh(0,0) + \xi_{P_3}^{}\, C_A^{\,4}\, \Hhh(0,0,0)
      \right]
\; . \;
\eeq

Obviously the N$^3$LO predictions (\ref{Pqg3DL}) and (\ref{Pgq3DL}), and the 
double-logarithmic part $\sim \ln^{\,4} \! \x1$ of Eq.~(\ref{Pqq3DSL}), exactly
follow the colour-factor pattern expected from the NLO and NNLO results (as
discussed below Eq.~(\ref{Pgg2DSL})). In our view this represents, for the time
being, an acceptably strong check of these predictions and thus of the 
underlying conjecture about the four-loop physical kernel matrix for the 
structure functions ($\,\Ftwo, F_\phi\,$). Furthermore we expect that a 
formal proof of this property, presumably first for subleading the 
$\x1^{0} \ln^{\,k}\! \x1$ terms corresponding to $N^{\,-1} \ln N$ contributions
in Mellin-$N$ space, can be obtained using, for instance, the recent 
path-integral approach of Ref.~\cite{Laenen:2008gt} or soft-collinear
effective theory (SCET) \cite{SCET}. For other recent research into 
$N^{\,-1} \ln N$ contributions to hard processes see Refs.~\cite{OneoverN}.

In fact, the $\x1^{0} \ln^{\,k}\! \x1$ contributions to the four-loop 
off-diagonal splitting function form the part of the above predictions which 
can be expected to become phenomenologically relevant first. These terms can
be obtained from Eqs.~(\ref{Pqg3DL}) and (\ref{Pgq3DL}) by removing all
$\widetilde{\rm H}$-functions (the expansions of which at $x=1$ generally start
with $\x1^m$ with $m \geq 1$) and setting $x=1$ in the remaining terms. For
the convenience of the reader we finally write down the resulting expressions,
\bea
\label{Pqg3DL0}
  P_{\rm qg}^{\,(3)}(x) &\! = \!&
%
     \ln^{\,6}\! \x1 \: \cdot \, 0
\nn \\[1mm] & & \mbox{\hspn}
     + \ln^{\,5}\! \x1 \* \Big[\,
            {22 \over 27}\: \* \caft \* \nf
          \,- \, {14 \over 27}\: \* \cafs \* \cf \* \nf
          \,- \, {4 \over 27}\: \* \cafs \* \nfs
          \Big]
\nn \\[1mm] & & \mbox{\hspn}
     + \ln^{\,4}\! \x1 \* \Big[\,
            \Big( \, {293 \over 27}
            \,-\, {80 \over 9}\: \* \z2\! \Big)\, \* \caft \* \nf
          \,+\, \Big( \, {4477 \over 162}
          \,-\, 8 \* \z2\! \Big)\, \* \cafs \* \cf \* \nf
\nn \\[1mm] & & \mbox{}
          \,-\, {13 \over 81}\: \* \caf \* \cfs \* \nf
          \,-\, {116 \over 81}\: \* \cafs \* \nfs
          \,+\, {17 \over 81}\: \* \caf \* \cf \* \nfs
          \,-\, {4 \over 81}\: \* \caf \* \nft
          \Big]
\nn \\[2mm] & & \mbox{\hspn}
      + {\cal O} \left( \ln^3 \! \x1 \right)
\:\: , \\[4mm] 
\label{Pgq3DL0}
  P_{\rm gq}^{\,(3)}(x) &\! = \!&
%
     \ln^{\,6}\! \x1 \: \cdot \, 0
\nn \\[1mm] & & \mbox{\hspn}
     + \ln^{\,5}\! \x1 \* \Big [ \:
            {70 \over 27}\: \* \caft \* \cf
          \,-\, {14 \over 27}\: \* \cafs \* \cfs 
          \,-\, {4 \over 27}\: \* \cafs \* \cf \* \nf
          \Big]
\nn \\[1mm] & & \mbox{\hspn}
     + \ln^{\,4}\! \x1 \* \Big [ \,
            \Big( \, {3280 \over 81}
            \,+\, {16 \over 9}\: \* \z2\! \Big)\, \* \caft \* \cf
          \,+\, \Big( \, {637 \over 18}
          \,-\, 8 \* \z2\! \Big)\, \* \cafs \* \cfs 
\nn \\[1mm] & & \mbox{}
          \,-\, {49 \over 81}\: \* \caf \* \cft 
          \,-\, {256 \over 27}\: \* \cafs \* \cf \* \nf
          \,+\, {17 \over 81}\: \* \caf \* \cfs \* \nf
          \,+\, {32 \over 81}\: \* \caf \* \cf \* \nfs
          \Big]
\nn \\[2mm] & & \mbox{\hspn}
      + {\cal O} \left( \ln^3 \! \x1 \right)
 \:\: .
\eea
 
\begin{figure}[p]
\vspace*{-1mm}
\centerline{\epsfig{file=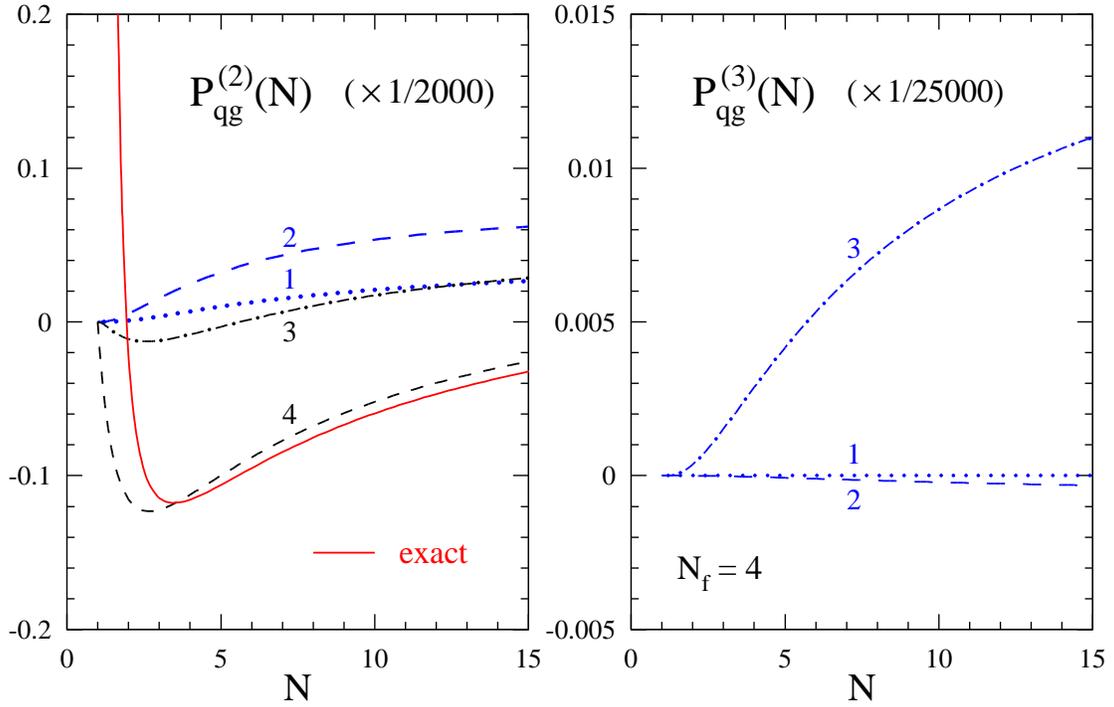,width=15.0cm,angle=0}}
\vspace{-2mm}
\caption{\label{pic:fig12}
 Large-$N$ expansions of the NNLO and N$^3$LO gluon-quark splitting functions
 $P_{\,\rm qg}^{\,(2)}(N)$ and $P_{\,\rm qg}^{\,(3)}(N)$, in the former case
 compared with the exact result of Ref.~\cite{MVV4}.
 The curves `1', `2', etc are successively including the leading logarithms,
 $N^{-1}\ln^{\,4} N$ (left) and $N^{-1}\ln^{\,6} N$ (right),
 next-to-leading logarithms, $N^{-1}\ln^{\,3} N$ (left) and $N^{-1}\ln^{\,5} N$
 (right), etc. All log-enhanced terms are shown at NNLO. Only the
 double-logarithmic contributions obtained in this article are~available at
 N$^3$LO.
 }
\end{figure}
\begin{figure}[p]
\centerline{\epsfig{file=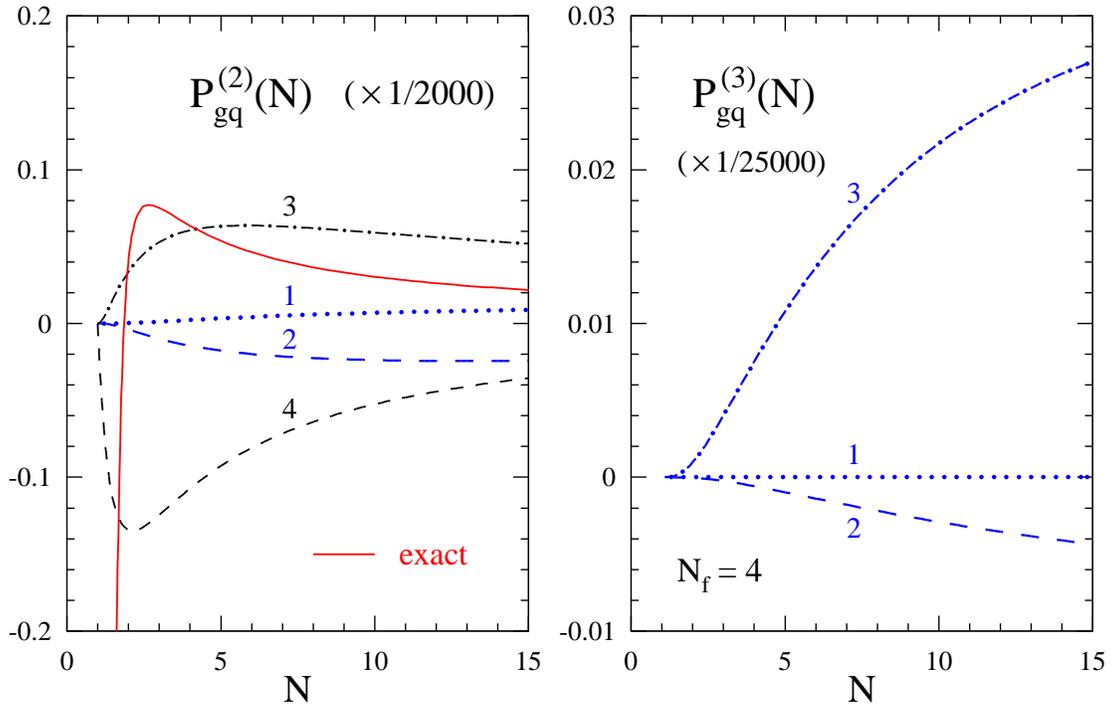,width=15.0cm,angle=0}}
\vspace{-2mm}
\caption{\label{pic:fig13}
 As Fig.~\ref{pic:fig12}, but for the NNLO (left) and N$^3$LO (right)
 quark-gluon splitting functions.
 }
\vspace{-1mm}
\end{figure}
\noindent
The only surprise in these results and Eqs.~(\ref{Pqg3DL}) and (\ref{Pgq3DL}) is
the absence, indicated explicitly~in Eqs.~(\ref{Pqg3DL0} and (\ref{Pgq3DL0}),
of the expected leading $\ln^{\,6} \! \x1$ terms. In general the -- related, 
recall Eqs.~(5.7/8) and Eqs.~(5.11/12) -- coefficients of the leading 
$\ln^{\,2n} \! \x1$ terms of $P_{\rm qg}^{\,(3)}$ and $P_{\rm gq}^{\,(3)}$
receive contributions with both signs from Eq.~(\ref{eq:Pexp}), which happen to 
cancel at $n=3$. 
Such `accidental' cancellations of leading coefficients at some orders do occur,
for a famous example see Refs.~\cite{BFKL}, and do not indicate any problem. 
We expect a non-vanishing coefficient of $\ln^{\,8} \! \x1$ at five loops.
Analogous comments apply to the (presumably related) absence of the expected 
$\ln^{\,5} \! \x1$ contribution to the diagonal fourth-order splitting 
functions in Eqs.~(\ref{Pqq3DSL}) and (\ref{Pgg3DSL}).

The Mellin moments of Eqs.~(\ref{Pqg3DL0}) and (\ref{Pgq3DL0}) and their NNLO 
counterparts in Eqs.\ (4.18) and (4.19) of Ref.~\cite{MVV4} are numerically 
illustrated in Figs.~12 and 13. 
Unlike analogous results for the non-singlet coefficient functions in
Refs.~\cite{MV3,MV5}, even all $\ln N$ enhanced terms do not provide a 
meaningful approximation at relevant values of $N$ for $P_{\,\rm gq}^{\,(2)}\!$,
and the double-logarithmic contributions (blue curves in the electronic version)
are nowhere near the full results in either case.  
Hence our predicted double-log terms are not phenomenologically useful without 
further four-loop information, e.g., in the form of an extension of 
Refs.~\cite{Mom3loop} to the next order. Indeed, given the fact that a first 
fourth-order calculation of a splitting-function moment has already been 
presented \cite{Baikov:2006ai}, such an extension should become feasible in 
the foreseeable future.
%
%
\setcounter{equation}{0}
\section{Discussion and outlook}
\label{sec:summary}
%
%
It is an old idea to replace the factorization-scheme dependent NLO (and now 
NNLO) quark and gluon distributions by suitable reference observables. One may 
expect to obtain smaller higher-order corrections to the coefficient functions 
for other observables in such a scheme, having absorbed some universal 
corrections entering all coefficient functions in, e.g., the usual \MSb\ scheme.
The predictions of (massless) perturbative QCD then depend on only one 
unphysical scale and scheme, for the renormalized strong coupling $\as$.  
Moreover physical constraints on the chosen observables, for instance on their 
positivity or endpoint behaviour, would directly apply to the unavoidable 
non-perturbative initial distributions for the scale evolution in this case.

An obvious quantity to replace the overall (flavour-singlet) quark density is
the (singlet part of the) standard structure function $\Ftwo$ in (say, 
charged- current) deep-inelastic scattering, i.e., one can define physical 
quark densities
$\tilde{q}_{\rm S}^{}$ by $\Ftwo \,=\, \tilde{q}_{\rm S}^{}$ to NLO and NNLO. 
Any such scheme is usually called a DIS scheme \cite{DISfact}. Of course, this
condition does not specify the lower row of the scheme-transformation matrix
$\ZV$ discussed in Section 2. The standard choice adopted in this respect,
used in parton-distribution analyses mainly in the 1990s, was to arbitrarily
use the coefficient functions for $\Ftwo$ also for this lower row. Consequently
the gluon distribution (which, in practical analyses, would profit most from 
being defined in terms of an observable) remains entirely unphysical in this
version of the DIS scheme, which is therefore mostly useful for quark-dominated
observables such as Drell-Yan lepton-pair production in $p\:\!\!\bar{p}$-%
collisions not too far from the threshold. 

Various alternatives to the above arbitrary choice -- which, given that 
gluon-dominated quantities usually receive larger higher-order corrections than
$\Ftwo$, `under-transforms' the gluon density -- have also been discussed 
already long ago, e.g., in Ref.~\cite{FP82}. The cleanest possibility is 
obviously the addition of a second observable which exactly mirrors the 
behaviour of $\Ftwo$, i.e., is directly sensitive only to the gluon 
distribution at LO. An example already mentioned in Ref.~\cite{FP82} is 
deep-inelastic scattering through the exchange of a scalar $\phi$ interacting 
with hadrons only via a $\,\phi\,G^{\,\mu\nu} G_{\mu\nu}$ coupling to the 
gluon field-strength tensor.
This interaction is realized in the Standard Model by the Higgs boson, for a
very heavy top quark and negligible couplings to all other flavours
\cite{HGGeff}.

In fact, as discussed in Section 2, the corresponding coefficient functions had
to be calculated to order $\as^{\,2}$ in the approach of Refs.~\cite{Mom3loop} 
to the determination of the complete set of third-order (NNLO) splitting 
functions, carried out for all-$N/$all-$x$ in Ref.~\cite{MVV4}. The numerous 
three-loop integrals for the latter were evaluated to order $\ep^{0}$ in 
dimensional regularization in order to obtain also the third-order coefficient 
functions for $\Ftwo$, $\FL$ and $\F3$ \cite{MVV5,MVV6,MVV10}. 
Hence also the three-loop coefficient functions for the $\phi$-exchange 
structure function $F_\phi$ could be calculated with a relatively minor extra 
effort. In Section 3 of this article we have presented and discussed these 
coefficient functions.

While, of course, Higgs-exchange DIS is of no experimental relevance at any
past or foreseeable lepton-nucleon facility, these results have already proven
useful in theoretical research as discussed above. Both the quark and gluon
coefficient functions for $F_\phi$ show, as expected, considerably larger 
higher-order corrections than their counterparts for $\Ftwo$. Their most
prominent feature is the double-logarithmic enhancement of the 
$1/x$-contributions dominating their small-$x$ behaviour, i.e., the occurrence
of terms up to $x^{\,-1} \ln^{\,2n} x$ in the $n$-loop coefficient functions
$c_{\phi,i}^{\,(n)}(x)$. These terms arise from using the large-$m_{\,\rm top}$
effective interaction outside its region of validity, see also Ref.~\cite
{Marzani:2008az} for a recent discussion in the context of Higgs production in
proton collisions.

The above double-logarithmic contributions also render the physical evolution
kernels for the system ($\,\Ftwo, F_\phi\,$), presented and discussed in 
Section 4, unstable within the physically interesting region of $x > 10^{\,-4}$
relevant to crucial high-scale processes at the LHC. Consequently this scheme
is unattractive for practical analyses of hard processes. There are various 
possibilities, theoretical and phenomenological, to improve upon this situation
which we, however, leave to future research. 
Already the present heavy-top results for the coefficient functions, however, 
can provide useful checks of, e.g., the order-dependence of fitted \MSb\ gluon 
densities by comparing the resulting shapes of the (in principle 
order-independent) quantity $F_\phi$ outside the small-$x$ region.

The above flavour-singlet evolution-kernel matrices share 
a theoretically very interesting feature with the (scalar) physical kernels for
a rather wide range of non-singlet observables \cite{MV3,MV5}: They show a
single-logarithmic enhancement at large values of the scaling variable $x$, 
i.e., terms of the form $\x1^a \ln^{\,k}\, \x1$ with $k > n$ do not contribute
to the N$^{\,n}$LO physical kernels for any value of $a$. This structure is in 
contrast to that of all coefficient functions and (pure-)$\,$singlet splitting 
functions. We consider this feature, which is already established to N$^3$LO 
for non-singlet structure functions, as a general property of the physical 
evolution kernels, and consequently conjecture that it holds to N$^3$LO also 
for the present ($\,\Ftwo, F_\phi\,$) singlet case.

Hence the double-logarithmic contributions from (a) the four-loop splitting 
functions and (b) the scheme transformation including lower orders and the 
three-loop coefficient functions have to cancel, i.e., the so far unknown 
corresponding terms in (a) can be inferred from our now complete results for 
(b). We have provided and discussed these predictions in Section 5. 
In particular, we note that the known double logarithms in the splitting 
functions to NNLO exhibit a particular colour structure in powers of 
$(\ca \!-\! \cf)$ which is non-trivially fulfilled also by our N$^3$LO 
predictions, a fact that we consider a significant check of their correctness.

As also discussed in Section 5, the predicted contributions to the splitting
functions $P_{\:\!ik}^{\,(3)}(x)$ alone are clearly insufficient for use in 
phenomenological applications. However they will become very useful in 
connection with future fourth-order calculations, such as the extension of
Ref.~\cite{Baikov:2006ch} to higher moments and the flavour-singlet sector. 
Already the corresponding three-loop predictions, had they been known at the
time, would have led to more accurate $x$-space approximation on the basis of
the moments calculated in Refs.~\cite{Mom3loop} than those actually obtained in 
Refs.~\cite{NV2,NV4}. In fact, this situation clearly illustrates the general
importance of calculations to at least NNLO: 
It usually requires two non-trivial orders to reliably identify most general 
structures -- for another example recall the `curious incident' of the 
coefficient of $\ln \x1$ in the three-loop diagonal splitting functions 
\cite{MVV3,MVV4} which triggered Ref.~\cite{DMS05} which in turn greatly 
assisted further developments including the determination of the NNLO splitting
functions for the final-state fragmentation case \cite{Mitov:2006ic,MV2}. 
We hope, accordingly, that also our present results will assist further 
developments in perturbative QCD, e.g., in the approach introduced in 
Ref.~\cite{Laenen:2008gt}.

{\sc Form} files of our main formulae, and {\sc Fortran} subroutines of the
$\phi$-exchange coefficient functions and the resulting physical evolution 
kernels can be obtained from the preprint server \ {\tt HTTP://arXiv.org} by 
downloading the source of this article. Furthermore they are available from the
authors upon request.
%
%
\vspace*{\fill}
\subsection*{Acknowledgments}
The work of G.S. has been funded by the UK Science \& Technology Facilities 
Council (STFC). 
S.M. acknowledges support by the Helmholtz Gemeinschaft under contract 
VH-NG-105 and in part by the Deutsche Forschungsgemeinschaft in 
Sonderforschungs\-be\-reich/Transregio~9. 
The~work of J.V. has been part of the research program of the Dutch Foundation 
for Fundamental Research of Matter (FOM). 
The research of A.V., who thanks S. Marzani for a useful discussion, has been 
supported by STFC under grant numbers PP/E007414/1 and ST/G00062X/1.

\newpage
%
%
\renewcommand{\theequation}{A.\arabic{equation}}
\setcounter{equation}{0}
\section*{Appendix A: The third-order coefficient functions for \boldmath
 $F_\phi$}
\label{sec:AppA}
%
\setlength{\baselineskip}{0.546cm}

In this appendix we present the very lengthy exact expressions for the 
third-order scalar-exchange coefficient functions 
$c^{\,(3)}_{\rm \phi\, , q}(x)$ and $c^{\,(3)}_{\rm \phi\, , g}(x)$. 
The notation is the same as used for the first- and second-order results given 
in Eqs.~(\ref{eq:cphiq1}) -- (\ref{eq:cphig2}) above. 
The result for the quark coefficient function reads
\bea
\label{eq:cphiq3}
 \lefteqn{c^{\,(3)}_{\rm \phi\, , q}(x) \; = \;
    \colour4colour{ \cf \* \cas } \, \*  \Big(
	    \pgq(x) \* (
          - 1652/15\: \* \H(1) \*  \zss\,
          + 240 \* \H(1,-3,0)
          - 160 \* \H(1,-2) \*  \z2 
          + 256 \* \H(1,-2,0,0)
 }
%
%
   \nn \\[-0.5mm] & & \mbox{}
          - 704/3\: \* \H(1,-2,-1,0)
          + 128/3\: \* \H(1,-2,2)
          + 712/3\: \* \H(1,0) \*  \z3 
          + 680/3\: \* \H(1,0,0) \*  \z2 
          + 32 \* \H(1,1,-2,0)
   \nn \\[0.5mm] & & \mbox{}
          - 376/3\: \* \H(1,0,0,0,0)
          + 80/3\: \* \H(1,1) \*  \z3 
          + 168 \* \H(1,1,0) \*  \z2 
          - 248/3\: \* \H(1,1,0,0,0)
          + 736/3\: \* \H(1,1,1) \*  \z2 
   \nn \\[0.5mm] & & \mbox{}
          - 896/3\: \* \H(1,1,1,0,0)
          - 1208/3\: \* \H(1,1,1,1,0)
          - 400 \* \H(1,1,1,1,1)
          - 1208/3\: \* \H(1,1,1,2)
          - 392 \* \H(1,1,2,0)
   \nn \\[0.5mm] & & \mbox{}
          - 1232/3\: \* \H(1,1,2,1)
          - 288 \* \H(1,1,3)
          + 776/3\: \* \H(1,2) \*  \z2 
          - 784/3\: \* \H(1,2,0,0)
          - 384 \* \H(1,2,1,0)
   \nn \\[0.5mm] & & \mbox{}
          - 1216/3\: \* \H(1,2,1,1)
          - 1192/3\: \* \H(1,2,2)
          - 1024/3\: \* \H(1,3,0)
          - 1144/3\: \* \H(1,3,1)
          - 272 \* \H(1,4)
          )
   \nn \\[0.5mm] & & \mbox{}
	  + \pgq( - x) \* (
          - 988/15\: \* \H(-1) \*  \zss\,
          - 992/3\: \* \H(-1,-3,0)
          + 1712/3\: \* \H(-1,-2) \*  \z2 
          - 432 \* \H(-1,-2,2)
   \nn \\[0.5mm] & & \mbox{}
          + 832/3\: \* \H(-1,-2,-1,0)
          - 1688/3\: \* \H(-1,-2,0,0)
          + 2240/3\: \* \H(-1,-1) \*  \z3 
          + 272 \* \H(-1,-1,-2,0)
   \nn \\[0.5mm] & & \mbox{}
          - 2120/3\: \* \H(-1,-1,-1) \*  \z2 
          - 944/3\: \* \H(-1,-1,-1,-1,0)
          + 2128/3\: \* \H(-1,-1,-1,0,0)
   \nn \\[0.5mm] & & \mbox{}
          + 1648/3\: \* \H(-1,-1,-1,2)
          + 2464/3\: \* \H(-1,-1,0) \*  \z2 
          - 1832/3\: \* \H(-1,-1,0,0,0)
          - 272 \* \H(-1,-1,2,0)
   \nn \\[0.5mm] & & \mbox{}
          - 992/3\: \* \H(-1,-1,2,1)
          - 2104/3\: \* \H(-1,-1,3)
          - 536 \* \H(-1,0) \*  \z3 
          - 1160/3\: \* \H(-1,0,0) \*  \z2 
   \nn \\[0.5mm] & & \mbox{}
          + 232 \* \H(-1,0,0,0,0)
          + 64/3\: \* \H(-1,2) \*  \z2 
          + 304/3\: \* \H(-1,2,0,0)
          + 224/3\: \* \H(-1,2,1,0)
          + 96 \* \H(-1,2,2)
   \nn \\[0.5mm] & & \mbox{}
          + 256/3\: \* \H(-1,2,1,1)
          + 632/3\: \* \H(-1,3,0)
          + 784/3\: \* \H(-1,3,1)
          + 1024/3\: \* \H(-1,4)
	      )
   \nn \\[0.5mm] & & \mbox{}
          - 35864/27\: \* (1 - 8/4483/x^2
          + 46675/53796/x + 7755/17932\: \* x
   \nn \\[0.5mm] & & \mbox{}
          + 3883/13449\: \* x^2\,) \* \H(-1,0)
          + 32/3\: \* (1 - 75/x - 56\: \* x) \* \H(2,1,1,1)
          + 80/3\: \* (1 - 162/5/x
   \nn \\[0.5mm] & & \mbox{}
          - 267/10\: \* x) \* \H(-2,0,0,0)
          + 208/3\: \* (1 - 48/13/x- 279/26\: \* x) \* \H(5)
          + 112/3\: \* (1 - 24/7/x
   \nn \\[0.5mm] & & \mbox{} 
          - 29/14\: \* x) \* \H(-3,2)
          - 208/3\: \* (1 - 36/13/x- 127/13\: \* x) \* \H(0,0,0) \*  \z2
   \nn \\[0.5mm] & & \mbox{}
          + 2368/15\: \* (1 - 349/148/x - 96/37\: \* x) \* \H(0) \*  \zss\,
          - 6112/9\: \* (1 - 841/382/x  + 85/764\: \* x 
   \nn \\[0.5mm] & & \mbox{}
          + 54/191\: \* x^2\,) \* \H(2) \*  \z2
          - 2620/9\: \* (1 - 1356/655/x + 187/262\: \* x + 44/655\: \* x^2\,) \* \H(1) \*  \z3 
   \nn \\[0.5mm] & & \mbox{}
          + 10040/81\: \* (1 - 16057/10040/x + 24191/10040\: \* x - 865/251\: \* x^2\,) \* \H(1,1)
   \nn \\[0.5mm] & & \mbox{}
          - 72928/27\: \* (1 - 55601/36464/x - 9461/36464\: \* x - 1697/9116\: \* x^2\,) \*  \z3 
   \nn \\[0.5mm] & & \mbox{}
          + 59026/81\: \* (1 - 86857/59026/x + 24793/59026\: \* x - 17892/29513\: \* x^2\,) \* \H(1,0)
   \nn \\[0.5mm] & & \mbox{}
          + 7976/9\: \* (1 - 1312/997/x - 229/1994\: \* x + 60/997\: \* x^2\,) \* \H(1,0,0,0)
   \nn \\[0.5mm] & & \mbox{}
          - 9376/9\: \* (1 - 382/293/x + 55/586\: \* x + 32/293\: \* x^2\,) \* \H(1,1) \*  \z2 
   \nn \\[0.5mm] & & \mbox{}
          - 2648/3\: \* (1 - 1289/993/x - 1/331\: \* x + 80/993\: \* x^2\,) \* \H(1,0) \*  \z2 
   \nn \\[0.5mm] & & \mbox{}
          + 11420/9\: \* (1 - 3704/2855/x + 257/2855\: \* x + 336/2855\: \* x^2\,) \* \H(1,1,0,0)
   \nn \\[0.5mm] & & \mbox{}
          + 12092/9\: \* (1 - 3614/3023/x + 107/3023\: \* x + 312/3023\: \* x^2\,) \* \H(1,1,2)
   \nn \\[0.5mm] & & \mbox{}
          + 11200/9\: \* (1 - 208/175/x - 73/1400\: \* x + 3/35\: \* x^2\,) \* \H(1,3)
          + 11240/9\: \* (1 - 1664/1405/x
   \nn \\[0.5mm] & & \mbox{}
          + 161/2810\: \* x + 148/1405\: \* x^2\,) \* \H(1,1,1,1)
          + 12076/9\: \* (1 - 3562/3019/x + 103/3019\: \* x 
   \nn \\[0.5mm] & & \mbox{}
          + 312/3019\: \* x^2\,) \* \H(1,2,1)
          + 1336\: \* (1 - 195/167/x + 6/167\: \* x + 52/501\: \* x^2\,) \* \H(1,1,1,0)
   \nn \\[0.5mm] & & \mbox{}
          + 13252/9\: \* (1 - 3852/3313/x - 53/3313\: \* x + 320/3313\: \* x^2\,) \* \H(1,2,0)
   \nn \\[0.5mm] & & \mbox{}
          + 1289237/243\: \* (1 - 1478082/1289237/x - 205504/1289237\: \* x
   \nn \\[0.5mm] & & \mbox{}
          + 167284/1289237\: \* x^2\,) \* \H(1)
          - 68764/27\: \* (1 - 38843/34382/x - 4969/34382\: \* x 
   \nn \\[0.5mm] & & \mbox{}
          + 432/17191\: \* x^2\,) \* \H(1) \*  \z2 
          + 1328/9\: \* (1 - 91/83/x - 89/166\: \* x + 8/83\: \* x^2\,) \* \H(1,-2,0)
   \nn \\[0.5mm] & & \mbox{}
          + 72760/27\: \* (1 - 19543/18190/x - 419/1819\: \* x + 9/9095\: \* x^2\,) \* \H(1,0,0)
   \nn \\[0.5mm] & & \mbox{}
          + 11480/9\: \* (1 - 1527/1435/x + 421/2870\: \* x + 268/1435\: \* x^2\,) \* \H(2,0,0)
   \nn \\[0.5mm] & & \mbox{}
          + 54620/27\: \* (1 - 27851/27310/x - 2458/13655\: \* x - 146/2731\: \* x^2\,) \* \H(1,2)
   \nn \\[0.5mm] & & \mbox{}
          + 48764/27\: \* (1 - 24351/24382/x - 4361/24382\: \* x - 782/12191\: \* x^2\,) \* \H(1,1,1)
   \nn \\[0.5mm] & & \mbox{}
          + 56864/27\: \* (1 - 28103/28432/x - 365/1777\: \* x - 383/7108\: \* x^2\,) \* \H(1,1,0)
   \nn \\[0.5mm] & & \mbox{}
          + 9412/9\: \* (1 - 2040/2353/x + 1046/2353\: \* x + 800/2353\: \* x^2\,) \* \H(4)
   \nn \\[0.5mm] & & \mbox{}
          + 13468/9\: \* (1 - 2866/3367/x + 557/3367\: \* x + 648/3367\: \* x^2\,) \* \H(2,2)
   \nn \\[0.5mm] & & \mbox{}
          + 13736/9\: \* (1 - 1431/1717/x + 623/3434\: \* x + 320/1717\: \* x^2\,) \* \H(2,1,1)
   \nn \\[0.5mm] & & \mbox{}
          + 4904/3\: \* (1 - 507/613/x + 247/1839\: \* x + 316/1839\: \* x^2\,) \* \H(2,1,0)
   \nn \\[0.5mm] & & \mbox{}
          - 9412/9\: \* (1 - 1932/2353/x + 956/2353\: \* x + 800/2353\: \* x^2\,) \* \H(0,0) \*  \z2 
   \nn \\[0.5mm] & & \mbox{}
          + 5200/3\: \* (1 - 523/650/x - 597/1300\: \* x) \*  \z5 
          + 12472/9\: \* (1 - 1128/1559/x
   \nn \\[0.5mm] & & \mbox{}
          + 511/1559\: \* x + 368/1559\: \* x^2\,) \* \H(3,0)
          + 15968/9\: \* (1 - 1095/1996/x + 1303/3992\: \* x 
   \nn \\[0.5mm] & & \mbox{}
          + 102/499\: \* x^2\,) \* \H(3,1)
          + 85748/27 \* (1 - 10598/21437/x - 3388/21437\: \* x
   \nn \\[0.5mm] & & \mbox{}
          + 70/1261\: \* x^2\,) \* \H(2,0)
          - 3896 \* (1 - 1888/4383/x + 200/4383\: \* x + 580/4383\: \* x^2\,) \* \H(0) \*  \z3 
   \nn \\[0.5mm] & & \mbox{}
          + 122288/27\: \* (1 - 3219/7643/x - 6949/61144\: \* x + 1455/15286\: \* x^2\,) \* \H(3)
   \nn \\[0.5mm] & & \mbox{}
          - 122288/27\: \* (1 - 12395/30572/x - 3535/61144\: \* x + 1455/15286\: \* x^2\,) \* \H(0) \*  \z2 
   \nn \\[0.5mm] & & \mbox{}
          - 160 \* (1 - 2/5/x - 1/10\: \* x) \* \H(2,-2,0)
          + 27832/9\: \* (1 - 8245/20874/x - 739/6958\: \* x 
   \nn \\[0.5mm] & & \mbox{}
          + 391/10437\: \* x^2\,) \* \H(2,1)
          + 1600/3\: \* (1 - 9/25/x - 61/200\: \* x) \* \H(-3,0,0)
   \nn \\[0.5mm] & & \mbox{}
          + 3047527/243\: \* (1 - 1062210/3047527/x - 139147/6095054\: \* x
   \nn \\[0.5mm] & & \mbox{}
          + 522652/3047527\: \* x^2\,) \* \H(0)
          + 170344/27\: \* (1 - 13635/42586/x - 4134/21293\: \* x 
   \nn \\[0.5mm] & & \mbox{}
          + 3017/21293\: \* x^2\,) \* \H(0,0,0)
          + 4904/3\: \* (1 - 166/613/x - 129/613\: \* x + 72/613\: \* x^2\,) \* \H(-2,-1,0)
   \nn \\[0.5mm] & & \mbox{}
          + 7796/3\: \* (1 - 270/1949/x + 121/1949\: \* x + 216/1949\: \* x^2\,) \* \H(-2) \*  \z2 
          + 480 \* (1 - 2/15/x
   \nn \\[0.5mm] & & \mbox{}
          - 11/10\: \* x) \* \H(0,0,0,0,0)
          - 31780/9\: \* (1 - 206/1589/x - 54/1135\: \* x + 928/7945\: \* x^2\,) \* \H(-2,0,0)
   \nn \\[0.5mm] & & \mbox{}
          - 381824/81\: \* (1 - 180471/1527296/x - 2124809/4581888\: \* x + 250877/572736\: \* x^2\,)
   \nn \\[0.5mm] & & \mbox{}
          + 2384/3\: \* (1 - 12/149/x - 25/298\: \* x) \* \H(-4,0)
          - 5344/3\: \* (1 - 13/167/x + 125/668\: \* x 
   \nn \\[0.5mm] & & \mbox{}
          + 18/167\: \* x^2\,) \* \H(-2,2)
          + 74912/27\: \* (1 - 481/18728/x - 1707/18728\: \* x
   \nn \\[0.5mm] & & \mbox{}
          + 1305/4682\: \* x^2\,) \* \H(-2,0)
          - 25544/9\: \* (1 + 54/3193/x - 45/3193\: \* x + 400/3193\: \* x^2\,) \* \H(-3,0)
   \nn \\[0.5mm] & & \mbox{}
          + 832 \* (1 + 1/6/x + 69/104\: \* x) \* \H(0,0) \*  \z3 
          + 9464/9\: \* (1 + 512/1183/x + 509/1183\: \* x 
   \nn \\[0.5mm] & & \mbox{}
          + 40/1183\: \* x^2\,) \* \H(-1,-1,2)
          - 4060/3\: \* (1 + 226/435/x + 13/35\: \* x + 136/3045\: \* x^2\,) \* \H(-1,-1) \*  \z2 
   \nn \\[0.5mm] & & \mbox{}
          - 1112/3\: \* (1 + 76/139/x + 347/278\: \* x) \*  \z2  \*  \z3 
          + 2800/3\: \* (1 + 4/7/x
          + 117/350\: \* x) \* \H(-2,-1,2)
   \nn \\[0.5mm] & & \mbox{}
          + 4732/3\: \* (1 + 748/1183/x + 789/2366\: \* x + 76/1183\: \* x^2\,) \* \H(-1) \*  \z3 
   \nn \\[0.5mm] & & \mbox{}
          - 3488/3\: \* (1 + 72/109/x + 85/218\: \* x) \* \H(-2,-1) \*  \z2
          - 7528/9\: \* (1 + 648/941/x
   \nn \\[0.5mm] & & \mbox{} 
          - 167/1882\: \* x - 268/941\: \* x^2\,) \* \H(0,0,0,0)
          - 16780/9\: \* (1 + 3232/4195/x
   \nn \\[0.5mm] & & \mbox{} 
          + 1274/4195\: \* x + 336/4195\: \* x^2\,) \* \H(-1,3)
          + 4648/3\: \* (1 + 65/83/x + 92/581\: \* x 
   \nn \\[0.5mm] & & \mbox{} 
          + 8/83\: \* x^2\,) \* \H(-1,-1,0,0)
          + 3176/3\: \* (1 + 314/397/x + 389/794\: \* x) \* \H(-2) \*  \z3 
   \nn \\[0.5mm] & & \mbox{} 
          - 7984/9\: \* (1 + 396/499/x + 319/998\: \* x + 44/499\: \* x^2\,) \* \H(-1,2,1)
   \nn \\[0.5mm] & & \mbox{} 
          + 20036/9\: \* (1 + 3982/5009/x + 1414/5009\: \* x + 416/5009\: \* x^2\,) \* \H(-1,0) \*  \z2 
   \nn \\[0.5mm] & & \mbox{} 
          + 186364/81\: \* (1 + 37486/46591/x - 6167/186364\: \* x + 22118/46591\: \* x^2\,) \*  \z2 
   \nn \\[0.5mm] & & \mbox{} 
          - 5432/9\: \* (1 + 558/679/x + 113/679\: \* x + 8/97\: \* x^2\,) \* \H(-1,-1,-1,0)
   \nn \\[0.5mm] & & \mbox{} 
          - 7240/9\: \* (1 + 752/905/x + 283/905\: \* x + 72/905\: \* x^2\,) \* \H(-1,2,0)
   \nn \\[0.5mm] & & \mbox{} 
          + 576 \* (1 + 71/81/x + 17/216\: \* x + 8/81\: \* x^2\,) \* \H(-1,-2,0)
          + 144 \* (1 + 8/9/x
   \nn \\[0.5mm] & & \mbox{} 
          + 37/54\: \* x) \* \H(-3,-1,0)
          - 19328/9\: \* (1 + 601/604/x + 751/4832\: \* x + 63/604\: \* x^2\,) \* \H(-1,0,0,0)
   \nn \\[0.5mm] & & \mbox{} 
          - 424 \* (1 + 160/159/x + 427/318\: \* x) \* \H(3,0,0)
          - 1376/3\: \* (1 + 44/43/x 
   \nn \\[0.5mm] & & \mbox{}
          + 53/86\: \* x) \* \H(-2,-1,-1,0) 
          - 293866/81\: \* (1 + 151137/146933/x + 8738/146933\: \* x 
   \nn \\[0.5mm] & & \mbox{}
          + 73960/146933\: \* x^2\,) \* \H(0,0)
          + 59324/27\: \* (1 + 15328/14831/x + 2398/14831\: \* x 
   \nn \\[0.5mm] & & \mbox{}
          + 4676/14831\: \* x^2\,) \* \H(-1,0,0)
          + 24536/27\: \* (1 + 3303/3067/x + 385/6134\: \* x 
   \nn \\[0.5mm] & & \mbox{}
          + 1178/3067\: \* x^2\,) \* \H(-1,2)
          - 704/3\: \* (1 + 25/22/x + 43/44\: \* x) \* \H(2,0,0,0)
   \nn \\[0.5mm] & & \mbox{}
          - 896/3\: \* (1 + 8/7/x + 11/14\: \* x) \* \H(-2,2,1)
          - 2024/3\: \* (1 + 292/253/x + 37/46\: \* x) \* \H(-2,3)
   \nn \\[0.5mm] & & \mbox{}
          - 38680/27\: \* (1 + 6051/4835/x + 823/19340\: \* x + 1759/4835\: \* x^2\,) \* \H(-1) \*  \z2 
   \nn \\[0.5mm] & & \mbox{}
          - 656/3\: \* (1 + 52/41/x + 75/82\: \* x) \* \H(-2,2,0)
          - 186364/81\: \* (1 + 121647/93182/x 
   \nn \\[0.5mm] & & \mbox{}
          + 40363/186364\: \* x + 22118/46591\: \* x^2\,) \* \H(2)
          + 2144/3\: \* (1 + 88/67/x 
          + 125/134\: \* x) \* \H(-2,0) \*  \z2
   \nn \\[0.5mm] & & \mbox{}
          - 1336/3\: \* (1 + 224/167/x + 415/334\: \* x) \* \H(2,1,0,0)
          + 240 \* (1 + 8/5/x + 11/10\: \* x) \* \H(-2,-2,0)
   \nn \\[0.5mm] & & \mbox{}
          + 1912/3\: \* (1 + 348/239/x + 485/478\: \* x) \* \H(-2,-1,0,0)
          + 408 \* (1 + 224/153/x 
   \nn \\[0.5mm] & & \mbox{}
          + 509/306\: \* x) \* \H(3) \*  \z2 
          - 28288/27\: \* (1 + 687/442/x + 53/7072\: \* x
          + 581/1768\: \* x^2\,) \* \H(-1,-1,0)
   \nn \\[0.5mm] & & \mbox{}
          + 512/3\: \* (1 + 53/32/x + 93/64\: \* x) \* \H(2) \*  \z3 
          - 1168/3\: \* (1 + 124/73/x + 271/146\: \* x) \* \H(3,1,0)
   \nn \\[0.5mm] & & \mbox{}
          + 880/3\: \* (1 + 102/55/x + 171/110\: \* x) \* \H(2,1) \*  \z2 
          - 336 \* (1 + 124/63/x + 13/6\: \* x) \* \H(3,2)
   \nn \\[0.5mm] & & \mbox{}
          - 632/3\: \* (1 + 168/79/x + 547/158\: \* x) \* \H(4,0)
          - 896/3\: \* (1 + 18/7/x + 37/14\: \* x) \* \H(3,1,1)
   \nn \\[0.5mm] & & \mbox{}
          - 616/3\: \* (1 + 204/77/x + 629/154\: \* x) \* \H(4,1)
          + 368/3\: \* (1 + 82/23/x + 119/46\: \* x) \* \H(2,0) \*  \z2 
   \nn \\[0.5mm] & & \mbox{}
          - 488/3\: \* (1 + 260/61/x + 421/122\: \* x) \* \H(2,2,0)
          + 104/3\: \* (1 + 72/13/x + 95/26\: \* x) \* \H(-3) \*  \z2 
   \nn \\[0.5mm] & & \mbox{}
          - 9986/45\: \* (1 + 27874/4993/x + 1529/4993\: \* x - 328/4993\: \* x^2\,) \*  \zss\,
   \nn \\[0.5mm] & & \mbox{}
          - 248/3\: \* (1 + 224/31/x  + 331/62\: \* x) \* \H(2,3)
          - 64 \* (1 + 73/6/x + 28/3\: \* x) \* \H(2,1,2)
   \nn \\[-0.5mm] & & \mbox{}
          - 160/3\: \* (1 + 73/5/x + 56/5\: \* x) \* \H(2,2,1)
          - 128/3\: \* (1 + 73/4/x + 14\: \* x) \* \H(2,1,1,0)
	   \Big)
   \nn \\[-0.5mm] & & \mbox{\hspn}
   + \colour4colour{ \cfs\, \*  \ca } \,   \*  \Big(
	   \pgq(x) \* (
          - 1856/15\: \* \H(1) \*  \zss\,
          - 176 \* \H(1,-3,0)
          + 264 \* \H(1,-2) \*  \z2 
          + 112 \* \H(1,-2,-1,0)
   \nn \\[-0.5mm] & & \mbox{}
          - 640/3\: \* \H(1,-2,0,0)
          - 208 \* \H(1,-2,2)
          - 32 \* \H(1,0) \*  \z3 
          + 640/3\: \* \H(1,0,0) \*  \z2 
          - 544/3\: \* \H(1,0,0,0,0)
   \nn \\[0.5mm] & & \mbox{}
          + 632/3\: \* \H(1,1) \*  \z3 
          - 464/3\: \* \H(1,1,-2,0)
          + 1400/3\: \* \H(1,1,0) \*  \z2 
          - 928/3\: \* \H(1,1,0,0,0)
   \nn \\[0.5mm] & & \mbox{}
          + 784/3\: \* \H(1,1,1) \*  \z2 
          - 320/3\: \* \H(1,1,1,0,0)
          - 40 \* \H(1,1,1,1,0)
          - 104 \* \H(1,1,1,2)
          - 120 \* \H(1,1,2,0)
   \nn \\[0.5mm] & & \mbox{}
          - 304/3\: \* \H(1,1,2,1)
          - 432 \* \H(1,1,3)
          + 832/3\: \* \H(1,2) \*  \z2 
          - 592/3\: \* \H(1,2,0,0)
          - 368/3\: \* \H(1,2,1,0)
   \nn \\[0.5mm] & & \mbox{}
          - 80 \* \H(1,2,1,1)
          - 488/3\: \* \H(1,2,2)
          - 176 \* \H(1,3,0)
          - 440/3\: \* \H(1,3,1)
          - 616/3\: \* \H(1,4)
	     )
   \nn \\[0.5mm] & & \mbox{}
	  + \pgq( - x) \* (
            1016/15\: \* \H(-1) \*  \zss\,
          + 192 \* \H(-1,-3,0)
          - 344/3\: \* \H(-1,-2) \*  \z2 
          - 688/3\: \* \H(-1,-2,-1,0)
   \nn \\[0.5mm] & & \mbox{}
          + 488/3\: \* \H(-1,-2,0,0)
          - 136 \* \H(-1,-1) \*  \z3 
          - 224 \* \H(-1,-1,-2,0)
          + 296/3\: \* \H(-1,-1,-1) \*  \z2
   \nn \\[0.5mm] & & \mbox{}
          + 944/3\: \* \H(-1,-1,-1,-1,0)
          - 736/3\: \* \H(-1,-1,-1,0,0)
          + 176/3\: \* \H(-1,-1,-1,2)
   \nn \\[0.5mm] & & \mbox{}
          - 112/3\: \* \H(-1,-1,0) \*  \z2 
          + 80 \* \H(-1,-1,0,0,0)
          + 112/3\: \* \H(-1,-1,2,0)
          + 160/3\: \* \H(-1,-1,2,1)
   \nn \\[0.5mm] & & \mbox{}
          + 8/3\: \* \H(-1,-1,3)
          - 448/3\: \* \H(-1,0) \*  \z3 
          - 320/3\: \* \H(-1,0,0) \*  \z2 
          + 96 \* \H(-1,0,0,0,0)
          - 232/3\: \* \H(-1,2) \*  \z2
   \nn \\[0.5mm] & & \mbox{}
          + 368/3\: \* \H(-1,2,0,0)
          + 128/3\: \* \H(-1,2,1,0)
          + 32/3\: \* \H(-1,2,1,1)
          + 64/3\: \* \H(-1,2,2)
          + 184/3\: \* \H(-1,3,0)
   \nn \\[0.5mm] & & \mbox{}
          + 64 \* \H(-1,3,1)
          + 344/3\: \* \H(-1,4)
	      )
          + 1280 \* (1 - 7/1080/x^2 + 139/640/x + 15683/17280\: \* x
   \nn \\[0.5mm] & & \mbox{}
          - 7/240\: \* x^2\,) \* \H(-1,0)
          + 742/27\: \* (1 - 13896/371/x + 59/2\: \* x - 144/53\: \* x^2\,) \*  \z3 
   \nn \\[0.5mm] & & \mbox{}
          + 904/9\: \* (1 - 543/113/x + 853/226\: \* x - 64/113\: \* x^2\,) \* \H(0) \*  \z3 
          - 128/3\: \* (1 - 15/8/x 
   \nn \\[0.5mm] & & \mbox{}
          - 1/4\: \* x) \* \H(-2,0,0,0)
          - 27938/27\: \* (1 - 22578/13969/x - 35941/27938\: \* x 
   \nn \\[0.5mm] & & \mbox{}
          - 720/13969\: \* x^2\,) \* \H(0) \*  \z2 
          + 27938/27\: \* (1 - 21120/13969/x - 27781/27938\: \* x
   \nn \\[0.5mm] & & \mbox{}
          - 720/13969\: \* x^2\,) \* \H(3)
          + 5032/9\: \* (1 - 936/629/x - 1715/1258\: \* x + 12/629\: \* x^2\,) \* \H(2,1,1)
   \nn \\[0.5mm] & & \mbox{}
          + 176 \* (1 - 16/11/x - 107/66\: \* x) \* \H(3,1,0)
          + 1120/3\: \* (1 - 46/35/x - 31/35\: \* x) \* \H(2,1,0,0)
   \nn \\[0.5mm] & & \mbox{}
          + 34460/27\: \* (1 - 10674/8615/x - 1888/8615\: \* x - 72/8615\: \* x^2\,) \* \H(2,0)
   \nn \\[0.5mm] & & \mbox{}
          + 784/3\: \* (1 - 8/7/x - 13/14\: \* x) \* \H(2,2,0)
          + 41896/27\: \* (1 - 5901/5237/x 
   \nn \\[0.5mm] & & \mbox{}
          - 4243/20948\: \* x + 288/5237\: \* x^2\,) \* \H(2,1)
          + 688/3\: \* (1 - 48/43/x - 125/86\: \* x) \* \H(3,2)
   \nn \\[0.5mm] & & \mbox{}
          + 43340/27\: \* (1 - 4745/4334/x - 3007/10835\: \* x + 24/2167\: \* x^2\,) \* \H(1,2)
   \nn \\[0.5mm] & & \mbox{}
          + 38416/27\: \* (1 - 20597/19208/x - 961/4802\: \* x + 30/2401\: \* x^2\,) \* \H(1,1,0)
   \nn \\[0.5mm] & & \mbox{}
          + 43012/27\: \* (1 - 23041/21506/x - 2357/10753\: \* x + 544/10753\: \* x^2\,) \* \H(1,1,1)
   \nn \\[0.5mm] & & \mbox{}
          + 6128/9\: \* (1 - 807/766/x - 605/766\: \* x + 30/383\: \* x^2\,) \* \H(2,1,0)
   \nn \\[0.5mm] & & \mbox{}
          + 78766/27\: \* (1 - 82071/78766/x - 21843/78766\: \* x + 20/39383\: \* x^2\,) \* \H(1,0)
   \nn \\[0.5mm] & & \mbox{}
          + 7304/9\: \* (1 - 927/913/x - 1483/1826\: \* x + 60/913\: \* x^2\,) \* \H(2,2)
          + 736/3\: \* (1 - 1/x
   \nn \\[0.5mm] & & \mbox{}
          - 35/46\: \* x) \* \H(2,1,1,0)
          + 8728/9\: \* (1 - 1073/1091/x - 436/1091\: \* x + 24/1091\: \* x^2\,) \* \H(1,2,0)
   \nn \\[0.5mm] & & \mbox{}
          + 274384/81\: \* (1 - 33685/34298/x - 20807/68596\: \* x - 331/34298\: \* x^2\,) \* \H(1,1)
   \nn \\[0.5mm] & & \mbox{}
          + 10688/9\: \* (1 - 653/668/x - 503/1336\: \* x + 3/167\: \* x^2\,) \* \H(1,1,0,0)
          + 768 \* (1 - 23/24/x
   \nn \\[0.5mm] & & \mbox{}
          - 143/288\: \* x + 1/72\: \* x^2\,) \* \H(1,1,1,0)
          + 1923029/486\: \* (1 - 1747009/1923029/x 
   \nn \\[0.5mm] & & \mbox{}
          - 795761/3846058\: \* x - 40576/1923029\: \* x^2\,) \* \H(1)
          + 9224/9\: \* (1 - 1003/1153/x 
   \nn \\[0.5mm] & & \mbox{}
          - 1105/2306\: \* x + 12/1153\: \* x^2\,) \* \H(1,2,1)
          - 286819/81\: \* (1 - 19056/22063/x
   \nn \\[0.5mm] & & \mbox{}
          - 181409/286819\: \* x + 360/286819\: \* x^2\,) \*  \z2 
          + 1262/27\: \* (1 - 540/631/x
   \nn \\[0.5mm] & & \mbox{}
          - 17713/1262\: \* x - 1600/631\: \* x^2\,) \* \H(0,0,0)
          + 7528/9\: \* (1 - 796/941/x - 583/941\: \* x
   \nn \\[0.5mm] & & \mbox{}
          - 4/941\: \* x^2\,) \* \H(1,1,1,1)
          + 8552/9\: \* (1 - 898/1069/x - 1039/2138\: \* x + 12/1069\: \* x^2\,) \* \H(1,1,2)
   \nn \\[0.5mm] & & \mbox{}
          + 8792/9\: \* (1 - 894/1099/x - 1945/2198\: \* x + 132/1099\: \* x^2\,) \* \H(3,1)
   \nn \\[0.5mm] & & \mbox{}
          + 848/3\: \* (1 - 42/53/x - 97/106\: \* x) \* \H(3,1,1)
          + 286819/81\: \* (1 - 225210/286819/x 
   \nn \\[0.5mm] & & \mbox{}
          - 87311/286819\: \* x + 360/286819\: \* x^2\,) \* \H(2)
          - 52040/27\: \* (1 - 4073/5204/x 
   \nn \\[0.5mm] & & \mbox{}
          - 4477/13010\: \* x + 36/6505\: \* x^2\,) \* \H(1) \*  \z2 
          - 480 \* (1 - 7/9/x - 31/60\: \* x) \* \H(2) \*  \z3 
   \nn \\[0.5mm] & & \mbox{}
          - 5800/9\: \* (1 - 112/145/x + 98/145\: \* x - 16/725\: \* x^2\,) \* \H(-1,-1,0)
   \nn \\[0.5mm] & & \mbox{}
          + 1023301/486\: \* (1 - 789822/1023301/x - 203321/1023301\: \* x
   \nn \\[0.5mm] & & \mbox{}
          - 6016/1023301\: \* x^2\,) \* \H(0)
          - 13160/9\: \* (1 - 181/235/x - 1627/3290\: \* x 
   \nn \\[0.5mm] & & \mbox{}
          + 24/1645\: \* x^2\,) \* \H(1,1) \*  \z2
          + 13972/9\: \* (1 - 2686/3493/x - 1709/6986\: \* x
   \nn \\[0.5mm] & & \mbox{}
          - 32/3493\: \* x^2\,) \* \H(1,0,0)
          + 1088/3\: \* (1 - 13/17/x - 43/68\: \* x) \* \H(2,2,1)
   \nn \\[0.5mm] & & \mbox{}
          + 1216/3\: \* (1 - 29/38/x - 11/19\: \* x) \* \H(2,1,2)
          - 1156 \* (1 - 1916/2601/x - 281/578\: \* x
   \nn \\[0.5mm] & & \mbox{}
          + 80/2601\: \* x^2\,) \* \H(1) \*  \z3 
          - 2368/3\: \* (1 - 47/74/x - 65/148\: \* x) \* \H(2,1) \*  \z2 
   \nn \\[0.5mm] & & \mbox{}
          + 3008/3\: \* (1 - 467/752/x - 293/752\: \* x + 1/47\: \* x^2\,) \* \H(1,0,0,0)
   \nn \\[0.5mm] & & \mbox{}
          + 12904/9\: \* (1 - 995/1613/x - 1553/3226\: \* x + 24/1613\: \* x^2\,) \* \H(1,3)
   \nn \\[0.5mm] & & \mbox{}
          + 4472/9\: \* (1 - 1719/2795/x - 4153/11180\: \* x + 348/2795\: \* x^2\,) \*  \zss\,
   \nn \\[0.5mm] & & \mbox{}
          - 10856/9\: \* (1 - 825/1357/x - 1729/2714\: \* x + 96/1357\: \* x^2\,) \* \H(2) \*  \z2 
   \nn \\[0.5mm] & & \mbox{}
          + 640/3\: \* (1 - 3/5/x - 2/5\: \* x) \* \H(-2,-1,2)
          + 1472/3\: \* (1 - 55/92/x - 13/23\: \* x) \* \H(2,0,0,0)
   \nn \\[0.5mm] & & \mbox{}
          - 13864/9\: \* (1 - 1028/1733/x - 1667/3466\: \* x + 36/1733\: \* x^2\,) \* \H(1,0) \*  \z2 
   \nn \\[0.5mm] & & \mbox{}
          + 10120/9\: \* (1 - 732/1265/x - 1469/2530\: \* x + 144/1265\: \* x^2\,) \* \H(3,0)
   \nn \\[0.5mm] & & \mbox{}
          - 528 \* (1 - 6/11/x - 68/99\: \* x) \* \H(3) \*  \z2 
          + 304 \* (1 - 10/19/x - 17/38\: \* x) \* \H(2,1,1,1)
   \nn \\[0.5mm] & & \mbox{}
          + 1088/3\: \* (1 - 71/136/x - 1/2\: \* x) \* \H(1,-2,0)
          + 8968/9\: \* (1 - 567/1121/x - 989/2242\: \* x
   \nn \\[0.5mm] & & \mbox{}
          + 72/1121\: \* x^2\,) \* \H(2,0,0)
          + 1360/3\: \* (1 - 42/85/x - 117/170\: \* x) \* \H(3,0,0)
   \nn \\[0.5mm] & & \mbox{}
          + 2315033/324\: \* (1 - 140715/330719/x - 306555/2315033\: \* x - 7104/2315033\: \* x^2\,)
   \nn \\[0.5mm] & & \mbox{}
          - 2560/3\: \* (1 - 17/40/x - 31/80\: \* x) \* \H(2,0) \*  \z2 
          + 2432/3\: \* (1 - 31/76/x - 29/76\: \* x) \* \H(2,3)
   \nn \\[0.5mm] & & \mbox{}
          - 14632/9\: \* (1 - 24/59/x - 2159/3658\: \* x + 240/1829\: \* x^2\,) \* \H(0,0) \*  \z2 
   \nn \\[0.5mm] & & \mbox{}
          + 1184/3\: \* (1 - 15/37/x - 19/74\: \* x) \* \H(2,-2,0)
          + 1472/3\: \* (1 - 37/92/x - 75/184\: \* x) \*  \z2  \*  \z3
   \nn \\[0.5mm] & & \mbox{}
          + 14632/9\: \* (1 - 690/1829/x - 1919/3658\: \* x + 240/1829\: \* x^2\,) \* \H(4)
   \nn \\[0.5mm] & & \mbox{}
          + 800/3\: \* (1 - 9/25/x - 4/25\: \* x) \* \H(-2,0) \*  \z2
          + 127235/81\: \* (1 - 44982/127235/x
   \nn \\[0.5mm] & & \mbox{}
          - 301739/254470\: \* x - 3352/127235\: \* x^2\,) \* \H(0,0)
          + 304/3\: \* (1 - 6/19/x + 26/19\: \* x) \* \H(0,0) \*  \z3 
   \nn \\[0.5mm] & & \mbox{}
          - 8084/9\: \* (1 - 514/2021/x + 1900/2021\: \* x - 48/2021\: \* x^2\,) \* \H(-1) \*  \z2
   \nn \\[0.5mm] & & \mbox{}
          - 2368/3\: \* (1 - 17/74/x + 41/148\: \* x + 3/37\: \* x^2\,) \* \H(-2,-1,0)
          + 2848/15\: \* (1 - 39/178/x
   \nn \\[0.5mm] & & \mbox{}
          - 407/712\: \* x) \* \H(0) \*  \zss\,
          - 928/3\: \* (1 - 6/29/x - 2/29\: \* x) \* \H(-2,3)
          + 7556/9\: \* (1 - 318/1889/x
   \nn \\[0.5mm] & & \mbox{}
          + 1750/1889\: \* x - 32/1889\: \* x^2\,) \* \H(-1,0,0)
          - 1120 \* (1 - 4/105/x - 11/420\: \* x 
   \nn \\[0.5mm] & & \mbox{}
          + 3/35\: \* x^2\,) \* \H(-2) \*  \z2 
          + 2704/3\: \* (1 - 7/338/x + 12/169\: \* x + 20/169\: \* x^2\,) \* \H(-2,0,0)
   \nn \\[0.5mm] & & \mbox{}
          + 576 \* (1 + 23/648/x + 235/216\: \* x - 2/81\: \* x^2\,) \* \H(-1,2)
          + 2176/3\: \* (1 + 9/136/x 
   \nn \\[0.5mm] & & \mbox{}
          - 13/68\: \* x + 3/34\: \* x^2\,) \* \H(-2,2)
          + 464 \* (1 + 2/29/x + 13/174\: \* x) \* \H(-3,0,0)
   \nn \\[0.5mm] & & \mbox{}
          - 1312/3\: \* (1 + 3/41/x + 19/164\: \* x) \* \H(-3) \*  \z2 
          + 1904/3\: \* (1 + 9/119/x + 20/119\: \* x 
   \nn \\[0.5mm] & & \mbox{}
          + 12/119\: \* x^2\,) \* \H(-3,0)
          - 1792/3\: \* (1 + 3/28/x + 11/112\: \* x) \* \H(-3,-1,0)
   \nn \\[0.5mm] & & \mbox{}
          - 1120/3\: \* (1 + 6/35/x + 33/35\: \* x) \* \H(5)
          + 1120/3\: \* (1 + 6/35/x + 34/35\: \* x) \* \H(0,0,0) \*  \z2 
   \nn \\[0.5mm] & & \mbox{}
          + 4576/9\: \* (1 + 243/1144/x + 85/143\: \* x - 4/143\: \* x^2\,) \* \H(-2,0)
          + 16 \* (1 + 1/3/x - 13/6\: \* x 
   \nn \\[0.5mm] & & \mbox{}
          + 4/3\: \* x^2\,) \* \H(-1,2,1)
          - 824 \* (1 + 109/309/x + 235/618\: \* x + 8/103\: \* x^2\,) \* \H(-1,-1,0,0)
   \nn \\[0.5mm] & & \mbox{}
          - 576 \* (1 + 31/72/x + 53/108\: \* x + 1/27\: \* x^2\,) \* \H(-1,-2,0)
          + 1024/3\: \* (1 + 111/256/x
   \nn \\[0.5mm] & & \mbox{}
         + 97/256\: \* x + 3/16\: \* x^2\,) \* \H(-1,0,0,0)
         - 480 \* (1 + 7/15/x + 3/10\: \* x) \* \H(-2,-2,0)
         - 64 \* \H(-2,2,1)
   \nn \\[0.5mm] & & \mbox{}
          - 432 \* (1 + 38/81/x + 31/162\: \* x + 4/81\: \* x^2\,) \* \H(-1,-1,2)
          - 312 \* (1 + 56/117/x 
          + 1/6\: \* x) \*  \z5 
   \nn \\[0.5mm] & & \mbox{}
          + 768 \* (1 + 1/2/x + 7/24\: \* x) \* \H(-2,-1,-1,0)
          + 944 \* (1 + 199/354/x 
          + 43/118\: \* x 
   \nn \\[0.5mm] & & \mbox{}        
          + 2/59\: \* x^2\,) \* \H(-1,-1) \*  \z2 
          - 836 \* (1 + 122/209/x + 149/418\: \* x
          + 32/627\: \* x^2\,) \* \H(-1) \*  \z3 
   \nn \\[0.5mm] & & \mbox{} 
          - 464 \* (1 + 18/29/x + 37/174\: \* x + 14/87\: \* x^2\,) \* \H(-1,0) \*  \z2 
          - 240 \* (1 + 83/90\: \* x) \* \H(0,0,0,0,0)
   \nn \\[0.5mm] & & \mbox{}
          + 1024 \* (1 + 41/64/x + 49/96\: \* x + 1/48\: \* x^2\,) \* \H(-1,-1,-1,0)
          + 1072/3\: \* (1 + 97/134/x
   \nn \\[0.5mm] & & \mbox{}
          + 9/67\: \* x + 12/67\: \* x^2\,) \* \H(-1,3)
          + 272/3\: \* (1 + 13/17/x + 2/17\: \* x + 4/17\: \* x^2\,) \* \H(-1,2,0)
   \nn \\[0.5mm] & & \mbox{}
          - 1024/3\: \* (1 + 15/16/x + 35/64\: \* x) \* \H(-2,-1,0,0)
          - 320/3\: \* (1 + 6/5/x + 59/20\: \* x) \* \H(4,0)
   \nn \\[0.5mm] & & \mbox{}
          - 160 \* (1 + 9/5/x + 21/20\: \* x) \* \H(-2) \*  \z3 
          + 512/3\: \* (1 + 15/8/x + 37/32\: \* x) \* \H(-2,-1) \*  \z2 
   \nn \\[0.5mm] & & \mbox{}
          - 64 \* (1 + 3/x + 67/12\: \* x) \* \H(4,1)
          + 13952/9\: \* (1 - 337/872\: \* x + 13/109\: \* x^2\,) \* \H(0,0,0,0)
   \nn \\[-0.5mm] & & \mbox{}
          - 160/3\: \* (1 - 1/10\: \* x) \* \H(-2,2,0)
          + 1952/3\: \* (1 + 1/61\: \* x) \* \H(-4,0)
          + 416/3\: \* (1 + 2/13\: \* x) \* \H(-3,2)
	   \Big)
   \nn \\[-0.5mm] & & \mbox{\hspn}
   + \colour4colour { \cft } \,  \*  \Big(
	    \pgq(x) \* (
            1288/15\: \* \H(1) \*  \zss\,
          + 640/3\: \* \H(1,-3,0)
          - 304 \* \H(1,-2) \*  \z2 
          - 160 \* \H(1,-2,-1,0)
   \nn \\[-0.5mm] & & \mbox{}
          + 256 \* \H(1,-2,0,0)
          + 224 \* \H(1,-2,2)
          + 400/3\: \* \H(1,0) \*  \z3 
          - 96 \* \H(1,0,0) \*  \z2 
          + 64 \* \H(1,0,0,0,0)
   \nn \\[0.5mm] & & \mbox{}
          - 352/3\: \* \H(1,1) \*  \z3 
          + 416/3\: \* \H(1,1,-2,0)
          - 944/3\: \* \H(1,1,0) \*  \z2 
          + 96 \* \H(1,1,0,0,0)
          - 368/3\: \* \H(1,1,1) \*  \z2
   \nn \\[0.5mm] & & \mbox{}
          - 608/3\: \* \H(1,1,1,0,0)
          - 400/3\: \* \H(1,1,1,1,0)
          - 80 \* \H(1,1,1,1,1)
          - 208/3\: \* \H(1,1,1,2)
          - 128 \* \H(1,1,2,0)
   \nn \\[0.5mm] & & \mbox{}
          - 64 \* \H(1,1,2,1)
          + 208 \* \H(1,1,3)
          - 96 \* \H(1,2) \*  \z2 
          - 272/3\: \* \H(1,2,0,0)
          - 400/3\: \* \H(1,2,1,0)
   \nn \\[0.5mm] & & \mbox{}
          - 272/3\: \* \H(1,2,1,1)
          - 80 \* \H(1,2,2)
          - 64 \* \H(1,3,0)
          - 80 \* \H(1,3,1)
          + 64 \* \H(1,4)
	      )
   \nn \\[0.5mm] & & \mbox{}
	  + \pgq( - x) \* (
            736/15\: \* \H(-1) \*  \zss\,
          - 832/3\: \* \H(-1,-3,0)
          + 400 \* \H(-1,-2) \*  \z2 
          + 352 \* \H(-1,-2,-1,0)
   \nn \\[0.5mm] & & \mbox{}
          - 1504/3\: \* \H(-1,-2,0,0)
          - 224 \* \H(-1,-2,2)
          + 464 \* \H(-1,-1) \*  \z3 
          + 352 \* \H(-1,-1,-2,0)
   \nn \\[0.5mm] & & \mbox{}
          - 448 \* \H(-1,-1,-1) \*  \z2 
          - 384 \* \H(-1,-1,-1,-1,0)
          + 704 \* \H(-1,-1,-1,0,0)
          + 256 \* \H(-1,-1,-1,2)
   \nn \\[0.5mm] & & \mbox{}
          + 1472/3\: \* \H(-1,-1,0) \*  \z2 
          - 1552/3\: \* \H(-1,-1,0,0,0)
          - 128 \* \H(-1,-1,2,0)
          - 128 \* \H(-1,-1,2,1)
   \nn \\[0.5mm] & & \mbox{}
          - 384 \* \H(-1,-1,3)
          - 112 \* \H(-1,0) \*  \z3 
          - 240 \* \H(-1,0,0) \*  \z2 
          + 448/3\: \* \H(-1,0,0,0,0)
          + 80 \* \H(-1,2) \*  \z2 
   \nn \\[0.5mm] & & \mbox{}
          - 96 \* \H(-1,2,0,0)
          + 64 \* \H(-1,3,0)
          + 64 \* \H(-1,3,1)
          + 208 \* \H(-1,4)
	   )
          - 6752/3\: \* (1 - 2/633/x^2 
   \nn \\[0.5mm] & & \mbox{}
          + 183/844/x + 4349/5064\: \* x) \* \H(-1,0)
          - 88/3\: \* (1 - 210/11/x - 21/2\: \* x) \*  \z5
   \nn \\[0.5mm] & & \mbox{}
          + 424/15\: \* (1 - 519/53/x + 4/53\: \* x) \*  \zss\,
          + 284/3\: \* (1 - 581/284/x + 887/284\: \* x) \* \H(1,1)
   \nn \\[0.5mm] & & \mbox{}
          + 1088/3\: \* (1 - 171/136/x - 65/136\: \* x) \* \H(1,2,1)
          + 920/3\: \* (1 - 57/46/x - 67/115\: \* x) \* \H(2,1)
   \nn \\[0.5mm] & & \mbox{}
          + 976/3\: \* (1 - 75/61/x - 101/244\: \* x) \* \H(1,1,1,1)
          + 1216/3\: \* (1 - 93/76/x 
   \nn \\[0.5mm] & & \mbox{}
          - 71/152\: \* x) \* \H(1,1,2)
          + 1340/3\: \* (1 - 358/335/x - 88/335\: \* x) \* \H(1,1,1)
   \nn \\[0.5mm] & & \mbox{}
          + 1856/3\: \* (1 - 30/29/x - 27/58\: \* x) \* \H(1,1,1,0)
          + 1136/3\: \* (1 - 72/71/x - 44/71\: \* x) \* \H(2,2)
   \nn \\[0.5mm] & & \mbox{}
          + 1436/3\: \* (1 - 364/359/x - 23/718\: \* x) \* \H(1,2)
          + 2036/3\: \* (1 - 504/509/x 
   \nn \\[0.5mm] & & \mbox{}
          - 193/1018\: \* x) \* \H(1,1,0)
          + 1424/3\: \* (1 - 84/89/x - 83/178\: \* x) \* \H(1,2,0)
   \nn \\[0.5mm] & & \mbox{}
          + 1600/3\: \* (1 - 9/10/x - 131/200\: \* x) \* \H(2,1,0)
          - 1664/3\: \* (1 - 93/104/x
   \nn \\[0.5mm] & & \mbox{}
          - 25/52\: \* x) \* \H(1,-2,0)
          + 1472/3\: \* (1 - 81/92/x - 187/368\: \* x) \* \H(2,1,1)
   \nn \\[0.5mm] & & \mbox{}
          + 916/3\: \* (1 - 180/229/x - 167/458\: \* x) \* \H(2,0)
          - 3265/6\: \* (1 - 493/653/x 
   \nn \\[0.5mm] & & \mbox{}
          - 4527/6530\: \* x) \* \H(1)
          + 896/3\: \* (1 - 3/4/x - 1/2\: \* x) \* \H(2,2,1)
          + 784/3\: \* (1 - 36/49/x 
   \nn \\[0.5mm] & & \mbox{}
          - 1/2\: \* x) \* \H(2,1,1,1)
          + 800/3\: \* (1 - 18/25/x - 27/50\: \* x) \* \H(2,1,2)
          + 1232/3\: \* (1 - 54/77/x 
   \nn \\[0.5mm] & & \mbox{}
          - 34/77\: \* x) \* \H(3,1)
          + 1216/3\: \* (1 - 12/19/x - 9/19\: \* x) \* \H(2,1,1,0)
          + 240 \* (1 - 3/5/x 
   \nn \\[0.5mm] & & \mbox{}
          - 3/5\: \* x) \* \H(3,0)
          + 1024/3\: \* (1 - 9/16/x - 15/32\: \* x) \* \H(2,2,0)
          - 70927/36\: \* (1 - 2098/3733/x 
   \nn \\[0.5mm] & & \mbox{}
          - 11754/70927\: \* x)
          + 1088/3\: \* (1 - 9/17/x - 1/2\: \* x) \* \H(3,2)
          - 1088/3\: \* (1 - 9/17/x 
   \nn \\[0.5mm] & & \mbox{}
          - 11/34\: \* x) \* \H(2,-2,0)
          + 368 \* (1 - 12/23/x - 1/2\: \* x) \* \H(3,1,1)
          + 424 \* (1 - 55/106/x 
   \nn \\[0.5mm] & & \mbox{}
          + 91/212\: \* x) \* \H(1,0)
          + 1312/3\: \* (1 - 18/41/x - 1/2\: \* x) \* \H(3,1,0)
          + 1232/3\: \* (1 - 24/77/x 
   \nn \\[0.5mm] & & \mbox{}
          - 13/22\: \* x) \* \H(1,1) \*  \z2 
          + 448 \* (1 - 2/7/x - 3/7\: \* x) \* \H(4,1)
          - 596 \* (1 - 36/149/x 
   \nn \\[0.5mm] & & \mbox{}
          + 271/447\: \* x) \* \H(0) \*  \z3 
          + 512/3\: \* (1 - 15/64/x - 77/128\: \* x) \* \H(1,1,0,0)
          + 832/3\: \* (1 - 3/13/x 
   \nn \\[0.5mm] & & \mbox{}
          - 3/26\: \* x) \* \H(2,1) \*  \z2 
          - 1000/3\: \* (1 - 27/125/x - 189/250\: \* x) \* \H(1,0,0)
          + 568 \* (1 - 15/71/x 
   \nn \\[0.5mm] & & \mbox{}
          - 121/213\: \* x) \* \H(1) \*  \z3 
          + 336 \* (1 - 4/21/x - 17/42\: \* x) \* \H(4,0)
          + 880/3\: \* (1 - 3/22/x 
   \nn \\[0.5mm] & & \mbox{}
          + 689/220\: \* x) \* \H(3)
          - 1592/3\: \* (1 - 24/199/x - 55/398\: \* x) \* \H(0,0) \*  \z3 
          + 976/3\: \* (1 - 6/61/x 
   \nn \\[0.5mm] & & \mbox{}
          - 33/122\: \* x) \* \H(2,1,0,0)
          + 2060/3\: \* (1 - 32/515/x - 861/1030\: \* x) \* \H(1) \*  \z2 
          - 288 \* (1 - 1/18/x 
   \nn \\[0.5mm] & & \mbox{}
          - 19/36\: \* x) \* \H(1,0,0,0)
          + 4352/3\: \* (1 + 3/68/x + 15/272\: \* x) \* \H(-3) \*  \z2 
          - 4192/3\: \* (1 + 6/131/x 
   \nn \\[0.5mm] & & \mbox{}
          + 21/262\: \* x) \* \H(-3,0,0)
          - 1304 \* (1 + 19/326/x + 1607/978\: \* x) \*  \z3 
          - 3584/3\: \* (1 + 9/112/x 
   \nn \\[0.5mm] & & \mbox{}
          + 33/224\: \* x) \* \H(-3,0)
          - 1056 \* (1 + 1/11/x + 5/33\: \* x) \* \H(-2,2)
          + 3712/3\: \* (1 + 3/29/x 
   \nn \\[0.5mm] & & \mbox{}
          + 15/116\: \* x) \* \H(-3,-1,0)
          - 7120/3\: \* (1 + 99/890/x + 121/445\: \* x) \* \H(-2,0)
          + 845 \* (1 + 8/65/x 
   \nn \\[0.5mm] & & \mbox{}
          + 39127/15210\: \* x) \* \H(0,0)
          + 1424/3\: \* (1 + 12/89/x - 9/178\: \* x) \* \H(2,0) \*  \z2
   \nn \\[0.5mm] & & \mbox{}
          + 5248/3\: \* (1 + 27/164/x + 167/656\: \* x) \* \H(-2) \*  \z2 
          - 736 \* (1 + 4/23/x + 9/46\: \* x) \* \H(-2,3)
   \nn \\[0.5mm] & & \mbox{}
          - 2720/3\: \* (1 + 3/17/x + 41/170\: \* x) \* \H(-2,0,0,0)
          + 176 \* (1 + 2/11/x - 5/22\: \* x) \* \H(2) \*  \z3 
   \nn \\[0.5mm] & & \mbox{}
          - 5392/3\: \* (1 + 63/337/x + 217/674\: \* x) \* \H(-2,0,0)
          + 960 \* (1 + 1/5/x + 2/9\: \* x) \* \H(-2,0) \*  \z2 
   \nn \\[0.5mm] & & \mbox{}
          + 1648/3\: \* (1 + 24/103/x - 273/412\: \* x) \* \H(1,0) \*  \z2 
          - 1264 \* (1 + 19/79/x + 62/79\: \* x) \* \H(-1,2)
   \nn \\[0.5mm] & & \mbox{}
          + 4160/3\: \* (1 + 18/65/x + 107/260\: \* x) \* \H(-2,-1,0)
          + 168 \* (1 + 1/3/x + 1079/504\: \* x) \* \H(2)
   \nn \\[0.5mm] & & \mbox{}
          + 4576/3\: \* (1 + 48/143/x + 81/286\: \* x) \* \H(-2,-1,0,0)
          + 7288/3\: \* (1 + 312/911/x 
   \nn \\[0.5mm] & & \mbox{}
          + 593/911\: \* x) \* \H(-1) \*  \z2 
          - 2528 \* (1 + 85/237/x + 151/237\: \* x) \* \H(-1,0,0)
          + 1152 \* (1 + 7/18/x 
   \nn \\[0.5mm] & & \mbox{}
          + 7/24\: \* x) \* \H(-2) \*  \z3 
          + 2432/3\: \* (1 + 15/38/x + 6/19\: \* x) \* \H(-2,-2,0)
          + 640 \* (1 + 2/5/x 
   \nn \\[0.5mm] & & \mbox{}
          + 3/10\: \* x) \* \H(-2,-1,2)
          - 224 \* (1 + 3/7/x + 4/7\: \* x) \* (\H(-1,2,0) + \H(-1,2,1) )
   \nn \\[0.5mm] & & \mbox{}
          - 1184 \* (1 + 16/37/x + 23/74\: \* x) \* \H(-2,-1) \*  \z2 
          + 6992/3\: \* (1 + 198/437/x 
   \nn \\[0.5mm] & & \mbox{}
          + 221/437\: \* x) \* \H(-1,-1,0)
          - 1088 \* (1 + 8/17/x + 11/34\: \* x) \* \H(-2,-1,-1,0)
   \nn \\[0.5mm] & & \mbox{}
          - 1088 \* (1 + 65/136/x + 55/102\: \* x) \* \H(-1,0,0,0)
          + 3472/3\: \* (1 + 108/217/x 
   \nn \\[0.5mm] & & \mbox{}
          + 249/434\: \* x) \* \H(-1,0) \*  \z2 
          + 256 \* (1 + 1/2/x + 19/48\: \* x) \* \H(3) \*  \z2 
          - 864 \* (1 + 1/2/x 
   \nn \\[0.5mm] & & \mbox{}
          + 7/12\: \* x) \* \H(-1,3)
          - 752/3\: \* (1 + 24/47/x + 17/94\: \* x) \* \H(2,3)
          + 368/3\: \* (1 + 12/23/x 
   \nn \\[0.5mm] & & \mbox{}
          - 5/46\: \* x) \*  \z2  \*  \z3 
          + 944/3\: \* (1 + 36/59/x - 19/118\: \* x) \* \H(2) \*  \z2 
          + 5984/3\: \* (1 + 21/34/x 
   \nn \\[0.5mm] & & \mbox{}
          + 395/748\: \* x) \* \H(-1,-1,0,0)
          + 928 \* (1 + 18/29/x + 31/58\: \* x) \* \H(-1,-1,2)
   \nn \\[0.5mm] & & \mbox{}
          + 1640 \* (1 + 138/205/x + 221/410\: \* x) \* \H(-1) \*  \z3 
          + 3424/3\: \* (1 + 147/214/x 
   \nn \\[0.5mm] & & \mbox{}
          + 55/107\: \* x) \* \H(-1,-2,0)
          - 1744 \* (1 + 75/109/x + 58/109\: \* x) \* \H(-1,-1) \*  \z2 
          - 832 \* \H(-3,2)
   \nn \\[0.5mm] & & \mbox{}
          - 880/3\: \* (1 + 42/55/x + 1173/220\: \* x) \* \H(0) \*  \z2 
          - 1632 \* (1 + 13/17/x 
   \nn \\[0.5mm] & & \mbox{}
          + 9/17\: \* x) \* \H(-1,-1,-1,0)
          - 256 \* (1 + 17/16/x - 51/64\: \* x) \* \H(1,3)
          - 236/3\: \* (1 + 72/59/x 
   \nn \\[0.5mm] & & \mbox{}
          + 365/59\: \* x) \* \H(0,0) \*  \z2 
          - 208/3\: \* (1 + 18/13/x + 105/52\: \* x) \* \H(2,0,0)
          - 272/5\: \* (1 + 40/17/x 
   \nn \\[0.5mm] & & \mbox{}
          + 2\: \* x) \* \H(0) \*  \zss\,
          - 168 \* (1 + 68/21/x + 20633/1512\: \* x) \*  \z2 
          + 2437/18\: \* (1 + 7972/2437/x 
   \nn \\[0.5mm] & & \mbox{}
          + 5058/2437\: \* x) \* \H(0)
          - 160/3\: \* (1 - 147/20\: \* x) \* \H(0,0,0,0)
          + 328 \* (1 - 21/82\: \* x) \* \H(5)
   \nn \\[0.5mm] & & \mbox{}
          - 416/3\: \* (1 - 5/26\: \* x) \* \H(2,0,0,0)
          - 328 \* (1 - 13/82\: \* x) \* \H(0,0,0) \*  \z2 
          - 2816/3\: \* (1 + 3/88\: \* x) \* \H(-4,0)
   \nn \\[0.5mm] & & \mbox{}
          - 192 \* (1 + 1/6\: \* x) \* ( \H(-2,2,0) + \H(-2,2,1) )
          + 120 \* (1 + 43/90\: \* x) \* \H(0,0,0,0,0)
   \nn \\[-0.5mm] & & \mbox{}
          + 1172/3\: \* (1 + 950/293\: \* x) \* \H(0,0,0)
          + 236/3\: \* (1 + 233/59\: \* x) \* \H(4)
          - 64 \* (1/x + 3/2\: \* x) \* \H(3,0,0)
          \Big)
   \nn \\[-0.5mm] & & \mbox{\hspn}
    +  \colour4colour { \cf \* \ca\, \* \nf } \, \*  \Big(
	    \pgq(x) \* (
            100/9\: \* \H(1) \*  \z3 
          + 160/9\: \* \H(1,-2,0)
          - 184/3\: \* \H(1,0) \*  \z2 
          + 688/9\: \* \H(1,0,0,0)
   \nn \\[-0.5mm] & & \mbox{}
          - 560/9\: \* \H(1,1) \*  \z2 
          + 544/9\: \* \H(1,1,0,0)
          + 128/3\: \* \H(1,1,1,0)
          + 376/9\: \* \H(1,1,1,1)
          + 448/9\: \* \H(1,1,2)
   \nn \\[0.5mm] & & \mbox{}
          + 512/9\: \* \H(1,2,0)
          + 416/9\: \* \H(1,2,1)
          + 560/9\: \* \H(1,3)
          )
	  + \pgq( - x) \* (
          - 20 \* \H(-1) \*  \z3 
          - 16 \* \H(-1,-2,0)
   \nn \\[0.5mm] & & \mbox{}
          + 80/3\: \* \H(-1,-1) \*  \z2 
          + 224/9\: \* \H(-1,-1,-1,0)
          - 32 \* \H(-1,-1,0,0)
          - 128/9\: \* \H(-1,-1,2)
   \nn \\[0.5mm] & & \mbox{}
          + 112/9\: \* \H(-1,0) \*  \z2 
          - 256/9\: \* \H(-1,0,0,0)
          - 128/9\: \* \H(-1,2,0)
          - 80/9\: \* \H(-1,2,1)
          - 104/9\: \* \H(-1,3)
	      )
   \nn \\[0.5mm] & & \mbox{}
          - 5744/27\: \* (1 + 2/359/x^2 + 1751/4308/x + 1049/1436\: \* x + 113/1077\: \* x^2\,) \* \H(-1,0)
   \nn \\[0.5mm] & & \mbox{}
          - 64/9\: \* (1 - 29/2/x - 23/2\: \* x) \* \H(2,2)
          - 176/9\: \* (1 - 63/11/x - 103/22\: \* x) \* \H(3,1)
   \nn \\[0.5mm] & & \mbox{}
          - 176/9\: \* (1 - 5/x - 7/2\: \* x) \* \H(2,1,1)
          - 15527/27\: \* (1 - 398483/139743/x 
   \nn \\[0.5mm] & & \mbox{}
          - 17114/139743\: \* x - 512/46581\: \* x^2\,)
          - 128/3\: \* (1 - 11/4/x - 5/3\: \* x) \* \H(2,1,0)
   \nn \\[0.5mm] & & \mbox{}
          + 448/9\: \* (1 - 5/2/x - 7/4\: \* x) \* \H(2) \*  \z2 
          - 512/9\: \* (1 - 5/2/x - 53/32\: \* x) \* \H(2,0,0)
   \nn \\[0.5mm] & & \mbox{}
          - 202498/243\: \* (1 - 183399/101249/x - 84161/202498\: \* x - 17064/101249\: \* x^2\,) \* \H(0)
   \nn \\[0.5mm] & & \mbox{}
          + 80/3\: \* (1 - 5/3/x + 7/15\: \* x) \* \H(0) \*  \z3 
          + 7376/27\: \* (1 - 1325/922/x - 1517/1844\: \* x 
   \nn \\[0.5mm] & & \mbox{}
          + 68/461\: \* x^2\,) \* \H(0) \*  \z2 
          - 7376/27\: \* (1 - 627/461/x - 1271/1844\: \* x + 68/461\: \* x^2\,) \* \H(3)
   \nn \\[0.5mm] & & \mbox{}
          - 110708/243\: \* (1 - 133181/110708/x - 26707/110708\: \* x - 6468/27677\: \* x^2\,) \* \H(1)
   \nn \\[0.5mm] & & \mbox{}
          + 11128/27\: \* (1 - 3301/2782/x - 1556/1391\: \* x - 10/1391\: \* x^2\,) \*  \z3 
   \nn \\[0.5mm] & & \mbox{}
          - 12256/27\: \* (1 - 1761/1532/x - 2001/3064\: \* x + 59/383\: \* x^2\,) \* \H(0,0,0)
   \nn \\[0.5mm] & & \mbox{}
          - 2560/9\: \* (1 - 533/480/x - 2/5\: \* x + 7/80\: \* x^2\,) \* \H(2,1)
          - 45940/81\: \* (1 - 24377/22970/x
   \nn \\[0.5mm] & & \mbox{}
          - 5353/22970\: \* x - 36/11485\: \* x^2\,) \* \H(1,0)
          - 9152/27\: \* (1 - 2407/2288/x - 667/2288\: \* x 
   \nn \\[0.5mm] & & \mbox{}
          + 1/22\: \* x^2\,) \* \H(1,2)
          - 9728/27\: \* (1 - 1261/1216/x - 467/1216\: \* x + 29/304\: \* x^2\,) \* \H(2,0)
   \nn \\[0.5mm] & & \mbox{}
          - 34412/81\: \* (1 - 1273/1229/x - 1474/8603\: \* x + 164/8603\: \* x^2\,) \* \H(1,1)
   \nn \\[0.5mm] & & \mbox{}
          - 86756/81\: \* (1 - 22113/21689/x - 7885/21689\: \* x + 1064/21689\: \* x^2\,) \* \H(0,0)
   \nn \\[0.5mm] & & \mbox{}
          - 13120/27\: \* (1 - 833/820/x - 577/1640\: \* x + 7/205\: \* x^2\,) \* \H(1,0,0)
   \nn \\[0.5mm] & & \mbox{}
          + 10624/27\: \* (1 - 84/83/x - 899/2656\: \* x + 29/664\: \* x^2\,) \* \H(1) \*  \z2 
   \nn \\[0.5mm] & & \mbox{}
          - 8264/27\: \* (1 - 2085/2066/x - 643/2066\: \* x + 44/1033\: \* x^2\,) \* \H(1,1,1)
          - 64/3\: \* (1 - 1/x 
   \nn \\[0.5mm] & & \mbox{}
          - x) \* \H(-2,2)
          - 2704/45\: \* (1 - 168/169/x - 185/338\: \* x) \*  \zss\,
          - 9776/27\: \* (1 - 2391/2444/x 
   \nn \\[0.5mm] & & \mbox{}
          - 823/2444\: \* x + 22/611\: \* x^2\,) \* \H(1,1,0)
          + 59024/81\: \* (1 - 3207/3689/x - 8753/29512\: \* x 
   \nn \\[0.5mm] & & \mbox{}
          + 303/3689\: \* x^2\,) \*  \z2 
          - 59024/81\: \* (1 - 11077/14756/x - 2459/29512\: \* x 
   \nn \\[0.5mm] & & \mbox{}
          + 303/3689\: \* x^2\,) \* \H(2)
          - 640/9\: \* (1 - 3/10/x - 3/10\: \* x) \* \H(-3,0)
          - 320/9\: \* (1 - 1/5/x 
   \nn \\[0.5mm] & & \mbox{}
          - 3/5\: \* x) \* \H(-2,0,0)
          + 1312/27\: \* (1 - 1/8/x + 83/82\: \* x + 9/82\: \* x^2\,) \* \H(-1) \*  \z2 
   \nn \\[0.5mm] & & \mbox{}
          - 1664/27\: \* (1 - 7/208/x + 79/104\: \* x + 1/4\: \* x^2\,) \* \H(-1,0,0)
          - 3392/27\: \* (1 + 71/424/x 
   \nn \\[0.5mm] & & \mbox{}
          + 123/424\: \* x + 13/106\: \* x^2\,) \* \H(-2,0)
          + 256/3\: \* (1 + 1/2/x + 1/8\: \* x) \* \H(-2,-1,0)
   \nn \\[0.5mm] & & \mbox{}
          + 1072/9\: \* (1 + 42/67/x + 143/134\: \* x) \* \H(0,0,0,0)
          + 2944/27\: \* (1 + 281/368/x + 29/46\: \* x 
   \nn \\[0.5mm] & & \mbox{}
          + 3/92\: \* x^2\,) \* \H(-1,-1,0)
          + 160/27\: \* (1 + 161/20/x - 5/2\: \* x - 3/5\: \* x^2\,) \* \H(-1,2)
   \nn \\[0.5mm] & & \mbox{}
          - 80/9\: \* (1 + 54/5/x + 113/10\: \* x) \* \H(0,0) \*  \z2 
          + 80/9\: \* (1 + 66/5/x + 137/10\: \* x) \* \H(4)
   \nn \\[-0.5mm] & & \mbox{}
          + 32/9\: \* (1 + 33/x + 55/2\: \* x) \* \H(3,0)
          + 64 \* (1 - 1/4\: \* x) \* \H(-2) \*  \z2 
	   \Big)
   \nn \\[-0.5mm] & & \mbox{\hspn}
   +  \colour4colour{ \cfs\, \* \nf } \, \* \Big(
	     \pgq(x) \* (
          - 128/3\: \* \H(1) \*  \z3 
          - 64/3\: \* \H(1,-2,0)
          - 8/9\: \* \H(1,0) \*  \z2 
          - 16/3\: \* \H(1,0,0,0)
   \nn \\[-0.5mm] & & \mbox{}
          + 104/9\: \* \H(1,1) \*  \z2 
          + 88/9\: \* \H(1,1,0,0)
          + 56/3\: \* \H(1,1,1,0)
          + 104/9\: \* \H(1,1,1,1)
          + 88/9\: \* \H(1,1,2)
   \nn \\[0.5mm] & & \mbox{}
          + 272/9\: \* \H(1,2,0)
          + 184/9\: \* \H(1,2,1)
          + 104/9\: \* \H(1,3)
	       )
	   + \pgq( - x) \* (
            112/3\: \* \H(-1) \*  \z3 
   \nn \\[0.5mm] & & \mbox{}
          + 128/3\: \* \H(-1,-2,0)
          - 128/3\: \* \H(-1,-1) \*  \z2 
          - 128/3\: \* \H(-1,-1,-1,0)
          + 224/3\: \* \H(-1,-1,0,0)
   \nn \\[0.5mm] & & \mbox{}
          + 64/3\: \* \H(-1,-1,2)
          + 64/3\: \* \H(-1,0) \*  \z2 
          - 32 \* \H(-1,0,0,0)
          - 32/3\: \* \H(-1,3)
	       )
   \nn \\[0.5mm] & & \mbox{}
	   +(2-x) \* (
          - 56/5\: \* \H(0) \*  \zss\,
          + 152/3\: \* \H(0,0) \*  \z3 
          + 88 \* \H(0,0,0) \*  \z2 
          - 120 \* \H(0,0,0,0,0)
          + 16 \* \H(3) \*  \z2 
   \nn \\[0.5mm] & & \mbox{}
          - 16 \* \H(3,0,0)
          - 16 \* \H(3,1,0)
          - 16 \* \H(3,1,1)
          - 16 \* \H(3,2)
          - 48 \* \H(4,0)
          - 48 \* \H(4,1)
          - 88 \* \H(5)
          )
   \nn \\[0.5mm] & & \mbox{}
	   + (2-x) \* (
            24 \*  \z5 
          - 32 \*  \z2  \*  \z3 
	       )
          - 112/9\: \* (1 - 25/7/x - 2/7\: \* x) \* \H(2,0,0)
          + 64/3\: \* (1 - 31/12/x
   \nn \\[0.5mm] & & \mbox{}
          + 13/4\: \* x - 1/3\: \* x^2\,) \* \H(-1,2)
          + 28030/81\: \* (1 - 16948/14015/x + 2234/14015\: \* x
   \nn \\[0.5mm] & & \mbox{}
          - 1488/14015\: \* x^2\,) \*  \z2 
          - 544/9\: \* (1 - 39/34/x - 7/68\: \* x) \* \H(2,1,1)
   \nn \\[0.5mm] & & \mbox{}
          - 28030/81\: \* (1 - 3168/2803/x - 9646/14015\: \* x - 1488/14015\: \* x^2\,) \* \H(2)
   \nn \\[0.5mm] & & \mbox{}
          - 608/9\: \* (1 - 21/19/x - 11/76\: \* x) \* \H(2,1,0)
          - 704/9\: \* (1 - 12/11/x - 17/88\: \* x) \* \H(2,2)
   \nn \\[0.5mm] & & \mbox{}
          + 1072/9\: \* (1 - 72/67/x - 53/268\: \* x) \* \H(0,0) \*  \z2 
          - 848/9\: \* (1 - 54/53/x + 19/106\: \* x) \* \H(3,1)
   \nn \\[0.5mm] & & \mbox{}
          - 880/9\: \* (1 - 54/55/x + 17/110\: \* x) \* \H(3,0)
          + 4640/27\: \* (1 - 142/145/x - 101/232\: \* x 
   \nn \\[0.5mm] & & \mbox{}
          + 9/145\: \* x^2\,) \* \H(1) \*  \z2 
          - 1072/9\: \* (1 - 60/67/x + 19/268\: \* x) \* \H(4)
          - 7712/27\: \* (1 - 823/964/x
   \nn \\[0.5mm] & & \mbox{}
          - 1087/1928\: \* x + 6/241\: \* x^2\,) \* \H(1,2)
          - 1624/9\: \* (1 - 139/174/x - 95/203\: \* x 
   \nn \\[0.5mm] & & \mbox{}
          + 32/609\: \* x^2\,) \* \H(1,0,0)
          - 6448/27\: \* (1 - 1279/1612/x - 445/806\: \* x + 12/403\: \* x^2\,) \* \H(1,1,1)
   \nn \\[0.5mm] & & \mbox{}
          - 7552/27\: \* (1 - 185/236/x - 1007/1888\: \* x + 3/118\: \* x^2\,) \* \H(1,1,0)
   \nn \\[0.5mm] & & \mbox{}
          - 9148/27\: \* (1 - 1790/2287/x - 1323/2287\: \* x - 192/2287\: \* x^2\,) \* \H(1,0)
   \nn \\[0.5mm] & & \mbox{}
          - 246517/243\: \* (1 - 185803/246517/x - 181615/493034\: \* x - 5280/246517\: \* x^2\,) \* \H(1)
   \nn \\[0.5mm] & & \mbox{}
          - 28300/81\: \* (1 - 5272/7075/x - 907/1415\: \* x - 576/7075\: \* x^2\,) \* \H(1,1)
   \nn \\[0.5mm] & & \mbox{}
          + 2528/9\: \* (1 - 43/79/x - 247/632\: \* x) \* \H(0) \*  \z3 
          - 19352/81\: \* (1 - 2481/4838/x 
   \nn \\[0.5mm] & & \mbox{}
          + 19385/19352\: \* x - 392/2419\: \* x^2\,) \* \H(0,0)
          - 585697/486\: \* (1 - 842416/1757091/x 
   \nn \\[0.5mm] & & \mbox{}
          - 129025/585697\: \* x - 9872/251013\: \* x^2\,)
          - 12872/27\: \* (1 - 705/1609/x 
   \nn \\[0.5mm] & & \mbox{}
          - 1111/3218\: \* x + 72/1609\: \* x^2\,) \* \H(2,0)
          - 12832/27\: \* (1 - 351/802/x - 139/401\: \* x 
   \nn \\[0.5mm] & & \mbox{}
          + 18/401\: \* x^2\,) \* \H(2,1)
          - 440/9\: \* (1 - 24/55/x + 89/110\: \* x) \* \H(0,0,0,0)
   \nn \\[0.5mm] & & \mbox{}
          - 254765/243\: \* (1 - 14591/36395/x - 103486/254765\: \* x - 2592/254765\: \* x^2\,) \* \H(0)
   \nn \\[0.5mm] & & \mbox{}
          + 18752/27\: \* (1 - 439/1172/x - 839/9376\: \* x + 12/293\: \* x^2\,) \* \H(0) \*  \z2 
   \nn \\[0.5mm] & & \mbox{}
          - 18752/27\: \* (1 - 189/586/x - 1343/9376\: \* x + 12/293\: \* x^2\,) \* \H(3)
   \nn \\[0.5mm] & & \mbox{}
          + 12176/27\: \* (1 - 239/761/x + 3553/6088\: \* x + 14/761\: \* x^2\,) \*  \z3 
   \nn \\[0.5mm] & & \mbox{}
          + 320/3\: \* (1 - 1/5/x - 3/10\: \* x) \* \H(-2,2)
          + 2528/9\: \* (1 - 61/474/x + 21/158\: \* x 
   \nn \\[0.5mm] & & \mbox{}
          - 10/237\: \* x^2\,) \* \H(-2,0)
          + 832/3\: \* (1 - 277/2808/x + 55/52\: \* x + 5/702\: \* x^2\,) \* \H(-1,0)
   \nn \\[0.5mm] & & \mbox{}
          + 320 \* (1 - 1/15/x - 1/10\: \* x) \* \H(-3,0)
          - 18896/27\: \* (1 - 15/1181/x - 151/4724\: \* x 
   \nn \\[0.5mm] & & \mbox{}
          + 52/1181\: \* x^2\,) \* \H(0,0,0)
          + 832/3\: \* (1 + 1/39/x - 3/26\: \* x) \* \H(-2,0,0)
          - 608/3\: \* (1 + 1/19/x 
   \nn \\[0.5mm] & & \mbox{}
          - 5/38\: \* x) \* \H(-2) \*  \z2 
          + 1504/9\: \* (1 + 19/282/x + 55/47\: \* x - 10/141\: \* x^2\,) \* \H(-1,0,0)
   \nn \\[0.5mm] & & \mbox{}
          - 1216/9\: \* (1 + 23/152/x + 175/152\: \* x - 3/38\: \* x^2\,) \* \H(-1) \*  \z2 
          - 192 \* (1 + 1/3/x 
   \nn \\[0.5mm] & & \mbox{}
          + 1/18\: \* x) \* \H(-2,-1,0)
          + 5096/45\: \* (1 + 246/637/x - 1/14\: \* x) \*  \zss\,
          - 2048/9\: \* (1 + 85/128/x 
   \nn \\[-0.5mm] & & \mbox{}
          + 97/128\: \* x - 1/32\: \* x^2\,) \* \H(-1,-1,0)
          - 160/9\: \* (1 + 3/x + 11/20\: \* x) \* \H(2) \*  \z2 
	   \Big)
   \nn \\[-0.5mm] & & \mbox{\hspn}
   +  \colour4colour { \cf \* \nfs } \, \*  \Big(
	    \pgq(x) \* (
            8/3\: \*  \z3 
          + 64/9\: \* \H(0) \*  \z2 
          - 16 \* \H(0,0,0)
          + 40/9\: \* \H(1) \*  \z2 
          - 64/9\: \* \H(1,0,0)
   \nn \\[-0.5mm] & & \mbox{}
          - 40/9\: \* \H(1,1,0)
          - 32/9\: \* \H(1,1,1)
          - 40/9\: \* \H(1,2)
          - 64/9\: \* \H(2,0)
          - 40/9\: \* \H(2,1)
          - 64/9\: \* \H(3)
          )
   \nn \\[0.5mm] & & \mbox{}
          + 13726/243\: \* (1 - 25291/13726/x - 5807/13726\: \* x)
          + 5440/81\: \* (1 - 959/680/x 
   \nn \\[0.5mm] & & \mbox{}
          - 59/136\: \* x) \* \H(0)
          + 1712/27\: \* (1 - 241/214/x - 101/214\: \* x) \* \H(0,0)
   \nn \\[0.5mm] & & \mbox{}
          + 1036/27\: \* (1 - 551/518/x - 239/518\: \* x) \* \H(1)
          - 848/27\: \* (1 - 1/x - 25/53\: \* x) \*  (\z2 
          - \H(1,0)
   \nn \\[-0.5mm] & & \mbox{}
          - \H(2))
          + 536/27\: \* (1 - 125/134/x - 61/134\: \* x) \* \H(1,1)
	   \Big)
\;\; .
\eea 
The corresponding expression for the gluon coefficient function is given by
\bea
\label{eq:cphig3}
 \lefteqn{c^{\,(3)}_{\rm \phi\, , g}(x) \; = \;
%
%
          \colour4colour{\cat}  \*  \Big(
          - 44228/27 \* (1 - 8/11057 / x^2 + 8977/11057/x + 11031/11057 \* x 
 }
%
%
   \nn \\[-0.5mm] & & \mbox{}
          + 8959/11057 \* x^2) \* \H(-1,0)
          - 7262/27 \* (1 - 19412/3631/x - 842/3631 \* x + 16705/3631 \* x^2)
          \*  \z3 
   \nn \\[0.5mm] & & \mbox{}
          + 12724/27 \* (1 - 48356/9543/x - 16641/6362 \* x + 67957/19086 \*
          x^2) \* \H(1,1)
   \nn \\[0.5mm] & & \mbox{}
          - 608/3 \* (1 - 275/57/x - 153/76 \* x + 869/228 \* x^2) \* \H(1,0)
          \*  \z2 
          + 7672/9 \* (1 - 59035/17262/x 
   \nn \\[0.5mm] & & \mbox{}
          - 739/411 \* x + 12371/4932 \* x^2) \* \H(1,0)
          + 544 \* (1 - 77/34/x - 113/34 \* x + 66/17 \* x^2) \* \H(2,1,1)
   \nn \\[0.5mm] & & \mbox{}
          + 408 \* (1 - 110/51/x - 383/306 \* x + 583/306 \* x^2) \* \H(1,0,0,0)
          + 5968/9 \* (1 - 759/373/x 
   \nn \\[0.5mm] & & \mbox{}
          - 1955/1492 \* x + 2563/1492 \* x^2) \* \H(1,3)
          - 6200/9 \* (1 - 1573/775/x - 941/775 \* x 
   \nn \\[0.5mm] & & \mbox{}
          + 1397/775 \* x^2) \* \H(1,1) \*  \z2 
          - 4616/3 \* (1 - 20489/10386/x - 1483/1154 \* x 
   \nn \\[0.5mm] & & \mbox{}
          + 18311/10386 \* x^2) \* \H(1) \*  \z2 
          + 8128/9 \* (1 - 913/508/x - 623/508 \* x + 803/508 \* x^2) \* \H(1,1,1,0)
   \nn \\[0.5mm] & & \mbox{}
          + 13448/9 \* (1 - 8963/5043/x 
          - 4355/3362 \* x + 7874/5043 \* x^2) \* \H(1,2)
   \nn \\[0.5mm] & & \mbox{}
          + 6544/9 \* (1 - 726/409/x - 4373/1636 \* x 
          + 5357/1636 \* x^2) \* \H(2,2)
          + 800 \* (1 - 44/25/x 
   \nn \\[0.5mm] & & \mbox{}
            - 61/50 \* x + 77/50 \* x^2) \* \H(1,1,1,1)
          + 8536/9 \* (1 - 169/97/x - 2597/2134 \* x 
   \nn \\[0.5mm] & & \mbox{}
            + 295/194 \* x^2) \* (\H(1,1,2) + \H(1,2,1))
          + 13960/9 \* (1 - 2943/1745/x - 871/698 \* x 
   \nn \\[0.5mm] & & \mbox{}
            + 5039/3490 \* x^2) \* \H(1,1,1)
          + 9832/9 \* (1 - 2013/1229/x - 1426/1229 \* x + 1826/1229 \* x^2) \*
          \H(1,2,0)
   \nn \\[0.5mm] & & \mbox{}
          + 137744/27 \* (1 - 13910/8609/x - 492137/413232 \* x +
          568297/413232 \* x^2) \* \H(1)
   \nn \\[0.5mm] & & \mbox{}
          + 15400/9 \* (1 - 9326/5775/x - 4593/3850 \* x + 7874/5775 \* x^2) \* \H(1,1,0)
   \nn \\[0.5mm] & & \mbox{}
          + 10904/9 \* (1 - 2189/1363/x - 1529/1363 \* x + 2013/1363 \* x^2) \* \H(1,1,0,0)
   \nn \\[0.5mm] & & \mbox{}
          + 16040/9 \* (1 - 3041/2005/x - 4621/4010 \* x + 5477/4010 \* x^2) \* \H(1,0,0)
   \nn \\[0.5mm] & & \mbox{}
          + 7304/9 \* (1 - 123/83/x - 1916/913 \* x + 248/83 \* x^2) \* \H(2,1,0)
          + 7544/9 \* (1 - 33/23/x 
   \nn \\[0.5mm] & & \mbox{}
          - 1931/943 \* x + 2541/943 \* x^2) \* \H(2,0,0)
          + 7784/9 \* (1 - 880/973/x - 4183/1946 \* x 
   \nn \\[0.5mm] & & \mbox{}
          + 1067/278 \* x^2) \* \H(3,1)
          + 2668/3 \* (1 - 583/667/x - 3995/2001 \* x + 7139/2001 \* x^2) \*
          \H(3,0)
   \nn \\[0.5mm] & & \mbox{}
          + 68572/27 \* (1 - 13324/17143/x - 599/553 \* x + 28492/17143 \*
          x^2) \* \H(2,0)
   \nn \\[0.5mm] & & \mbox{}
          + 62572/27 \* (1 - 10990/15643/x - 14213/15643 \* x + 27766/15643 \* x^2) \* \H(2,1)
   \nn \\[0.5mm] & & \mbox{}
          + 788 \* (1 - 1111/1773/x - 4202/1773 \* x + 8558/1773 \* x^2) \* \H(4)
          + 9506/3 \* (1 - 23900/42777/x 
   \nn \\[0.5mm] & & \mbox{}
          - 13907/14259 \* x + 25964/14259 \* x^2) \* \H(3)
          - 788 \* (1 - 935/1773/x - 4160/1773 \* x 
   \nn \\[0.5mm] & & \mbox{}
          + 8558/1773 \* x^2) \* \H(0,0) \*  \z2 
          - 9506/3 \* (1 - 20684/42777/x - 4463/6111 \* x 
   \nn \\[0.5mm] & & \mbox{}
          + 25964/14259 \* x^2) \* \H(0) \*  \z2 
          + 3184/9 \* (1 - 88/199/x - 301/398 \* x + 253/398 \* x^2) \*
          \H(1,-2,0)
   \nn \\[0.5mm] & & \mbox{}
          + 959401/81 \* (1 - 306237/959401/x - 230274/959401 \* x +
          3294515/2878203 \* x^2) \* \H(0)
   \nn \\[0.5mm] & & \mbox{}
          + 146410/27 \* (1 - 16284/73205/x - 57243/73205 \* x + 101818/73205 \* x^2) \* \H(0,0,0)
   \nn \\[0.5mm] & & \mbox{}
          - 22004/9 \* (1 - 957/5501/x - 1450/5501 \* x + 9372/5501 \* x^2) \*
          \H(0) \*  \z3
   \nn \\[0.5mm] & & \mbox{} 
          - 16168/9 \* (1 - 341/2021/x + 619/2021 \* x + 1661/2021 \* x^2) \* \H(-2,2)
   \nn \\[0.5mm] & & \mbox{}
          + 23116/9 \* (1 - 891/5779/x + 998/5779 \* x + 4752/5779 \* x^2) \* \H(-2) \*  \z2 
   \nn \\[0.5mm] & & \mbox{}
          - 10360/3 \* (1 - 517/3885/x - 13/2590 \* x + 7337/7770 \* x^2) \* \H(-2,0,0)
   \nn \\[0.5mm] & & \mbox{}
          + 1544 \* (1 - 209/1737/x - 80/579 \* x + 1430/1737 \* x^2) \* \H(-2,-1,0)
          + 63584/27 \* (1 - 201/1987/x 
   \nn \\[0.5mm] & & \mbox{}
          - 655/1987 \* x + 1959/1987 \* x^2) \* \H(-2,0)
          - 15722/3 \* (1 - 31615/1910223/x - 477635/848988 \* x 
   \nn \\[0.5mm] & & \mbox{}
          + 3411235/7640892 \* x^2)
          - 7568/3 \* (1 + 4/129/x + 7/946 \* x + 89/86 \* x^2) \* \H(-3,0)
   \nn \\[0.5mm] & & \mbox{}
          + 1344 \* (1 + 143/378/x + 263/252 \* x + 253/756 \* x^2) \* \H(-1,-1,2)
          - 3464/9 \* (1 + 198/433/x 
   \nn \\[0.5mm] & & \mbox{}
          + 2302/433 \* x - 2970/433 \* x^2) \* \H(0,0,0,0)
          - 14432/9 \* (1 + 39/82/x + 3871/3608 \* x 
   \nn \\[0.5mm] & & \mbox{}
          + 133/328 \* x^2) \* \H(-1,-1) \*  \z2 
          + 16532/9 \* (1 + 2321/4133/x + 4363/4133 \* x + 2101/4133 \* x^2) \* \H(-1) \*  \z3 
   \nn \\[0.5mm] & & \mbox{}
          - 2072 \* (1 + 1441/2331/x + 788/777 \* x + 1408/2331 \* x^2) \* \H(-1,3)
          + 22792/9 \* (1 + 169/259/x 
   \nn \\[0.5mm] & & \mbox{}
          + 5765/5698 \* x + 331/518 \* x^2) \* \H(-1,0) \*  \z2 
          - 2960/3 \* (1 + 121/185/x + x + 121/185 \* x^2) \* \H(-1,2,1)
   \nn \\[0.5mm] & & \mbox{}
          - 2384/3 \* (1 + 99/149/x + x + 99/149 \* x^2) \* \H(-1,2,0)
          + 5552/3 \* (1 + 275/347/x + 1497/1388 \* x 
   \nn \\[0.5mm] & & \mbox{}
          + 1001/1388 \* x^2) \* \H(-1,-1,0,0)
          - 302995/81 \* (1 + 244674/302995/x + 313876/302995 \* x 
   \nn \\[0.5mm] & & \mbox{}
          + 19086/43285 \* x^2) \* \H(0,0)
          - 6776/3 \* (1 + 72/77/x + 157/154 \* x + 141/154 \* x^2) \*
          \H(-1,0,0,0)
   \nn \\[0.5mm] & & \mbox{}
          + 4364/9 \* (1 + 1045/1091/x - 532/1091 \* x - 506/1091 \* x^2) \* \H(1) \*  \z3 
          - 4672/9 \* (1 + 143/146/x 
   \nn \\[0.5mm] & & \mbox{}
          + 715/584 \* x + 451/584 \* x^2) \* \H(-1,-1,-1,0)
          + 5104/9 \* (1 + 30/29/x + 368/319 \* x 
   \nn \\[0.5mm] & & \mbox{}
          + 26/29 \* x^2) \* \H(-1,-2,0)
          + 1848 \* (1 + 7127/6237/x + 2074/2079 \* x + 7127/6237 \* x^2) \*
          \H(-1,0,0)
   \nn \\[0.5mm] & & \mbox{}
          + 172045/81 \* (1 + 243046/172045/x + 207847/172045 \* x -
          26002/172045 \* x^2) \*  \z2 
   \nn \\[0.5mm] & & \mbox{}
          + 5888/9 \* (1 + 3557/2208/x + x + 3557/2208 \* x^2) \* \H(-1,2)
          - 172045/81 \* (1 + 70154/34409/x 
   \nn \\[0.5mm] & & \mbox{}
          + 340219/172045 \* x - 26002/172045 \* x^2) \* \H(2)
          - 2096/3 \* (1 + 9677/4716/x + 261/262 \* x 
   \nn \\[0.5mm] & & \mbox{}
          + 9677/4716 \* x^2) \* \H(-1) \*  \z2 
          - 17752/45 \* (1 + 7172/2219/x + 6750/2219 \* x - 6061/4438 \* x^2)
          \*  \zss
   \nn \\[0.5mm] & & \mbox{}
          - 800/9 \* (1 + 2563/300/x + 47/50 \* x + 2563/300 \* x^2) \* \H(-1,-1,0)
          + 404/9 \* (1 + 3113/101/x 
   \nn \\[0.5mm] & & \mbox{}
          + 4613/101 \* x - 3927/101 \* x^2) \* \H(2) \*  \z2 
          - 2752/3 \* (1 - 13/43 \* x) \* \H(-3,-1,0)
   \nn \\[0.5mm] & & \mbox{}
          + 6464/3 \* (1 - 16/101 \* x) \* \H(-3,0,0)
          + 5440/3 \* (1 - 7/85 \* x) \* \H(-4,0)
          - 4384/3 \* (1 - 11/137 \* x) \* \H(-3) \*  \z2 
   \nn \\[0.5mm] & & \mbox{}
          + 1856 \* (1 - 3/58 \* x) \*  \z5 
          + 3008/3 \* (1 + 1/47 \* x) \* \H(-3,2)
          - 1760 \* (1 + 131/165 \* x) \*  \z2  \*  \z3 
   \nn \\[0.5mm] & & \mbox{}
          + 6656/3 \* (1 + 11/13 \* x) \* \H(0,0) \*  \z3 
          - 1472 \* (1 + 85/69 \* x) \* \H(3,1,0)
          - 1344 \* (1 + 9/7 \* x) \* \H(3,1,1)
   \nn \\[0.5mm] & & \mbox{}
          - 4288/3 \* (1 + 87/67 \* x) \* \H(3,2)
          - 3616/3 \* (1 + 157/113 \* x) \* \H(3,0,0)
          - 5152/3 \* (1 + 35/23 \* x) \* \H(4,1)
   \nn \\[0.5mm] & & \mbox{}
          - 4384/3 \* (1 + 221/137 \* x) \* \H(4,0)
          + 2912/3 \* (1 + 23/13 \* x) \* \H(3) \*  \z2 
          + 1344 \* (1 + 2 \* x) \* \H(0,0,0) \*  \z2 
   \nn \\[0.5mm] & & \mbox{}
          - 1344 \* (1 + 19/9 \* x) \* \H(5)
          - 2336/15 \* (1 + 452/73 \* x) \* \H(0) \*  \zss
          - 1920 \* x \* \H(0,0,0,0,0)
   \nn \\[0.5mm] & & \mbox{}
	  + \pgg( - x) \* (
           504 \*  \z5 
          + 1876/3 \*  \z3 
          + 6464/81 \*  \z2 
          - 1888/3 \*  \z2  \*  \z3 
          + 5038/45 \*  \zss
          + 1376/3 \* \H(-4,0)
   \nn \\[0.5mm] & & \mbox{}
          - 1184 \* \H(-3) \*  \z2 
          - 832 \* \H(-3,-1,0)
          + 3872/9 \* \H(-3,0)
          + 1120 \* \H(-3,0,0)
          + 768 \* \H(-3,2)
          - 1712 \* \H(-2) \*  \z3 
   \nn \\[0.5mm] & & \mbox{}
          - 616 \* \H(-2) \*  \z2 
          - 1984/3 \* \H(-2,-2,0)
          + 1712 \* \H(-2,-1) \*  \z2 
          + 800 \* \H(-2,-1,-1,0)
          - 5104/9 \* \H(-2,-1,0)
   \nn \\[0.5mm] & & \mbox{}
          - 5408/3 \* \H(-2,-1,0,0)
          - 1312 \* \H(-2,-1,2)
          + 536 \* \H(-2,0)
          - 1808 \* \H(-2,0) \*  \z2 
          + 704 \* \H(-2,0,0)
   \nn \\[0.5mm] & & \mbox{}
          + 4144/3 \* \H(-2,0,0,0)
          + 2992/9 \* \H(-2,2)
          + 1376/3 \* \H(-2,2,0)
          + 544 \* \H(-2,2,1)
          + 4576/3 \* \H(-2,3)
   \nn \\[0.5mm] & & \mbox{}
          - 7436/9 \* \H(-1) \*  \z3 
          - 2144/3 \* \H(-1) \*  \z2 
          + 1528/15 \* \H(-1) \*  \zss
          - 832 \* \H(-1,-3,0)
          + 800 \* \H(-1,-2,-1,0)
   \nn \\[0.5mm] & & \mbox{}
          + 1712 \* \H(-1,-2) \*  \z2 
          - 5192/9 \* \H(-1,-2,0)
          - 5408/3 \* \H(-1,-2,0,0)
          - 1312 \* \H(-1,-2,2)
          + 6448/3 \* \H(-1,-1) \*  \z3 
   \nn \\[0.5mm] & & \mbox{}
          + 7568/9 \* \H(-1,-1) \*  \z2 
          + 800 \* \H(-1,-1,-2,0)
          - 2112 \* \H(-1,-1,-1) \*  \z2 
          - 768 \* \H(-1,-1,-1,-1,0)
   \nn \\[0.5mm] & & \mbox{}
          + 6688/9 \* \H(-1,-1,-1,0)
          + 2336 \* \H(-1,-1,-1,0,0)
          + 1728 \* \H(-1,-1,-1,2)
          - 2144/3 \* \H(-1,-1,0)
   \nn \\[0.5mm] & & \mbox{}
          + 7648/3 \* \H(-1,-1,0) \*  \z2 
          - 8624/9 \* \H(-1,-1,0,0)
          - 2096 \* \H(-1,-1,0,0,0)
          - 1408/3 \* \H(-1,-1,2)
   \nn \\[0.5mm] & & \mbox{}
          - 2176/3 \* \H(-1,-1,2,0)
          - 2432/3 \* \H(-1,-1,2,1)
          - 6496/3 \* \H(-1,-1,3)
          + 12928/81 \* \H(-1,0)
   \nn \\[0.5mm] & & \mbox{}
          - 4784/3 \* \H(-1,0) \*  \z3 
          - 1936/3 \* \H(-1,0) \*  \z2 
          + 2680/3 \* \H(-1,0,0)
          - 4400/3 \* \H(-1,0,0) \*  \z2 
   \nn \\[0.5mm] & & \mbox{}
          + 5500/9 \* \H(-1,0,0,0)
          + 2864/3 \* \H(-1,0,0,0,0)
          + 1072/3 \* \H(-1,2)
          + 48 \* \H(-1,2) \*  \z2 
          + 704/3 \* \H(-1,2,0)
   \nn \\[0.5mm] & & \mbox{}
          + 256 \* \H(-1,2,0,0)
          + 704/3 \* \H(-1,2,1)
          + 704/3 \* \H(-1,2,1,0)
          + 192 \* \H(-1,2,1,1)
          + 704/3 \* \H(-1,2,2)
   \nn \\[0.5mm] & & \mbox{}
          + 528 \* \H(-1,3)
          + 672 \* \H(-1,3,0)
          + 2336/3 \* \H(-1,3,1)
          + 1328 \* \H(-1,4)
          + 120 \* \H(0)
          + 5360/27 \* \H(0) \*  \z2 
   \nn \\[0.5mm] & & \mbox{}
          + 3212/9 \* \H(0) \*  \z3 
          + 904/15 \* \H(0) \*  \zss
          - 6464/81 \* \H(0,0)
          + 1712/3 \* \H(0,0) \*  \z3 
          + 1804/9 \* \H(0,0) \*  \z2 
          - 268 \* \H(0,0,0)
   \nn \\[0.5mm] & & \mbox{}
          + 432 \* \H(0,0,0) \*  \z2 
          - 1364/9 \* \H(0,0,0,0)
          - 208 \* \H(0,0,0,0,0)
          - 496/3 \* \H(2) \*  \z3 
          - 836/9 \* \H(2) \*  \z2 
          - 112 \* \H(2,0) \*  \z2 
   \nn \\[0.5mm] & & \mbox{}
          - 176/9 \* \H(2,0,0)
          + 64/3 \* \H(2,0,0,0)
          - 304/3 \* \H(2,1) \*  \z2 
          + 88/9 \* \H(2,1,0)
          + 64/3 \* \H(2,1,0,0)
          - 88/9 \* \H(2,2)
   \nn \\[0.5mm] & & \mbox{}
          + 64/3 \* \H(2,3)
          - 536/3 \* \H(3)
          - 608/3 \* \H(3) \*  \z2 
          - 308/3 \* \H(3,0)
          - 352/3 \* \H(3,0,0)
          - 352/3 \* \H(3,1)
   \nn \\[0.5mm] & & \mbox{}
          - 352/3 \* \H(3,1,0)
          - 96 \* \H(3,1,1)
          - 352/3 \* \H(3,2)
          - 1628/9 \* \H(4)
          - 784/3 \* \H(4,0)
          - 976/3 \* \H(4,1)
          - 400 \* \H(5)
	      )
   \nn \\[0.5mm] & & \mbox{}
	  + \pgg(x) \* (
          - 1616486/729
          - 480 \*  \z5 
          + 21994/9 \*  \z3 
          + 170666/81 \*  \z2 
          - 2896/3 \*  \z2  \*  \z3 
   \nn \\[0.5mm] & & \mbox{}
          - 25102/45 \*  \zss
          + 1184/3 \* \H(-4,0)
          - 960 \* \H(-3) \*  \z2 
          - 640 \* \H(-3,-1,0)
          + 3520/9 \* \H(-3,0)
          + 896 \* \H(-3,0,0)
   \nn \\[0.5mm] & & \mbox{}
          + 640 \* \H(-3,2)
          - 2144/3 \* \H(-2) \*  \z3 
          - 572/3 \* \H(-2) \*  \z2 
          - 544/3 \* \H(-2,-2,0)
          + 752 \* \H(-2,-1) \*  \z2 
   \nn \\[0.5mm] & & \mbox{}
          + 608/3 \* \H(-2,-1,-1,0)
          - 616/3 \* \H(-2,-1,0)
          - 2048/3 \* \H(-2,-1,0,0)
          - 1952/3 \* \H(-2,-1,2)
   \nn \\[0.5mm] & & \mbox{}
          + 13400/27 \* \H(-2,0)
          - 2320/3 \* \H(-2,0) \*  \z2 
          + 2552/9 \* \H(-2,0,0)
          + 1312/3 \* \H(-2,0,0,0)
          + 88 \* \H(-2,2)
   \nn \\[0.5mm] & & \mbox{}
          + 544/3 \* \H(-2,2,0)
          + 608/3 \* \H(-2,2,1)
          + 2048/3 \* \H(-2,3)
          - 191858/81 \* \H(0)
          + 11792/9 \* \H(0) \*  \z3 
   \nn \\[0.5mm] & & \mbox{}
          + 6656/3 \* \H(0) \*  \z2 
          - 2408/5 \* \H(0) \*  \zss
          - 130514/81 \* \H(0,0)
          + 2224/3 \* \H(0,0) \*  \z3 
          + 11264/9 \* \H(0,0) \*  \z2 
   \nn \\[0.5mm] & & \mbox{}
          - 10952/9 \* \H(0,0,0)
          + 688 \* \H(0,0,0) \*  \z2 
          - 5060/9 \* \H(0,0,0,0)
          - 272 \* \H(0,0,0,0,0)
          - 192268/81 \* \H(1)
   \nn \\[0.5mm] & & \mbox{}
          + 9196/9 \* \H(1) \*  \z3 
          + 17236/9 \* \H(1) \*  \z2 
          - 296 \* \H(1) \*  \zss
          + 1664/3 \* \H(1,-3,0)
          - 400 \* \H(1,-2) \*  \z2 
   \nn \\[0.5mm] & & \mbox{}
          - 1696/3 \* \H(1,-2,-1,0)
          + 3080/9 \* \H(1,-2,0)
          + 1792/3 \* \H(1,-2,0,0)
          + 352/3 \* \H(1,-2,2)
          - 176374/81 \* \H(1,0)
   \nn \\[0.5mm] & & \mbox{}
          + 2032/3 \* \H(1,0) \*  \z3 
          + 12760/9 \* \H(1,0) \*  \z2 
          - 47888/27 \* \H(1,0,0)
          + 688 \* \H(1,0,0) \*  \z2 
          - 9196/9 \* \H(1,0,0,0)
   \nn \\[0.5mm] & & \mbox{}
          - 1456/3 \* \H(1,0,0,0,0)
          - 19246/9 \* \H(1,1)
          + 240 \* \H(1,1) \*  \z3 
          + 10736/9 \* \H(1,1) \*  \z2 
          + 32 \* \H(1,1,-2,0)
   \nn \\[0.5mm] & & \mbox{}
          - 19484/9 \* \H(1,1,0)
          + 640 \* \H(1,1,0) \*  \z2 
          - 10384/9 \* \H(1,1,0,0)
          - 592 \* \H(1,1,0,0,0)
          - 2040 \* \H(1,1,1)
          + 768 \* \H(1,1,1) \*  \z2 
   \nn \\[0.5mm] & & \mbox{}
          - 14696/9 \* \H(1,1,1,0)
          - 1216 \* \H(1,1,1,0,0)
          - 4400/3 \* \H(1,1,1,1)
          - 1152 \* \H(1,1,1,1,0)
          - 960 \* \H(1,1,1,1,1)
   \nn \\[0.5mm] & & \mbox{}
          - 1152 \* \H(1,1,1,2)
          - 14080/9 \* \H(1,1,2)
          - 1280 \* \H(1,1,2,0)
          - 1152 \* \H(1,1,2,1)
          - 1024 \* \H(1,1,3)
          - 20452/9 \* \H(1,2)
   \nn \\[0.5mm] & & \mbox{}
          + 880 \* \H(1,2) \*  \z2 
          - 14432/9 \* \H(1,2,0)
          - 3296/3 \* \H(1,2,0,0)
          - 1584 \* \H(1,2,1)
          - 1280 \* \H(1,2,1,0)
   \nn \\[0.5mm] & & \mbox{}
          - 1152 \* \H(1,2,1,1)
          - 1280 \* \H(1,2,2)
          - 13816/9 \* \H(1,3)
          - 3488/3 \* \H(1,3,0)
          - 1216 \* \H(1,3,1)
          - 2480/3 \* \H(1,4)
   \nn \\[0.5mm] & & \mbox{}
          - 177130/81 \* \H(2)
          + 1568/3 \* \H(2) \*  \z3 
          + 3520/3 \* \H(2) \*  \z2 
          + 96 \* \H(2,-2,0)
          - 16424/9 \* \H(2,0)
   \nn \\[0.5mm] & & \mbox{}
          + 752 \* \H(2,0) \*  \z2 
          - 8360/9 \* \H(2,0,0)
          - 1616/3 \* \H(2,0,0,0)
          - 6656/3 \* \H(2,1)
          + 880 \* \H(2,1) \*  \z2 
          - 14432/9 \* \H(2,1,0)
   \nn \\[0.5mm] & & \mbox{}
          - 3296/3 \* \H(2,1,0,0)
          - 14344/9 \* \H(2,1,1)
          - 1280 \* \H(2,1,1,0)
          - 1152 \* \H(2,1,1,1)
          - 1280 \* \H(2,1,2)
          - 13112/9 \* \H(2,2)
   \nn \\[0.5mm] & & \mbox{}
          - 1184 \* \H(2,2,0)
          - 1280 \* \H(2,2,1)
          - 3104/3 \* \H(2,3)
          - 60440/27 \* \H(3)
          + 2432/3 \* \H(3) \*  \z2 
          - 3784/3 \* \H(3,0)
   \nn \\[0.5mm] & & \mbox{}
          - 832 \* \H(3,0,0)
          - 14344/9 \* \H(3,1)
          - 3680/3 \* \H(3,1,0)
          - 1280 \* \H(3,1,1)
          - 3680/3 \* \H(3,2)
          - 11440/9 \* \H(4)
   \nn \\[0.5mm] & & \mbox{}
          - 2704/3 \* \H(4,0)
          - 3568/3 \* \H(4,1)
          - 720 \* \H(5)
	      )
	  + (1-x) \* (
          - 1504/3 \* \H(-2) \*  \z3 
          - 1088/3 \* \H(-2,-2,0)
   \nn \\[0.5mm] & & \mbox{}
          + 384 \* \H(-2,-1) \*  \z2 
          + 896/3 \* \H(-2,-1,-1,0)
          - 2656/3 \* \H(-2,-1,0,0)
          - 704/3 \* \H(-2,-1,2)
          - 2560/3 \* \H(-2,0) \*  \z2 
   \nn \\[0.5mm] & & \mbox{}
          + 3584/3 \* \H(-2,0,0,0)
          + 704/3 \* \H(-2,2,0)
          + 256 \* \H(-2,2,1)
          + 1952/3 \* \H(-2,3)
	      )
	  + (1+x) \* (
            896/3 \* \H(2) \*  \z3 
   \nn \\[0.5mm] & & \mbox{}
          - 128/3 \* \H(2,-2,0)
          + 512 \* \H(2,0) \*  \z2 
          - 1600/3 \* \H(2,0,0,0)
          + 2240/3 \* \H(2,1) \*  \z2 
          - 3296/3 \* \H(2,1,0,0)
   \nn \\[0.5mm] & & \mbox{}
          - 896 \* \H(2,1,1,0)
          - 768 \* \H(2,1,1,1)
          - 896 \* \H(2,1,2)
          - 992 \* \H(2,2,0)
          - 896 \* \H(2,2,1)
          - 2144/3 \* \H(2,3)
          )
   \nn \\ & & \mbox{}
          + \delta(1 - x) \* \Big\{14160613/2916 + 4048/9 \*  \z5  -
          90632/27 \*  \z3  + 464/3 \*  \z3 ^2 - 196718/81 \*  \z2  
   \nn \\[-1mm] & & \mbox{}
          + 836 \*  \z2  \*  \z3  
          + 22501/27 \*  \zss - 15644/105 \*  \zts \Big\}
	   \Big)
   \nn \\[-0.5mm] & & \mbox{\hspn}
       +  \colour4colour{\cas \* \nf}  \*  \Big(
          - 15128/27 \* (1 - 4/1891 / x^2 + 375/1891/x + 1893/1891 \* x + 381/1891 \* x^2) \* \H(-1,0)
   \nn \\[-0.5mm] & & \mbox{}
          - 631/27 \* (1 - 264040/5679/x - 45427/1893 \* x + 132160/5679 \* x^2) \* \H(1)
   \nn \\[0.5mm] & & \mbox{}
          - 428/9 \* (1 - 848/107/x - 481/107 \* x + 534/107 \* x^2) \* \H(1,2)
          - 416/9 \* (1 - 102/13/x - 215/52 \* x 
   \nn \\[0.5mm] & & \mbox{}
          + 127/26 \* x^2) \* \H(1,1,1)
          - 92 \* (1 - 2692/621/x - 491/207 \* x + 1786/621 \* x^2) \* \H(1,1,0)
   \nn \\[0.5mm] & & \mbox{}
          + 26362/81 \* (1 - 34234/13181/x - 62089/13181 \* x + 31748/13181 \*
          x^2) \*  \z2 
   \nn \\[0.5mm] & & \mbox{}
          - 3284/27 \* (1 - 2048/821/x - 1513/821 \* x + 2594/821 \* x^2) \* \H(2,1)
          - 9536/27 \* (1 - 5465/2384/x 
   \nn \\[0.5mm] & & \mbox{}
          - 9027/4768 \* x + 857/596 \* x^2) \* \H(1,0)
          - 26362/81 \* (1 - 29734/13181/x - 39373/13181 \* x 
   \nn \\[0.5mm] & & \mbox{}
          + 31748/13181 \* x^2) \* \H(2)
          + 1228/9 \* (1 - 668/307/x - 681/307 \* x + 360/307 \* x^2) \* \H(1)
          \*  \z2 
   \nn \\[0.5mm] & & \mbox{}
          - 460/3 \* (1 - 2128/1035/x - 31/15 \* x + 944/345 \* x^2) \* \H(2,0)
          - 452/3 \* (1 - 656/339/x - 571/339 \* x 
   \nn \\[0.5mm] & & \mbox{}
          + 442/339 \* x^2) \* \H(1,0,0)
          - 10538/27 \* (1 - 30380/15807/x - 9705/5269 \* x + 18482/15807 \*
          x^2) \* \H(1,1)
   \nn \\[0.5mm] & & \mbox{}
          - 13184/27 \* (1 - 599/412/x - 22057/6592 \* x + 3555/1648 \* x^2) \* \H(0,0)
          + 1600/9 \* (1 - 9/10/x + x 
   \nn \\[0.5mm] & & \mbox{}
          - 87/100 \* x^2) \* \H(-1,-1,0)
          + 64/3 \* (1 - 2/3/x - 1/8 \* x + 37/24 \* x^2) \* \H(1,0,0,0)
          - 248/3 \* (1 - 16/31/x 
   \nn \\[0.5mm] & & \mbox{}
          - 17/31 \* x + 30/31 \* x^2) \* \H(1,0) \*  \z2 
          - 218/45 \* (1 - 48/109/x - 1782/109 \* x - 804/109 \* x^2) \*  \zss
   \nn \\[0.5mm] & & \mbox{}
          + 3472/27 \* (1 - 185/434/x - 53/7 \* x + 85/31 \* x^2) \*  \z3 
          - 1168/3 \* (1 - 226/657/x + 218/219 \* x 
   \nn \\[0.5mm] & & \mbox{}
          - 443/1314 \* x^2) \* \H(-1,0,0)
          + 776/9 \* (1 - 28/97/x - 109/194 \* x + 71/97 \* x^2) \* \H(1,3)
   \nn \\[0.5mm] & & \mbox{}
          - 1672/9 \* (1 - 3/11/x + 206/209 \* x) \* \H(0) \*  \z3 
          + 3968/9 \* (1 - 61/248/x + x - 119/496 \* x^2) \* \H(-1) \*  \z2 
   \nn \\[0.5mm] & & \mbox{}
          - 2312/9 \* (1 - 64/289/x + 266/289 \* x - 42/289 \* x^2) \* \H(-1,-1,-1,0)
          - 120 \* (1 - 28/135/x + 14/15 \* x 
   \nn \\[0.5mm] & & \mbox{}
          - 19/135 \* x^2) \* \H(-1,0,0,0)
          + 976/3 \* (1 - 12/61/x + 225/244 \* x - 15/122 \* x^2) \*
          \H(-1,-1,0,0)
   \nn \\[0.5mm] & & \mbox{}
          + 1664/9 \* (1 - 5/26/x + 191/208 \* x - 3/26 \* x^2) \* \H(-1,-2,0)
          - 2524/9 \* (1 - 112/631/x + 584/631 \* x 
   \nn \\[0.5mm] & & \mbox{}
          - 66/631 \* x^2) \* \H(-1,-1) \*  \z2 
          + 2440/9 \* (1 - 43/305/x + 569/610 \* x - 23/305 \* x^2) \* \H(-1)
          \*  \z3 
   \nn \\[0.5mm] & & \mbox{}
          + 152 \* (1 - 8/57/x + 53/57 \* x - 4/57 \* x^2) \* \H(-1,-1,2)
          - 700/3 \* (1 - 2/15/x + 761/525 \* x 
   \nn \\[0.5mm] & & \mbox{}
          + 4/35 \* x^2) \* \H(0,0) \*  \z2 
          + 904/3 \* (1 - 34/339/x + 38/113 \* x - 20/113 \* x^2) \*
          \H(-2,-1,0)
   \nn \\[0.5mm] & & \mbox{}
          - 776/9 \* (1 - 9/97/x 
          + 241/97 \* x + 71/97 \* x^2) \* \H(2) \*  \z2 
          + 1268/9 \* (1 - 28/317/x + 304/317 \* x 
   \nn \\[0.5mm] & & \mbox{}
          - 14/317 \* x^2) \*
          \H(-1,0) \*  \z2 
          - 352 \* (1 - 8/99/x + x - 8/99 \* x^2) \* \H(-1,2)
          + 700/3 \* (1 - 38/525/x 
   \nn \\[0.5mm] & & \mbox{}
          + 977/525 \* x + 4/35 \* x^2) \* \H(4)
          + 1520/9 \* (1 - 6/95/x + 40/19 \* x) \* \H(0,0,0,0)
          - 968/3 \* (1 - 16/363/x 
   \nn \\[0.5mm] & & \mbox{}
          - 36/121 \* x - 70/363 \* x^2) \* \H(-3,0)
          - 4808/27 \* (1 - 26/601/x + 740/601 \* x - 630/601 \* x^2) \* \H(-2,0)
   \nn \\[0.5mm] & & \mbox{}
          + 2056/9 \* (1 - 10/257/x + 116/257 \* x 
          + 32/257 \* x^2) \* \H(2,1,0)
          + 712/3 \* (1 - 3/89/x + 677/534 \* x 
   \nn \\[0.5mm] & & \mbox{}
          + 31/267 \* x^2) \* \H(3,0)
          + 548/9 \* (1 - 4/137/x - 52/137 \* x + 90/137 \* x^2) \* \H(1,1,2)
          + 2368/9 \* (1 - 1/37/x 
   \nn \\[0.5mm] & & \mbox{}
          + 817/592 \* x + 53/296 \* x^2) \* \H(3,1)
          - 1028/3 \* (1 + 4/257/x + 35/257 \* x - 158/771 \* x^2) \*
          \H(-2,0,0)
   \nn \\[0.5mm] & & \mbox{}
          + 740/9 \* (1 + 4/185/x - 23/37 \* x + 64/185 \* x^2) \* \H(1,2,0)
          + 644/9 \* (1 + 4/161/x - 76/161 \* x 
   \nn \\[0.5mm] & & \mbox{}
          + 82/161 \* x^2) \* \H(1,2,1)
          + 2132/9 \* (1 + 16/533/x + 368/533 \* x + 2/13 \* x^2) \* \H(2,2)
   \nn \\[0.5mm] & & \mbox{}
          - 412/3 \* (1 + 4/103/x 
          + 99/103 \* x + 8/103 \* x^2) \* \H(-1,3)
          + 3164/9 \* (1 + 6/113/x 
          + 4/113 \* x 
   \nn \\[0.5mm] & & \mbox{}
          - 152/791 \* x^2) \* \H(-2) \*
          \z2 
          - 836/9 \* (1 + 12/209/x - 271/209 \* x - 4/11 \* x^2) \*
          \H(1,1,0,0)
   \nn \\[0.5mm] & & \mbox{}
          + 952/9 \* (1 + 2/17/x + 571/238 \* x 
          + 6/17 \* x^2) \* \H(2,0,0)
          - 1808/9 \* (1 + 19/113/x - 43/226 \* x 
   \nn \\[0.5mm] & & \mbox{}
          - 23/113 \* x^2) \* \H(-2,2)
          - 160/3 \* (1 + 1/3/x 
          + x + 1/3 \* x^2) \* \H(-1,2,1)
          - 40 \* (1 + 4/9/x + x 
   \nn \\[0.5mm] & & \mbox{}
          + 4/9 \* x^2) \* \H(-1,2,0)
          + 608/9 \* (1 + 17/38/x - 107/76 \* x 
          - 33/38 \* x^2) \* \H(1,1) \*  \z2 
   \nn \\[0.5mm] & & \mbox{}
          + 496/9 \* (1 + 77/62/x - 56/31 \* x - 63/31 \* x^2) \* \H(1) \*
          \z3 
          + 208 \* x \* \H(0,0) \*  \z3 
          - 32 \* x \* \H(0,0,0,0,0)
   \nn \\[0.5mm] & & \mbox{}
          + 16306/81 \* (1 + 52501/8153/x + 78053/16306 \* x - 22866/8153 \* x^2) \* \H(0)
   \nn \\[0.5mm] & & \mbox{}
          + 544/27 \* (1 + 309/34/x 
          + 1667/34 \* x - 1903/68 \* x^2) \* \H(0,0,0)
   \nn \\[0.5mm] & & \mbox{}
          + 28211/243 \* (1 
          + 1582888/84633/x 
          + 475721/56422 \* x 
          - 1639871/169266 \* x^2)
   \nn \\[0.5mm] & & \mbox{}
          - 88/27 \* (1 + 804/11/x + 6137/22 \* x 
          - 1624/11 \* x^2) \* \H(0)
          \*  \z2 
          + 88/27 \* (1 + 830/11/x 
   \nn \\[0.5mm] & & \mbox{}
          + 4657/22 \* x - 1624/11 \* x^2) \* \H(3)
          - 1600/9 \* (1 - 23/25 \* x 
          + 7/100 \* x^2) \* \H(1,-2,0)
   \nn \\[0.5mm] & & \mbox{}
          + 240 \* (1 - 11/15 \* x) \* \H(-3,-1,0)
          + 824/9 \* (1 - 62/103 \* x + 40/103 \* x^2) \* \H(1,1,1,0)
   \nn \\[0.5mm] & & \mbox{}
          + 64 \* (1 - 1/2 \* x 
          + 1/2 \* x^2) \* \H(1,1,1,1)
          - 704/3 \* (1 - 17/44 \* x) \* \H(-3,0,0)
          + 728/3 \* (1 - 19/91 \* x) \* \H(-3) \*  \z2 
   \nn \\[0.5mm] & & \mbox{}
          - 656/3 \* (1 - 1/41 \* x) \* \H(-4,0)
          - 368/3 \* (1 + 7/23 \* x) \* \H(-3,2)
          - 80 \* (1 + 31/75 \* x) \* \H(0) \*  \zss
   \nn \\[0.5mm] & & \mbox{}
          + 224 \* (1 + 4/7 \* x + 1/7 \* x^2) \* \H(2,1,1)
          + 376/3 \* (1 + 31/47 \* x) \* \H(3) \*  \z2 
          - 232/3 \* (1 + 21/29 \* x) \* \H(3,0,0)
   \nn \\[0.5mm] & & \mbox{}
          - 48 \* (1 + 11/9 \* x) \* \H(5)
          + 48 \* (1 + 4/3 \* x) \* \H(0,0,0) \*  \z2 
          - 232/3 \* (1 + 65/29 \* x) \*  \z2  \*  \z3 
          - 56/3 \* (1 + 27/7 \* x) \*  \z5
   \nn \\[0.5mm] & & \mbox{} 
	  + \pgg( - x) \* (
          - 280/3 \*  \z3 
          - 896/81 \*  \z2 
          - 916/45 \*  \zss
          - 704/9 \* \H(-3,0)
          + 112 \* \H(-2) \*  \z2 
          + 928/9 \* \H(-2,-1,0)
   \nn \\[0.5mm] & & \mbox{} 
          - 80 \* \H(-2,0)
          - 128 \* \H(-2,0,0)
          - 544/9 \* \H(-2,2)
          + 1352/9 \* \H(-1) \*  \z3 
          + 320/3 \* \H(-1) \*  \z2 
          + 944/9 \* \H(-1,-2,0)
   \nn \\[0.5mm] & & \mbox{} 
          - 1376/9 \* \H(-1,-1) \*  \z2 
          - 1216/9 \* \H(-1,-1,-1,0)
          + 320/3 \* \H(-1,-1,0)
          + 1568/9 \* \H(-1,-1,0,0)
   \nn \\[0.5mm] & & \mbox{} 
          + 256/3 \* \H(-1,-1,2)
          - 1792/81 \* \H(-1,0)
          + 352/3 \* \H(-1,0) \*  \z2 
          - 400/3 \* \H(-1,0,0)
          - 1000/9 \* \H(-1,0,0,0)
   \nn \\[0.5mm] & & \mbox{} 
          - 160/3 \* \H(-1,2)
          - 128/3 \* \H(-1,2,0)
          - 128/3 \* \H(-1,2,1)
          - 96 \* \H(-1,3)
          - 48 \* \H(0)
          - 584/9 \* \H(0) \*  \z3 
   \nn \\[0.5mm] & & \mbox{} 
          - 800/27 \* \H(0) \*  \z2 
          + 896/81 \* \H(0,0)
          - 328/9 \* \H(0,0) \*  \z2 
          + 40 \* \H(0,0,0)
          + 248/9 \* \H(0,0,0,0)
          + 152/9 \* \H(2) \*  \z2 
   \nn \\[0.5mm] & & \mbox{} 
          + 32/9 \* \H(2,0,0)
          - 16/9 \* \H(2,1,0)
          + 16/9 \* \H(2,2)
          + 80/3 \* \H(3)
          + 56/3 \* \H(3,0)
          + 64/3 \* \H(3,1)
          + 296/9 \* \H(4)
	      )
   \nn \\[0.5mm] & & \mbox{} 
	  + \pgg(x) \* (
            1234307/1458
          - 8608/27 \*  \z3 
          - 55712/81 \*  \z2 
          + 4852/45 \*  \zss
          - 640/9 \* \H(-3,0)
   \nn \\[0.5mm] & & \mbox{} 
          + 104/3 \* \H(-2) \*  \z2 
          + 112/3 \* \H(-2,-1,0)
          - 2000/27 \* \H(-2,0)
          - 464/9 \* \H(-2,0,0)
          - 16 \* \H(-2,2)
   \nn \\[0.5mm] & & \mbox{} 
          + 68830/81 \* \H(0)
          - 992/9 \* \H(0) \*  \z3 
          - 4672/9 \* \H(0) \*  \z2 
          + 44684/81 \* \H(0,0)
          - 2048/9 \* \H(0,0) \*  \z2 
   \nn \\[0.5mm] & & \mbox{} 
          + 2776/9 \* \H(0,0,0)
          + 920/9 \* \H(0,0,0,0)
          + 67730/81 \* \H(1)
          - 520/9 \* \H(1) \*  \z3 
          - 4544/9 \* \H(1) \*  \z2 
          - 560/9 \* \H(1,-2,0)
   \nn \\[0.5mm] & & \mbox{} 
          + 56284/81 \* \H(1,0)
          - 2320/9 \* \H(1,0) \*  \z2 
          + 12128/27 \* \H(1,0,0)
          + 1672/9 \* \H(1,0,0,0)
          + 6212/9 \* \H(1,1)
   \nn \\[0.5mm] & & \mbox{} 
          - 1952/9 \* \H(1,1) \*  \z2 
          + 4672/9 \* \H(1,1,0)
          + 1888/9 \* \H(1,1,0,0)
          + 512 \* \H(1,1,1)
          + 2672/9 \* \H(1,1,1,0)
   \nn \\[0.5mm] & & \mbox{} 
          + 800/3 \* \H(1,1,1,1)
          + 2560/9 \* \H(1,1,2)
          + 5024/9 \* \H(1,2)
          + 2624/9 \* \H(1,2,0)
          + 288 \* \H(1,2,1)
          + 2512/9 \* \H(1,3)
   \nn \\[0.5mm] & & \mbox{} 
          + 56608/81 \* \H(2)
          - 640/3 \* \H(2) \*  \z2 
          + 1360/3 \* \H(2,0)
          + 1520/9 \* \H(2,0,0)
          + 1616/3 \* \H(2,1)
          + 2624/9 \* \H(2,1,0)
   \nn \\[0.5mm] & & \mbox{} 
          + 2608/9 \* \H(2,1,1)
          + 2384/9 \* \H(2,2)
          + 14096/27 \* \H(3)
          + 688/3 \* \H(3,0)
          + 2608/9 \* \H(3,1)
          + 2080/9 \* \H(4)
	      )
   \nn \\[0.5mm] & & \mbox{} 
	  + (1-x) \* (
            488/3 \* \H(-2) \*  \z3 
          + 368/3 \* \H(-2,-2,0)
          - 544/3 \* \H(-2,-1) \*  \z2 
          - 544/3 \* \H(-2,-1,-1,0)
   \nn \\[0.5mm] & & \mbox{} 
          + 216 \* \H(-2,-1,0,0)
          + 272/3 \* \H(-2,-1,2)
          + 224/3 \* \H(-2,0) \*  \z2 
          - 80 \* \H(-2,0,0,0)
          + 16/3 \* \H(-2,2,0)
   \nn \\[0.5mm] & & \mbox{} 
          - 152/3 \* \H(-2,3)
          + 16/3 \* \H(3,1,0)
          - 16/3 \* \H(3,2)
	      )
	  + (1+x) \* (
            464/3 \* \H(2) \*  \z3 
          - 224/3 \* \H(2,-2,0)
   \nn \\[0.5mm] & & \mbox{} 
          + 176/3 \* \H(2,0) \*  \z2 
          - 32 \* \H(2,0,0,0)
          + 304/3 \* \H(2,1) \*  \z2 
          - 232/3 \* \H(2,1,0,0)
          + 32/3 \* \H(2,1,1,0)
          - 32/3 \* \H(2,1,2)
   \nn \\ & & \mbox{} 
          + 40/3 \* \H(2,2,0)
          - 104/3 \* \H(2,3)
          - 24 \* \H(4,0)
          - 40 \* \H(4,1)
	      )
          + \delta(1 - x) \* \Big\{ - 1988899/972 
   \nn \\[-1mm] & & \mbox{}
          + 1064/9 \*  \z5  + 6794/27 \*
          \z3  + 70774/81 \*  \z2  - 24 \*  \z2  \*  \z3  
          - 28526/135 \*  \zss \Big\}
          \Big)
   \nn \\[-0.5mm] & & \mbox{\hspn}
       +  \colour4colour{\cf  \*  \ca \* \nf}   \*  \Big(
            11528/9 \* (1 - 4/1441 / x^2 + 122/1441/x + 1384/1441 \* x + 69/1441 \* x^2) \* \H(-1,0)
   \nn \\[-0.5mm] & & \mbox{}
          - 56/9 \* (1 - 309/7/x - 78/7 \* x + 50 \* x^2) \* \H(1,1,0)
          - 340/27 \* (1 - 1338/85/x - 643/17 \* x 
   \nn \\[0.5mm] & & \mbox{}
          + 3304/85 \* x^2) \* \H(2,0)
          - 68/3 \* (1 - 1450/153/x - 41/17 \* x + 98/9 \* x^2) \* \H(1,1,1)
   \nn \\[0.5mm] & & \mbox{}
          + 2362/27 \* (1 - 2384/1181/x + 3209/1181 \* x + 7616/1181 \* x^2) \* \H(0) \*  \z2 
   \nn \\[0.5mm] & & \mbox{}
          - 2362/27 \* (1 - 2256/1181/x - 1843/1181 \* x + 7616/1181 \* x^2) \* \H(3)
          + 800/3 \* (1 - 67/75/x + x 
   \nn \\[0.5mm] & & \mbox{}
          - 67/75 \* x^2) \* \H(-1,-1) \*  \z2 
          - 169601/243 \* (1 - 121601/169601/x - 89143/169601 \* x 
   \nn \\[0.5mm] & & \mbox{}
          + 276912/169601 \* x^2) \* \H(0)
          - 3952/9 \* (1 - 141/247/x + 265/247 \* x - 123/247 \* x^2) \* \H(-1,-1,0)
   \nn \\[0.5mm] & & \mbox{}
          - 904/3 \* (1 - 58/113/x + x - 58/113 \* x^2) \* \H(-1) \*  \z3 
          + 6128/9 \* (1 - 466/1149/x + 847/766 \* x 
   \nn \\[0.5mm] & & \mbox{}
          - 689/2298 \* x^2) \* \H(-1,0,0)
          - 1400/3 \* (1 - 212/525/x + x - 212/525 \* x^2) \* \H(-1,-1,0,0)
   \nn \\[0.5mm] & & \mbox{}
          - 6712/9 \* (1 - 259/839/x + 911/839 \* x - 187/839 \* x^2) \* \H(-1) \*  \z2 
          + 544 \* (1 - 46/153/x + x 
   \nn \\[0.5mm] & & \mbox{}
          - 46/153 \* x^2) \* \H(-1,-1,-1,0)
          - 1136/3 \* (1 - 20/71/x + x - 20/71 \* x^2) \* \H(-1,-2,0)
   \nn \\[0.5mm] & & \mbox{}
          - 23150/27 \* (1 - 2664/11575/x - 5048/11575 \* x + 9784/11575 \* x^2) \* \H(0,0,0)
   \nn \\[0.5mm] & & \mbox{}
          + 4736/9 \* (1 - 59/296/x + 323/296 \* x - 4/37 \* x^2) \* \H(-1,2)
          - 1712/3 \* (1 - 20/107/x + 66/107 \* x 
   \nn \\[0.5mm] & & \mbox{}
          - 68/321 \* x^2) \* \H(-2,-1,0)
          - 664 \* (1 - 44/249/x - 20/83 \* x - 100/747 \* x^2) \* \H(-2) \*  \z2 
   \nn \\[0.5mm] & & \mbox{}
          + 1136/3 \* (1 - 12/71/x - 63/71 \* x - 16/213 \* x^2) \* \H(-2,2)
          + 656 \* (1 - 44/369/x + 1/246 \* x 
   \nn \\[0.5mm] & & \mbox{}
          - 26/369 \* x^2) \* \H(-2,0,0)
          + 1904/3 \* (1 - 8/119/x - 39/119 \* x - 12/119 \* x^2) \* \H(-3,0)
   \nn \\[0.5mm] & & \mbox{}
          + 4384/9 \* (1 - 8/411/x + 421/548 \* x - 475/822 \* x^2) \* \H(-2,0)
          + 1184/3 \* (1 + 4/111/x - x 
   \nn \\[0.5mm] & & \mbox{}
          - 4/111 \* x^2) \* \H(1,-2,0)
          + 13612/45 \* (1 + 352/3403/x - 2282/3403 \* x - 776/3403 \* x^2) \*
          \zss
   \nn \\[0.5mm] & & \mbox{}
          + 22804/27 \* (1 + 4697/34206/x - 9101/11402 \* x - 4828/17103 \* x^2) \* \H(1,1)
   \nn \\[0.5mm] & & \mbox{}
          + 20020/27 \* (1 + 159/910/x - 7283/10010 \* x - 156/385 \* x^2) \* \H(1,0)
   \nn \\[0.5mm] & & \mbox{}
          + 384329/243 \* (1 + 455021/2305974/x - 19253/69878 \* x - 486427/1152987 \* x^2)
   \nn \\[0.5mm] & & \mbox{}
          + 3140/9 \* (1 + 192/785/x + 1919/785 \* x) \* \H(0,0,0,0)
          - 78058/81 \* (1 + 10992/39029/x 
   \nn \\[0.5mm] & & \mbox{}
          - 68023/39029 \* x - 9282/39029 \* x^2) \*  \z2 
          + 2960/9 \* (1 + 12/37/x + 104/185 \* x - 54/185 \* x^2) \* \H(3,1)
   \nn \\[0.5mm] & & \mbox{}
          - 4496/9 \* (1 + 92/281/x - 83/562 \* x - 76/281 \* x^2) \* \H(2) \*  \z2 
          + 3976/9 \* (1 + 172/497/x 
   \nn \\[0.5mm] & & \mbox{}
          - 13/497 \* x - 152/497 \* x^2) \* \H(2,0,0)
          + 2752/9 \* (1 + 15/43/x + 197/344 \* x - 12/43 \* x^2) \* \H(3,0)
   \nn \\[0.5mm] & & \mbox{}
          + 3100/9 \* (1 + 288/775/x + 1339/775 \* x - 168/775 \* x^2) \* \H(4)
          + 78058/81 \* (1 + 15384/39029/x 
   \nn \\[0.5mm] & & \mbox{}
          - 18199/39029 \* x - 9282/39029 \* x^2) \* \H(2)
          - 1112/3 \* (1 + 68/139/x - x - 68/139 \* x^2) \* \H(1,1) \*  \z2 
   \nn \\[0.5mm] & & \mbox{}
          - 3100/9 \* (1 + 384/775/x + 1807/775 \* x - 168/775 \* x^2) \* \H(0,0) \*  \z2 
          + 1928/9 \* (1 + 124/241/x 
   \nn \\[0.5mm] & & \mbox{}
          + 115/241 \* x - 84/241 \* x^2) \* \H(2,2)
          + 344/3 \* (1 + 24/43/x + x + 24/43 \* x^2) \* \H(-1,0,0,0)
   \nn \\[0.5mm] & & \mbox{}
          + 2024/9 \* (1 + 148/253/x + 97/253 \* x - 100/253 \* x^2) \* \H(2,1,0)
          + 664/3 \* (1 + 148/249/x - x 
   \nn \\[0.5mm] & & \mbox{}
          - 148/249 \* x^2) \* \H(1,3)
          + 1648/9 \* (1 + 64/103/x + 34/103 \* x - 40/103 \* x^2) \* \H(2,1,1)
   \nn \\[0.5mm] & & \mbox{}
          + 280 \* (1 + 40/63/x - x - 40/63 \* x^2) \* \H(1,1,0,0)
          - 640/3 \* (1 + 5/6/x - x - 5/6 \* x^2) \* \H(1,0) \*  \z2 
   \nn \\[0.5mm] & & \mbox{}
          - 760/3 \* (1 + 50/57/x - x - 50/57 \* x^2) \* \H(1) \*  \z3 
          + 152/3 \* (1 + 52/57/x + x + 52/57 \* x^2) \* \H(-1,3)
   \nn \\[0.5mm] & & \mbox{}
          + 4468/27 \* (1 + 1062/1117/x + 2947/1117 \* x - 2918/1117 \* x^2) \* \H(2,1)
          + 296/3 \* (1 + 112/111/x 
   \nn \\[0.5mm] & & \mbox{}
          - x - 112/111 \* x^2) \* \H(1,1,2)
          - 2888/9 \* (1 + 1114/1083/x - 306/361 \* x - 1369/1083 \* x^2) \* \H(1) \*  \z2 
   \nn \\[0.5mm] & & \mbox{}
          + 224/3 \* (1 + 22/21/x + x + 22/21 \* x^2) \* \H(-1,2,1)
          + 31550/81 \* (1 + 16644/15775/x 
   \nn \\[0.5mm] & & \mbox{}
          - 9451/3155 \* x - 2784/15775 \* x^2) \* \H(0,0)
          + 136 \* (1 + 200/153/x - x - 200/153 \* x^2) \* \H(1,0,0,0)
   \nn \\[0.5mm] & & \mbox{}
          + 8 \* (1 + 4/3/x - x - 4/3 \* x^2) \* ( 22/3 \* \H(1,1,1,1)
          + 7 \* \H(1,2,1) )
          + 160/3 \* (1 + 4/3/x + x + 4/3 \* x^2) \* \H(-1,2,0)
   \nn \\[0.5mm] & & \mbox{}
          - 164 \* (1 + 592/369/x + 455/123 \* x - 64/123 \* x^2) \* \H(0) \*
          \z3 
          + 1792/9 \* (1 + 11/6/x 
   \nn \\[0.5mm] & & \mbox{}
          - 143/224 \* x - 1475/672 \* x^2) \* \H(1,0,0)
          + 152/3 \* (1 + 112/57/x - x 
          - 112/57 \* x^2) \* \H(1,1,1,0)
   \nn \\[0.5mm] & & \mbox{}
          + 48 \* (1 + 2/x - x - 2 \* x^2) \* \H(1,2,0)
          + 304/3 \* (1 + 691/342/x - 41/114 \* x 
          - 500/171 \* x^2) \* \H(1,2)
   \nn \\[0.5mm] & & \mbox{}
          - 4262/27 \* (1 + 6900/2131/x - 16385/2131 \* x - 1952/2131 \* x^2)
          \*  \z3 
   \nn \\[0.5mm] & & \mbox{}
          + 15112/81 \* (1 + 33767/7556/x - 4061/7556 \* x - 29945/7556 \* x^2) \* \H(1)
          + 16/3 \* (1 + 88/3/x 
   \nn \\[0.5mm] & & \mbox{}
          + x + 88/3 \* x^2) \* \H(-1,-1,2)
          - 256/3 \* (1 - 155/16 \* x) \* \H(5)
          + 256/3 \* (1 - 149/16 \* x) \* \H(0,0,0) \*  \z2 
   \nn \\[0.5mm] & & \mbox{}
          - 480 \* (1 - 47/30 \* x) \* \H(0,0,0,0,0)
          - 544 \* (1 - 14/17 \* x) \* \H(-3,-1,0)
          + 304 \* (1 - 9/19 \* x) \* \H(-3,0,0)
   \nn \\[0.5mm] & & \mbox{}
          - 400 \* (1 - 8/25 \* x) \* \H(-3) \*  \z2 
          + 1024/3 \* (1 + 3/32 \* x) \* \H(-4,0)
          - 448 \* (1 + 31/168 \* x) \*  \z5 
   \nn \\[0.5mm] & & \mbox{}
          + 128 \* (1 + 3/4 \* x) \* \H(-3,2)
          + 560/3 \* (1 + 437/350 \* x) \* \H(0) \*  \zss
          - 1280/3 \* (1 + 107/80 \* x) \* \H(3) \*  \z2 
   \nn \\[0.5mm] & & \mbox{}
          + 384 \* (1 + 25/18 \* x) \* \H(3,0,0)
          + 1232/3 \* (1 + 138/77 \* x) \*  \z2  \*  \z3 
          + 560/3 \* (1 + 71/35 \* x) \* \H(3,1,0)
   \nn \\[0.5mm] & & \mbox{}
          + 464/3 \* (1 + 65/29 \* x) \* \H(3,2)
          + 416/3 \* (1 + 31/13 \* x) \* \H(3,1,1)
          - 496/3 \* (1 + 163/31 \* x) \* \H(0,0) \*  \z3 
   \nn \\[0.5mm] & & \mbox{}
          + 112 \* (1 + 39/7 \* x) \* \H(4,1)
          + 208/3 \* (1 + 109/13 \* x) \* \H(4,0)
	  + \pgg( - x) \* (
          - 224/5 \*  \zss
          - 128/3 \* \H(-3,0)
   \nn \\[0.5mm] & & \mbox{}
          - 24 \* \H(0)
          - 128/3 \* \H(0) \*  \z3 
          - 64/3 \* \H(0,0) \*  \z2 
          + 64/3 \* \H(0,0,0,0)
	      )
	  + \pgg(x) \* (
            7810/27
          - 2152/9 \*  \z3 
   \nn \\[0.5mm] & & \mbox{}
          - 100/3 \*  \z2 
          + 32/15 \*  \zss
          - 256/3 \* \H(-2) \*  \z2 
          + 128/3 \* \H(-2,0,0)
          + 256/3 \* \H(-2,2)
          + 2020/9 \* \H(0)
   \nn \\[0.5mm] & & \mbox{}
          - 64 \* \H(0) \*  \z3 
          - 176/9 \* \H(0) \*  \z2 
          + 24 \* \H(0,0)
          + 64 \* \H(0,0) \*  \z2 
          - 64/3 \* \H(0,0,0,0)
          + 598/3 \* \H(1)
          - 128 \* \H(1) \*  \z3 
   \nn \\[0.5mm] & & \mbox{}
          + 32 \* \H(1,0)
          + 32 \* \H(1,1)
          + 100/3 \* \H(2)
          - 176/9 \* \H(2,0)
          + 176/9 \* \H(3)
          - 128/3 \* \H(4)
	   )
   \nn \\[0.5mm] & & \mbox{}
    + (1-x) \* (
          - 1120/3 \* \H(-2) \*  \z3 
          - 992/3 \* \H(-2,-2,0)
          + 1408/3 \* \H(-2,-1) \*  \z2 
          + 1472/3 \* \H(-2,-1,-1,0)
   \nn \\[0.5mm] & & \mbox{}
          - 496 \* \H(-2,-1,0,0)
          - 224 \* \H(-2,-1,2)
          - 128/3 \* \H(-2,0) \*  \z2 
          - 16/3 \* \H(-2,0,0,0)
          - 64 \* \H(-2,2,0)
   \nn \\[0.5mm] & & \mbox{}
          - 64 \* \H(-2,2,1)
          - 16 \* \H(-2,3)
	)
    + (1+x) \* (
          - 128/3 \* \H(-1,0) \*  \z2 
          - 416 \* \H(2) \*  \z3 
          + 640/3 \* \H(2,-2,0)
   \nn \\[0.5mm] & & \mbox{}
          - 992/3 \* \H(2,0) \*  \z2 
          + 880/3 \* \H(2,0,0,0)
          - 1232/3 \* \H(2,1) \*  \z2 
          + 1072/3 \* \H(2,1,0,0)
          + 400/3 \* \H(2,1,1,0)
   \nn \\[0.5mm] & & \mbox{}
          + 352/3 \* \H(2,1,1,1)
          + 496/3 \* \H(2,1,2)
          + 128 \* \H(2,2,0)
          + 112 \* \H(2,2,1)
          + 272 \* \H(2,3)
          )
   \nn \\ & & \mbox{}
          + \delta(1 - x) \* \Big\{ - 411137/324 + 200 \*  \z5  + 6664/9 \*  \z3  +
          2128/9 \*  \z2  - 128 \*  \z2  \*  \z3  + 176/45 \*  \zss \Big\}
    \Big)
   \nn \\[-0.5mm] & & \mbox{\hspn}
       +  \colour4colour{\cfs \* \nf}  \*  \Big(
          - 12016/9 \* (1 - 2/2253 / x^2 + 5/1502/x + 4505/4506 \* x + 3/751 \* x^2) \* \H(-1,0)
   \nn \\[-0.5mm] & & \mbox{}
          - 16/3 \* (1 - 13/x - 5 \* x + 17 \* x^2) \* \H(1,0,0)
          + 44/3 \* (1 - 32/11/x + 70/11 \* x - 128/33 \* x^2) \* \H(0,0,0,0)
   \nn \\[0.5mm] & & \mbox{}
          - 256/3 \* (1 - 19/12/x + x - 19/12 \* x^2) \* \H(-1,0,0,0)
          + 592/3 \* (1 - 172/111/x + x 
   \nn \\[0.5mm] & & \mbox{}
          - 172/111 \* x^2) \* \H(-1,-1,0,0)
          - 296/3 \* (1 - 56/37/x - 94/37 \* x - 112/111 \* x^2) \* \H(0) \*
          \z3 
   \nn \\[0.5mm] & & \mbox{}
          + 452/3 \* (1 - 280/339/x - 137/113 \* x + 352/339 \* x^2) \* \H(1) \*  \z2 
          + 704/3 \* (1 - 20/33/x + x 
   \nn \\[0.5mm] & & \mbox{}
          - 20/33 \* x^2) \* \H(-1,-2,0)
          - 320 \* (1 - 8/15/x + x - 8/15 \* x^2) \* \H(-1,-1,-1,0)
          - 192 \* (1 - 4/9/x - 13/6 \* x 
   \nn \\[0.5mm] & & \mbox{}
          - 2/3 \* x^2) \* \H(-2,2)
          + 352 \* (1 - 4/11/x - 10/11 \* x - 20/33 \* x^2) \* \H(-2) \*  \z2 
          - 592/3 \* (1 - 12/37/x 
   \nn \\[0.5mm] & & \mbox{}
          + 19/74 \* x - 14/37 \* x^2) \* \H(0,0) \*  \z2 
          + 320 \* (1 - 4/15/x + 3/5 \* x - 8/15 \* x^2) \* \H(-2,-1,0)
   \nn \\[0.5mm] & & \mbox{}
          + 1424/3 \* (1 - 70/267/x + 79/89 \* x - 100/267 \* x^2) \* \H(-1,-1,0)
          - 1216/3 \* (1 - 4/19/x - 21/76 \* x 
   \nn \\[0.5mm] & & \mbox{}
          - 37/57 \* x^2) \* \H(-2,0,0)
          + 712 \* (1 - 122/801/x + 257/267 \* x - 152/801 \* x^2) \* \H(-1)
          \*  \z2 
   \nn \\[0.5mm] & & \mbox{}
          + 592/3 \* (1 - 4/37/x + 51/74 \* x - 14/37 \* x^2) \* \H(4)
          - 1424/3 \* (1 - 26/267/x + x - 26/267 \* x^2) \* \H(-1,2)
   \nn \\[0.5mm] & & \mbox{}
          - 1216/3 \* (1 - 2/19/x - 4/19 \* x - 26/57 \* x^2) \* \H(-3,0)
          - 9104/9 \* (1 - 48/569/x + 554/569 \* x 
   \nn \\[0.5mm] & & \mbox{}
          - 63/569 \* x^2) \* \H(-1,0,0)
          - 136 \* (1 - 4/51/x - x + 4/51 \* x^2) \* \H(1,3)
          + 2141/9 \* (1 + 264/2141/x 
   \nn \\[0.5mm] & & \mbox{}
          + 42253/6423 \* x - 9776/6423 \* x^2) \* \H(0,0)
          + 680/3 \* (1 + 16/85/x + 77/85 \* x - 28/85 \* x^2) \* \H(3,1)
   \nn \\[0.5mm] & & \mbox{}
          + 48631/162 \* (1 + 9513/48631/x - 1270/48631 \* x - 58494/48631 \* x^2)
          + 200 \* (1 + 16/75/x 
   \nn \\[0.5mm] & & \mbox{}
          + 24/25 \* x - 8/25 \* x^2) \* \H(3,0)
          + 7442/27 \* (1 + 836/3721/x + 773/7442 \* x - 18254/11163 \* x^2)
          \* \H(0)
   \nn \\[0.5mm] & & \mbox{}
          + 136 \* (1 + 4/17/x - x - 4/17 \* x^2) \* \H(1,1) \*  \z2 
          + 488/3 \* (1 + 16/61/x + 49/61 \* x - 52/183 \* x^2) \* \H(2,1,0)
   \nn \\[0.5mm] & & \mbox{}
          + 200 \* (1 + 4/15/x + 58/75 \* x - 56/225 \* x^2) \* \H(2,1,1)
          + 216 \* (1 + 8/27/x + 7/9 \* x - 20/81 \* x^2) \* \H(2,2)
   \nn \\[0.5mm] & & \mbox{}
          - 56 \* (1 + 8/21/x + 33/7 \* x + 4/7 \* x^2) \* \H(2) \*  \z2 
          + 376/3 \* (1 + 68/141/x - x - 68/141 \* x^2) \* \H(1,0) \*  \z2 
   \nn \\[0.5mm] & & \mbox{}
          + 64/3 \* (1 + 1/2/x + 39/8 \* x - 1/2 \* x^2) \* \H(2,0,0)
          + 7141/27 \* (1 + 11065/21423/x + 1340/7141 \* x 
   \nn \\[0.5mm] & & \mbox{}
          - 36508/21423 \* x^2) \* \H(1)
          + 220 \* (1 + 284/495/x + 17/33 \* x - 472/495 \* x^2) \* \H(2,0)
   \nn \\[0.5mm] & & \mbox{}
          + 78 \* (1 + 272/351/x + 911/117 \* x - 1312/351 \* x^2) \* \H(3)
          - 78 \* (1 + 272/351/x + 445/39 \* x 
   \nn \\[0.5mm] & & \mbox{}
          - 1312/351 \* x^2) \* \H(0) \*  \z2 
          + 416/3 \* (1 + 71/78/x + 29/52 \* x - 22/13 \* x^2) \* \H(2,1)
          + 152/3 \* (1 + 52/57/x 
   \nn \\[0.5mm] & & \mbox{}
          - x - 52/57 \* x^2) \* \H(1,1,1,0)
          - 112 \* (1 + 20/21/x + x + 20/21 \* x^2) \* (2 \* \H(-1,-1,2) - \H(-1,3) )
   \nn \\[0.5mm] & & \mbox{}
          + 64 \* (1 + 1/x - x - x^2) \* \H(1,2,0)
          + 856/9 \* (1 + 374/321/x + 1379/428 \* x - 1321/321 \* x^2) \* \H(2)
   \nn \\[0.5mm] & & \mbox{}
          - 856/9 \* (1 + 389/321/x + 22157/1284 \* x - 1321/321 \* x^2) \*  \z2 
          + 224/3 \* (1 + 10/7 \* x) \*  \z5 
   \nn \\[0.5mm] & & \mbox{}
          + 8 \* (1 + 4/3/x - x - 4/3 \* x^2) \* (\H(1,1,0,0) + 14/3 \*
          \H(1,1,1,1) + 7 \* \H(1,2,1) )
          - 368/3 \* (1 + 100/69/x + x 
   \nn \\[0.5mm] & & \mbox{}
          + 100/69 \* x^2) \* \H(-1,0) \*  \z2 
          + 820/9 \* (1 + 316/205/x - 49/205 \* x - 472/205 \* x^2) \*
          \H(1,1,0)
          - 576 \* \H(-4,0)
   \nn \\[0.5mm] & & \mbox{}
          - 128/3 \* (1 + 11/6/x - x - 11/6 \* x^2) \* \H(1,0,0,0)
          + 812/9 \* (1 + 384/203/x - 59/203 \* x 
   \nn \\[0.5mm] & & \mbox{}
          - 528/203 \* x^2) \* \H(1,1,1)
          + 260/3 \* (1 + 28/13/x - 21/65 \* x - 184/65 \* x^2) \* \H(1,2)
          + 24 \* (1 + 20/9/x 
   \nn \\[0.5mm] & & \mbox{}
          - x - 20/9 \* x^2) \* \H(1,1,2)
          + 686/9 \* (1 + 3208/1029/x + 241/343 \* x - 4960/1029 \* x^2) \*
          \H(1,0)
   \nn \\[0.5mm] & & \mbox{}
          + 64 \* (1 + 14/3/x + x + 14/3 \* x^2) \* \H(-1,-1) \*  \z2 
          + 188/15 \* (1 + 808/141/x + 577/47 \* x + 44/47 \* x^2) \*  \zss
   \nn \\[0.5mm] & & \mbox{}
          + 40/3 \* (1 + 28/3/x - x - 28/3 \* x^2) \* \H(1) \*  \z3 
          + 50/9 \* (1 + 88/5/x - 4609/25 \* x + 1144/25 \* x^2) \*  \z3 
   \nn \\[0.5mm] & & \mbox{}
          + 70/3 \* (1 + 424/45/x + 667/105 \* x - 5284/315 \* x^2) \* \H(1,1)
          - 8 \* (1 + 92/3/x + x + 92/3 \* x^2) \* \H(-1) \*  \z3 
   \nn \\[0.5mm] & & \mbox{}
          + 32/3 \* (1 - 11 \* x) \*  \z2  \*  \z3 
          - 304/3 \* (1 - 14/19 \* x) \* \H(0) \*  \zss
          + 1600/3 \* (1 - 16/25 \* x) \* \H(-3,-1,0)
   \nn \\[0.5mm] & & \mbox{}
          - 608 \* (1 - 13/57 \* x) \* \H(-3,0,0)
          + 1568/3 \* (1 - 10/49 \* x) \* \H(-3) \*  \z2 
          - 712/3 \* (1 - 15/89 \* x) \* \H(0,0) \*  \z3 
   \nn \\[0.5mm] & & \mbox{}
          + 320/3 \* (1 + 1/10 \* x) \* \H(3) \*  \z2 
          - 160/3 \* (1 + 1/5 \* x) \* \H(3,0,0)
          - 256 \* (1 + 1/4 \* x) \* \H(-3,2)
   \nn \\[0.5mm] & & \mbox{}
          - 7120/9 \* (1 + 159/445 \* x - 20/89 \* x^2) \* \H(-2,0)
          + 946/9 \* (1 + 4570/473 \* x - 912/473 \* x^2) \* \H(0,0,0)
   \nn \\[0.5mm] & & \mbox{}
	  + \pgg( - x) \* (
            224/5 \*  \zss
          + 128/3 \* \H(-3,0)
          + 128/3 \* \H(0) \*  \z3 
          + 64/3 \* \H(0,0) \*  \z2 
          - 64/3 \* \H(0,0,0,0)
	      )
   \nn \\[0.5mm] & & \mbox{}
	  + \pgg(x) \* (
          - 1
          - 128/15 \*  \zss
          + 256/3 \* \H(-2) \*  \z2 
          - 128/3 \* \H(-2,0,0)
          - 256/3 \* \H(-2,2)
          - 64 \* \H(0,0) \*  \z2 
   \nn \\[0.5mm] & & \mbox{}
          - 64 \* \H(0) \*  \z3 
          + 64/3 \* \H(0,0,0,0)
          + 128/3 \* \H(4)
	      )
	  + (1-x) \* (
            336 \* \H(-2) \*  \z3 
          + 896/3 \* \H(-2,-2,0)
          - 384 \* \H(-2,-1) \*  \z2 
   \nn \\[0.5mm] & & \mbox{}
          - 384 \* \H(-2,-1,-1,0)
          + 1504/3 \* \H(-2,-1,0,0)
          + 192 \* \H(-2,-1,2)
          + 544/3 \* \H(-2,0) \*  \z2 
          - 640/3 \* \H(-2,0,0,0)
   \nn \\[0.5mm] & & \mbox{}
          - 96 \* \H(-2,3)
          - 256 \* \H(1,-2,0)
	      )
	  + (1+x) \* (
          - 616/3 \* \H(0,0,0) \*  \z2 
          + 120 \* \H(0,0,0,0,0)
          + 560/3 \* \H(2) \*  \z3 
          + 616/3 \* \H(5)
   \nn \\[0.5mm] & & \mbox{}
          - 128 \* \H(2,-2,0)
          + 144 \* \H(2,0) \*  \z2 
          - 128 \* \H(2,0,0,0)
          + 368/3 \* \H(2,1) \*  \z2 
          - 16/3 \* \H(2,1,0,0)
          + 80 \* \H(2,1,1,0)
   \nn \\[0.5mm] & & \mbox{}
          + 224/3 \* \H(2,1,1,1)
          + 208/3 \* \H(2,1,2)
          + 320/3 \* \H(2,2,0)
          + 112 \* \H(2,2,1)
          - 176/3 \* \H(2,3)
          + 128 \* \H(3,1,0)
   \nn \\ & & \mbox{}
          + 448/3 \* \H(3,1,1)
          + 160 \* \H(3,2)
          + 176 \* \H(4,0)
          + 640/3 \* \H(4,1)
          )
          + \delta(1 - x) \* \Big\{751/9 - 400 \*  \z5  + 740/3 \*  \z3 \Big\}
	   \Big)
   \nn \\[-0.5mm] & & \mbox{\hspn}
       +  \colour4colour{\ca \* \nfs}   \*  \Big(
           4498/81 \* (1 + 1128/2249/x + 131/2249 \* x + 524/2249 \* x^2) \*  (\z2 - \H(2))
   \nn \\[-0.5mm] & & \mbox{}
          - 1286/27 \* (1 + 1024/1929/x - 99/643 \* x + 524/1929 \* x^2) \* \H(1,1)
          - 1264/27 \* (1 + 269/474/x 
   \nn \\[0.5mm] & & \mbox{}
          - 47/316 \* x + 59/237 \* x^2) \* \H(1,0)
          - 3524/81 \* (1 + 642/881/x + 428/881 \* x + 56/881 \* x^2) \* \H(0,0)
   \nn \\[0.5mm] & & \mbox{}
          - 217/3 \* (1 + 18169/17577/x + 104/837 \* x - 262/17577 \* x^2) \* \H(1)
          - 2170/27 \* (1 + 1208/1085/x 
   \nn \\[0.5mm] & & \mbox{}
          + 214/651 \* x + 1/9765 \* x^2) \* \H(0)
          - 1112/27 \* (1 - 68/139 \* x + 69/139 \* x^2) \*  \z3 
   \nn \\[0.5mm] & & \mbox{}
          - 22751/243 \* (1 + 34312/22751/x + 17705/45502 \* x - 10817/45502 \* x^2)
   \nn \\[0.5mm] & & \mbox{}
          + 16/3 \* (1 - 1/2 \* x + 1/2 \* x^2) \* ( 16/3 \* \H(1) \*  \z2 
          - 10/3 \* \H(1,0,0)
          - 4 \* \H(1,1,0)
          - 14/3 \* \H(1,1,1)
          - 16/3 \* \H(1,2) )
   \nn \\[0.5mm] & & \mbox{}
          - 704/27 \* (1 - 19/44 \* x + 21/44 \* x^2) \* \H(2,1)
          - 544/27 \* (1 - 11/34 \* x + 15/34 \* x^2) \* \H(2,0)
   \nn \\[0.5mm] & & \mbox{}
          + 568/27 \* (1 - 19/71 \* x + 30/71 \* x^2) \* (\H(0) \*  \z2 - \H(3))
	  + \pgg(x) \* (
          - 138493/1458
          - 296/27 \*  \z3 
   \nn \\[0.5mm] & & \mbox{}
          + 584/9 \*  \z2 
          - 6740/81 \* \H(0)
          + 256/9 \* \H(0) \*  \z2 
          - 56 \* \H(0,0)
          - 176/9 \* \H(0,0,0)
          - 6652/81 \* \H(1)
          + 304/9 \* \H(1) \*  \z2 
   \nn \\[0.5mm] & & \mbox{}
          - 584/9 \* \H(1,0)
          - 256/9 \* \H(1,0,0)
          - 584/9 \* \H(1,1)
          - 272/9 \* \H(1,1,0)
          - 32 \* \H(1,1,1)
          - 304/9 \* \H(1,2)
   \nn \\[0.5mm] & & \mbox{}
          - 584/9 \* \H(2)
          - 256/9 \* \H(2,0)
          - 32 \* \H(2,1)
          - 256/9 \* \H(3)
	      )
          - 200/27 \* \H(0,0,0) \* (1+x)
   \nn \\ & & \mbox{}
          + \delta(1 - x) \* \Big\{ 174769/972 + 404/9 \*  \z3  - 6740/81 \*  \z2  +
          608/45 \*  \zss \Big\}
          \Big)
   \nn \\[-0.5mm] & & \mbox{\hspn}
       +  \colour4colour{\cf \* \nfs}   \*  \Big(
          - 184/45 \* (1+x) \*  \zss
          + 2684/27 \* (1 + 96/671/x + 515/671 \* x - 16/671 \* x^2) \* (\H(0) \*  \z2 - \H(3) )
   \nn \\[-0.5mm] & & \mbox{}
          + 3092/27 \* (1 + 136/773/x + 71/773 \* x + 244/773 \* x^2) \*  \z3 
          - 3508/27 \* (1 + 168/877/x 
   \nn \\[0.5mm] & & \mbox{}
          + 826/877 \* x - 56/877 \* x^2) \* \H(0,0,0)
          - 1406/3 \* (1 + 38524/170829/x - 3974/6327 \* x 
   \nn \\[0.5mm] & & \mbox{}
          - 53617/170829 \* x^2)
          - 1240/27 \* (1 + 36/155/x + 134/155 \* x - 16/155 \* x^2) \* \H(2,1)
   \nn \\[0.5mm] & & \mbox{}
          - 68290/243 \* (1 + 1941/6829/x + 4931/34145 \* x 
          - 11746/34145 \* x^2) \* \H(0)
   \nn \\[0.5mm] & & \mbox{}
          - 14240/81 \* (1 + 57/178/x + 1177/1780 \* x - 211/890 \* x^2) \*
          \H(0,0)
          - 1016/27 \* (1 + 48/127/x 
   \nn \\[0.5mm] & & \mbox{}
          + 166/127 \* x - 32/127 \* x^2) \* \H(2,0)
          + 56/9 \* (1 + 8/21/x - 9/7 \* x - 20/21 \* x^2) \* (\H(1) \* \z2 -
          \H(1,2) )
   \nn \\[0.5mm] & & \mbox{}
          + 8020/81 \* (1 + 948/2005/x + 230/401 \* x - 160/401 \* x^2) \*  (\z2 - \H(2) )
   \nn \\[0.5mm] & & \mbox{}
          - 3316/27 \* (1 + 9397/14922/x - 887/1658 \* x - 6143/7461 \* x^2) \* \H(1)
          - 32 \* (1 + 157/162/x 
   \nn \\[0.5mm] & & \mbox{}
          - 7/9 \* x - 211/162 \* x^2) \* \H(1,0)
          - 8/9 \* (1 + 4/3/x - x - 4/3 \* x^2) \* ( 7 \* \H(1,0,0) + 4 \* \H(1,1,1) )
   \nn \\[0.5mm] & & \mbox{}
          - 24 \* (1 + 346/243/x - 7/9 \* x - 400/243 \* x^2) \* \H(1,1)
          - 8/3 \* (1 + 32/9/x - 1/3 \* x - 20/9 \* x^2) \* \H(1,1,0)
   \nn \\[0.5mm] & & \mbox{}
	  + \pgg(x) \* (
          - 350/9
          + 112/3 \*  \z3 
          - 100/9 \* \H(0)
          + 32/9 \* \H(0) \*  \z2 
          - 20/3 \* \H(1)
          + 32/9 \* \H(2,0)
          - 32/9 \* \H(3)
	      )
   \nn \\[0.5mm] & & \mbox{}
	  + (1+x) \* (
            40/3 \* \H(0) \*  \z3 
          + 328/9 \* \H(0,0) \*  \z2 
          - 536/9 \* \H(0,0,0,0)
          + 80/9 \* \H(2) \*  \z2 
          - 112/9 \* \H(2,0,0)
   \nn \\[0.5mm] & & \mbox{}
          - 80/9 \* \H(2,1,0)
          - 64/9 \* \H(2,1,1)
          - 80/9 \* \H(2,2)
          - 208/9 \* \H(3,0)
          - 176/9 \* \H(3,1)
          - 328/9 \* \H(4)
          )
   \nn \\ & & \mbox{}
          + \delta(1 - x) \* \Big\{ 28945/162 - 1144/9 \*  \z3  - 100/9 \*  \z2  -
          32/45 \*  \zss \Big\}
	   \Big)
   \nn \\[-0.5mm] & & \mbox{\hspn}
       +   \colour4colour{\nft}  \*  \Big(
          32/9 \* (1 - 1/2 \* x + 1/2 \* x^2) \* ( 25/18 
          - \z2 
          + 5/3 \* \H(0)
          + \H(0,0)
          + 5/3 \* \H(1)
          + \H(1,0)
          + \H(1,1)
          + \H(2) )
   \nn \\[-0.5mm] & & \mbox{}
	  + 16/9 \* \pgg(x) \* (
	    25/18
          - \z2 
          + 5/3 \* \H(0)
          + \H(0,0)
          + 5/3 \* \H(1)
          + \H(1,0)
          + \H(1,1)
          + \H(2)
          )
   \nn \\ & & \mbox{}
          + \delta(1 - x) \* \Big\{ - 1000/729 + 80/27 \*  \z2 \Big\}
	   \Big)
\;\; .
\eea
It may be worthwhile to note that, unlike the third-order coefficient functions
for gauge-boson exchange DIS \cite{MVV6,MVV10}, Eqs.~(\ref{eq:cphiq3}) and
(\ref{eq:cphig3}) do not involve additional special functions including terms
of the form $(1 \!\pm\! x)^{-n}\, {\rm H}_{\:\!\vec{w}}(x)$ with $n>1$.
%
%
\setlength{\baselineskip}{0.55cm}
\renewcommand{\theequation}{B.\arabic{equation}}
\setcounter{equation}{0}
\section*{Appendix B: The third-order physical kernels for \boldmath
 $( \Ftwo , \: F_\phi ) $}
\label{sec:AppB}
\setlength{\baselineskip}{0.543cm}
%
%
In this second appendix we write down the exact expressions, again using the
notations introduced in Section 3, for the matrix elements of the NNLO 
contribution to the physical evolution kernel for the system 
($\,\Ftwo, F_\phi\,$) of singlet structure functions. 
The coefficients in the expansion in powers of of the strong coupling are 
normalized according to Eq.~(\ref{eq:Kabexp}). 

As the corresponding NLO quantity before, we present $K_{\,22}^{\,(2)}$ via
the decomposition (\ref{eq:K22nsps}) into its non-singlet and pure-singlet
components. The former quantity, which is of direct interest for analyses of
data on, e.g., the difference of the neutron and proton structure functions, 
is given by
\bea
 \label{eq:Kns2}
 \lefteqn{K^{\,(2)}_{\,2,\rm ns}(x)\,  = }
%
%
   \nn \\ & & \mbox{\hspn}
       \colour4colour{ \cas \*  \cf}  \*  \Big(
         \pqq(x)  \*  (
            50689/162\:
          - 20/3\: \*  \z3 
          - 1504/9\: \*  \z2 
          + 192/5\: \*  \zss
          + 16 \* \H(-3,0)
          + 64 \* \H(-2,-1,0)
   \nn \\[-0.5mm] & & \mbox{}
          + 64 \* \H(-2,0)
          - 16 \* \H(-2,0,0)
          + 32 \* \H(-2,2)
          + 10585/27\: \* \H(0)
          + 64 \* \H(0) \*  \z3 
          - 212/3\: \* \H(0) \*  \z2 
   \nn \\[0.5mm] & & \mbox{}
          + 1988/9\: \* \H(0,0)
          + 256/3\: \* \H(0,0,0)
          - 16 \* \H(0,0,0,0)
          + 4649/27\: \* \H(1)
          + 144 \* \H(1) \*  \z3 
          - 88 \* \H(1) \*  \z2 
   \nn \\[0.5mm] & & \mbox{}
          + 96 \* \H(1,-2,0)
          + 484/9\: \* \H(1,0)
          - 16 \* \H(1,0) \*  \z2 
          + 44/3\: \* \H(1,0,0)
          - 48 \* \H(1,0,0,0)
          + 484/9\: \* \H(1,1)
   \nn \\[0.5mm] & & \mbox{}
          - 88/3\: \* \H(1,1,0)
          - 64 \* \H(1,1,0,0)
          + 88/3\: \* \H(1,2)
          + 64 \* \H(1,3)
          + 968/9\: \* \H(2)
          - 32 \* \H(2,0,0)
          + 44 \* \H(3)
          + 16 \* \H(4)
          )
   \nn \\[0.5mm] & & \mbox{}
       + \pqq(-x)  \*  (
            176/3\: \*  \z3 
          + 536/9\: \*  \z2 
          - 16 \*  \zss
          - 16 \* \H(-3,0)
          + 128 \* \H(-2) \*  \z2 
          + 352/3\: \* \H(-2,0)
   \nn \\[0.5mm] & & \mbox{}
          - 64 \* \H(-2,0,0)
          - 128 \* \H(-2,2)
          + 192 \* \H(-1) \*  \z3 
          - 176/3\: \* \H(-1) \*  \z2 
          - 256 \* \H(-1,-1) \*  \z2 
          - 352/3\: \* \H(-1,-1,0)
   \nn \\[0.5mm] & & \mbox{}
          + 128 \* \H(-1,-1,0,0)
          + 256 \* \H(-1,-1,2)
          + 1072/9\: \* \H(-1,0)
          + 176 \* \H(-1,0) \*  \z2 
          + 176 \* \H(-1,0,0)
   \nn \\[0.5mm] & & \mbox{}
          - 48 \* \H(-1,0,0,0)
          - 128 \* \H(-1,3)
          - 124/3\: \* \H(0)
          - 64 \* \H(0) \*  \z3 
          + 80/3\: \* \H(0) \*  \z2 
          - 536/9\: \* \H(0,0)
          - 48 \* \H(0,0) \*  \z2 
   \nn \\[0.5mm] & & \mbox{}
          - 256/3\: \* \H(0,0,0)
          + 16 \* \H(0,0,0,0)
          + 32 \* \H(2) \*  \z2 
          + 32 \* \H(4)
          )
       + (1+x)  \*  (
          - 96 \* \H(-1,2)
          + 112 \* \H(2)
          - 8 \* \H(2) \*  \z2 
   \nn \\[0.5mm] & & \mbox{}
          - 4 \* \H(2,0,0)
          + 48 \* \H(3)
          + 12 \* \H(4)
          )
       + (1-x)  \*  (
            8 \* \H(-3,0)
          - 8 \* \H(-2) \*  \z2 
          - 16 \* \H(-2,-1,0)
          + 8 \* \H(-2,0,0)
          )
   \nn \\[0.5mm] & & \mbox{}
       + 16/3\: \* (1 - 43 \* x)  \*  \H(-1,-1,0)
       + 8/3\: \* (1 + 43 \* x)  \*  (
            \H(1) \*  \z2 
          - \H(1,0,0)
          )
       - 4 \*  \z2  \* (3 + 5 \* x)  \*  \H(0,0)
   \nn \\[0.5mm] & & \mbox{}
       + 2 \*  \zss \* (5 + 3 \* x)
       - 8 \*  \z2  \* (6 + 11 \* x)  \*  \H(0)
       + 8/3\: \* (13 - 15 \* x)  \*  \H(-2,0)
       - 8/3\: \* (19 - 25 \* x)  \*  \H(-1,0,0)
   \nn \\[0.5mm] & & \mbox{}
       - 4/3\: \*  \z3  \* (27 - 91 \* x)
       + 8/3\: \*  \z2  \* (37 - 7 \* x)  \*  \H(-1)
       - 8/15\: \*  \z2  \* (210 - 145 \* x + 198 \* x^3)
   \nn \\[0.5mm] & & \mbox{}
       + 11/9\: \* (229 - 467 \* x)  \*  \H(1)
       + 8/15\: \* (575 + 22 \* x^{-2} + 355 \* x 
       - 198 \* x^3)  \*  \H(-1,0)
       + 1/45\: \* (2619  
   \nn \\[0.5mm] & & \mbox{}
       - 528 \* x^{-1} - 33241 \* x
       + 4752 \* x^2)  \*  \H(0)
       + 1/270\: \* (64993 + 3168 \* x^{-1} 
       - 283643 \* x + 28512 \* x^2)
   \nn \\[0.5mm] & & \mbox{}
       - 8/15\: \* (145 \* x - 198 \* x^3)  \*  \H(0,0)
       + x  \*  (
          - 16 \* \H(0) \*  \z3 
          + 40 \* \H(0,0,0)
          + 8 \* \H(0,0,0,0)
          )
   \nn \\[-0.5mm] & & \mbox{}
       + (20/27\: \* (827 + 54 \*  \z5 )
          - 1544/3\: \*  \z3 
          + 22286/27\: \*  \z2 
          - 1592/15\: \*  \zss
          )\, \* \delta(1 - x)
              \Big)
   \nn \\[-0.5mm] & & \mbox{\hspn}
+  \colour4colour{ \ca  \*  \cfs } \* \Big(
         \pqq(x)  \*  (
            11/2\:
          - 664/3\: \*  \z3 
          - 276/5\: \*  \zss
          - 16 \* \H(-3,0)
          - 48 \* \H(-2) \*  \z2 
          - 224 \* \H(-2,-1,0)
   \nn \\[-0.5mm] & & \mbox{}
          - 128 \* \H(-2,0)
          + 80 \* \H(-2,0,0)
          - 64 \* \H(-2,2)
          - 317/2\: \* \H(0)
          - 136 \* \H(0) \*  \z3 
          - 188/3\: \* \H(0) \*  \z2 
          - 74 \* \H(0,0)
   \nn \\[0.5mm] & & \mbox{}
          - 64 \* \H(0,0) \*  \z2 
          - 4/3\: \* \H(0,0,0)
          + 80 \* \H(0,0,0,0)
          - 384 \* \H(1) \*  \z3 
          + 176/3\: \* \H(1) \*  \z2 
          - 256 \* \H(1,-2,0)
   \nn \\[0.5mm] & & \mbox{}
          + 1468/9\: \* \H(1,0)
          - 32 \* \H(1,0) \*  \z2 
          + 584/3\: \* \H(1,0,0)
          + 176 \* \H(1,0,0,0)
          + 352/3 \* \H(1,1,0)
          + 128 \* \H(1,1,0,0)
   \nn \\[0.5mm] & & \mbox{}
          + 176/3\: \* \H(1,2)
          - 128 \* \H(1,3)
          + 1072/9\: \* \H(2)
          - 32 \* \H(2) \*  \z2 
          + 440/3\: \* \H(2,0)
          + 80 \* \H(2,0,0)
          + 88 \* \H(2,1)
   \nn \\[0.5mm] & & \mbox{}
          + 128 \* \H(3)
          + 16 \* \H(3,0)
          + 16 \* \H(4)
          )
       + \pqq(-x)  \*  (
          - 244/3\: \*  \z3 
          - 1072/9\: \*  \z2 
          + 4 \*  \zss
          + 80 \* \H(-3,0)
   \nn \\[0.5mm] & & \mbox{}
          - 512 \* \H(-2) \*  \z2 
          - 64 \* \H(-2,-1,0)
          - 776/3\: \* \H(-2,0)
          + 336 \* \H(-2,0,0)
          + 480 \* \H(-2,2)
          - 672 \* \H(-1) \*  \z3 
   \nn \\[0.5mm] & & \mbox{}
          + 208/3\: \* \H(-1) \*  \z2 
          - 64 \* \H(-1,-2,0)
          + 896 \* \H(-1,-1) \*  \z2 
          + 704/3\: \* \H(-1,-1,0)
          - 576 \* \H(-1,-1,0,0)
   \nn \\[0.5mm] & & \mbox{}
          - 896 \* \H(-1,-1,2)
          - 2144/9\: \* \H(-1,0)
          - 672 \* \H(-1,0) \*  \z2 
          - 376 \* \H(-1,0,0)
          + 272 \* \H(-1,0,0,0)
          + 48 \* \H(-1,2)
   \nn \\[0.5mm] & & \mbox{}
          + 32 \* \H(-1,2,0)
          + 512 \* \H(-1,3)
          + 284/3\: \* \H(0)
          + 232 \* \H(0) \*  \z3 
          - 124/3\: \* \H(0) \*  \z2 
          + 1072/9\: \* \H(0,0)
   \nn \\[0.5mm] & & \mbox{}
          + 208 \* \H(0,0) \*  \z2 
          + 620/3\: \* \H(0,0,0)
          - 80 \* \H(0,0,0,0)
          - 112 \* \H(2) \*  \z2 
          - 24 \* \H(3)
          - 16 \* \H(3,0)
          - 160 \* \H(4)
          )
   \nn \\[0.5mm] & & \mbox{}
       + (1+x)  \*  (
            384 \* \H(-1,2)
          + 16 \* \H(2) \*  \z2 
          - 44/3\: \* \H(2,0)
          + 32 \* \H(2,0,0)
          - 48 \* \H(4)
          )
       + (1-x)  \*  (
            16 \* \H(-2) \*  \z2
   \nn \\[0.5mm] & & \mbox{}
          + 32 \* \H(-2,-1,0)
          - 48 \* \H(-2,0,0)
          + 224/3\: \* \H(1,0)
          )
       + 16/3\: \* (1 + 43 \* x)  \*  \H(1,0,0)
       - 32 \* (2 - x)  \*  \H(-3,0)
   \nn \\[0.5mm] & & \mbox{}
       - 16 \*  \z3  \* (3 - 2 \* x)  \*  \H(0)
       + 16 \*  \z2  \* (3 + 5 \* x)  \*  \H(0,0)
       - 64/3\: \* (5 - 17 \* x)  \*  \H(-1,-1,0)
       - 32/3\: \*  \z2  \* (5 + 17 \* x)  \*  \H(1)
   \nn \\[0.5mm] & & \mbox{}
       - 8 \*  \zss \* (7 + 6 \* x)
       - 44/3\: \* (7 + 25 \* x)  \*  \H(0,0,0)
       - 8/3\: \* (11 - 75 \* x)  \*  \H(-2,0)
       - 16 \* (22 + 35 \* x)  \*  \H(2)
   \nn \\[0.5mm] & & \mbox{}
       + 32/3\: \* (29 + 7 \* x)  \*  \H(-1,0,0)
       - 32/3\: \*  \z2  \* (41 + 19 \* x)  \*  \H(-1)
       - 44/3\: \* (49 - 72 \* x)  \*  \H(1)
   \nn \\[0.5mm] & & \mbox{}
       - 8/3\: \* (65 + 101 \* x)  \*  \H(3)
       + 8/3\: \*  \z2  \* (65 + 176 \* x)  \*  \H(0)
       + 4/3\: \*  \z3  \* (99 + 61 \* x)
       - 32/15\: \* (280 
   \nn \\[0.5mm] & & \mbox{}
       + 11 \* x^{-2}
       + 170 \* x - 99 \* x^3)  \*  \H(-1,0)
       + 16/15\: \*  \z2  \* (330 + 185 \* x + 198 \* x^3)
       - 2/45\: \* (635 + 7115 \* x 
   \nn \\[0.5mm] & & \mbox{}
       + 4752 \* x^3)  \*  \H(0,0)
       + 1/90\: \* (799 + 2112 \* x^{-1} + 89679 \* x - 19008 \* x^2)  \*  \H(0)
   \nn \\[0.5mm] & & \mbox{}
       - 1/90\: \* (10817 + 2112 \* x^{-1} - 29627 \* x + 19008 \* x^2)
          - 64 \* x \* \H(0,0,0,0)
       +  ( 2/3\: \* (47 + 180 \*  \z5 )
   \nn \\[-0.5mm] & & \mbox{}
          + 2296/3\: \*  \z3 
          - 1235/3\: \*  \z2 
          + 16 \*  \z2  \*  \z3 
          - 856/15\: \*  \zss
          )\, \* \delta(1 - x)
                  \Big)
   \nn \\[-0.5mm] & & \mbox{\hspn}
+  \colour4colour{ \cft } \* \Big(
         \pqq(x)  \*  (
            72/5\: \*  \zss
          - 32 \* \H(-3,0)
          + 96 \* \H(-2) \*  \z2 
          + 192 \* \H(-2,-1,0)
          - 96 \* \H(-2,0,0)
          - 3 \* \H(0)
   \nn \\[-0.5mm] & & \mbox{}
          + 16 \* \H(0) \*  \z3 
          - 24 \* \H(0) \*  \z2 
          + 26 \* \H(0,0)
          - 32 \* \H(0,0,0,0)
          + 192 \* \H(1) \*  \z3 
          + 128 \* \H(1,-2,0)
          - 96 \* \H(1,0,0)
   \nn \\[0.5mm] & & \mbox{}
          - 64 \* \H(1,0,0,0)
          + 64 \* \H(1,2,0)
          + 128 \* \H(1,3)
          - 48 \* \H(2,0)
          + 32 \* \H(2,0,0)
          + 64 \* \H(2,1,0)
          + 64 \* \H(2,2)
          + 64 \* \H(3,0)
   \nn \\[0.5mm] & & \mbox{}
          + 64 \* \H(3,1)
          + 32 \* \H(4)
          )
       + \pqq(-x)  \*  (
          - 72 \*  \z3 
          + 56 \*  \zss
          - 96 \* \H(-3,0)
          + 512 \* \H(-2) \*  \z2 
          + 128 \* \H(-2,-1,0)
   \nn \\[0.5mm] & & \mbox{}
          + 48 \* \H(-2,0)
          - 416 \* \H(-2,0,0)
          - 448 \* \H(-2,2)
          + 576 \* \H(-1) \*  \z3 
          + 96 \* \H(-1) \*  \z2 
          + 128 \* \H(-1,-2,0)
   \nn \\[0.5mm] & & \mbox{}
          - 768 \* \H(-1,-1) \*  \z2 
          + 640 \* \H(-1,-1,0,0)
          + 768 \* \H(-1,-1,2)
          + 640 \* \H(-1,0) \*  \z2 
          + 48 \* \H(-1,0,0)
   \nn \\[0.5mm] & & \mbox{}
          - 352 \* \H(-1,0,0,0)
          - 96 \* \H(-1,2)
          - 64 \* \H(-1,2,0)
          - 512 \* \H(-1,3)
          - 24 \* \H(0)
          - 208 \* \H(0) \*  \z3 
          - 24 \* \H(0) \*  \z2 
   \nn \\[0.5mm] & & \mbox{}
          - 224 \* \H(0,0) \*  \z2 
          - 72 \* \H(0,0,0)
          + 96 \* \H(0,0,0,0)
          + 96 \* \H(2) \*  \z2 
          + 48 \* \H(3)
          + 32 \* \H(3,0)
          + 192 \* \H(4)
          )
   \nn \\[0.5mm] & & \mbox{}
       + (1+x)  \*  (
            296/5\: \*  \zss
          + 480 \* \H(-1) \*  \z2 
          + 192 \* \H(-1,-1,0)
          - 32 \* \H(-1,0)
          - 416 \* \H(-1,0,0)
          - 384 \* \H(-1,2)
   \nn \\[0.5mm] & & \mbox{}
          - 48 \* \H(2,0,0)
          - 32 \* \H(3,0)
          - 16 \* \H(4)
          )
       + (1-x)  \*  (
          - 62
          + 64 \* \H(-2,0,0)
          + 16 \* \H(0,0) \*  \z2 
          - 48 \* \H(0,0,0,0)
   \nn \\[0.5mm] & & \mbox{}
          + 560 \* \H(1)
          + 96 \* \H(1) \*  \z2 
          - 16 \* \H(1,0)
          )
       - 80 \* (1 + 3 \* x)  \*  \H(-2,0)
       + 8 \* (3 - 5 \* x)  \*  \H(2,0)
       + 32 \* (3 - x)  \*  \H(-3,0)
   \nn \\[0.5mm] & & \mbox{}
       + 16 \* (3 + 20 \* x)  \*  \H(0,0,0)
       - (3 + 539 \* x)  \*  \H(0)
       - 24 \*  \z3  \* (5 + 27 \* x)
       - 2 \* (9 - 191 \* x)  \*  \H(0,0)
   \nn \\[0.5mm] & & \mbox{}
       + 16 \* (9 + 13 \* x)  \*  \H(3)
       - 16 \*  \z2  \* (9 + 28 \* x)  \*  \H(0)
       + 4 \* (77 + 85 \* x)  \*  \H(2)
       - 4 \*  \z2  \* (77 + 93 \* x)
       + 96 \*  \z3  \* \H(0)
   \nn \\[-0.5mm] & & \mbox{}
       + (  1/2\: \* (29 - 480 \*  \z5 )
          + 68 \*  \z3 
          + 18 \*  \z2 
          - 32 \*  \z2  \*  \z3 
          + 288/5\: \*  \zss
          )\, \* \delta(1 - x)
                \Big)
   \nn \\[-0.5mm] & & \mbox{\hspn}
+  \colour4colour{ \ca  \*  \cf  \*  \nf } \* \Big(
         \pqq(x)  \*  (
          - 7531/81\:
          - 64/3\: \*  \z3 
          + 48 \*  \z2 
          - 16 \* \H(-2,0)
          - 3536/27\: \* \H(0)
          + 32/3\: \* \H(0) \*  \z2
   \nn \\[-0.5mm] & & \mbox{}
          - 608/9\: \* \H(0,0)
          - 40/3\: \* \H(0,0,0)
          - 1552/27\: \* \H(1)
          + 16 \* \H(1) \*  \z2 
          - 176/9\: \* \H(1,0)
          - 8/3\: \* \H(1,0,0)
   \nn \\[0.5mm] & & \mbox{}
          - 176/9\: \* \H(1,1)
          + 16/3\: \* \H(1,1,0)
          - 16/3\: \* \H(1,2)
          - 352/9\: \* \H(2)
          - 8 \* \H(3)
          )
   \nn \\[0.5mm] & & \mbox{}
       + \pqq(-x)  \*  (
          - 32/3\: \*  \z3 
          - 80/9\: \*  \z2 
          - 64/3\: \* \H(-2,0)
          + 32/3\: \* \H(-1) \*  \z2 
          + 64/3\: \* \H(-1,-1,0)
   \nn \\[0.5mm] & & \mbox{}
          - 160/9\: \* \H(-1,0)
          - 32 \* \H(-1,0,0)
          + 16/3\: \* \H(0)
          - 8/3\: \* \H(0) \*  \z2 
          + 80/9\: \* \H(0,0)
          + 40/3\: \* \H(0,0,0)
          )
   \nn \\[0.5mm] & & \mbox{}
       - 8 \* (1+x)  \*  \H(2)
       + 16/3\: \* (1 - 5 \* x)  \*  (
            \H(1) \*  \z2 
          - \H(1,0,0)
          )
       + 16/3\: \* (1 + 5 \* x)  \*  (
            \H(-1) \*  \z2 
          + 2 \* \H(-1,-1,0)
   \nn \\[0.5mm] & & \mbox{}
          - \H(-1,0,0)
          )
       - 16/45\: \* (3 - 6 \* x^{-1} - 497 \* x + 54 \* x^2)  \*  \H(0)
       + 8/15\: \*  \z2  \* (15 - 55 \* x + 36 \* x^3)
   \nn \\[0.5mm] & & \mbox{}
       - 16/15\: \* (55 + 2 \* x^{-2} + 35 \* x - 18 \* x^3)  \*  \H(-1,0)
       - 4/9\: \* (73 - 269 \* x)  \*  \H(1)
       - 1/135\: \* (7303 + 288 \* x^{-1} 
   \nn \\[0.5mm] & & \mbox{}
       - 43013 \* x + 2592 \* x^2)
       + 8/15\: \* (55 \* x - 36 \* x^3)  \*  \H(0,0)
       - 160/3\: \*  \z3  \* x
       - 32/3\: \* \H(-2,0)
       - (
            5516/27\:
   \nn \\[-0.5mm] & & \mbox{}
          - 224/3\: \*  \z3 
          + 7216/27\: \*  \z2 
          - 296/15\: \*  \zss
          )\, \* \delta(1 - x)
                \Big)
   \nn \\[-0.5mm] & & \mbox{\hspn}
+  \colour4colour{ \cfs \*  \nf } \* \Big(
         \pqq(x)  \*  (
          - 67/3\:
          + 208/3\: \*  \z3 
          + 32 \* \H(-2,0)
          + 19 \* \H(0)
          + 80/3\: \* \H(0) \*  \z2 
          + 12 \* \H(0,0)
   \nn \\[-0.5mm] & & \mbox{}
          - 32/3\: \* \H(0,0,0)
          - 4 \* \H(1)
          - 32/3\: \* \H(1) \*  \z2 
          - 232/9\: \* \H(1,0)
          - 80/3\: \* \H(1,0,0)
          - 64/3\: \* \H(1,1,0)
          - 32/3\: \* \H(1,2)
   \nn \\[0.5mm] & & \mbox{}
          - 160/9\: \* \H(2)
          - 80/3\: \* \H(2,0)
          - 16 \* \H(2,1)
          - 32 \* \H(3)
          )
       + \pqq(-x)  \*  (
            64/3\: \*  \z3 
          + 160/9\: \*  \z2 
   \nn \\[0.5mm] & & \mbox{}
          + 128/3\: \* \H(-2,0)
          - 64/3\: \* \H(-1) \*  \z2 
          - 128/3\: \* \H(-1,-1,0)
          + 320/9\: \* \H(-1,0)
          + 64 \* \H(-1,0,0)
          - 32/3\: \* \H(0)
   \nn \\[0.5mm] & & \mbox{}
          + 16/3\: \* \H(0) \*  \z2 
          - 160/9\: \* \H(0,0)
          - 80/3\: \* \H(0,0,0)
          )
       + 8/3\: \* (1+x)  \*  (
          - 2 \* \H(0) \*  \z2 
          + 7 \* \H(0,0,0)
          +  \H(2,0)
   \nn \\[0.5mm] & & \mbox{}
          + 2 \* \H(3)
          )
       - 32/3\: \*  (1-x)  \*  \H(1,0)
       - 32/3\: \* (1 - 5 \* x)  \*  (
            \H(1) \*  \z2 
          - \H(1,0,0)
          )
       + 16 \* (1 + 3 \* x)  \*  \H(2)
   \nn \\[0.5mm] & & \mbox{}
       - 32/3\: \*  (1 + 5 \* x)  \*  (
            \H(-1) \*  \z2 
          + 2 \* \H(-1,-1,0)
          - \H(-1,0,0)
          )
       - 16/15\: \*  \z2  \* (15 - 25 \* x + 36 \* x^3)
   \nn \\[0.5mm] & & \mbox{}
       + 8/3\: \* (19 - 42 \* x)  \*  \H(1)
       + 32/15\: \* (55 + 2 \* x^{-2} + 35 \* x - 18 \* x^3)  \*  \H(-1,0)
       + 4/45\: \* (85 - 95 \* x 
   \nn \\[0.5mm] & & \mbox{}
       + 432 \* x^3)  \*  \H(0,0)
       + 8/45\: \* (134 + 24 \* x^{-1} - 269 \* x + 216 \* x^2)
       - 1/45\: \* (209 + 192 \* x^{-1} + 4929 \* x 
   \nn \\[0.5mm] & & \mbox{}
       - 1728 \* x^2)  \*  \H(0)
       + 320/3\: \*  \z3  \* x
       + 64/3\: \* \H(-2,0)
       - (
            239/6\:
          + 400/3\: \*  \z3 
          - 146/3\: \*  \z2 
   \nn \\[-0.5mm] & & \mbox{}
          - 208/15\: \*  \zss
          )\, \* \delta(1 - x)
              \Big)
   \nn \\[-0.5mm] & & \mbox{\hspn}
+  \colour4colour{ \cf  \*  \nfs } \* \Big(
         \pqq(x)  \*  (
            470/81\:
          - 32/9\: \*  \z2 
          + 268/27\: \* \H(0)
          + 16/3\: \* \H(0,0)
          + 116/27\: \* \H(1)
          + 16/9\: \* \H(1,0)
   \nn \\[-0.5mm] & & \mbox{}
          + 16/9\: \* \H(1,1)
          + 32/9\: \* \H(2)
          )
       - 4/9\: \* (1 + 13 \* x)  \*  \H(1)
       - 4/9\: \* (3 + 23 \* x)  \*  \H(0)
       + 2/27\: \* (29 - 295 \* x)
   \nn \\[-0.5mm] & & \mbox{}
       + (
            406/27\:
          + 536/27\: \*  \z2 
          )\, \* \delta(1 - x)
              \Big) 
\:\: .
\eea
The corresponding pure-singlet component reads
\bea
\label{eq:Kps2}
  \lefteqn{ K^{\,(2)}_{\,22, \rm ps}(x)\, \:  =  }
%
%
   \nn \\ & & \mbox{\hspn}
       \colour4colour{ \ca  \*  \cf  \*  \nf } \* \Big(
         8 \* (1+x)  \*  (
            4 \* \H(2) \*  \z2 
          - 7 \* \H(2,0,0)
          - 4 \* \H(3,0)
          - \H(4)
          )
       + 8 \* (1-x)  \*  (
            4 \* \H(-2) \*  \z2 
          + 8 \* \H(-2,-1,0)
   \nn \\[-0.5mm] & & \mbox{}
          + 4 \* \H(-2,0,0)
          + \H(0,0) \*  \z2 
          )
       + 32 \*  \z3  \* (2 + x)  \*  \H(0)
       + 64/3\: \* (3 - x^{-1} + 3 \* x - x^2)  \*  \H(-1,-1,0)
   \nn \\[0.5mm] & & \mbox{}
       + 32/3\: \*  \z2  \* (3 + x^{-1} - 3 \* x - x^2)  \*  \H(1)
       - 32/3\: \* (3 + 2 \* x^{-1} + 3 \* x + 2 \* x^2)  \*  \H(-1,2)
       - 16/3\: \* (3 + 8 \* x^{-1} 
   \nn \\[0.5mm] & & \mbox{}
       + 3 \* x + 8 \* x^2)  \*  \H(-1,0,0)
       + 16 \* (3 - x)  \*  \H(-3,0)
       + 16 \* (4 - 3 \* x)  \*  \H(0,0,0,0)
       + 8/3\: \* (5 - 13 \* x + 8 \* x^2)  \*  \H(2,0)
   \nn \\[0.5mm] & & \mbox{}
       + 32/3\: \*  \z2  \* (6 + x^{-1} + 6 \* x + x^2)  \*  \H(-1)
       - 8/3\: \*  \z3  \* (11 - 32 \* x^{-1} + 89 \* x + 28 \* x^2)
   \nn \\[0.5mm] & & \mbox{}
       + 4/5\: \*  \zss \* (11 + 31 \* x)
       - 4/3\: \*  \z2  \* (15 - 16 \* x^{-1} + 15 \* x + 16 \* x^2)  \*  \H(0)
       + 4/3\: \* (15 + 16 \* x^{-1} - 51 \* x 
   \nn \\[0.5mm] & & \mbox{}
       + 16 \* x^2)  \*  \H(3)
       - 8 \* (15 + 32 \* x)  \*  \H(0,0,0)
       - 4/9\: \*  \z2  \* (21 - 368 \* x^{-1} - 363 \* x + 160 \* x^2)
   \nn \\[0.5mm] & & \mbox{}
       + 4/9\: \* (21 + 104 \* x^{-1} + 21 \* x + 160 \* x^2)  \*  \H(2)
       - 8/3\: \* (27 - 16 \* x^{-1} + 33 \* x + 24 \* x^2)  \*  \H(-2,0)
   \nn \\[0.5mm] & & \mbox{}
       - 4/3\: \* (39 + 16 \* x^{-1} - 39 \* x - 16 \* x^2)  \*  \H(1,0,0)
       - 4/3\: \* (49 - 32 \* x^{-1} - x - 16 \* x^2)  \*  \H(1,0)
   \nn \\[0.5mm] & & \mbox{}
       - 4/9\: \* (57 - 127 \* x^{-1} - 48 \* x + 118 \* x^2)  \*  \H(1)
       + 32/9\: \* (84 + 59 \* x^{-1} + 48 \* x + 23 \* x^2)  \*  \H(-1,0)
   \nn \\[0.5mm] & & \mbox{}
       + 4/9\: \* (743 + 28 \* x^{-1} - 874 \* x + 744 \* x^2)  \*  \H(0,0)
       + 1/27\: \* (8115 - 15971 \* x^{-1} + 8688 \* x - 832 \* x^2)
   \nn \\[-0.5mm] & & \mbox{}
       - 2/27\: \* (11373 + 820 \* x^{-1} + 1860 \* x - 1396 \* x^2)  \*  \H(0)
              \Big)
   \nn \\[-0.5mm] & & \mbox{\hspn}
+  \colour4colour{  \cfs \*  \nf } \* \Big(
         16 \* (1+x)  \*  (
          - 3 \* \H(-1) \*  \z2 
          - 6 \* \H(-1,-1,0)
          + 3 \* \H(-1,0,0)
          + 6 \* \H(0,0) \*  \z2 
          - 4 \* \H(0,0,0,0)
          - 2 \* \H(2) \*  \z2 
   \nn \\[-0.5mm] & & \mbox{}
          + 2 \* \H(2,0,0)
          - 2 \* \H(3,0)
          - 6 \* \H(4)
          )
       + 16 \* (1-x)  \*  (
          - 2 \* \H(-2) \*  \z2 
          - 4 \* \H(-2,-1,0)
          + 2 \* \H(-2,0,0)
          + 2 \* \H(0) \*  \z3 
   \nn \\[0.5mm] & & \mbox{}
          - 3 \* \H(1) \*  \z2 
          + 3 \* \H(1,0,0)
          )
       - 16/3\: \* (3 - 4 \* x^{-1} - 8 \* x^2)  \*  \H(2,0)
       + 16/3\: \*  \z3  \* (3 + 4 \* x^{-1} + 36 \* x)
   \nn \\[0.5mm] & & \mbox{}
       + 16/3\: \* (3 - 7 \* x + 24 \* x^2)  \*  \H(0,0,0)
       - 16/3\: \* (9 - 4 \* x^{-1} + 12 \* x - 16 \* x^2)  \*  \H(3)
       + 16/3\: \*  \z2  \* (9 - 4 \* x^{-1} 
   \nn \\[0.5mm] & & \mbox{}
       + 16 \* x - 16 \* x^2)  \*  \H(0)
       + 32/3\: \* (9 + 2 \* x)  \*  \H(-2,0)
       + 16/5\: \*  \zss \* (11 + 6 \* x)
       - 32/3\: \* (12 - 5 \* x^{-1} - 12 \* x 
   \nn \\[0.5mm] & & \mbox{}
       + 5 \* x^2)  \*  \H(1,0)
       - 8/9\: \* (234 - 80 \* x^{-1} - 27 \* x + 32 \* x^2)  \*  \H(2)
       + 16/45\: \* (585 + 4 \* x^{-2} + 625 \* x 
   \nn \\[0.5mm] & & \mbox{}
       - 36 \* x^3)  \*  \H(-1,0)
       - 4/9\: \* (657 - 149 \* x^{-1} - 450 \* x - 58 \* x^2)  \*  \H(1)
       + 8/45\: \*  \z2  \* (1170 - 400 \* x^{-1} 
   \nn \\[0.5mm] & & \mbox{}
       + 1115 \* x + 160 \* x^2 - 72 \* x^3)
       - 2/45\: \* (1484 - 628 \* x^{-1} - 7911 \* x - 1588 \* x^2)  \*  \H(0)
   \nn \\[0.5mm] & & \mbox{}
       - 4/45\: \* (3015 + 1825 \* x + 920 \* x^2 - 144 \* x^3)  \*  \H(0,0)
       - 1/45\: \* (6931 - 5184 \* x^{-1} - 7391 \* x 
   \nn \\[-0.5mm] & & \mbox{}
       + 5644 \* x^2)
       + 64 \* \H(-3,0)
         \Big)
   \nn \\[-0.5mm] & & \mbox{\hspn}
   +  \colour4colour{  \cf  \*  \nfs } \* \Big(
         16/3\: \* (1+x)  \*  (
          - 2 \*  \z3 
          + 3 \* \H(0,0,0)
          - \H(2,0)
          )
       - 64/9\: \* (3 + x^{-1} + 3 \* x + x^2)  \*  \H(-1,0)
   \nn \\[-0.5mm] & & \mbox{}
       - 8/9\: \* (3 + 4 \* x^{-1} - 3 \* x - 4 \* x^2)  \*  \H(1,0)
       + 8/9\: \* (23 - 12 \* x^{-1} + 95 \* x - 20 \* x^2)  \*  \H(0,0)
   \nn \\[0.5mm] & & \mbox{}
       + 4/27\: \* (351 - 212 \* x^{-1} + 114 \* x - 240 \* x^2)  \*  \H(0)
       + 2/81\: \* (5139 - 2155 \* x^{-1} - 3420 \* x + 436 \* x^2)
   \nn \\[-0.5mm] & & \mbox{}
       - 64/9\: \*  \z2  \* (x^{-1} + 3 \* x)
            \Big)
\:\: .
\eea
The off-diagonal contributions $K^{\,(2)}_{\,\rm qg}$ and 
$K^{\,(2)}_{\,\rm gq}$ can be written as
\bea
 \label{eq:Kqg2}
 \lefteqn{ K^{\,(2)}_{\rm 2\phi}(x)\,  =  }
%
%
   \nn \\ & & \mbox{\hspn}
        \colour4colour{ \cas\, \* \nf } \, \*  \Big(
          8\, \* \pqg(x)\,  \*  [
          - 11\, \* \z3\, \* \H(1)\,
          - 8\, \* \z2\, \* \Hh(1,0)\,
          + 4\, \* \Hhhh(1,0,0,0)\,
          - 8\, \* \z2\, \* \Hh(1,1)\,
          + 12\, \* \Hhh(1,1,0)\,
          + 4\, \* \Hhhh(1,1,0,0)\,
   \nn \\[-0.5mm] & & \mbox{}
          + 14/3\: \* \Hh(1,2)\,
          + 4\, \* \Hh(1,3)\,
          + 61/3\: \* \Hh(2,1)\,
          + 4\, \* \Hhh(2,1,0)\,
          + 4\, \* \Hh(2,2)\,
          + 8\, \* \Hh(3,1)\,
          ]\,
       + 4\, \* \pqg(-x)\,  \*  [
          - 16\, \* \Hh(-2,2)\,
   \nn \\[0.5mm] & & \mbox{}
          - 6\, \* \z3\, \* \H(-1)\,
          - 16\, \* \Hhh(-1,-2,0)\,
          + 8\, \* \z2\, \* \Hh(-1,-1)\,
          + 32\, \* \Hhhh(-1,-1,-1,0)\,
          + 4\, \* \Hhhh(-1,-1,0,0)\,
          + 12\, \* \z2\, \* \Hh(-1,0)\,
   \nn \\[0.5mm] & & \mbox{}
          - 8\, \* \Hhhh(-1,0,0,0)\,
          - 20\, \* \Hh(-1,3)\,
          ]\,
       + 8/3\: \* \z2\, \* (- 16\, \* x^{-1} + 1\, + 11\, \* x + 18\, \* x^2)\, \* \H(-1)\,
   \nn \\[0.5mm] & & \mbox{}
       + 32\, \* (1\, - 4\, \* x - 6\, \* x^2)\,  \*  \Hh(-1,2)\,
       + 8\, \* (1\, - 2\, \* x)\,  \*  [
          - 7\, \* \z3\, \* \H(1)\,
          - 4\, \* \z2\, \* \Hh(1,0)\,
          + 2\, \* \Hhhh(1,0,0,0)\,
          - 4\, \* \z2\, \* \Hh(1,1)\,
   \nn \\[0.5mm] & & \mbox{}
          + 2\, \* \Hhhh(1,1,0,0)\,
          + 2\, \* \Hh(1,3)\,
          ]\,
       + 8\, \* (1\, + 2\, \* x)\,  \*  [
           4\, \* \Hh(-2,2)\,
          - 7\, \* \z3\, \* \H(-1)\,
          - 4\, \* \Hhh(-1,-2,0)\,
          + 8\, \* \z2\, \* \Hh(-1,-1)\,
   \nn \\[0.5mm] & & \mbox{}
          + 8\, \* \Hhhh(-1,-1,-1,0)\,
          - 6\, \* \Hhhh(-1,-1,0,0)\,
          - 4\, \* \z2\, \* \Hh(-1,0)\,
          + 2\, \* \Hhhh(-1,0,0,0)\,
          + 2\, \* \Hh(-1,3)\,
          ]\,
          - 8/3\: \* (1\, + 345\, \* x
   \nn \\[0.5mm] & & \mbox{}
       - 36\, \* x^2)\, \* \Hhh(0,0,0)\,
       + 64\, \* \z3\, \* (2\, + 9\, \* x)\,  \*  \H(0)\,
       - 8\, \* (3\, + 2\, \* x)\,  \*  \Hh(-3,0)\,
       - 12\, \* (3\, + 14\, \* x - 8\, \* x^2)\,  \*  \H(4)\,
   \nn \\[0.5mm] & & \mbox{}
       + 16\, \* (4\, - 15\, \* x)\,  \*  \Hhhh(0,0,0,0)\,
       - 8\, \* (5\, + 6\, \* x + 16\, \* x^2)\,  \*  \Hhh(-2,0,0)\,
       + 8\, \* \z2\, \* (7\, + 38\, \* x)\,  \*  \H(2)\,
       + 4\, \* \z2\, \* (9
   \nn \\[0.5mm] & & \mbox{}
       +  38\, \* x - 24\, \* x^2)\,  \*  \Hh(0,0)\,
       + 16\, \* (11\, - 30\, \* x + 8\, \* x^2)\,  \*  \Hhh(-2,-1,0)\,
       + 16/3\: \* (- 16\, \* x^{-1} + 13\, - 37\, \* x
   \nn \\[0.5mm] & & \mbox{}
       - 54\, \* x^2)\,  \*  \Hhh(-1,-1,0)\,
       - 8/3\: \* (- 8\, \* x^{-1} + 15\, + 420\, \* x - 436\, \* x^2)\,  \*  \H(1)\,
       + 8\, \* \z2\, \* (15\, - 22\, \* x
   \nn \\[0.5mm] & & \mbox{}
       + 24\, \* x^2)\,  \*  \H(-2)\,
       + 4/3\: \* (32\, \* x^{-1} + 17\, - 691\, \* x + 308\, \* x^2)\,  \*  \H(3)\,
       - 4/3\: \* \z2\, \* (17\, - 441\, \* x + 308\, \* x^2)\,  \*  \H(0)\,
   \nn \\[0.5mm] & & \mbox{}
       + 8/3\: \* (17\,  - 184\, \* x + 116\, \* x^2)\,  \*  \Hh(2,0)\,
       - 8/3\: \* (17\, + 127\, \* x + 150\, \* x^2)\,  \*  \Hhh(-1,0,0)\,
       + 8/3\: \* \z2\, \* (16\, \* x^{-1}
   \nn \\[0.5mm] & & \mbox{}
       + 23\, + 65\, \* x
       - 82\, \* x^2)\,  \*  \H(1)\,
       - 4\, \* (25\, + 90\, \* x + 8\, \* x^2)\,  \*  \Hhh(2,0,0)\,
       + 8/3\: \* (37\, - 222\, \* x + 210\, \* x^2)\,  \*  \Hh(1,1)\,
   \nn \\[0.5mm] & & \mbox{}
       - 8/3\: \* (- 16\, \* x^{-1} + 46\, + 125\, \* x - 10\, \* x^2)\,  \*  \Hh(-2,0)\,
       - 4/3\: \* (32\, \* x^{-1} + 99\, + 159\, \* x - 280\, \* x^2)\,  \*  \Hhh(1,0,0)\,
   \nn \\[0.5mm] & & \mbox{}
       + 8/9\: \* (244\, \* x^{-1} + 99\, - 264\, \* x - 179\, \* x^2)\,  \*  \Hh(-1,0)\,
       + 4/3\: \* \z3\, \* (80\, \* x^{-1} + 133\, + 134\, \* x - 286\, \* x^2)
   \nn \\[0.5mm] & & \mbox{}
       + 2/5\: \* \z2^2\, \* (149 + 174\, \* x + 352\, \* x^2)
       - 2/9\: \* (- 192\, \* x^{-1} + 163\, + 2758\, \* x - 2824\, \* x^2)\,  \*  \Hh(1,0)\,
   \nn \\[0.5mm] & & \mbox{}
       - 4/9\: \* (- 200\, \* x^{-1}  + 375
       + 2514\, \* x - 1852\, \* x^2)\,  \*  \H(2)\,
       + 4/9\: \* \z2\, \* (288\, \* x^{-1} + 375\, + 1986\, \* x
   \nn \\[0.5mm] & & \mbox{}
       - 1852\, \* x^2)
       - 2/9\: \* (372\, \* x^{-1} + 1589
       + 6362\, \* x - 2308\, \* x^2)\,  \*  \H(0)\,
       + 2/9\: \* (56\, \* x^{-1} + 2039\,
   \nn \\[0.5mm] & & \mbox{}
       - 4942\, \* x + 6340\, \* x^2)\,  \*  \Hh(0,0)\,
       + 1/27\: \* (- 19297\, \* x^{-1}
       + 25545\, - 58161\, \* x + 46270\, \* x^2)
   \nn \\[-0.5mm] & & \mbox{}
       - 64\, \*  ( (3\, \* x - x^2)\,  \*  \Hh(3,0)\,
       - x^2\,  \*  \Hhh(-1,-1,2)\, 
       + \Hhh(1,1,0)\,
       + \Hh(1,2)\,
       + 2\, \* \Hh(2,1)\, )\,
                \Big)
   \nn \\[-0.5mm] & & \mbox{\hspn}
       + \colour4colour{ \cfs\, \* \nf } \, \*  \Big(
          16\, \* \pqg(x)\,  \*  [
          - 4\, \* \z2\, \* \Hh(0,0)\,
          - 9\, \* \z3\, \* \H(1)\,
          - 8\, \* \Hhh(1,-2,0)\,
          - 10\, \* \z2\, \* \Hh(1,0)\,
          + 4\, \* \Hhhh(1,0,0,0)\,
          - 4\, \* \z2\, \* \Hh(1,1)\,
   \nn \\[-0.5mm] & & \mbox{}
          - 3\, \* \Hhh(1,1,0)\,
          + 2\, \* \Hhhh(1,1,0,0)\,
          - 3\, \* \Hh(1,2)\,
          + 4\, \* \Hhh(1,2,0)\,
          + 6\, \* \Hh(1,3)\,
          - 4\, \* \z2\, \* \H(2)\,
          + 4\, \* \Hh(3,0)\,
          + 4\, \* \H(4)\,
          ]\,
   \nn \\[0.5mm] & & \mbox{}
       + 16\, \* \pqg(-x)\,  \*  [
          - 8\, \* \z2\, \* \H(-2)\,
          - 8\, \* \Hhh(-2,-1,0)\,
          + 4\, \* \Hhh(-2,0,0)\,
          - 7\, \* \z3\, \* \H(-1)\,
          + 8\, \* \z2\, \* \Hh(-1,-1)\,
   \nn \\[0.5mm] & & \mbox{}
          + 8\, \* \Hhhh(-1,-1,-1,0)\,
          + 2\, \* \Hhhh(-1,-1,0,0)\,
          - 4\, \* \Hhh(-1,-1,2)\,
          - 6\, \* \z2\, \* \Hh(-1,0)\,
          + 2\, \* \Hh(-1,3)\,
          + 4\, \* \Hhhh(0,0,0,0)\,
          ]\,
   \nn \\[0.5mm] & & \mbox{}
       + 16\, \* (1\, - 2\, \* x)\,  \*  [
          - 7\, \* \z3\, \* \H(1)\,
          - 4\, \* \z2\, \* \Hh(1,0)\,
          + 2\, \* \Hhhh(1,0,0,0)\,
          - 4\, \* \z2\, \* \Hh(1,1)\,
          + 2\, \* \Hhhh(1,1,0,0)\,
          + 2\, \* \Hh(1,3)\,
          - 2\, \* \Hh(3,0)\,
   \nn \\[0.5mm] & & \mbox{}
          - 2\, \* \H(4)\,
          ]\,
       + 16\, \* (1\, + 2\, \* x)\,  \*  [
          - 7\, \* \z3\, \* \H(-1)\,
          - 4\, \* \Hhh(-1,-2,0)\,
          + 8\, \* \z2\, \* \Hh(-1,-1)\,
          + 8\, \* \Hhhh(-1,-1,-1,0)\,
   \nn \\[0.5mm] & & \mbox{}
          - 6\, \* \Hhhh(-1,-1,0,0)\,
          - 4\, \* \Hhh(-1,-1,2)\,
          - 4\, \* \z2\, \* \Hh(-1,0)\,
          + 2\, \* \Hhhh(-1,0,0,0)\,
          + 2\, \* \Hh(-1,3)\,
          - 3\, \* \Hhhh(0,0,0,0)\,
          ]\,
   \nn \\[0.5mm] & & \mbox{}
       - 16\, \* \z3\, \* (1\, - 34\, \* x + 4\, \* x^2)\,  \*  \H(0)\,
       + 128\, \* (1\, + x^2)\,  \*  \Hh(-2,2)\,
       + 8\, \* (3\, - 2\, \* x - 12\, \* x^2)\,  \*  \Hh(2,0)\,
       - 32\, \* (3\,
   \nn \\[0.5mm] & & \mbox{}
       - 2\, \* x + 8\, \* x^2)\,  \*  \Hh(-3,0)\,
       - 32/5\: \* \z2\, \* (5\, - 80\, \* x
       - 15\, \* x^2\, - 24\, \* x^3)\,  \*  \H(0)\,
       - 8/5\: \* \z2^2\, \* (5\, + 90\, \* x - 24\, \* x^2)
   \nn \\[0.5mm] & & \mbox{}
       - 16\, \* (7\, + 2\, \* x - 12\, \* x^2)\,  \*  \Hhh(1,0,0)\,
       + 8\, \* (13\, - 42\, \* x
       + 34\, \* x^2)\,  \*  \Hh(1,1)\,
   \nn \\[0.5mm] & & \mbox{}
       + 16/15\: \* \z2\, \* (- x^{-2}\, + 15\, + 70\, \* x - 180\, \* x^2\, + 36\, \* x^3)\,  \*  \H(1)\,
       - 16/3\: \* \z3\, \* (15\, - 175\, \* x - 36\, \* x^2\,
   \nn \\[0.5mm] & & \mbox{}
       - 36\, \* x^3)
       + 32/15\: \* (15\, - 55\, \* x - 45\, \* x^2\, - 36\, \* x^3)\,  \*  \H(3)\,
       + 4\, \* (19\, - 68\, \* x + 52\, \* x^2)\,  \*  \Hh(1,0)\,
   \nn \\[0.5mm] & & \mbox{}
       + 32/15\: \* (- x^{-2}\,
       + 30 - 160\, \* x - 270\, \* x^2\, - 36\, \* x^3)\,  \*  \Hhh(-1,-1,0)\,
       - 32/15\: \* (- x^{-2}\, + 30\, + 50\, \* x 
   \nn \\[0.5mm] & & \mbox{}
       - 45\, \* x^2\, - 36\, \* x^3)\,  \*  \Hhh(-1,0,0)\,
       - 16/5\: \* \z2\, \* (x^{-2}\, + 50\, + 80\, \* x + 90\, \* x^2\, + 36\, \* x^3)\,  \*  \H(-1)\,
   \nn \\[0.5mm] & & \mbox{}
       + 32/15\: \* (x^{-2}\, + 90\, + 40\, \* x + 36\, \* x^3)\,  \*  \Hh(-1,2)\,
       - 32/15\: \* (150\, - 185\, \* x - 270\, \* x^2\, - 36\, \* x^3)\,  \*  \Hh(-2,0)\,
   \nn \\[0.5mm] & & \mbox{}
       - 4/15\: \* (8\, \* x^{-1} + 161\, + 1986\, \* x - 732\, \* x^2)\,  \*  \H(2)\,
       - 1/45\: \* (64\, \* x^{-1} + 217\, + 3528\, \* x 
   \nn \\[0.5mm] & & \mbox{}
       - 13320\, \* x^2)\,  \*  \H(0)\,
       + 2/45\: \* (- 48\, \* x^{-1} + 249\, + 2494\, \* x - 1728\, \* x^2
   \nn \\[0.5mm] & & \mbox{}
       - 2880\, \* x^3)\,  \*  \Hh(0,0)\,
       + 4/45\: \* \z2\, \* (48\, \* x^{-1} + 483\, + 1426\, \* x - 2196\, \* x^2\, + 1440\, \* x^3)
   \nn \\[0.5mm] & & \mbox{}
       - 2/15\: \* (- 16\, \* x^{-1} + 1849 + 756\, \* x - 2604\, \* x^2)\,  \*  \H(1)\,
       - 1/45\: \* (- 64\, \* x^{-1} + 2572\, + 13953\, \* x
   \nn \\[0.5mm] & & \mbox{}
       - 16596\, \* x^2)
       - 8/45\: \* ( - 20\, \* x^{-2}
       - 12\, \* x^{-1}
       + 3651+ 2266\, \* x - 1602\, \* x^2\, - 720\, \* x^3)\, \*  \Hh(-1,0)\,
   \nn \\[0.5mm] & & \mbox{}
       - 64\, \* (4\, \* x - x^2)\,  \*  \Hhh(2,0,0)\,
       - 16/15\: \* (85\, \* x 
       - 90\, \* x^2\, + 72\, \* x^3)\,  \*  \Hhh(0,0,0)\,
       - 192\, \* x^2\,  \*  \Hhh(-2,0,0)\,
   \nn \\[-0.5mm] & & \mbox{}
       - 32\, \* (2\, \* \z2\, \* \H(-2)\,
       - \z2\, \* \Hh(0,0)\,
       - 2\, \* \Hhh(1,1,0)\,
       - 2\, \* \Hh(1,2)\, )
       + 128\, \* x  \*  [
          - 2\, \* \Hhh(-2,-1,0)\,
          + \z2\, \* \H(2)\,
          ]\,
               \Big)
   \nn \\[-0.5mm] & & \mbox{\hspn}
       +  \colour4colour{ \ca\, \* \cf\, \* \nf } \, \*  \Big(
        8\, \* \pqg(x)\,  \*  [
           23\, \* \z3\, \* \H(1)\,
          + 12\, \* \Hhh(1,-2,0)\,
          + 22\, \* \z2\, \* \Hh(1,0)\,
          - 4\, \* \Hhhh(1,0,0,0)\,
          + 16\, \* \z2\, \* \Hh(1,1)\,
   \nn \\[-0.5mm] & & \mbox{}
          + 7/3\: \* \Hhh(1,1,0)\,
          + 29/3\: \* \Hh(1,2)\,
          - 8\, \* \Hh(1,3)\,
          - 12\, \* \Hh(2,1)\,
          + 4\, \* \Hhh(2,1,0)\,
          + 4\, \* \Hh(2,2)\,
          ]\,
   \nn \\[0.5mm] & & \mbox{}
       + 16\, \* \pqg(-x)\,  \*  [
           13\, \* \z2\, \* \H(-2)\,
          + 8\, \* \Hhh(-1,-2,0)\,
          + 9\, \* \Hhhh(-1,-1,0,0)\,
          + 10\, \* \Hhh(-1,-1,2)\,
          + 12\, \* \z2\, \* \Hh(-1,0)\,
   \nn \\[0.5mm] & & \mbox{}
          - 6\, \* \Hhhh(-1,0,0,0)\,
          - 2\, \* \Hhh(-1,2,0)\,
          - 5\, \* \Hh(-1,3)\,
          - 4\, \* \z3\, \* \H(0)\,
          ]\,
       + 24\, \* (1\, - 2\, \* x)\,  \*  [
           7\, \* \z3\, \* \H(1)\,
          + 4\, \* \z2\, \* \Hh(1,0)\,
   \nn \\[0.5mm] & & \mbox{}
          - 2\, \* \Hhhh(1,0,0,0)\,
          + 4\, \* \z2\, \* \Hh(1,1)\,
          - 2\, \* \Hhhh(1,1,0,0)\,
          - 2\, \* \Hh(1,3)\,
          ]\,
       + 16\, \* (1\, + 2\, \* x)\,  \*  [
           3\, \* \z2\, \* \H(-2)\,
          + 6\, \* \Hhh(-1,-2,0)\,
   \nn \\[0.5mm] & & \mbox{}
          + 9\, \* \Hhhh(-1,-1,0,0)\,
          + 6\, \* \Hhh(-1,-1,2)\,
          + 6\, \* \z2\, \* \Hh(-1,0)\,
          - 3\, \* \Hhhh(-1,0,0,0)\,
          - 3\, \* \Hh(-1,3)\,
          - 2\, \* \Hh(3,0)\,
          ]\,
   \nn \\[0.5mm] & & \mbox{}
       - 32\, \* (1\, + 3\, \* x)\,  \*  \Hhhh(0,0,0,0)\,
       + 64\, \* (1\, + 14\, \* x + 5\, \* x^2)\,  \*  \Hhh(-2,-1,0)\,
       + 8\, \* \z2\, \* (3\, - 2\, \* x - 4\, \* x^2)\,  \*  \Hh(0,0)\,
   \nn \\[0.5mm] & & \mbox{}
       - 8\, \* (3\, + 10\, \* x - 4\, \* x^2)\,  \*  \H(4)\,
       - 96\, \* \z2\, \* (5\, + 10\, \* x + 6\, \* x^2)\,  \*  \Hh(-1,-1)\,
       + 4/15\: \* (5\, + 1660\, \* x + 840\, \* x^2\,
   \nn \\[0.5mm] & & \mbox{}
       + 288\, \* x^3)\,  \*  \H(3)\,
       - 4/15\: \* \z2\, \* (5\, + 2960\, \* x + 840\, \* x^2\, + 576\, \* x^3)\,  \*  \H(0)\,
       - 32\, \* (7\, + 2\, \* x + 8\, \* x^2)\,  \*  \Hh(-2,2)\,
   \nn \\[0.5mm] & & \mbox{}
       - 64\, \* (7\, + 14\, \* x + 8\, \* x^2)\,  \*  \Hhhh(-1,-1,-1,0)\,
       - 16\, \* (7\, + 26\, \* x + 10\, \* x^2)\,  \*  \Hhh(-2,0,0)\,
       + 16\, \* (13\, - 6\, \* x
   \nn \\[0.5mm] & & \mbox{}
       + 8\, \* x^2)\,  \*  \Hh(-3,0)\,
       + 8\, \* (17\, + 38\, \* x + 16\, \* x^2)\,  \*  \Hhh(2,0,0)\,
       + 16\, \* \z3\, \* (25\, + 50\, \* x + 29\, \* x^2)\,  \*  \H(-1)\,
   \nn \\[0.5mm] & & \mbox{}
       - 4/3\: \* (- 16\, \* x^{-1} + 33\, - 42\, \* x
       - 60\, \* x^2)\,  \*  \Hh(2,0)\,
       - 8/15\: \* \z2\, \* (- 2\, \* x^{-2} + 60\, \* x^{-1} + 35\, + 370\, \* x
   \nn \\[0.5mm] & & \mbox{}
       - 465\, \* x^2\, + 72\, \* x^3)\,  \*  \H(1)\,
       + 4/5\: \* \z2^2\, \* (59\,
       + 314\, \* x - 120\, \* x^2)
       + 16/15\: \* (2\, \* x^{-2} + 60\, \* x^{-1} + 110\,
   \nn \\[0.5mm] & & \mbox{}
       + 660\, \* x + 755\, \* x^2\, + 72\, \* x^3)\,  \*  \Hhh(-1,-1,0)\,
       - 4/3\: \* (131\, - 654\, \* x + 582\, \* x^2)\,  \*  \Hh(1,1)\,
       - 4/3\: \* \z3\, \* (175\,
   \nn \\[0.5mm] & & \mbox{}
       + 934\, \* x + 114\, \* x^2\, + 144\, \* x^3)
       + 4/15\: \* (195\, + 550\, \* x
       + 1800\, \* x^2\, + 288\, \* x^3)\,  \*  \Hhh(0,0,0)\,
   \nn \\[0.5mm] & & \mbox{}
       - 16/15\: \* (2\, \* x^{-2} + 40\, \* x^{-1} + 205\, + 355\, \* x + 320\, \* x^2\, + 72\, \* x^3)\,  \*  \Hhh(-1,0,0)\,
   \nn \\[0.5mm] & & \mbox{}
       + 4/3\: \* (16\, \* x^{-1} + 235\, - 116\, \* x - 72\, \* x^2)\,  \*  \Hhh(1,0,0)\,
       + 4/15\: \* (128\, \* x^{-1} + 261\, + 3881\, \* x
   \nn \\[0.5mm] & & \mbox{}
       - 2142\, \* x^2)\,  \*  \H(2)\,
       + 16/15\: \* (265\, - 325\, \* x - 755\, \* x^2 - 72\, \* x^3)\,  \*  \Hh(-2,0)\,
       + 2/15\: \* (784\, \* x^{-1} + 399\,
   \nn \\[0.5mm] & & \mbox{}
       + 7156\, \* x - 8064\, \* x^2)\,  \*  \H(1)\,
       - 16/15\: \* (2\, \* x^{-2} + 20\, \* x^{-1} + 420 + 350\, \* x + 65\, \* x^2\, + 72\, \* x^3)\,  \*  \Hh(-1,2)\,
   \nn \\[0.5mm] & & \mbox{}
       - 1/45\: \* (- 848\, \* x^{-1} + 695\, - 25786\, \* x
       + 30216\, \* x^2)\,  \*  \H(0)\,
       - 4/45\: \* \z2\, \* (408\, \* x^{-1}+ 783\,
   \nn \\[0.5mm] & & \mbox{}
       + 2531\, \* x - 6426\, \* x^2\, + 2016\, \* x^3)
       - 2/9\: \* (- 256\, \* x^{-1} + 831\, - 3264\, \* x + 2596\, \* x^2)\,  \*  \Hh(1,0)\,
   \nn \\[0.5mm] & & \mbox{}
       - 4/45\: \* (- 24\, \* x^{-1} + 927\, + 1897\, \* x + 4726\, \* x^2\,
       - 2016\, \* x^3)\,  \*  \Hh(0,0)\,
       + 8/15\: \* \z2\, \* (6\, \* x^{-2} + 100\, \* x^{-1}
   \nn \\[0.5mm] & & \mbox{}
       + 950\, + 1360\, \* x + 885\, \* x^2\, + 216\, \* x^3)\, \* \H(-1)\,
       - 1/45\: \* (- 4732\, \* x^{-1} + 3042\, - 41977\, \* x
   \nn \\[0.5mm] & & \mbox{}
       + 46292\, \* x^2)
       + 8/45\: \* (- 28\, \* x^{-2} - 12\, \* x^{-1} + 4281
       + 4556\, \* x - 522\, \* x^2\, - 1008\, \* x^3)\, \* \Hh(-1,0)\,
   \nn \\[-0.5mm] & & \mbox{}
       - 96\, \* \z2\, \* (4\, \* x - x^2)\,  \*  \H(2)\,
       - 800\, \* \z3\, \* x  \*  \H(0)\,
       + 128\, \* \Hh(2,1)\,
                \Big)
   \nn \\[-0.5mm] & & \mbox{\hspn}
       +  \colour4colour{ \ca\, \* \nfs } \, \*       \Big(
        8/3\: \* \pqg(x)\,  \*  [
          - 3\, \* \z2\, \* \H(1)\,
          + 6\, \* \Hh(1,1)\,
          + 4\, \* \Hh(1,2)\,
          + 2\, \* \Hh(2,1)\,
          + 2\, \* \H(3)\,
          ]\,
       + 8/3\: \* \pqg(-x)\,  \*  [
           \z2\, \* \H(-1)\,
   \nn \\[-0.5mm] & & \mbox{}
          + 2\, \* \Hhh(-1,-1,0)\,
          + 3\, \* \Hhh(-1,0,0)\,
          ]\,
       + 8/3\: \* (1\, + 2\, \* x)\,  \*  [
          - 2\, \* \Hh(-2,0)\,
          + \H(-1)\, \* \z2
          + 2\, \* \Hhh(-1,-1,0)\,
          - \Hhh(-1,0,0)\,
          ]\,
   \nn \\[0.5mm] & & \mbox{}
       + 8/3\: \* \z3\, \* (1\, - 46\, \* x + 10\, \* x^2)
       + 8/3\: \* \z2\, \* (1\, - 2\, \* x)\,  \*  \H(1)\,
       - 16/3\: \* (1\, + 4\, \* x)\,  \*  \Hh(2,0)\,
   \nn \\[0.5mm] & & \mbox{}
       - 16/3\: \* \z2\, \* (1\,   + 2\, \* x^2)\,  \*  \H(0)\,
       + 8\, \* (2\, - 3\, \* x + x^2)\,  \*  \H(1)\,
       + 8/9\: \* (- 4\, \* x^{-1} + 5\, + 38\, \* x + 39\, \* x^2)\,  \*  \Hh(0,0)\,
   \nn \\[0.5mm] & & \mbox{}
       + 16/9\: \* (- 2\, \* x^{-1} + 6\, - 15\, \* x
       + 11\, \* x^2)\,  \*  \H(2)\,
       - 16/9\: \* \z2\, \* (2\, \* x^{-1} + 6\, + 3\, \* x + 11\, \* x^2)
   \nn \\[0.5mm] & & \mbox{}
       + 8/9\: \* (- 4\, \* x^{-1} + 8\, - 46\, \* x + 53\, \* x^2)\,  \*  \Hh(1,0)\,
       - 16/9\: \* (4\, \* x^{-1} + 15\, + 18\, \* x + 4\, \* x^2)\,  \*  \Hh(-1,0)\,
   \nn \\[0.5mm] & & \mbox{}
       - 4/27\: \* (114\, \* x^{-1} + 69\, + 90\, \* x - 278\, \* x^2)\,  \*  \H(0)\,
       + 2/81\: \* (- 1328\, \* x^{-1} + 3105\, - 3114\, \* x
   \nn \\[-0.5mm] & & \mbox{}
       + 3524\, \* x^2)
       + 16/3\: \* (x^2\,  \*  \Hhh(1,0,0)\,
       - \Hhh(0,0,0)\,
       - 2\, \* \Hh(1,1)\, )\,
                  \Big)
   \nn \\[-0.5mm] & & \mbox{\hspn}
       + \colour4colour{ \cf\, \* \nfs } \, \*  \Big(
        8/3\: \* \pqg(x)\,  \*  [
          - 2\, \* \Hhh(1,0,0)\,
          - 3\, \* \Hh(1,1)\,
          + 2\, \* \Hhh(1,1,0)\,
          - 2\, \* \Hh(1,2)\,
          - 3\, \* \H(2)\,
          + 3\, \* \Hh(2,0)\,
          - 2\, \* \H(3)\,
          ]\,
   \nn \\[-0.5mm] & & \mbox{}
       + 32/3\: \* \pqg(-x)\,  \*  [
          - \z2\, \* \H(-1)\,
          - 2\, \* \Hhh(-1,-1,0)\,
          + \Hhh(-1,0,0)\,
          ]\,
       + (1\, - 2\, \* x)\,  \*  [
          - 32\, \* \Hhhh(0,0,0,0)\,
   \nn \\[0.5mm] & & \mbox{}
          - 16/3\: \* \H(1)\, \* \z2
          ]\,
       + 16/3\: \* \z2\, \* (1\, - 14\, \* x + 2\, \* x^2)\,  \*  \H(0)\,
       - 4/9\: \* (8\, \* x^{-1} + 3\, + 48\, \* x - 44\, \* x^2)\,  \*  \Hh(1,0)\,
   \nn \\[0.5mm] & & \mbox{}
       - 16\, \* (3\, - 11\, \* x
       - 2\, \* x^2)\,  \*  \Hhh(0,0,0)\,
       + 32/3\: \* (5\, - 6\, \* x + 4\, \* x^2)\,  \*  \Hh(-2,0)\,
       + 16/3\: \* \z3\, \* (7\, - 6\, \* x)
   \nn \\[0.5mm] & & \mbox{}
       - 8/45\: \* \z2\, \* (40\, \* x^{-1} + 15\, + 10\, \* x 
       - 90\, \* x^2\, - 72\, \* x^3)
       + 16/45\: \* (x^{-2} - 20\, \* x^{-1} + 225\, + 40\, \* x
   \nn \\[0.5mm] & & \mbox{}
       - 155\, \* x^2\, + 36\, \* x^3)\,  \*  \Hh(-1,0)\,
       - 8/45\: \* (40\, \* x^{-1}
       + 885\, + 470\, \* x
       + 460\, \* x^2\, + 72\, \* x^3)\,  \*  \Hh(0,0)\,
   \nn \\[0.5mm] & & \mbox{}
       - 4/135\: \* (712\, \* x^{-1} + 5304\, - 6771\, \* x - 9078\, \* x^2)\,  \*  \H(0)\,
       - 1/405\: \* (9796\, \* x^{-1} + 7731
   \nn \\[0.5mm] & & \mbox{}
       - 151506\, \* x + 105494\, \* x^2)
       - 8/3\: \* (2\, \* x^{-1} + 6\, \* x - 7\, \* x^2)\,  \*  \H(1)\,
       + 32/3\: \* ( \Hh(1,1)\,
       + \H(2)\, )
   \nn \\[-0.5mm] & & \mbox{}
       + 16/3\: \* x^2\,  \*  [
           2\, \* \z2\, \* \H(-1)\,
          + 4\, \* \Hhh(-1,-1,0)\,
          - 2\, \* \Hhh(-1,0,0)\,
          - 2\, \* \Hhh(1,0,0)\,
          - \Hh(2,0)\,
          ]\,
               \Big)
   \nn \\[-0.5mm] & & \mbox{\hspn}
       +  \colour4colour{  \nft } \, \*  \Big(
        8/9\: \* \pqg(x)\,  \*  [
          - \Hh(0,0)\,
          - \Hh(1,0)\,
          ]\,
       + 8/27\: \* ( 2\, \* x^{-1} - 3\, + 6\, \* x)\,  \*  \H(0)\,
       + 4/81\: \* ( 13\, \* x^{-1} - 45\,
   \nn \\[-0.5mm] & & \mbox{}
       + 81\, \* x
       - 49\, \* x^2)
               \Big)
\eea
and
\bea
 \label{eq:Kgq2}
 \lefteqn{ K^{\,(2)}_{\,\phi 2}(x)\,  =  }
%
%
   \nn \\ & & \mbox{\hspn}
      \colour4colour{ \cas\, \*  \cf } \, \*  \Big(
          8\, \*  \pgq(x)\, \*   [
          - 8\, \*  \Hhh(1,-2,0)\,
          + 2\, \*  \Hhhh(1,0,0,0)\,
          + 8\, \*  \z2\, \*  \Hh(1,1)\,
          + 13/3\: \*  \Hhh(1,1,0)\,
          - 4\, \*  \Hhhh(1,1,0,0)\,
          + 13/3\: \*  \Hh(1,2)\,
   \nn \\[-0.5mm] & & \mbox{}
          + 8\, \*  \Hhh(1,2,0)\,
          + 4\, \*  \Hh(1,3)\,
          ]\,\,
       + 16\, \*  \pgq(-x)\, \*   [
          - 4\, \*  \Hh(-2,2)\,
          + 8\, \*  \Hhh(-1,-2,0)\,
          - 14\, \*  \z2\, \*  \Hh(-1,-1)\,
          - 8\, \*  \Hhhh(-1,-1,-1,0)\,
   \nn \\[0.5mm] & & \mbox{}
          + 13\, \*  \Hhhh(-1,-1,0,0)\,
          + 10\, \*  \Hhh(-1,-1,2)\,
          + 7\, \*  \z2\, \*  \Hh(-1,0)\,
          - 5\, \*  \Hhhh(-1,0,0,0)\,
          - 2\, \*  \Hhh(-1,2,0)\,
          - 5\, \*  \Hh(-1,3)\,
          ]\,\,
   \nn \\[0.5mm] & & \mbox{}
       + 24\, \*  (2\, - x)\,\*   (
          - \Hhh(1,1,0)\,
          - \Hh(1,2)\,
          )
       - 104\, \*  \z3\, \*  (- 2\, \*  x^{-1} + 2\, - x)\,\*   \H(1)\,
       - 168\, \*  \z3\, \*  (2\, \*  x^{-1} + 2\, + x)\,\*   \H(-1)\,
   \nn \\[0.5mm] & & \mbox{}
       + 8\, \*  (32\, \*  x^{-1} + 2\, + 25\, \*  x)\,\*   \Hhh(-2,0,0)\,
       + - 4\, \*  \z2\, \*  (2\, + 11\, \*  x)\,\*   \Hh(0,0)\,
       + 4\, \*  (2\, + 13\, \*  x)\,\*   \H(4)\,
       - 16\, \*  \z3\, \*  (- 8\, \*  x^{-1}
   \nn \\[0.5mm] & & \mbox{}
       + 4\, - 9\, \*  x)\,\*   \H(0)\,
       + 32\, \*  (2\, \*  x^{-1} + 4\, + 3\, \*  x)\,\*   \Hh(3,0)\,
       - 8\, \*  \z2\, \*  (- 8\, \*  x^{-1} + 6\, - 5\, \*  x)\,\*   \H(2)\,
       - 16\, \*  (8\, \*  x^{-1} + 6\,
   \nn \\[0.5mm] & & \mbox{}
       + 5\, \*  x)\,\*   \Hhh(-2,-1,0)\,
       - 8\, \*  (16\, - 15\, \*  x)\,\*   \Hhhh(0,0,0,0)\,
       + 4\, \*  (18\, + x)\,\*   \Hhh(2,0,0)\,
       - 8\, \*  \z2\, \*  (24\, \*  x^{-1} + 30\, + 13\, \*  x)\,\*   \H(-2)\,
   \nn \\[0.5mm] & & \mbox{}
       - 8\, \*  (34\, - x)\,\*   \Hh(-3,0)\,
       - 4/3\: \*  \z2\, \*  (- 6\, \*  x^{-1} + 56\, - 7\, \*  x + 16\, \*  x^2)\,\*   \H(1)\,
       - 16/3\: \*  (- 20\, \*  x^{-1} + 65\, - 7\, \*  x
   \nn \\[0.5mm] & & \mbox{}
       + 12\, \*  x^2)\,\*   \Hh(2,0)\,
       - 4\, \*  (- 92\, \*  x^{-1} + 75\, - 32\, \*  x)\,\*   \Hhh(1,0,0)\,
       + 16/3\: \*  (75\, \*  x^{-1} + 106\, + 50\, \*  x + 8\, \*  x^2)\,\*   \Hh(-1,2)\,
   \nn \\[0.5mm] & & \mbox{}
       + 8/3\: \*  (22\, \*  x^{-1} + 110\, + 19\, \*  x - 24\, \*  x^2)\,\*   \Hhh(0,0,0)\,
       - 8/3\: \*  (58\, \*  x^{-1} + 144\, + 51\, \*  x + 16\, \*  x^2)\,\*   \Hhh(-1,-1,0)\,
   \nn \\[0.5mm] & & \mbox{}
       + 4/3\: \*  \z2\, \*  (- 44\, \*  x^{-1} + 187\, + 145\, \*  x + 32\, \*  x^2)\,\*   \H(0)\,
       - 4/3\: \*  (44\, \*  x^{-1} + 187\, + 63\, \*  x + 32\, \*  x^2)\,\*   \H(3)\,
   \nn \\[0.5mm] & & \mbox{}
       - 4/9\: \*  (- 17\, \*  x^{-1} + 242\, - 121\, \*  x)\,\*   \Hh(1,1)\,
       - 2/5\: \*  \z2^2\, \*  (384\, \*  x^{-1} + 242\, + 247\, \*  x)
       + 8/3\: \*  (- 44\, \*  x^{-1}
   \nn \\[0.5mm] & & \mbox{}
       + 353\, + 41\, \*  x + 48\, \*  x^2)\,\*   \Hh(-2,0)\,
       + 8/3\: \*  \z3\, \*  (- 347\, \*  x^{-1} + 444\, - 24\, \*  x + 16\, \*  x^2)
       - 4/3\: \*  \z2\, \*  (358\, \*  x^{-1}
   \nn \\[0.5mm] & & \mbox{}
       + 568\, + 251\, \*  x + 48\, \*  x^2)\,\*   \H(-1)\,
       + 8/9\: \*  \z2\, \*  (- 269\, \*  x^{-1} + 704\, + 24\, \*  x + 96\, \*  x^2)
       + 4/3\: \*  (632\, \*  x^{-1}
   \nn \\[0.5mm] & & \mbox{}
       + 724\, + 215\, \*  x + 64\, \*  x^2)\,\*   \Hhh(-1,0,0)\,
       - 4/9\: \*  (671\, \*  x^{-1} + 1408\, + 205\, \*  x + 192\, \*  x^2)\,\*   \H(2)\,
   \nn \\[0.5mm] & & \mbox{}
       - 4/9\: \*  (1209\, \*  x^{-1}
       + 1484\, + 157\, \*  x + 472\, \*  x^2)\,\*   \Hh(-1,0)\,
       + 4/27\: \*  (- 5419\, \*  x^{-1} + 1687\, + 709\, \*  x 
   \nn \\[0.5mm] & & \mbox{}
       + 1446\, \*  x^2)\,\*   \H(1)\,
       - 2/9\: \*  (- 1946\, \*  x^{-1} + 2190\, - 1269\, \*  x - 400\, \*  x^2)\,\*   \Hh(1,0)\,
       - 2/9\: \*  (- 1892\, \*  x^{-1}
   \nn \\[0.5mm] & & \mbox{}
       + 10782\, - 1667\, \*  x
       + 1632\, \*  x^2)\,\*   \Hh(0,0)\,
       + 1/27\: \*  (28529\, \*  x^{-1} + 27762\, + 10188\, \*  x + 15400\, \*  x^2)\,\*   \H(0)\,
   \nn \\[-0.5mm] & & \mbox{}
       - 1/54\: \*  (- 223601\, \*  x^{-1}
       + 172270\, - 72332\, \*  x - 7424\, \*  x^2)
       + 64\, \*  \Hh(-2,2)\,
                \Big)
   \nn \\[-0.5mm] & & \mbox{\hspn}
   +  \colour4colour{ \ca\, \*  \cfs } \, \*  \Big(
        8\, \*  \pgq(x)\, \*   [
           4\, \*  \Hhh(1,-2,0)\,
          - 14\, \*  \z2\, \*  \Hh(1,0)\,
          + 10\, \*  \Hhhh(1,0,0,0)\,
          - 16\, \*  \z2\, \*  \Hh(1,1)\,
          + 11/3\: \*  \Hhh(1,1,0)\,
   \nn \\[-0.5mm] & & \mbox{}
          + 16\, \*  \Hhhh(1,1,0,0)\,
          + 11/3\: \*  \Hh(1,2)\,
          + 8\, \*  \Hh(1,3)\,
          + 10\, \*  \Hhh(2,0,0)\,
          + 5\, \*  \Hh(2,1)\,
          + 4\, \*  \Hhh(2,1,0)\,
          + 4\, \*  \Hh(2,2)\,
          ]\,
   \nn \\[0.5mm] & & \mbox{}
       + 16\, \*  \pgq(-x)\, \*   [
          - 8\, \*  \Hhh(-1,-2,0)\,
          + 14\, \*  \z2\, \*  \Hh(-1,-1)\,
          + 16\, \*  \Hhhh(-1,-1,-1,0)\,
          - 7\, \*  \Hhhh(-1,-1,0,0)\,
          - 6\, \*  \Hhh(-1,-1,2)\,
   \nn \\[0.5mm] & & \mbox{}
          - 2\, \*  \z2\, \*  \Hh(-1,0)\,
          - \Hhhh(-1,0,0,0)\,
          - \Hh(-1,3)\,
          ]\,
       + 16\, \*  (2\, - x)\,\*   (
          - 3\, \*  \Hh(2,1)\,
          - \Hh(3,0)\,
          )
       - 72\, \*  \z2\, \*  (2\, + x)\,\*   \Hh(0,0)\,
   \nn \\[0.5mm] & & \mbox{}
       + 264\, \*  \z3\, \*  (- 2\, \*  x^{-1} + 2\,  - x)\,\*   \H(1)\,
       + 184\, \*  \z3\, \*  (2\, \*  x^{-1} + 2\, + x)\,\*   \H(-1)\,
       + 16\, \*  (4\, + 3\, \*  x)\,\*   \Hhhh(0,0,0,0)\,
   \nn \\[0.5mm] & & \mbox{}
       - 32\, \*  \z3\, \*  (3\, \*  x^{-1} + 5\, + 9\, \*  x)\,\*   \H(0)\,
       + 24\, \*  \z2\, \*  (4\, \*  x^{-1} + 6\, + 5\, \*  x)\,\*   \H(-2)\,
       - 8\, \*  (12\, \*  x^{-1} + 6\, + 11\, \*  x)\,\*   \Hhh(-2,0,0)\,
   \nn \\[0.5mm] & & \mbox{}
       + 16\, \*  (12\, \*  x^{-1} + 10\, + 15\, \*  x)\,\*   \Hhh(-2,-1,0)\,
       + 8\, \*  \z2\, \*  (- 20\, \*  x^{-1} + 18\, - 19\, \*  x)\,\*   \H(2)\,
       - 32/3\: \*  (11\, \*  x^{-1} + 18\,
   \nn \\[0.5mm] & & \mbox{}
       + 9\, \*  x + 2\, \*  x^2)\,\*   \Hh(-1,2)\,
       + 8\, \*  (18\, + 7\, \*  x)\,\*   \H(4)\,
       - 8\, \*  (6\, \*  x^{-1} + 44\, + 17\, \*  x + 8\, \*  x^2)\,\*   \Hh(-2,0)\,
   \nn \\[0.5mm] & & \mbox{}
       + 16/3\: \*  \z3\, \*  (- 18\, \*  x^{-1} + 56\, - 49\, \*  x + 6\, \*  x^2)
       + 8\, \*  (30\, \*  x^{-1} + 60\,
       + 13\, \*  x + 8\, \*  x^2)\,\*   \Hhh(-1,-1,0)\,
   \nn \\[0.5mm] & & \mbox{}
       - 4/5\: \*  \z2^2\, \*  (- 44\, \*  x^{-1} + 72\, + 27\, \*  x)
       - 8/3\: \*  (- 45\, \*  x^{-1} + 94\, - 17\, \*  x + 8\, \*  x^2)\,\*   \Hhh(1,0,0)\,
   \nn \\[0.5mm] & & \mbox{}
       - 2/3\: \*  (- 210\, \*  x^{-1} + 116\, - 23\, \*  x)\,\*   \Hh(1,0)\,
       - 4\, \*  (134\, - 67\, \*  x + 16\, \*  x^2)\,\*   \Hhh(0,0,0)\,
       - 4/3\: \*  (- 96\, \*  x^{-1} + 142\,
   \nn \\[0.5mm] & & \mbox{}
       - 95\, \*  x)\,\*   \Hh(2,0)\,
       + 4/15\: \*  (16\, \*  x^{-2} - 135\, \*  x^{-1} + 180\, + 205\, \*  x + 36\, \*  x^3)\,\*   \Hh(-1,0)\,
       - 4/3\: \*  (134\, \*  x^{-1} + 204\,
   \nn \\[0.5mm] & & \mbox{}
       + 63\, \*  x
       + 32\, \*  x^2)\,\*   \Hhh(-1,0,0)\,
       + 4/3\: \*  \z2\, \*  (- 134\, \*  x^{-1} + 224\, - 61\, \*  x + 24\, \*  x^2)\,\*   \H(1)\,
       + 4/3\: \*  \z2\, \*  (178\, \*  x^{-1}
   \nn \\[0.5mm] & & \mbox{}
       + 324\, + 111\, \*  x + 40\, \*  x^2)\,\*   \H(-1)\,
       - 8/9\: \*  (- 1043\, \*  x^{-1} + 420\, - 201\, \*  x + 65\, \*  x^2)\,\*   \H(1)\,
   \nn \\[0.5mm] & & \mbox{}
       + 4/3\: \*  \z2\, \*  (- 256\, \*  x^{-1} + 556\, - 299\, \*  x
       + 48\, \*  x^2)\,\*   \H(0)\,
       - 4/3\: \*  (- 220\, \*  x^{-1} + 556\, - 197\, \*  x + 48\, \*  x^2)\,\*   \H(3)\,
   \nn \\[0.5mm] & & \mbox{}
       - 4/9\: \*  (- 1497\, \*  x^{-1} + 607\, - 683\, \*  x
       - 208\, \*  x^2)\,\*   \H(2)\,
       + 4/45\: \*  \z2\, \*  (- 7890\, \*  x^{-1} + 3035\, - 2800\, \*  x
   \nn \\[0.5mm] & & \mbox{}
       - 1040\, \*  x^2\, + 108\, \*  x^3)
       + 2/45\: \*  (4450\,
       + 2785\, \*  x + 2080\, \*  x^2\,
       - 216\, \*  x^3)\,\*   \Hh(0,0)\,
       - 2/45\: \*  (- 7269\, \*  x^{-1} 
   \nn \\[0.5mm] & & \mbox{}
       + 10022\, - 6078\, \*  x + 1756\, \*  x^2)\,\*   \H(0)\,
       - 1/90\: \*  (- 52479\, \*  x^{-1} + 133506\, - 13606\, \*  x - 3656\, \*  x^2)
   \nn \\[-0.5mm] & & \mbox{}
       + 68\, \*  x^{-1} \*   \Hh(1,1)\,
       + 16\, \*  x \*   (- \Hh(-3,0)\, + 6\, \*  \Hhh(2,0,0)\,)
       - 64\, \*  \Hh(-2,2)\,
               \Big)
   \nn \\[-0.5mm] & & \mbox{\hspn}
    + \colour4colour{ \cft } \, \* \Big(
        8\, \* \pgq(x)\, \*  [
           14\, \* \z3\, \* \H(1)\,
          + 8\, \* \z2\, \* \Hh(1,0)\,
          - 4\, \* \Hhhh(1,0,0,0)\,
          + 8\, \* \z2\, \* \Hh(1,1)\,
          - 4\, \* \Hhhh(1,1,0,0)\,
          - 4\, \* \Hh(1,3)\,
          + 3\, \* \Hh(2,0)\,
   \nn \\[-0.5mm] & & \mbox{}
          + 4\, \* \Hhh(2,1,0)\,
          + 4\, \* \Hh(2,2)\,
          + 4\, \* \Hh(3,0)\,
          + 8\, \* \Hh(3,1)\,
          ]\,\,
       + 16\, \* \pgq(-x)\, \*  [
           7\, \* \z3\, \* \H(-1)\,
          + 4\, \* \Hhh(-1,-2,0)\,
          - 8\, \* \z2\, \* \Hh(-1,-1)\,
   \nn \\[0.5mm] & & \mbox{}
          - 8\, \* \Hhhh(-1,-1,-1,0)\,
          + 6\, \* \Hhhh(-1,-1,0,0)\,
          + 4\, \* \Hhh(-1,-1,2)\,
          + 4\, \* \z2\, \* \Hh(-1,0)\,
          - 2\, \* \Hhhh(-1,0,0,0)\,
          - 2\, \* \Hh(-1,3)\,
          ]\,\,
   \nn \\[0.5mm] & & \mbox{}
       + 64\, \* (1\, + x\,)\, \*  \Hh(-1,2)\,
       + 16\, \* (2\, - x)\,\*  [
       + 2\, \* \Hh(2,0)\,
       + \H(3)\,
       - \Hh(3,0)\,
       - 3\, \* \H(4)\,
          ]\,\,
       - 16\, \* (2\, + x)\,\*  \Hhhh(0,0,0,0)\,
   \nn \\[0.5mm] & & \mbox{}
       - 16\, \* \z2\, \* (- 6\, \* x^{-1} + 2\, - 11\, \* x)\,\*  \H(0)\,
       + 16\, \* (2\, - 9\, \* x)\,\*  \Hhh(0,0,0)\,
       + 16\, \* (2\, - 5\, \* x)\,\*  \Hhh(2,0,0)\,
       + 32\, \* \z3\, \* ( 3\, \* x^{-1}
   \nn \\[0.5mm] & & \mbox{}
       + 3\, + 8\, \* x)
       + 32\, \* \z3\, \* (- 2\, \* x^{-1} + 4\, + 5\, \* x)\,\*  \H(0)\,
       + 24\, \* (2\, \* x^{-1} + 4\, + 3\, \* x)\,\*  \Hhh(-1,0,0)\,
       + 32\, \* (3\, \* x^{-1} + 4\,
   \nn \\[0.5mm] & & \mbox{}
       + 5\, \* x)\,\*  \Hh(-2,0)\,
       + 8\, \* (4\, - x)\,\*  \Hhh(1,0,0)\,
       + 32\, \* (4\, + x)\,\*  \Hh(-3,0)\,
       - 16\, \* \z2\, \* (6\, - 5\, \* x)\,\*  \H(2)\,
       + 16\, \* \z2\, \* (6\, - x)\,\*  \Hh(0,0)\,
   \nn \\[0.5mm] & & \mbox{}
       - 16\, \* (6\, \* x^{-1} + 8\, + 5\, \* x)\,\*  \Hhh(-1,-1,0)\,
       - 8\, \* \z2\, \* (8\, - 5\, \* x)\,\*  \H(1)\,
       - 32\, \* (4\, \* x^{-1} + 10\, + 7\, \* x)\,\*  \Hhh(-2,-1,0)\,
   \nn \\[0.5mm] & & \mbox{}
       + 16\, \* (4\, \* x^{-1} + 14\, + 7\, \* x)\,\*  \Hhh(-2,0,0)\,
       - 8/5\: \* \z2^2\, \* (- 16\, \* x^{-1} + 16\, + x)
       - 8\, \* \z2\, \* (6\, \* x^{-1} + 16\,
   \nn \\[0.5mm] & & \mbox{}
       + 13\, \* x)\,\*  \H(-1)\,
       - 4\, \* (- 14\, \* x^{-1} + 18\, + x)\,\*  \Hh(1,0)\,
       - 4\, \* (- 10\, \* x^{-1} + 18\, + 25\, \* x)\,\*  \H(2)\,
       - 16\, \* \z2\, \* (4\, \* x^{-1} + 18\,
   \nn \\[0.5mm] & & \mbox{}
       + 7\, \* x)\,\*  \H(-2)\,
       - 2\, \* (+ 27\, \* x^{-1} + 42\, - 16\, \* x)\,\*  \H(1)\,
       + 8/15\: \* (- 16\, \* x^{-2} + 285\, \* x^{-1} + 180\, - 85\, \* x
   \nn \\[0.5mm] & & \mbox{}
       - 36\, \* x^3)\,\*  \Hh(-1,0)\,
       + 4/15\: \* \z2\, \* (420\, \* x^{-1} + 270\, + 205\, \* x - 72\, \* x^3)
       - 4/15\: \* (360\, + 85\, \* x - 72\, \* x^3)\,\*  \Hh(0,0)\,
   \nn \\[0.5mm] & & \mbox{}
       - 2/15\: \* (716\, \* x^{-1}
       + 2132\, + 597\, \* x - 144\, \* x^2)\,\*  \H(0)\,
       + 2/15\: \* (- 2734\, \* x^{-1} + 2716\, + 369\, \* x + 144\, \* x^2)
   \nn \\[-0.5mm] & & \mbox{}
          + 16\, \* x^{-1} \*  (2\, \* \Hh(1,1)\,
          + 3\, \* \Hhh(1,1,0)\,
          + 3\, \* \Hh(1,2)\,
          + 6\, \* \Hh(2,1)\,
          )
          + 128\, \* \Hh(-2,2)\,
                 \Big)
   \nn \\[-0.5mm] & & \mbox{\hspn}
     + \colour4colour{ \cfs\, \* \nf} \, \* \Big(
        16/3\: \* \pgq(x)\, \*  [
           7\, \* \z2\, \* \H(0)\,
          - 2\, \* \z2\, \* \H(1)\,
          + 2\, \* \Hhh(1,0,0)\,
          - \Hhh(1,1,0)\,
          - \Hh(1,2)\,
          - 3\, \* \Hh(2,0)\,
          - 3\, \* \Hh(2,1)\,
          - 7\, \* \H(3)\,
          ]\,\,
   \nn \\[-0.5mm] & & \mbox{}
       + 16\, \* \pgq(-x)\, \*  [
          - \z2\, \* \H(-1)\,
          - 2\, \* \Hhh(-1,-1,0)\,
          + \Hhh(-1,0,0)\,
          ]\,\,
       - 32\, \* (1+x)\,\*  \Hh(-1,0)\,
       + 8/3\: \* (2\, - x)\,\*  [
          - 2\, \* \z2\, \* \H(0)\,
   \nn \\[0.5mm] & & \mbox{}
          + 12\, \* \Hhhh(0,0,0,0)\,
          + \Hh(2,0)\,
          + 2\, \* \H(3)\,
          ]\,\,
       + 4/3\: \* (- 18\, \* x^{-1} + 4\, - x)\,\*  \Hh(1,0)\,
       + 8\, \* (8\, + 3\, \* x)\,\*  \Hhh(0,0,0)\,
   \nn \\[0.5mm] & & \mbox{}
       + 4/3\: \* (- 91\, \* x^{-1}
       + 30\, - 7\, \* x)\,\*  \H(1)\,
       - 16/3\: \* \z3\, \* (- 26\, \* x^{-1} + 38\, - 13\, \* x)
       + 8/9\: \* (- 161\, \* x^{-1} + 110\,
   \nn \\[0.5mm] & & \mbox{}
       - 67\, \* x)\,\*  \H(2)\,
       - 8/9\: \* \z2\, \* (- 161\, \* x^{-1} + 110\, - 31\, \* x)
       + 4/9\: \* (104\, \* x^{-1} + 506\, + 47\, \* x)\,\*  \Hh(0,0)\,
   \nn \\[0.5mm] & & \mbox{}
       + 4/27\: \* (55\, \* x^{-1} + 1401
       - 198\, \* x - 112\, \* x^2)\,\*  \H(0)\,
       + 1/27\: \* (- 5779\, \* x^{-1} + 13788\, - 8223\, \* x 
   \nn \\[-0.5mm] & & \mbox{}
       + 304\, \* x^2)
       - 24\, \* x^{-1} \*  \Hh(1,1)\,
       - 64\, \* \Hh(-2,0)\,
              \Big)
   \nn \\[-0.5mm] & & \mbox{\hspn}
   + {\colour4colour \ca\, \* \cf\, \* \nf } \, \* \Big(
        4/9\: \* \pgq(x)\, \*  [
          - 12\, \* \z2\, \* \H(0)\,
          + 42\, \* \z2\, \* \H(1)\,
          - 90\, \* \Hhh(1,0,0)\,
          - 17\, \* \Hh(1,1)\,
          - 24\, \* \Hhh(1,1,0)\,
          - 24\, \* \Hh(1,2)\,
          ]\,\,
   \nn \\[-0.5mm] & & \mbox{}
       + 8/3\: \* \pgq(-x)\, \*  [
          - \z2\, \* \H(-1)\,
          + 6\, \* \Hhh(-1,-1,0)\,
          + 7\, \* \Hhh(-1,0,0)\,
          + 4\, \* \Hh(-1,2)\,
          + 2\, \* \H(3)\,
          ]\,\,
       + 32/3\: \* (- 3\, \* x^{-1} + 1\,
   \nn \\[0.5mm] & & \mbox{}
       - 2\, \* x)\,\*  \Hh(2,0)\,
       + 32/3\: \* (- 2\, \* x^{-1} + 1\, - 2\, \* x)\,\*  \Hh(-2,0)\,
       + 12\, \* (2\, - x)\,\*  \Hh(1,1)\,
       + 8/9\: \* (- 45\, \* x^{-1} + 4\, + 17\, \* x
   \nn \\[0.5mm] & & \mbox{}
       + 8\, \* x^2)\,\*  \Hh(-1,0)\,
       + 16\, \* \z3\, \* (- 4\, \* x^{-1} + 4\, - 3\, \* x)
       - 32/3\: \* (x^{-1} + 5\, + 4\, \* x)\,\*  \Hhh(0,0,0)\,
       + 8/9\: \* (- 17\, \* x^{-1}
   \nn \\[0.5mm] & & \mbox{}
       + 160\,
       - 26\, \* x + 16\, \* x^2)\,\*  \H(2)\,
       - 8/9\: \* \z2\, \* (28\, \* x^{-1} + 160\, - 43\, \* x + 16\, \* x^2)
       + 8/9\: \* (- 212\, \* x^{-1} + 232\,
   \nn \\[0.5mm] & & \mbox{}
       - 101\, \* x
       + 4\, \* x^2)\,\*  \Hh(1,0)\,
       + 8/9\: \* (- 229\, \* x^{-1} + 368\, - 201\, \* x + 44\, \* x^2)\,\*  \Hh(0,0)\,
       + 4/27\: \* (26\, \* x^{-1} + 418\,
   \nn \\[0.5mm] & & \mbox{}
       - 407\, \* x
       - 192\, \* x^2)\,\*  \H(1)\,
       + 4/27\: \* (- 2579\, \* x^{-1} + 4572\, - 1050\, \* x + 112\, \* x^2)\,\*  \H(0)\,
   \nn \\[-0.5mm] & & \mbox{}
       + 2/81\: \* (- 21551\, \* x^{-1}
       + 11598\,
       - 12585\, \* x 
       - 6364\, \* x^2)
       - 32/3\: \* x \*  \H(3)\,
             \Big)
   \nn \\[-0.5mm] & & \mbox{\hspn}
      + \colour4colour{ \cf\, \* \nfs } \, \*  \Big(
        8/27\: \*  \pgq(x)\,  \*   [
           39\, \*  \Hh(0,0)\,
          + 29\, \*  \H(1)\,
          + 15\, \*  \Hh(1,0)\,
          + 6\, \*  \Hh(1,1)\,
          + 12\, \*  \H(2)\,
          ]\,\,
       + 32/9\: \*  \z2\, \*  (- 2\, \*  x^{-1} + 2\,
   \nn \\[-0.5mm] & & \mbox{}
       - x)
       - 4/9\: \*  (- 119\, \*  x^{-1} + 106\, - 52\, \*  x)\,  \*   \H(0)\,
       - 2/27\: \*  (- 1003\, \*  x^{-1} + 814\, - 422\, \*  x)
              \Big)
 \;\; .
\eea
Finally the second diagonal NNLO entry is given by
\bea
 \label{eq:Kgg2}
 \lefteqn{ K^{\,(2)\,}_{\, \phi\phi}(x)\,  =  }
%
%
   \nn \\ & & \mbox{\hspn}
      \colour4colour{ \cat } \, \* \Big(
        \pgg(x)\,  \*  [
           18974/27\:
          - 440\, \* \z3\,
          - 3008/9\: \* \z2\,
          - 24/5\: \* \zss\,
          - 64\, \* \Hh(-3,0)\,
          + 96\, \* \z2\, \* \H(-2)\,
   \nn \\[-0.5mm] & & \mbox{}
          + 64\, \* \Hhh(-2,-1,0)\,
          - 352/3\: \* \Hh(-2,0)\,
          - 64\, \* \Hhh(-2,0,0)\,
          - 64\, \* \Hh(-2,2)\,
          + 11185/27\: \* \H(0)\,
          - 112\, \* \z3\, \* \H(0)\,
   \nn \\[0.5mm] & & \mbox{}
          - 880/3\: \* \z2\, \* \H(0)\,
          + 3008/9\: \* \Hh(0,0)\,
          - 128\, \* \z2\, \* \Hh(0,0)\,
          + 440/3\: \* \Hhh(0,0,0)\,
          + 64\, \* \Hhhh(0,0,0,0)\,
          + 9782/27\: \* \H(1)\,
   \nn \\[0.5mm]  & &\mbox{}
          - 96\, \* \H(1)\, \* \z3\,
          - 176/3\: \* \z2\, \* \H(1)\,
          - 64\, \* \Hhh(1,-2,0)\,
          + 1360/3\: \* \Hh(1,0)\,
          - 96\, \* \z2\, \* \Hh(1,0)\,
          + 616/3\: \* \Hhh(1,0,0)\,
   \nn \\[0.5mm]  & &\mbox{}
          + 128\, \* \Hhhh(1,0,0,0)\,
          + 968/9\: \* \Hh(1,1)\,
          + 176\, \* \Hhh(1,1,0)\,
          + 128\, \* \Hhhh(1,1,0,0)\,
          + 176\, \* \Hh(1,2)\,
          + 128\, \* \Hhh(1,2,0)\,
   \nn \\[0.5mm]  & &\mbox{}
          + 128\, \* \Hh(1,3)\,
          + 1360/3\: \* \H(2)\,
          - 64\, \* \z2\, \* \H(2)\,
          + 176\, \* \Hh(2,0)\,
          + 160\, \* \Hhh(2,0,0)\,
          + 176\, \* \Hh(2,1)\,
          + 128\, \* \Hhh(2,1,0)\,
   \nn \\[0.5mm]  & &\mbox{}
          + 128\, \* \Hh(2,2)\,
          + 968/3\: \* \H(3)\,
          + 160\, \* \Hh(3,0)\,
          + 128\, \* \Hh(3,1)\,
          + 128\, \* \H(4)\,
          ]\,
       + 8\, \* \pgg(-x)\,  \*  [
          - 77/3\: \* \z3\,
   \nn \\[0.5mm]  & &\mbox{}
          - 134/9\: \* \z2\,
          + 11\, \* \zss\,
          - 8\, \* \Hh(-3,0)\,
          + 32\, \* \z2\, \* \H(-2)\,
          + 16\, \* \Hhh(-2,-1,0)\,
          - 22\, \* \Hh(-2,0)\,
          - 36\, \* \Hhh(-2,0,0)\,
   \nn \\[0.5mm]  & &\mbox{}
          - 24\, \* \Hh(-2,2)\,
          + 24\, \* \z3\, \* \H(-1)\,
          + 88/3\: \* \z2\, \* \H(-1)\,
          + 16\, \* \Hhh(-1,-2,0)\,
          - 32\, \* \z2\, \* \Hh(-1,-1)\,
          + 88/3\: \* \Hhh(-1,-1,0)\,
   \nn \\[0.5mm]  & &\mbox{}
          + 48\, \* \Hhhh(-1,-1,0,0)\,
          + 32\, \* \Hhh(-1,-1,2)\,
          - 268/9\: \* \Hh(-1,0)\,
          + 36\, \* \z2\, \* \Hh(-1,0)\,
          - 110/3\: \* \Hhh(-1,0,0)\,
   \nn \\[0.5mm]  & &\mbox{}
          - 32\, \* \Hhhh(-1,0,0,0)\,
          - 44/3\: \* \Hh(-1,2)\,
          - 8\, \* \Hhh(-1,2,0)\,
          - 32\, \* \Hh(-1,3)\,
          - 10\, \* \z3\, \* \H(0)\,
          - 11\, \* \z2\, \* \H(0)\,
          + 134/9\: \* \Hh(0,0)\,
   \nn \\[0.5mm]  & &\mbox{}
          - 16\, \* \z2\, \* \Hh(0,0)\,
          + 11\, \* \Hhh(0,0,0)\,
          + 8\, \* \Hhhh(0,0,0,0)\,
          + 4\, \* \z2\, \* \H(2)\,
          + 22/3\: \* \H(3)\,
          + 4\, \* \Hh(3,0)\,
          + 16\, \* \H(4)\,
          ]\,
   \nn \\[0.5mm]  & &\mbox{}
       + 32\, \* (1+x)\,  \*  [
          - 2\, \* \z2\, \* \H(2)\,
          + 5\, \* \Hhh(2,0,0)\,
          + 8\, \* \Hh(3,0)\,
          + 13\, \* \H(4)\,
          ]\,
       + 64\, \* (1-x)\,  \*  [
          - \z2\, \* \H(-2)\,
          - 2\, \* \Hhh(-2,-1,0)\,
   \nn \\[0.5mm]  & &\mbox{}
          - 3\, \* \Hhh(-2,0,0)\,
          ]\,
       - 16\, \* (- 11\, \* x^{-1}  + 3\, + 3\, \* x - 11\, \* x^2)\, \*  \Hhh(-1,-1,0)\,
       - 8\, \* \z2\, \* (11\, \* x^{-1} + 3\, - 3\, \* x
   \nn \\[0.5mm]  & &\mbox{}
       - 11\, \* x^2)\, \*  \H(1)\,
       - 128\, \* \z3\, \* (3\, + x)\, \*  \H(0)\,
       + 8\, \* (12\, \* x^{-1} + 5\, - 5\, \* x - 12\, \* x^2)\, \*  \Hh(1,0)\,
       - 64\, \* (5\, - x)\, \*  \Hh(-3,0)\,
   \nn \\[0.5mm]  & &\mbox{}
       - 32\, \* \z2\, \* (13\, + 11\, \* x)\, \*  \Hh(0,0)\,
       + 16/3\: \* (19\, + 117\, \* x - 132\, \* x^2)\, \*  \Hhh(0,0,0)\,
       - 16/3\: \* (- 22\, \* x^{-1} + 21\,
   \nn \\[0.5mm]  & &\mbox{}
       - 51\, \* x + 66\, \* x^2)\, \*  \Hh(2,0)\,
       - 8\, \* (- 33\, \* x^{-1} + 25\,
       - 25\, \* x + 33\, \* x^2)\, \*  \Hhh(1,0,0)\,
       + 16\, \* (11\, \* x^{-1} + 25\, + 25\, \* x
   \nn \\[0.5mm]  & &\mbox{}
       + 11\, \* x^2)\, \*  \Hh(-1,2)\,
       - 16/5\: \* \zss\, \* (33\, + 43\, \* x)
       - 8\, \* \z2\, \* (11\, \* x^{-1} + 53\, + 53\, \* x + 11\, \* x^2)\, \*  \H(-1)\,
   \nn \\[0.5mm]  & &\mbox{}
       + 32/3\: \* (44\, \* x^{-1} + 57\, + 57\, \* x + 44\, \* x^2)\, \*  \Hhh(-1,0,0)\,
       + 8\, \* \z3\, \* (- 55\, \* x^{-1} + 59\, + 22\, \* x + 66\, \* x^2)
   \nn \\[0.5mm]  & &\mbox{}
       + 16/3\: \* (- 22\, \* x^{-1} + 62\, + 19\, \* x + 77\, \* x^2)\, \*  \Hh(-2,0)\,
       + 8/3\: \* \z2\, \* (- 33\, \* x^{-1} + 75\, - 7\, \* x + 231\, \* x^2)\, \*  \H(0)\,
   \nn \\[0.5mm]  & &\mbox{}
       - 8/3\: \* (11\, \* x^{-1} + 75\, - 45\, \* x + 231\, \* x^2)\, \*  \H(3)\,
       + 8/9\: \* (- 134\, \* x^{-1} + 137\, - 316\, \* x - 229\, \* x^2)\, \*  \H(2)\,
   \nn \\[0.5mm]  & &\mbox{}
       - 8/9\: \* \z2\, \* (631\, \* x^{-1} + 137\, + 461\, \* x - 229\, \* x^2)
       - 8/3\: \* (255\, \* x^{-1} + 259\, + 259\, \* x + 255\, \* x^2)\, \*  \Hh(-1,0)\,
   \nn \\[0.5mm]  & &\mbox{}
       - 8/9\: \* (- 242\, \* x^{-1} + 1593\, - 642\, \* x + 2217\, \* x^2)\, \*  \Hh(0,0)\,
       + 2/27\: \* (- 3774\, \* x^{-1} + 8194\, - 6797\, \* x
   \nn \\[0.5mm]  & &\mbox{}
       + 5105\, \* x^2)\, \*  \H(1)\,
       + 1/27\: \* (23364\, \* x^{-1} + 65830\, + 21391\, \* x
       + 53565\, \* x^2)\, \*  \H(0)\,
   \nn \\[0.5mm]  & &\mbox{}
       - 2/81\: \* (- 129517\, \* x^{-1} + 83236\, - 96986\, \* x + 107847\, \* x^2)
       + 448\, \* x \* \Hhhh(0,0,0,0)\,
       - 1/486\: \* (592399\, 
   \nn \\[-0.5mm]  & &\mbox{}
       + 38880\, \* \z5\,)\, \* \delta \x1 \,
       +  (
           1988/3\: \* \z3\,
          + 11185/27\: \* \z2\,
          - 16\, \* \z2\, \* \z3\,
          - 88\, \* \zss\,
          )\, \* \delta \x1 \,
                \Big)
   \nn \\[-0.5mm]  & &\mbox{\hspn}
     + \colour4colour{ \cas\, \* \nf } \, \* \Big(
        2\, \* \pgg(x)\,  \*  [
          - 10429/81\:
          + 24\, \* \z3\,
          + 48\, \* \z2\,
          + 32/3\: \* \Hh(-2,0)\,
          - 2281/27\: \* \H(0)\,
          + 80/3\: \* \z2\, \* \H(0)\,
   \nn \\[-0.5mm]  & &\mbox{}
          - 48\, \* \Hh(0,0)\,
          - 40/3\: \* \Hhh(0,0,0)\,
          - 1882/27\: \* \H(1)\,
          + 16/3\: \* \z2\, \* \H(1)\,
          - 512/9\: \* \Hh(1,0)\,
          - 56/3\: \* \Hhh(1,0,0)\,
   \nn \\[0.5mm]  & &\mbox{}
          - 176/9\: \* \Hh(1,1)\,
          - 16\, \* \Hhh(1,1,0)\,
          - 16\, \* \Hh(1,2)\,
          - 512/9\: \* \H(2)\,
          - 16\, \* \Hh(2,0)\,
          - 16\, \* \Hh(2,1)\,
          - 88/3\: \* \H(3)\,
          ]\,
   \nn \\[0.5mm]  & &\mbox{}
       + 16/9\: \* \pgg(-x)\,  \*  [
           21\, \* \z3\,
          + 10\, \* \z2\,
          + 18\, \* \Hh(-2,0)\,
          - 24\, \* \z2\, \* \H(-1)\,
          - 24\, \* \Hhh(-1,-1,0)\,
          + 20\, \* \Hh(-1,0)\,
   \nn \\[0.5mm]  & &\mbox{}
          + 30\, \* \Hhh(-1,0,0)\,
          + 12\, \* \Hh(-1,2)\,
          + 9\, \* \z2\, \* \H(0)\,
          - 10\, \* \Hh(0,0)\,
          - 9\, \* \Hhh(0,0,0)\,
          - 6\, \* \H(3)\,
          ]\,
       + 8\, \* (1+x)\,  \*  [
          - 6\, \* \z2\, \* \H(2)\,
   \nn \\[0.5mm]  & &\mbox{}
          - 4\, \* \Hh(2,0)\,
          + 3\, \* \Hhh(2,0,0)\,
          + 3\, \* \H(4)\,
          ]\,
       + 48\, \* (1-x)\,  \*  [
          - \z2\, \* \H(-2)\,
          - 2\, \* \Hhh(-2,-1,0)\,
          + \Hhh(-2,0,0)\,
          ]\,
   \nn \\[0.5mm]  & &\mbox{}
       + 32/3\: \* (x^{-1} + 3\,
       + 3\, \* x + x^2)\, \*  \Hh(-1,2)\,
       + 2/27\: \* (374\, \* x^{-1} + 3\, - 807\, \* x - 1100\, \* x^2)\, \*  \H(1)\,
   \nn \\[0.5mm]  & &\mbox{}
       - 8\, \* \z2\, \* (3\, + 5\, \* x)\, \*  \Hh(0,0)\,
       + 16\, \* \z3\, \* (- x^{-1} + 5\, + 5\, \* x)
       + 16\, \* (5\, - x)\, \*  \Hh(-3,0)\,
       - 16\, \* (- 2\, \* x^{-1} + 7\, + 7\, \* x
   \nn \\[0.5mm]  & &\mbox{}
       - 2\, \* x^2)\, \*  \Hhh(-1,-1,0)\,
       - 8\, \* \z2\, \* (2\, \* x^{-1} + 7\, - 7\, \* x - 2\, \* x^2)\, \*  \H(1)\,
       + 40/3\: \* (- 2\, \* x^{-1} + 17\, + 17\, \* x
   \nn \\[0.5mm]  & &\mbox{}
       - 2\, \* x^2)\, \*  \Hh(-1,0)\,
       - 8/3\: \* (20\, + 51\, \* x)\, \*  \Hhh(0,0,0)\,
       + 4/3\: \* (4\, \* x^{-1} + 21\, - 21\, \* x
       - 4\, \* x^2)\, \*  \Hhh(1,0,0)\,
   \nn \\[0.5mm]  & &\mbox{}
       + 12/5\: \* \zss\, \* (23\, + 13\, \* x)
       + 16/9\: \* (- 35\, \* x^{-1} + 27\, - 27\, \* x + 35\, \* x^2)\, \*  \Hh(1,0)\,
       + 8/3\: \* (- 4\, \* x^{-1}
       + 27\, 
   \nn \\[0.5mm]  & &\mbox{}
       + 27\, \* x - 4\, \* x^2)\, \*  \Hhh(-1,0,0)\,
       - 4/9\: \* (88\, \* x^{-1} + 32\, + 761\, \* x - 396\, \* x^2)\, \*  \Hh(0,0)\,
       - 8/3\: \* \z2\, \* (- 2\, \* x^{-1} + 33\,
   \nn \\[0.5mm]  & &\mbox{}
       + 33\, \* x - 2\, \* x^2)\, \*  \H(-1)\,
       + 4/3\: \* \z2\, \* (- 4\, \* x^{-1} + 39\, + 71\, \* x + 12\, \* x^2)\, \*  \H(0)\,
       - 4/3\: \* (- 4\, \* x^{-1} + 39\, + 81\, \* x
   \nn \\[0.5mm]  & &\mbox{}
       + 12\, \* x^2)\, \*  \H(3)\,
       + 8/3\: \* (47\, - 5\, \* x - 12\, \* x^2)\, \*  \Hh(-2,0)\,
       + 4/9\: \* (- 52\, \* x^{-1} + 145\, + 193\, \* x + 140\, \* x^2)\, \*  \H(2)\,
   \nn \\[0.5mm]  & &\mbox{}
       - 4/9\: \* \z2\, \* (8\, \* x^{-1} + 145\, - 317\, \* x + 140\, \* x^2)
       - 2/27\: \* (2008\, \* x^{-1} + 586\, + 3196\, \* x - 1757\, \* x^2)\, \*  \H(0)\,
   \nn \\[0.5mm]  & &\mbox{}
       - 2/81\: \* (7616\, \* x^{-1} + 13155\, - 5298\, \* x + 2473\, \* x^2)
       + 16\, \* x  \*  [
          - 6\, \* \z3\, \* \H(0)\,
          + \Hhhh(0,0,0,0)\,
          ]\,
   \nn \\[-0.5mm]  & &\mbox{}
       +  (
           89027/162\:
          - 80/3\: \* \z3\,
          - 4562/27\: \* \z2\,
          + 16\, \* \zss\,
          )\, \* \delta(1-x)\,
               \Big)
   \nn \\[-0.5mm]  & &\mbox{\hspn}
    + \colour4colour{ \cfs\, \* \nf} \, \* \Big(
         32\, \* (1+x)\,  \*  [
           \z2\, \* \Hh(0,0)\,
          - \Hhhh(0,0,0,0)\,
          - \z2\, \* \H(2)\,
          + \Hhh(2,0,0)\,
          - \Hh(3,0)\,
          - \H(4)\,
          ]\,
   \nn \\[-0.5mm]  & &\mbox{}
       + 32\, \* (1-x)\,  \*  [
          - \z2\, \* \H(-2)\,
          - 2\, \* \Hhh(-2,-1,0)\,
          + \Hhh(-2,0,0)\,
          ]\,
       - 16\, \* \z3\, \* (4\, \* x^{-1} + 1\, + 4\, \* x)
   \nn \\[0.5mm]  & &\mbox{}
       - 32\, \* \z3\, \* (1\, + 3\, \* x)\, \*  \H(0)\,
       + (- 4\, \* x^{-1} + 3\, + 3\, \* x
          - 4\, \* x^2)\, \*  [
          - 16/3\: \* \z2\, \* \H(-1)\,
          - 32/3\: \* \Hhh(-1,-1,0)\,
   \nn \\[0.5mm]  & &\mbox{}
          + 16/3\: \* \Hhh(-1,0,0)\,
          ]\,
       + (4\, \* x^{-1} + 3\,
          - 3\, \* x - 4\, \* x^2)\, \*  [- 16/3\: \* \z2\, \* \H(1)\,
          + 16/3\: \* \Hhh(1,0,0)\,
          ]\,
   \nn \\[0.5mm]  & &\mbox{}
       + 16/5\: \* \zss\, \* (3\, - 2\, \* x)
       + 32/3\: \* (3\, - 2\, \* x - 4\, \* x^2)\, \*  \Hh(-2,0)\,
       - 16/3\: \* (3\, + 2\, \* x)\, \*  \Hhh(0,0,0)\,
       - 16/3\: \* (4\, \* x^{-1}
   \nn \\[0.5mm]  & &\mbox{}
       + 6\, + 3\, \* x)\, \*  \Hh(2,0)\,
       + 16/3\: \* \z2\, \* (4\, \* x^{-1} + 9\, - 10\, \* x)\, \*  \H(0)\,
       - 16/3\: \* (4\, \* x^{-1} + 9\, - 6\, \* x)\, \*  \H(3)\,
   \nn \\[0.5mm]  & &\mbox{}
       + 8/9\: \* (- 44\, \* x^{-1} + 27\, - 27\, \* x + 44\, \* x^2)\, \*  \Hh(1,0)\,
       + 64/45\: \* (- x^{-2} + 105\,
       + 95\, \* x + 9\, \* x^3)\, \*  \Hh(-1,0)\,
   \nn \\[0.5mm]  & &\mbox{}
       - 4/45\: \* (135\, + 1835\, \* x - 440\, \* x^2\, + 144\, \* x^3)\, \*  \Hh(0,0)\,
       + 4/27\: \* (- 691\, \* x^{-1} + 675\,
       - 54\, \* x
   \nn \\[0.5mm]  & &\mbox{}
       + 70\, \* x^2)\, \*  \H(1)\,
       - 2/135\: \* (1884\, \* x^{-1} + 1368\, + 6933\, \* x + 164\, \* x^2)\, \*  \H(0)\,
       + - 1/135\: \* (7632\, \* x^{-1}
   \nn \\[0.5mm]  & &\mbox{}
       + 4807\, - 3427\, \* x
       - 9012\, \* x^2)
       - 8/9\: \* (80\, \* x^{-1} + 27\, \* x)\, \*  \H(2)\,
       + 8/45\: \* \z2\, \* (400\, \* x^{-1} + 895\, \* x + 72\, \* x^3)
   \nn \\[-0.5mm]  & &\mbox{}
       + 64\, \* \Hh(-3,0)\,
       + \delta(1-x)\,
             \Big)
   \nn \\[-0.5mm]  & &\mbox{\hspn}
     + \colour4colour{ \ca\, \* \cf\, \* \nf } \, \* \Big(
        8\, \* \pgg(x)\,  \*  [
          - 22/3\:
          + 4\, \* \z3\,
          - \H(0)\,
          - \H(1)\,
          ]\,
       + 16\, \* (1+x)\,  \*  [
           6\, \* \z2\, \* \Hh(0,0)\,
          + 7\, \* \z2\, \* \H(2)\,
          - 7\, \* \Hhh(2,0,0)\,
   \nn \\[-0.5mm]  & &\mbox{}
          - 2\, \* \Hh(3,0)\,
          - 6\, \* \H(4)\,
          ]\,
       + 16\, \* (1-x)\,  \*  [
           7\, \* \H(-2)\, \* \z2\,
          + 14\, \* \Hhh(-2,-1,0)\,
          - 3\, \* \Hhh(-2,0,0)\,
          ]\,
       + 128\, \* \z3\, \* (1\, + 2\, \* x)\, \*  \H(0)\,
   \nn \\[0.5mm]  & &\mbox{}
       + 64\, \* (2\, - 3\, \* x)\, \*  \Hhhh(0,0,0,0)\,
       - 64/3\: \* (x^{-1} + 3\, + 3\, \* x + x^2)\, \*  \Hh(-1,2)\,
       - 4/3\: \* (32\, \* x^{-1} + 11\, - 59\, \* x
   \nn \\[0.5mm]  & &\mbox{}
       + 16\, \* x^2)\, \*  \Hh(1,0)\,
       - 16/3\: \* (4\, \* x^{-1} + 15\, - 18\, \* x - 4\, \* x^2)\, \*  \H(3)\,
       + 16/3\: \* \z2\, \* (15\, - 3\, \* x - 4\, \* x^2)\, \*  \H(0)\,
   \nn \\[0.5mm]  & &\mbox{}
       - 8/3\: \* (8\, \* x^{-1} + 23\, - 7\, \* x)\, \*  \Hh(2,0)\,
       - 8/3\: \* \z3\, \* (- 20\, \* x^{-1} + 25\, + 43\, \* x + 8\, \* x^2)
       - 16/3\: \* (4\, \* x^{-1} + 30\,
   \nn \\[0.5mm]  & &\mbox{}
       - 15\, \* x - 12\, \* x^2)\, \*  \Hh(-2,0)\,
       - 8/3\: \* (20\, \* x^{-1} + 33\, - 33\, \* x - 20\, \* x^2)\, \*  \Hhh(1,0,0)\,
       - 8/3\: \* (43\, + 31\, \* x)\, \*  \Hhh(0,0,0)\,
   \nn \\[0.5mm]  & &\mbox{}
       + 16/3\: \* (- 16\, \* x^{-1} + 45\, + 45\, \* x - 16\, \* x^2)\, \*  \Hhh(-1,-1,0)\,
       - 8/3\: \* (4\, \* x^{-1} + 45\, + 45\, \* x + 4\, \* x^2)\, \*  \Hhh(-1,0,0)\,
   \nn \\[0.5mm]  & &\mbox{}
       + 8/3\: \* \z2\, \* (16\, \* x^{-1} + 45\, - 45\, \* x - 16\, \* x^2)\, \*  \H(1)\,
       + 8/3\: \* \z2\, \* (- 8\, \* x^{-1} + 69\, + 69\, \* x - 8\, \* x^2)\, \*  \H(-1)\,
   \nn \\[0.5mm]  & &\mbox{}
       + 16/9\: \* \z2\, \* (87\,
       - 216\, \* x + 14\, \* x^2)\,
       - 16/9\: \* (87\, - 3\, \* x  + 14\, \* x^2)\, \*  \H(2)\,
       - 16/3\: \* (95\, + 71\, \* x
   \nn \\[0.5mm]  & &\mbox{}
       - 24\, \* x^2)\, \*  \Hh(-1,0)\,
       + 4/9\: \* (- 116\, \* x^{-1}
       + 335\, + 1223\, \* x + 360\, \* x^2)\, \*  \Hh(0,0)\,
       - 4/27\: \* (- 241\, \* x^{-1}
   \nn \\[0.5mm]  & &\mbox{}
       + 2196\,
       - 2223\, \* x + 268\, \* x^2)\, \*  \H(1)\,
       - 2/27\: \* (1284\, \* x^{-1}
       + 2463\, - 2286\, \* x + 8464\, \* x^2)\, \*  \H(0)\,
   \nn \\[0.5mm]  & &\mbox{}
       - 1/81\: \* (24445\, \* x^{-1}
       + 5079\, + 31074\, \* x - 52282\, \* x^2)
       - 8\, \* ( 7\, \* \zss\, + 12\, \* \Hh(-3,0)\, )
       + (
           465/2\:
   \nn \\[-0.5mm]  & &\mbox{}
          - 176\, \* \z3\,
          - 8\, \* \z2\,
          )\, \* \delta(1-x)\,
               \Big)
   \nn \\[-0.5mm]  & &\mbox{\hspn}
     + \colour4colour{ \ca\, \* \nfs} \, \* \Big(
        4/81\: \* \pgg(x)\,  \*  [
           571
          - 144\, \* \z2\,
          + 417\, \* \H(0)\,
          + 144\, \* \Hh(0,0)\,
          + 318\, \* \H(1)\,
          + 144\, \* \Hh(1,0)\,
          + 72\, \* \Hh(1,1)\,
   \nn \\[-0.5mm]  & &\mbox{}
          + 144\, \* \H(2)\,
          ]\,
       + 32/9\: \* (1+x)\,  \*  \Hh(0,0)\,
       + 32/9\: \* (2\, - x + x^2)\, \*  [
          - \z2\,
          + \H(2)\,
          ]\,
       + 2/27\: \* (52\, \* x^{-1} + 225\, 
   \nn \\[0.5mm]  & &\mbox{}
       - 81\, \* x
       + 80\, \* x^2)\, \*  \H(1)\,
       + 2/27\: \* (192\, \* x^{-1} + 404\, + 5\, \* x + 146\, \* x^2)\, \*  \H(0)\,
       + 2/81\: \* (1540\, \* x^{-1} + 891\,
   \nn \\[-0.5mm]  & &\mbox{}
       + 303\, \* x
       - 202\, \* x^2)
       + (
          - 1139/18\:
          - 16\, \* \z3\,
          + 556/27\: \* \z2\,
          )\, \* \delta(1-x)\,
                 \Big)
   \nn \\[-0.5mm]  & &\mbox{\hspn}
    + \colour4colour{ \cf\, \* \nfs} \, \* \Big(
        4\, \* \pgg(x)\,
       + 16/3\: \* (1+x)\,  \*  [
           2\, \* \z3\,
          + 5\, \* \Hhh(0,0,0)\,
          + \Hh(2,0)\,
          ]\,
       + 8/9\: \* (4\, \* x^{-1} + 3\, - 3\, \* x
   \nn \\[-0.5mm]  & &\mbox{}
       - 4\, \* x^2)\, \*  \Hh(1,0)\,
       + 8/9\: \* (20\, \* x^{-1} + 47\, + 5\, \* x - 4\, \* x^2)\, \*  \Hh(0,0)\,
       + 4/9\: \* (100\, \* x^{-1} + 155\, + 2\, \* x)\, \*  \H(0)\,
   \nn \\[-0.5mm]  & &\mbox{}
       + 2/27\: \* (1079\, \* x^{-1}
       + 309\, - 1314\, \* x + 178\, \* x^2)
       + (
          - 43
          + 32\, \* \z3\,
          )\, \* \delta(1-x)\,
                 \Big)
   \nn \\[-0.5mm]  & &\mbox{\hspn}
     + \colour4colour{ \nft } \, \* \Big(
       8/81\: \*  \pgg(x)\,  \*  [
          - 10
          - 9\, \* \H(0)\,
          - 6\, \* \H(1)\,
          ]\,
       + 8/81\: \* (2\, - x + x^2)\, \*  [
          - 10
          - 9\, \* \H(0)\,
          - 6\, \* \H(1)\,
          ]\,
   \nn \\[-0.5mm] & &\mbox{}
       + 8\, \* (
           25/243\:
          - 1/9\: \* \z2\,
          )\, \* \delta \x1 \,
                \Big)
 \:\: .
\eea
%
%
\setlength{\baselineskip}{0.55cm}
\renewcommand{\theequation}{C.\arabic{equation}}
\setcounter{equation}{0}
\section*{Appendix C: The gluon coefficient function for \boldmath 
 $\Ftwo$ at large $x$}
\label{sec:AppC}
%
%
In this final appendix we provide the large-$x$ coefficients of the gluon
coefficient function $c_{\rm 2\, , g}^{\,(n)}(x)$ for the photon-exchange 
structure function $\Ftwo\,$, which were not written down in Ref.~\cite{MVV6} 
for brevity. As their counterparts for $c_{\rm \phi\, , q}^{\,(n)}(x)$ in 
Eqs.~(\ref{eq:cphiq2L3}) -- (\ref{eq:cphiq3L1}), these coefficients contribute 
to the large-$x$ behaviour of the off-diagonal physical kernels 
$K_{\,2\phi}^{\,(n)}$ and $K_{\,\phi 2}^{\,(n)}$ presented above.

\noindent
The second-order coefficients are (recall $L_1 \equiv \ln \x1\:\!$)
\bea
\label{eq:c2g2L3}
 c_{\rm 2\, , g}^{\,(2)} \Big|_{\,L_1^3} &\! = \! &
          {2 \over 3}\: \* \ca\, \* \nf\,
        + {10 \over 3}\: \* \cf\, \* \nf\,
 \; ,
 \\[0.5mm]
\label{eq:c2g2L2}
 c_{\rm 2\, , g}^{\,(2)} \Big|_{\,L_1^2} &\! = \! &
         - 4\, \* \ca\, \* \nf\,
         - 9\, \* \cf\, \* \nf\,
 \; ,
 \\[0.5mm]
\label{eq:c2g2L1}
 c_{\rm 2\, , g}^{\,(2)} \Big|_{\,L_1} &\! = \! &
           (14\, - 8\, \* \z2\,)\, \* \ca\, \* \nf\,
         - (2\, + 16\, \* \z2\,)\, \* \cf\, \* \nf\,
 \; .
\eea
The corresponding three-loop results read
\bea
\label{eq:c2g3L5}
 c_{\rm 2\, , g}^{\,(3)} \Big|_{\,L_1^5} &\! = \! &
          {2 \over 3}\: \* \cas\, \* \nf\,
        + {10 \over 3}\: \* \cfs\, \* \nf\,
 \; ,
 \\[1mm]
 \label{eq:c2g3L4}
 c_{\rm 2\, , g}^{\,(3)} \Big|_{\,L_1^4} &\! = \! & \mbox{}
         - {293 \over 54}\: \* \cas\, \* \nf\,
         - {206 \over 27}\: \* \cf\, \* \ca\, \* \nf\,
         - {83 \over 6}\: \* \cfs\, \* \nf\,
         + {17 \over 27}\: \* \cf\, \* \nfs\,
         + {7 \over 27}\: \* \ca\, \* \nfs\,
 \; ,
 \\[1mm]
 \label{eq:c2g3L3}
 c_{\rm 2\, , g}^{\,(3)} \Big|_{\,L_1^3} &\! = \! &
           \Big( \, {3056 \over 81}\:
         - {136 \over 9}\: \* \z2 \Big)\, \* \cas\, \* \nf\,
         + \Big( \, {3442 \over 81}\:
         - {136 \over 9}\: \* \z2 \Big)\, \* \cf\, \* \ca\, \* \nf\,
   \nn \\[0.5mm] & & \mbox{}
         + \Big( \, {127 \over 9}\:
         - {376 \over 9}\: \* \z2 \Big)\, \* \cfs\, \* \nf\,
         - {496 \over 81}\: \* \cf\, \* \nfs\,
         - {152 \over 81}\: \* \ca\, \* \nfs\,
 \; ,
 \\[2.5mm]
 \label{eq:c2g3L2}
 c_{\rm 2\, , g}^{\,(3)} \Big|_{\,L_1^2} &\! = \! & \mbox{}
         - \Big( \, {12043 \over 162}\:
         - {232 \over 3}\: \* \z2\, 
         - {52 \over 3}\: \* \z3
           \Big)\, \* \cf\, \* \ca\, \* \nf\,
   \nn \\[0.5mm] & & \mbox{}
         - \Big( \, {13789 \over 81}\:
         - {652 \over 9}\: \* \z2\, 
         - {176 \over 3}\: \* \z3
           \Big)\, \* \cas\, \* \nf\,
         + \Big( \, {1096 \over 81}\:
         - {40 \over 9}\: \* \z2 
           \Big)\, \* \ca\, \* \nfs\,
   \nn \\[1mm] & & \mbox{}
         - \Big( \, {205 \over 6}\:
         - 142\, \* \z2\, 
         - 72\, \* \z3
           \Big)\, \* \cfs\, \* \nf\,
         + \Big( \, {1565 \over 81}\:
         - {16 \over 3}\: \* \z2 
           \Big)\, \* \cf\, \* \nfs\,
 \; ,
 \\[2.5mm]
 \label{eq:c2g3L1}
 c_{\rm 2\, , g}^{\,(3)} \Big|_{\,L_1} &\! = \! & \mbox{}
         - \Big( \, {32968 \over 243}\:
         + {2380 \over 9}\: \* \z2\,
         - {1250 \over 9}\: \* \z3\,
         - {64 \over 5}\: \* \zss \Big)\, \* \cf\, \* \ca\, \* \nf\,
   \nn \\[0.5mm] & & \mbox{}
         + \Big( \, {69526 \over 243}\:
         - {6052 \over 27}\: \* \z2\,
         - {952 \over 9}\: \* \z3\,
         + 50\, \* \zss \Big)\, \* \cas\, \* \nf\,
   \nn \\[0.5mm] & & \mbox{}
         + \Big( \, {254 \over 3}\:
         - {46 \over 3}\: \* \z2\,
         - 456\, \* \z3\,
         + {632 \over 5}\: \* \zss \Big)\, \* \cfs\, \* \nf\,
   \nn \\[0.5mm] & & \mbox{}
         - \Big( \, {10798 \over 243}\:
         - {496 \over 27}\: \* \z2\, 
         - {16 \over 9}\: \* \z3
           \Big)\, \* \ca\, \* \nfs\,
         + \Big( \, {3076 \over 243}\:
         + {328 \over 9}\: \* \z2\, 
         + {64 \over 9}\: \* \z3
           \Big)\, \* \cf\, \* \nfs\,
 \; . \qquad
\eea
Note the close similarity between Eqs.~(\ref{eq:cphiq2L3}), 
(\ref{eq:cphiq3L5}), (\ref{eq:c2g2L3}) and (\ref{eq:c2g3L5}) for the leading 
logarithms at NNLO and N$^3$LO which may point to yet another general structure.
%
%
{\footnotesize
\setlength{\baselineskip}{0.5cm}

}

\end{document}